\documentclass[12pt,12pt]{article}
\usepackage{amsmath,amssymb}
\usepackage{slashed} 

\font\mathbf cmbxti10 at 12pt
\font\mathbfsy cmmib10 at 12pt

\catcode`\@=11
\newdimen\ex@
\def\beq{\begin{equation}}
\def\eeq{\end{equation}}
\def\beqa{\begin{eqnarray}}
\def\eeqa{\end{eqnarray}}
\newcommand{\ba}{\begin{eqnarray}}
\newcommand{\ea}{\end{eqnarray}}
\newcommand\BA{\begin{array}}
\newcommand\EA{\end{array}}

\begin{document}
\def\thefootnote{\fnsymbol{footnote}}

\title{\bf
$\!\!\!\!\!\!\!\!\!\!\!\!\!\!\!\!$
A NEW DESCRIPTION OF MOTION OF THE\\
FERMIONIC SO(2N+2) TOP IN THE\\
CLASSICAL LIMIT UNDER THE\\
QUASI-ANTICOMMUTATION-RELATION APPROXIMATION
\footnotemark[1]\\[-0.4cm]}
\author{
$\!\!\!\!\!\!\!\!\!\!\!\!\!\!\!\!$
SEIYA NISHIYAMA$\!$\footnotemark[2]~,
JO\~AO DA PROVID\^{E}NCIA$\!$\footnotemark[3]~
and CONSTAN\c{C}A PROVID\^{E}NCIA\footnotemark[4]
\\ \\[-0.5cm]
Centro de F\'\i sica Computacional,
Departamento de F\'\i sica,\\
Universidade de Coimbra,
P-3004-516 Coimbra, Portugal\footnotemark[2] 
\\ \\[-0.6cm]
{\it Dedicated to the Memory of Hideo Fukutome}}

\maketitle

\vspace{0.1mm}
\footnotetext[1]{A preliminary version of
this work has been presented by S. Nishiyama
at the YITP Workshop \\
~~~~~YITP-W-10-02 on
{\it Development of Quantum Fields Theory and String Theory 2010},\\
~~~~~Yukawa Institute for Theoretical Physics, Kyoto University, Kyoto, Japan, 20-24 July, 2010.
}
\footnotetext[2]{
E-mail address: seikoceu@khe.biglobe.ne.jp;nisiyama@teor.fis.uc.pt;}
\footnotetext[3]{
E-mail address: providencia@teor.fis.uc.pt}
\footnotetext[4]{
E-mail address: cp@teor.fis.uc.pt}

%%%%%%%%%%%%%%%%%
%               %
%  0  Abstract  %
%               % 
%%%%%%%%%%%%%%%%%

\vspace{-1.1cm}

\begin{abstract}
%\vskip-0.1cm
$\!\!\!\!$The boson images of fermion $SO(2N \!\!+\!\! 1)$ Lie operators
have been given together with those of $SO(2N \!+\! 2)$ ones.
The $SO(2N \!\!+\!\! 1)$ Lie operators are generators of rotation
in the $(2N \!\!+\!\! 1)$-dimensional Euclidian space
($N$: number of single-particle states of the fermions).
The images of fermion annihilation-creation operators
must satisfy the canonical anti-commutation relations,
when they operate on a spinor subspace.
In the regular representation space we use 
a boson Hamiltonian with Lagrange multipliers
to select out the spinor subspace.
Based on these facts,
a new description of a fermionic $SO(2N \!+\! 2)$ top is proposed.
From the Heisenberg equations of motions for the boson operators,
we get the $SO(2N \!+\! 1)$ self-consistent field (SCF)
Hartree-Bogoliubov (HB) equation for the classical stationary motion of the fermion top.
Decomposing an $SO(2N \!\!+\!\! 1)$ matrix
into matrices describing paired and unpaired modes of fermions,
we obtain a new form of the $SO(2N \!\!+\!\! 1)$ SCF equation
with respect to the paired-mode amplitudes.
To demonstrate the effectiveness of the new description
based on the bosonization theory,
the extended HB eigenvalue equation is applied to
a superconducting toy-model
which consists of a particle-hole plus BCS type interaction.
It is solved to reach an interesting and exciting solution
which is not found in the traditional HB eigenvalue equation,
due to the unpaired-mode effects.
To complete the new description,
the Lagrange multipliers
must be determined in the classical limit.
For this aim
a quasi anti-commutation-relation approximation is proposed.
Only if a certain relation between an $SO(2N \!\!+\!\! 1)$ parameter $z$
and the $N$ is satisfied,
unknown parameters $k$ and $l$ in the Lagrange multipliers can be determined
withuout any inconcistency.
\vskip0.1cm

$Keywords$: Fermion $SO(2N \!\!+\!\! 1)$ Lie operators;
fermion top;
extended Hartree-Bogoliubov theory;
\end{abstract}

\newpage

%%%%%%%%%%%%%%%%%%%%%%
%                    %
%  1  Introduction   %
%                    %
%%%%%%%%%%%%%%%%%%%%%%

\def\thesection{\arabic{section}}
\setcounter{equation}{0}
\renewcommand{\theequation}{\arabic{section}.\arabic{equation}}

\section{Introduction}

~~
The time dependent Hartree-Bogoliubov (TDHB) theory is the leading standard approximation in the many-body theoretical description of a superconducting fermion system
\cite{Bogo.58}-
%\cite{RS.80},
\cite{BlaizotRipka.86}.
The HB wave function (WF) for the fermion system represents
Bose condensate states of fermion pairs.
It is a good approximation for the ground state of the fermion system with a short-range pairing interaction
that produces two-body correlations which is taken into account as a spontaneous Bose condensation of the fermion pairs.
The fermion number-nonconservation of the HB WF is a consequence of the spontaneous Bose condensation of fermion pairs which causes a coherence in phases of superconducting (Bose condensed) fermions.
The $SO(2N)$ Lie algebra of the fermion pair operators contains
the $U(N)$ Lie algebra as a subalgebra.
Here the $SO(2N)$ and the $U(N)$ denote the special orthogonal group 
of $2N$ dimensions and the unitary group of $N$ dimensions,
respectively
($N$: number of single-particle states of the fermions).
The canonical transformation of the fermion operators generated by
the Lie operators in the $SO(2N)$ Lie algebra induces the
generalized Bogoliubov transformation for the fermions
\cite{Bogo.59}
to vanish the {\it dangerous} term for arbitrary pairs.
The TDHB equation is derived from the classical Euler-Lagrange
equation of motion for the $\frac{SO(2N)}{U(N)}$ coset variables
\cite{Nishi.81}.
Usually the solutions of this equation provides the ground state of an even fermion system.
For the odd fermion system,
we must consider a one quasiparticle state
on the HB WF
in which paired and unpaired states are not treated
in an equal manner.
In this sense
for such a system we have no TD self-consistent field (SCF) theory
with the same power for the mean field approximation (MFA) 
as the TDHB theory.

One of the most challenging problems in current studies of condensed matter physics is to obtain a theory suitable for the
description of collective motions with large amplitudes
in fermion systems with strong collective correlations.
For providing a general microscopic framework for a unified SCF description for Bose- and Fermi- type collective excitations in those systems, 
a many-body theory has been proposed by
Fukutome, Yamamura and one of the present authors (S.N.) 
based on
the $SO(2N\!\!+\!\!1)$ Lie algebra of the fermion operators
\cite{FYN.77}.
An induced representation
of an $SO(2N\!\!+\!\!1)$ group has been obtained from
a group extension of the
$SO(2N)$ Bogoliubov transformation for fermions to the $SO(2N\!\!+\!\!1)$
transformation group. We start with the fact that the set of the
fermion operators consisting of the creation-annihilation and the pair
operators forms a larger Lie algebra, the $SO(2N\!\!+\!\!1)$ Lie algebra.
The fermion Lie operators, when operating
on the integral representation of the $SO(2N\!\!+\!\!1)$ WF,
are mapped into the regular representation of the $SO(2N\!\!+\!\!1)$ group and are represented by Bose operators.
The Bose images of the fermion Lie operators
are expressed by closed first order differential forms. The
creation-annihilation operators themselves as well as the pair 
operators are given by the finite Schwinger type boson representation
\cite{Sch.65,YN.76}.

Embedding the $SO(2\!N\!\!+\!\!1)$ group into
an $SO(2\!N\!\!+\!\!2)$ group and using the boson images of 
$SO(2\!N\!\!+\!\!2)$ Lie operators,
we have developed an extended TDHB (ETDHB) theory
\cite{Nishi.98}
and extended supersymmetric $\sigma$-model
\cite{SJCF.08,SJCF.11}.
Particularly in the ETDHB theory,
which is obtained from the Heisenberg equation of motion
for the boson operators,
paired and unpaired modes are treated on an equal footing.
$\!$A static extended Hartree-Bogoliubov (EHB) theory is derived easily from the ETDHB theory.
$\!$The EHB theory applicable to both even and odd fermion systems is
a SCF theory with the same power for the MFA
as the usual HB theory for even fermion systems.
Based on these facts,
a new description of a fermionic $SO(2N \!+\! 2)$ top is proposed.
We start from the Hamiltonian of the fermion system which includes, however, the Lagrange multipliers needed
to select the physical spinor subspace.
The EHB equation is written in terms of appropriate variables 
representing the paired and the unpaired modes.
The EHB eigenvalue equation is solved by 
a method parallel to the two-step diagonalization method
for the usual HB eigenvalue equation
\cite{Ba.61}.
We obtain a new eigenvalue involving unpaired-mode effects
in contrast to the usual HB theory, which is unable to describe
the unpaired modes.
Through the coordinate transformations
for space fixed and body fixed coordinate frames,
the fermion $SO(2N \!\!+\!\! 1)$ Lie operators are expressed
in terms of the quasiparticle expectation values
($c$-number) of them and the quasiparticle
$SO(2N \!\!+\!\! 1)$ Lie operators
(quantum mechanical fluctuations).
To treat the quantum mechanical fluctuations,
we make successive coordinate transformations.
The fluctuating Hamiltonian with the Lagrange multipliers 
is given up to the first order in the quasi-particle
$SO(2N \!\!+\!\! 1)$ Lie operators.
We found that Sawada's eigenmode method
\cite{Sawada.57,Sawada.60}
is not appropriate to obtain
the excitation energy due to the Hamiltonian because it contains the
unpaired-mode amplitudes in a particular form.
However, with the help of the traditional Bogoliubov's approach
\cite{Bogo.59},
we were able to derive a very simple expression for the excitation energy due to the fluctuating Hamiltonian.
The unknown parameters in the Lagrange multipliers terms must be determined. 
The anticommutators of the fermion Lie operators given in the first order differentials satisfy exactly the anticommutation relations,
when they operate on the
$SO(2N \!\!+\!\! 1)$ HB WF.
However, this fact plays no role to determine
the parameters.
A determination of the parameters is possible if we demand instead that expectation values of the anticommutators by an approximate
$SO(2N \!\!+\!\! 1)$ HB WF satisfy the anticommutation relations, i.e.,
the quasi anticommutation relation approximation for the fermions
\cite{Nishi.98,Nishi.96}.
Under the approximation,
the determination could be made successfully.
This plays crucial roles for a unified self-consistent description of Bose-Fermi type collective excitations at regions very near 
$z \!\!=\!\! 1$ (the case of non existence of unpaired modes)
and $z \!\!=\!\! 0$ (the case of largest contribution from unpaired modes)
since they are the necessary and sufficient conditions to determine the unknown parameters
in the Langrange multipliers.
This means a great step to towards verification of the validity of
the present new description.
The above quantum fluctuations can also be studied
in the framework of quantum group symmetry formalism
\cite{TripodiLima.97}-\cite{BonatsosDaskoloyannis.99}.
Finally,
we point out that
it is an important and interesting problem
to attempt a group theoretical approach to
formation of the Lax pair of the $SO(2N \!\!+\!\! 2)$ top
\cite{OlshanetskyPerelomov.81}-\cite{ReymanShansky.94}.

This paper is organized as follows.
In Sec. 2,
we recapitulate briefly the induced representation of the
$SO(2N\!+\!1)$ canonical transformation group,
the embedding of the $SO(2N\!+\!1)$ group into
an $SO(2N\!+\!2)$ one and the introduction of the
$\frac{SO(2N\!+\!2)}{U(N\!+\!1)}$ coset variables.
In Sec. 3,
we give a brief sketch of the derivation of the extended TDHB equation from the Heisenberg equation of motion for the boson operators.
The extended TDHB theory is just the TDSCF theory.
In Sec. 4,
a static extended HB equation is derived from the extended TDHB equation.
The EHB eigenvalue equation is solved with
the two-step diagonalization method.
In Sec. 5,
energies of classical motion and quantum mechanical fluctuation of the fermion $SO(2N\!\!+\!\!1)$ top are given.
Sections 6 and 7 are devoted to a determination of the unknown parameters
in the Lagrange multipliers by the quasi-anticommutation relation approximation for the fermions.
Finally, in the last section, we give some concluding remarks and
further perspectives.
In particular
we will attempt a group theoretical approach to
the formation of the Lax pair of the $SO(2N\!\!+\!\!2)$ top.
In App. A,
we give expressions for differential $SO(2N\!\!+\!\!2)$ Lie operators
on the coset spaces.
Further
we provide useful formulas to evaluate the expectation values of
differential annihilation-creation operators.
We characterize the action of
the corresponding operators on the $SO(2N\!\!+\!\!1)$ wave function.
Throughout this paper, we use the summation convention over
repeated indices
unless otherwise stated.

\newpage

%%%%%%%%%%%%%%%%%%%%%%%%%%%%%%%%%%%%%%%%%%%%%%%%%%%%%%%%%%%%%%%%%%%%%%%%%%%%
%                                                                          %
%  2  Brief review of Bogoliubov transformation generated by SO(2N+1)      %
%                                                                          %
%  Lie algebra of fermion operators and its embedding into SO(2N+2) group  %
%                                                                          %
%%%%%%%%%%%%%%%%%%%%%%%%%%%%%%%%%%%%%%%%%%%%%%%%%%%%%%%%%%%%%%%%%%%%%%%%%%%%

\def\thesection{\arabic{section}}
\setcounter{equation}{0}
\renewcommand{\theequation}{\arabic{section}.\arabic{equation}}

\section{Brief review of Bogoliubov transformation generated by SO(2N+1) Lie algebra of fermion operators and its embedding into SO(2N+2) group}

\def\bra#1{{<\!#1\,|}}
\def\ket#1{{|\,#1\!>}}
~~~
We consider a fermion system with $N$ single-particle states.
Let $c_\alpha$ and $c_\alpha^\dagger$, $\alpha \!=\!1,\ldots,N$, be the
annihilation and creation operators of the fermion, respectively.
The $SO(2N\!+\!1)$ canonical transformation $U(G)$ is generated by the fermion $SO(2N\!+\!1)$ Lie operators consisting of the set
$\{
c_\alpha, c_\alpha^\dagger,
E^\alpha_{~\beta }
=
c_\alpha^\dagger c_\beta -\frac{1}{2} \!\cdot\! \delta_{\alpha\beta },
E_{\alpha\beta }
=
c_\alpha c_\beta ,
E^{\alpha\beta }
=
c_\alpha^\dagger c_\beta^\dagger;
~\alpha,\beta = 1,\ldots,N
\}$.
The $U(G)$ induces an inhomogeneous linear transformation (TR) (for $z \!\neq\! 1$) in the space spanned by the fermion annihilation and creation operators (not a linear one)
involving a $q$-number gauge and is specified by an
$SO(2N+1)$ matrix $G$ as
\ba
U(G)(c,c^\dagger,\frac{1}{\sqrt{2}})U^\dagger(G)
\!=\!
(c,c^\dagger,\frac{1}{\sqrt{2}})(z \!-\! \rho)G ~,
\label{UGtrans}
\ea
\vskip-1.6\bigskipamount
\ba
G
\!\equiv\!\!
\left[ \!
\BA{ccc}
A&B^\star&-X^\star\\
B&A^\star&X\\
Y&
-Y^\star&Z
\EA \!
\right] ,
X\!\equiv\!\frac{x}{\sqrt{2}},
Y\!\equiv\!\frac{y}{\sqrt{2}},
Z\!\equiv\!z,~
G^\dagger G\!=\!GG^\dagger \!=\! 1_{2N+1}, 
\label{matG}
\ea
\vskip-1.6\bigskipamount
\ba
U(G)U(G')\!=\!U(GG')~,~U(G^{-1})\!=\!U^{-1}(G)\!=\!U^\dagger(G)~,~
U(1) \!=\! {\hbox {\bf 1}}~,
\label{UGproperty}
\ea
where
$(c,c^\dagger,\frac{1}{\sqrt{2}})$ is
a $(2N\!+\!1)$-dimensional row vector
$\bigl((c_\alpha),(c^\dagger_\alpha),\frac{1}{\sqrt{2}}\bigr)$ and
$A\!=\!(A^\alpha_{~i})$ and
$B\!=\!(B_{\alpha i})$ are $N \!\times\! N$ matrices.
The symbol $\star$ denotes the complex conjugation.
The operator $\rho$ in the gauge factor $z \!-\! \rho$ is defined as
$\rho
\!=\!
x_\alpha c_\alpha^\dagger \!-\! x^\star_\alpha c_\alpha$
and satisfies 
$\rho^2
\!=\!
-x_\alpha^\star x_\alpha
\!=\!
z^2 \!-\! 1$.

When $z \!=\! 1$,
then $G$ becomes
essentially an $SO(2N)$ matrix $g$
given by
\ba
g
\!=\!
\left[ \!
\BA{cc}
a&b^\star\\
b&a^\star
\EA \!
\right] ,~
g^\dagger g \!=\! g g^\dagger \!=\! 1_{2N} .
\label{matrixg}
\ea
The HB $(SO(2N))$ WF $\ket g$ is generated as
$\ket g \!=\! U(g)\,\ket 0$,
where $\ket 0$ is the vacuum satisfying
$c_\alpha\ket 0 \!=\! 0$.
The WF $\ket g$ is expressed as
\ba 
\ket g
\!=\!
\bra0\,U(g)\,\ket0\exp(\frac{1}{2} \!\cdot\! q_{\alpha \beta }c^\dagger_\alpha 
c^\dagger_\beta)\,\ket 0~,~q \!=\! ba^{-1}
\!=\!
-q^{\mbox{\scriptsize T}} ,
\label{ketg}
\ea
\vskip-1.8\bigskipamount
\ba
\bra0\,U(g)\,\ket 0\,
\!=\!
\left[\strut\det(1 \!-\! q^\star q)\right]^{-\frac{1}{4}}
e^{i\frac{\tau }{2}} ,
\tau
\!\equiv\!
\frac{i}{2} \!\cdot\! \ln \!
\left[
\frac{\det({a}^*)}{\det({a})}
\right] ,
\label{0Ug0}
\ea
where $\det$ represents determinant and
the symbol ${\mbox{\scriptsize T}}$ denotes
the transposition.
By $a \!=\! (a^\alpha_{~i})$ and $b \!=\! (b_{\alpha i})$
we denote $N \!\times\! N$ matrices satisfying 
the orthonormalization condition for the HB amplitudes $a$ and $b$
\cite{Bogo.59}.
By $q$ a $N \!\times\! N$ matrix is denoted which is a variable of
the $\frac{SO(2N)}{U(N)}$ coset space.
On the other hand, the $SO(2N\!+\!1)$ WF
$| G \!\! >$ is generated as
\cite{Fu.77}-\cite{Doba.82}
$\ket G \!=\! U(G)\,\ket 0$
and can be expressed as
\ba
\ket G
\!=\!
\bra0\,U(G)\,\ket0\,(1 \!+\! r_\alpha c^\dagger_\alpha)
\exp(\frac{1}{2} \!\cdot\! q_{\alpha\beta }
c_\alpha^\dagger c_\beta^\dagger)\,
\ket0,
\,\,\,\,
r_\alpha
\!=\!
\frac{1}{1\!+\!z}(x_\alpha \!+\! q_{\alpha\beta }x^\star_\beta)\,,
\label{ketG}
\ea
\vspace{-0.7cm}
\ba
\bra0\, U(G)\,\ket0
\!=\!
\sqrt{\frac{1\!+\!z}{2}}
\left[\strut\det(1 \!-\! q^\star q)\right]^{-\frac{1}{4}}
e^{i\frac{\tau }{2}}~.
\label{0UG0}
\ea

The $SO(2N\!+\!1)$ group is embedded into an $SO(2N\!+\!2)$ group.
The embedding leads to a unified formulation of the $SO(2N\!+\!1)$
regular representation in which paired and unpaired modes are
treated in an equal manner
\cite{Fu.77}-\cite{Fu.81}.
We define
$(\!N\!\!+\!\!1\!) \!\!\times\!\! (\!N\!\!+\!\!1\!)$ matrices
$\cal A$ and $\cal B$ as\\[-12pt]
\ba
{\cal A}
\!\equiv\!\!
\left[ \!\!
\BA{cc}
A&
-{\displaystyle \frac{x^\star }{2}}\\
{\displaystyle \frac{y}{2}}&{\displaystyle \frac{1\!+\!z}{2}}
\EA \!\!
\right] ,~
{\cal B}
\!\equiv\!\!
\left[ \!\!
\BA{cc}
B&
{\displaystyle \frac{x}{2}}\\
-{\displaystyle \frac{y}{2}}&{\displaystyle \frac{1\!-\!z}{2}}
\EA \!\!
\right] ,~
A
\!=\!
a \!-\! {\displaystyle \frac{x^\star y}{2(1\!+\!z)}},~
B
\!=\!
b \!+\!{\displaystyle \frac{xy}{2(1\!+\!z)}},~
y
\!=\!
x^{\mbox{\scriptsize T}} a \!-\! x^\dagger b .
\label{matAandB}
\ea
Using the orthonormalization
(\ref{matG}),
the matrices $\cal A$ and $\cal B$ are shown to satisfy the orthonormalization condition
for the $(N\!+\!1)$-dimensional HB amplitudes
and to form an $SO(2N\!+\!2)$ matrix
$\cal G$. The representations for $\cal G$ and for
$\frac{SO(2N+2)}{U(N+1)}$ coset variables $\cal Q$ are given as\\[-12pt]
\ba
{\cal G}
\!=\!
\left[
\BA{cc}
\cal A&\cal B^\star\\
\cal B& \cal A^\star
\EA
\right]~,~
{\cal G^\dagger G=G G^\dagger={}}1_{2N+2}~,~
{\cal Q}
\!=\!{\cal B A}^{-1}
\!=\!
\left[
\BA{cc}
q&r\\
-r^{\mbox{\scriptsize T}}&0
\EA
\right]
\!=\!
-{\cal Q}^{\mbox{\scriptsize T}} ,
\label{matcalG}
\ea
which shows that the $SO(2N\!+\!1)$ variables $q_{\alpha\beta }$ and $r_\alpha$ are just the independent variables of the
$\frac{SO(2N+2)}{U(N+1)}$ coset space.
The paired mode and the unpaired mode
variables $q_{\alpha\beta }$ and $r_\alpha$, respectively,
in the $SO(2N\!+\!1)$ algebra are unified as the paired variables
in the $SO(2N\!+\!2)$ algebra.
We denote the $(N\!+\!1)$-th dimension of the matrices 
$\cal A$ and $\cal B$ by
the index 0 and use the indices $p,q,\ldots$ running over 0 and 
the single-particle states $\alpha,\beta,\ldots$.

\def\erw#1{{<\!#1\!>_G}}
$\!\!\!$Expectation values of the fermion
$\!SO ( 2 \! N \!\!+\!\! 1 \!)\!$ Lie operators, i.e.,
the generators of rotation
in $(2N \!+\! 1)$-dimensional Euclidian space,
with respect to $| G \!\! >$ are given as
\ba
\left.
\BA{ll}
&\erw{E^\alpha_{~\beta }
\!+\!
{\displaystyle \frac{1}{2}\delta_{\alpha \beta }}}
\!=\!
R_{\alpha \beta }
\!=\!
{\displaystyle \frac{1}{2}}
\left( \!
B^\star_{\alpha i} B_{\beta i}
\!-\!
A^\alpha _{~i} A^{\beta \star }_{~i}
\right)
\!+\!
{\displaystyle \frac{1}{2}}\delta_{\alpha \beta } ,\\
\\[-10pt]
&\erw{E_{\alpha \beta }}
\!=\!
-K_{\alpha \beta }
\!=\!
{\displaystyle \frac{1}{2}}
\left( \!
A^{\alpha \star }_{~i} B_{\beta i}
\!-\!
B_{\alpha i} A^{ \beta \star }_{~i}
\right),~~
\erw{E^{\alpha \beta }}
\!=\!
K^\star_{\alpha \beta }~,\\
\\[-10pt]
&\erw{c_\alpha }
\!=\!
K_{\alpha 0} \!-\! R^\star_{\alpha 0}
\!=\!
{\displaystyle \frac{1}{2}}
\left(
A^{\alpha \star }_{~i} y_{i}
\!+\!
B_{\alpha i} y^{\star }_{i}
\right),~~
\erw{c^\dagger_\alpha }
\!=\!
K^\star_{\alpha 0} \!-\! R_{\alpha 0} .
\EA
\right\}
\label{expectG}
\ea
The unified matrices $R_{\!pq}$ and $K_{\!pq}$ are expressed
in terms of the
$\frac{SO(2N\!+\!2)}{U(N\!+\!1)}$ coset variable $Q_{\!pq}$ as\\[-16pt]
\ba
R_{pq}
\!=\!
-\left[\strut Q^\star Q\,(1 \!-\! Q^\star Q)^{-1}\right]_{pq}~,~
K_{pq}
\!=\!
\left[\strut Q\,(1 \!-\! Q^\star Q)^{-1}\right]_{pq}~.
\label{matRandK}
\ea
The expectation value of a two-body operator is given as
\\[-12pt]
\ba
\erw{E^{\alpha\gamma }E_{\delta\beta }}
\!=\!
R_{\alpha\beta }R_{\gamma\delta }-
R_{\alpha\delta }R_{\gamma\beta }-
K^\star_{\alpha\gamma }K_{\delta\beta }~.
\label{2bodtexpectG}
\ea

Let the Hamiltonian of the fermion system
under consideration be\\[-14pt]
\ba
H
\!=\!
h_{\alpha\beta }
\left( \! E^\alpha_{~\beta } \!+\! \frac{1}{2}\delta_{\alpha\beta } \! \right)
\!+\!
\frac{1}{4}[\alpha\beta|\gamma\delta]
E^{\alpha\gamma }E_{\delta\beta } ,
\label{Hamiltonian}
\ea
which is expressed in terms of the generators of rotation.
We call such a sytem a fermion top.
The matrix $h_{\alpha\beta }$ related to a single-particle hamiltonian includes a chemical
potential and
$
[\alpha\beta|\gamma\delta]
\!=\!
-
[\alpha\delta|\gamma\beta]
\!=\!
[\gamma\delta|\alpha\beta]
\!=\!
[\beta\alpha|\delta\gamma]^\star
$
are anti-symmetrized matrix
elements of an interaction potential.
Parallel to calculations by the usual HB factorization method
(See Refs. \cite{RS.80} and \cite{BlaizotRipka.86}),
the expectation value of $H$ with respect to $| G \!\! >$
is calculated as\\[-14pt]
\ba
\BA{l}
\erw H
\!=\!
h_{\alpha\beta }\erw{E^\alpha_{~\beta }
\!+\!
{\displaystyle \frac{1}{2}}\delta_{\alpha\beta }}\\
\\[-10pt]
{} 
\!+\!
{\displaystyle \frac{1}{2}}[\alpha\beta|\gamma\delta] \!
\left\{ \!
\erw{E^\alpha_{~\beta }
\!+\!
{\displaystyle \frac{1}{2}\delta_{\alpha\beta }}}
\erw{E^\gamma_{~\delta }
\!+\!
{\displaystyle \frac{1}{2}}\delta_{\gamma\delta }}
\!+\!
{\displaystyle \frac{1}{2}}\erw{E^{\alpha\gamma }}
\erw{E_{\delta\beta }}
\right\} .
\EA
\label{HexpectG}
\ea
Then the $\erw H$
(\ref{HexpectG})
represents approximately the energy of  classical motion of the fermion top.

\newpage

%%%%%%%%%%%%%%%%%%%%%%%%%%%%%%%%%%%%%%%%%%%%%%%%%%%%%%%
%                                                     %
%  3  Boson images of fermion SO(2N+1) Lie operators  %
%                                                     %
%         and equations of motions for bosons         %
%                                                     %
%%%%%%%%%%%%%%%%%%%%%%%%%%%%%%%%%%%%%%%%%%%%%%%%%%%%%%%

\def\thesection{\arabic{section}}
\setcounter{equation}{0}
\renewcommand{\theequation}{\arabic{section}.\arabic{equation}}

\section{Boson images of fermion SO(2N+1) Lie operators and equations of motions for bosons}

~~~According to Fukutome
\cite{Fu.77}-\cite{Fu.81},
the fermion $SO(2N \!\!+\!\! 2)$ Lie operators
$\{E^p_{~q},E_{pq},E^{pq}\}$
are constructed by the fermion $SO(2N)$ Lie operators
$\{E^\alpha _{~\beta },E_{\alpha \beta },E^{\alpha \beta }\}$
and the operators\\[-20pt]
\ba
\left.
\BA{ll}
&E^{\alpha 0}
\!=\!
c_\alpha^\dagger P_+
\!=\!
-P_-c_\alpha^\dagger ,~
P_\pm
\!=\!
{\displaystyle \frac{1}{2}} \! \left(1 \pm (-1)^n \right) ,~
n \!\equiv\! c_\alpha^\dagger c_\alpha , \\
\\[-14pt]
&E^\alpha_{~0}
\!=\!
c^\dagger_\alpha P_ ,~
E^0_{~\alpha }
\!=\!
c_\alpha P_+ ,~
E^0_{~0}
\!=\!
{\displaystyle \frac{1}{2}} \! \left(P_- \!-\! P_+ \right) ,~
E_{\alpha 0}
\!=\!
\alpha P_-
\!=\!
P_+c_\alpha .
\EA
\right\}
\ea\\[-12pt]
The boson images
$\hbox{\mathbfsy\char'42}^p_{~q}$ etc. of them
are given in the following forms
\cite{SJCF.08}, \cite{SJCF.11}:\\[-20pt]
\ba
\left.
\BA{ll}
&\hbox{\mathbfsy\char'42}^p_{~q}
\!=\!
{\cal B}^\star _{pr}
{\displaystyle \frac{\partial }{\partial {\cal B}^\star _{qr}}}
\!-\!
{\cal B}_{qr}
{\displaystyle \frac{\partial }{\partial {\cal B}_{pr}}}
\!-\!
{\cal A}^{q \star }_{~r}
{\displaystyle \frac{\partial }{\partial {\cal A}^{p \star }_{~r}}}
\!+\!
{\cal A}^p_{~r}
{\displaystyle \frac{\partial }{\partial {\cal A}^q_{~r}}} 
\!=\!
\hbox{\mathbfsy\char'42}^{q \dag }_{~p} ,~~
\hbox{\mathbfsy\char'42}^{p \dag }_{~q}
\!=\!
-
\hbox{\mathbfsy\char'42}^{p \star }_{~q} ,\\
\\[-14pt]
&\hbox{\mathbfsy\char'42}_{pq}
\!=\!
{\cal A}^{p \star }_{~r}
{\displaystyle \frac{\partial }{\partial {\cal B}^\star _{qr}}}
\!-\!
{\cal B}_{qr}
{\displaystyle \frac{\partial }{\partial {\cal A}^p_{~r}}}
\!-\!
{\cal A}^{q \star }_{~r}
{\displaystyle \frac{\partial }{\partial {\cal B}^\star _{pr}}}
\!+\!
{\cal B}_{pr}
{\displaystyle \frac{\partial }{\partial {\cal A}^q_{~r}}} 
\!=\!
\hbox{\mathbfsy\char'42}^{qp \dag } ,~~
\hbox{\mathbfsy\char'42}_{pq }^\dag
\!=\!
-
\hbox{\mathbfsy\char'42}^\star _{pq } ,~~
\hbox{\mathbfsy\char'42}_{pq}
\!=\!
-
\hbox{\mathbfsy\char'42}_{qp} .
\EA
\right\}
\label{Lieopepsilon}
\ea\\[-12pt]
Using
(\ref{Lieopepsilon}),
the images of the fermion $SO(2N\!+\!1)$ Lie operators are expressed
\cite{Fu.77}-\cite{Fu.81}
as\\[-20pt]
\ba
\BA{c}
\hbox{\mathbf E}^\alpha_{~\beta }
\!=\!
\hbox{\mathbfsy\char'42}^\alpha_{~\beta~ } ,~
\hbox{\mathbf E}_{\alpha\beta }
\!=\!
\hbox{\mathbfsy\char'42}_{\alpha\beta } ,~
\hbox{\mathbf E}^{\alpha\beta }
\!=\!
\hbox{\mathbfsy\char'42}^{\alpha\beta } ,~
\hbox{\mathbf c}_\alpha
\!=\!
\hbox{\mathbfsy\char'42}_{0\alpha } 
\!-\!
\hbox{\mathbfsy\char'42}^0_{~\alpha } ,~
\hbox{\mathbf c}^\dagger_\alpha
\!=\!
\hbox{\mathbfsy\char'42}^{\alpha 0}
\!-\!
\hbox{\mathbfsy\char'42}_{~0}^\alpha ,
\EA
\label{imageLieop}
\ea\\[-20pt]
which are rewritten as\\[-18pt]
\ba
\!\!\!\!\!\!\!\!
\left.
\BA{ll}
&\hbox{\mathbf E}^\alpha_{~\beta }
\!=\!
{\displaystyle \frac{1}{2}}
[\hbox{\mathbf c}^\dagger_\alpha,\hbox{\mathbf c}_\beta ]
\!=\!
\hbox{\boldmath ${\cal B}$}^\dagger_{\alpha \widetilde{r}}
\hbox{\boldmath ${\cal B}$}_{\beta \widetilde{r}}
\!-\!
\hbox{\boldmath ${\cal A}$}^{\beta~\dagger }_{~\widetilde{r}}
\hbox{\boldmath ${\cal A}$}^\alpha_{~\widetilde{r}},
\left(
\hbox{\boldmath ${\cal A}$}^{\alpha }_{~r\!+\!N\!+\!1}
\!\equiv\!
\hbox{\boldmath ${\cal B}$}^\star_{\alpha r},
\hbox{\boldmath ${\cal B}$}_{\alpha r\!+\!N\!+\!1}
\!\equiv\!
\hbox{\boldmath ${\cal A}$}^{\alpha \star }_{~r}
\right),\\
\\[-12pt]
&\hbox{\mathbf E}_{\alpha \beta }
\!=\!
{\displaystyle \frac{1}{2}}
[\hbox{\mathbf c}_\alpha,\hbox{\mathbf c}_\beta ]
\!=\!
\hbox{\boldmath ${\cal A}$}^{\alpha~\dagger }_{~\widetilde{r}}
\hbox{\boldmath ${\cal B}$}_{\beta \widetilde{r}}
\!-\!
\hbox{\boldmath ${\cal A}$}^{\beta~\dagger }_{~\widetilde{r}}
\hbox{\boldmath ${\cal B}$}_{\alpha \widetilde{r}},~
\hbox{\mathbf E}^{\alpha\beta }
\!=\!
-\hbox{\mathbf E}_{\alpha\beta }^\dagger
\!=\!
\hbox{\mathbf E}_{\alpha\beta }^\star,~\\
\\[-12pt]
&\hbox{\mathbf c}_\alpha
\!=\!
\sqrt{2}
\left(
\hbox{\boldmath ${\cal A}$}^{\alpha\dagger }_{~\widetilde{r}}
\hbox{\boldmath ${\cal Y}$}_{\widetilde{r}}
\!+\!
\hbox{\boldmath ${\cal Y}$}^\dagger_{\widetilde{r}}
\hbox{\boldmath ${\cal B}$}_{\alpha \widetilde{r}}
\right) ,~
\hbox{\boldmath ${\cal Y}$}_{\widetilde{r}}
\!\equiv\!
{\displaystyle \frac{1}{\sqrt 2}}
\left(
\hbox{\boldmath ${\cal A}$}^{0}_{~\widetilde{r}}
\!-\!
\hbox{\boldmath ${\cal B}$}^\dagger _{0 \widetilde{r}}
\right) ,~
\hbox{\mathbf c}^\dagger_\alpha
\!=\!
-\hbox{\mathbf c}^\star_\alpha , \\
\\[-12pt]
&\hbox{\boldmath ${\cal A}$}^{\alpha }_{~0}
\!\equiv\!
-\hbox{\boldmath ${\cal X}$}^\star_{\!\alpha },~
\hbox{\boldmath ${\cal A}$}^{\alpha }_{~N\!+\!1}
\!\equiv\!
\hbox{\boldmath ${\cal X}$}_{\!\alpha },~
\hbox{\boldmath ${\cal B}$}_{\alpha 0}
\!\equiv\!
\hbox{\boldmath ${\cal X}$}_{\!\alpha },~
\hbox{\boldmath ${\cal B}$}_{\alpha N\!+\!1}
\!\equiv\!
-\hbox{\boldmath ${\cal X}$}^\star_{\!\alpha }, \\
\\[-10pt]
&\hbox{\boldmath ${\cal Y}$}_{r\!+\!N\!+\!1}
\!\equiv\!
-\hbox{\boldmath ${\cal Y}$}^\star_{r},~
\hbox{\boldmath ${\cal Y}$}_0
\!\equiv\!
\hbox{\boldmath ${\cal Z}$} ,~
\hbox{\boldmath ${\cal Y}$}_{\!N\!+\!1}
\!\equiv\!
-\hbox{\boldmath ${\cal Z}$} ,
(r \!=\! 0, \ldots, N,
\widetilde{r} \!=\! 0, \ldots, N, N\!+\!1,
\ldots, 2N\!+\!1) ,
\EA \!\!
\right\}
\label{bosonimageLieop}
\ea
where
$\hbox{\boldmath ${\cal A}$}^p_{~q}$
and
$\hbox{\boldmath ${\cal A}$}^{p\star }_{~q}$,
etc.
are defined by all the variables
${\cal A}^p_{~q}$
and
${\cal A}^{p\star }_{~q}$,
etc., in ${\cal G}$
(\ref{matcalG})
and their partial differentials as\\[-16pt]
\ba
\left.
\BA{ll}
&
\left.
\BA{c}
\hbox{\boldmath ${\cal A}^p_{~q}$}\\
\\[-10pt]
\hbox{\boldmath ${\cal A}^{\mbox{\scriptsize T \!\!p}}_{~~q}$} \!\!
\EA \!
\right\}
\stackrel{\mathrm{def}}{=}
{\displaystyle \frac{1}{\sqrt 2}} \!
\left( \!
{\cal A}^p_{~q}
\!\pm\!
{\displaystyle \frac{\partial }{\partial {\cal A}^{\star p}_{~~q}}} \!
\right),~
\left.
\BA{c}
\hbox{\boldmath ${\cal A}^{\dagger p}_{~~q}$}\\
\\[-10pt]
\hbox{\boldmath ${\cal A}^{\star p}_{~~q}$} \!\!
\EA \!
\right\}
\stackrel{\mathrm{def}}{=}
{\displaystyle \frac{1}{\sqrt 2}} \!
\left( \!
{\cal A}^{\star p}_{~~q}
\!\mp\!
{\displaystyle \frac{\partial }{\partial {\cal A}^p_{~q}}} \!
\right),\\
\\[-4pt]
&[\hbox{\boldmath ${\cal A}^p_{~q}$},
\hbox{\boldmath ${\cal A}$}^{\dagger r}_{~~s} ]
\!=\!
[\hbox{\boldmath ${\cal A}$}^{\star p}_{~~q},
\hbox{\boldmath ${\cal A}$}^{\mbox{\scriptsize T \!\!r}}_{~~s} ]
\!=\!
\delta^{pr} \delta_{qs} , ~~
\hbox{\boldmath ${\cal A}^p_{~q}$}|~)
\!=\!
\hbox{\boldmath ${\cal A}$}^{\star p}_{~~q}|~)
\!=\!
0 , \\
\\[-4pt]
&[\hbox{\boldmath ${\cal A}^p_{~q}$},
\hbox{\boldmath ${\cal A}$}^{\star r}_{~~s} ]
\!=\!
[\hbox{\boldmath ${\cal A}^p_{~q}$},
\hbox{\boldmath ${\cal A}$}^{\mbox{\scriptsize T \!\!r}}_{~~s} ]
\!=\!
{\bf 0},~
[\hbox{\boldmath ${\cal A}$}^{\dagger p}_{~~q},
\hbox{\boldmath ${\cal A}$}^{\mbox{\scriptsize T \!\!r}}_{~~s} ]
\!=\!
[\hbox{\boldmath ${\cal A}$}^{\dagger p}_{~~q},
\hbox{\boldmath ${\cal A}$}^{\star r}_{~~s} ]
\!=\!
{\bf 0} .
\EA
\right\}
\label{bosonops}
\ea
Similar definitions hold for ${\cal B}$ in order to define
the boson operators 
$\hbox{\boldmath ${\cal B}$}_{pq}$
and
$\hbox{\boldmath ${\cal B}$}^{\star }_{pq}$,
etc.
Here we have used the same symbols as those used in Refs.
\cite{SJCF.08} and \cite{SJCF.11}.

The $SO(2N \!\!+\!\! 1)$ Lie operators are the generators of rotation
in the $(2N \!\!+\!\! 1)$-dimensional Euclidian space.
The fermion top has two kinds of coordinate TRs, namely,
the TR for space fixed coordinate frame
(left TR)
and
the TR for body fixed coordinate frame
(right TR).
To discuss the $c$-number limit of the bosonized fermion
$SO(2N \!\!+\!\! 2)$ Lie operators,
we make a coordinate TR bringing the body fixed coordinate to the frame referred to a quasi-particle:
\ba
\!\!\!\!
\left.
\BA{ll}
&\hbox{\mathbf \boldmath ${\cal A}$}^{p}_{~\widetilde{q}}
\!=\!
{\cal A}^{p}_{~\widetilde{r}}
\hbox{\boldmath ${\cal O}$}_{\widetilde{r}\widetilde{q}},~
\hbox{\mathbf \boldmath ${\cal B}$}_{p \widetilde{q}}
\!=\!
{\cal B}_{p \widetilde{r}}
\hbox{\boldmath ${\cal O}$}_{\widetilde{r}\widetilde{q}},~
(\widetilde{q}, \widetilde{r}, \!=\! 0, \ldots, N,
N \!+\! 1, \ldots, 2 N \!+\! 1) ,\\
\\[-8pt]
&\hbox{\boldmath ${\cal O}$}_{r\widetilde{p}}
\!=\!
\widetilde{\hbox{\boldmath ${\cal A}$}}^{r}_{~\widetilde{p}},~
\hbox{\boldmath ${\cal O}$}_{r\!+\!N\!+\!1,\widetilde{p}}
\!=\!
\widetilde{\hbox{\boldmath ${\cal B}$}}_{r\widetilde{p}},~
(\mbox{boson operators in the quasi-particle}) , \\
\\[-8pt]
&\widetilde{\hbox{\boldmath ${\cal A}$}}^{p}_{~r\!+\!N\!+\!1}
\!\equiv\!
\widetilde{\hbox{\boldmath ${\cal B}$}}^\star_{p r},
\widetilde{\hbox{\boldmath ${\cal A}$}}^{p}_{~0}
\!\equiv\!
-\widetilde{\hbox{\boldmath ${\cal X}$}}^\star_{p},~
\widetilde{\hbox{\boldmath ${\cal A}$}}^{p}_{~N\!+\!1}
\!\equiv\!
\widetilde{\hbox{\boldmath ${\cal X}$}}_{p},~
\widetilde{\hbox{\boldmath ${\cal B}$}}_{p, r\!+\!N\!+\!1}
\!\equiv\!
\widetilde{\hbox{\boldmath ${\cal A}$}}^{p\star }_{~r} ,
\widetilde{\hbox{\boldmath ${\cal B}$}}_{p 0}
\!\equiv\!
\widetilde{\hbox{\boldmath ${\cal X}$}}_{p},
\widetilde{\hbox{\mathbf B}}_{p, N\!+\!1}
\!\equiv\!
-\widetilde{\hbox{\boldmath ${\cal X}$}}^\star_{p}, \\
\\[-8pt]
&\widetilde{\hbox{\mathbf Y}}_{r\!+\!N\!+\!1}
\!\equiv\!
-\widetilde{\hbox{\boldmath ${\cal Y}$}}^\star_{r},~
\widetilde{\hbox{\boldmath ${\cal Y}$}}_0
\!\equiv\!
\widetilde{\hbox{\boldmath ${\cal Z}$}} ,~
\widetilde{\hbox{\boldmath ${\cal Y}$}}_{\!N\!+\!1}
\!\equiv\!
-\widetilde{\hbox{\boldmath ${\cal Z}$}} .
\EA
\right\} \!
\label{bosonrotation}
\ea
Though we use the same symbols parameters, 
${\cal A}^p_{~\widetilde{r}}$ etc. in
(\ref{bosonrotation})
are not identical with those involved in the variable ${\cal G}$.
They are coefficients of a time dependent coordinate transformation
for the ${\cal G}$.
The coordinate transformation is represented in the matrix form as
\\[-16pt]
\ba
\left[ \!
\BA{cc}
\hbox{\boldmath ${\cal A}$}^p_{~q}&
\hbox{\boldmath ${\cal B}$}^\star _{p q}\\
\\[-6pt]
\hbox{\boldmath ${\cal B}$}_{p q}&
\hbox{\boldmath ${\cal A}$}^{p\star }_{~q}
\EA \!
\right] \!
\!=\!
\left[ \!
\BA{cc}
{\cal A}^p_{~r}&{\cal B}^\star _{p r}\\
\\[-6pt]
{\cal B}_{p r}&{\cal A}^{p\star }_{~r}
\EA \!
\right] \!
\left[ \!
\BA{cc}
\widetilde{\hbox{\boldmath ${\cal A}$}}^r_{~q}&
\widetilde{\hbox{\boldmath ${\cal B}$}}^\star_{r q}\\
\\[-6pt]
\widetilde{\hbox{\boldmath ${\cal B}$}}_{r q}&
\widetilde{\hbox{\boldmath ${\cal A}$}}^{r\star }_{~q}
\EA \!
\right] \! .
\label{bosonrotationmat}
\ea\\[-12pt]
The bosonized quasiparticle
$SO(2N \!+\! 1)$ Lie operators with indices $i$ and $j$
$(i,j \!=\! 1, \ldots, N)$
are constructed from the operators in
(\ref{bosonrotation})
in the same way as
(\ref{bosonimageLieop})
and the operator
$\hbox{\boldmath ${\cal W}$}^{\alpha }_{\widetilde{p}}$
is introduced in the following form:\\[-16pt]
\ba
\left.
\BA{ll}
&\hbox{\mathbf E}{}^{\!~i}_{~j}
\!=\!
{\displaystyle \frac{1}{2}}
[\hbox{\mathbf d}^\dagger_i,\hbox{\mathbf d}_j]
\!=\!
\widetilde{\hbox{\boldmath ${\cal B}$}}^\dagger_{i \widetilde{r}}
\widetilde{\hbox{\boldmath ${\cal B}$}}_{j \widetilde{r}}
\!-\!
\widetilde{\hbox{\boldmath ${\cal A}$}}^{j~\dagger }_{~\widetilde{r}}
\widetilde{\hbox{\boldmath ${\cal A}$}}^i_{~\widetilde{r}},~\\
\\[-8pt]
&\hbox{\mathbf E}_{i j}
\!=\!
{\displaystyle \frac{1}{2}}
[\hbox{\mathbf d}_i,\hbox{\mathbf d}_j]
\!=\!
\widetilde{\hbox{\boldmath ${\cal A}$}}^{i~\dagger }_{~\widetilde{r}}
\widetilde{\hbox{\boldmath ${\cal B}$}}_{j \widetilde{r}}
\!-\!
\widetilde{\hbox{\boldmath ${\cal A}$}}^{j~\dagger }_{~\widetilde{r}}
\widetilde{\hbox{\boldmath ${\cal B}$}}_{i \widetilde{r}},~
\hbox{\mathbf E}^{i j}
\!=\!
-\hbox{\mathbf E}_{i j}^\dagger
\!=\!
\hbox{\mathbf E}_{i j}^\star, \\
\\[-6pt]
&\hbox{\mathbf d}_i
\!=\!
\sqrt{2}
\left( \widetilde{\hbox{\boldmath ${\cal A}$}}^{i\dagger }_{~~\widetilde{r}}
\widetilde{\hbox{\boldmath ${\cal Y}$}}_{\widetilde{r}}
\!+\!
\widetilde{\hbox{\boldmath ${\cal Y}$}}^\dagger_{\widetilde{r}}
\widetilde{\hbox{\boldmath ${\cal B}$}}_{i \widetilde{r}}
\right) ,~
\widetilde{\hbox{\boldmath ${\cal Y}$}}_{\widetilde{r}}
\!\equiv\!
{\displaystyle \frac{1}{\sqrt 2}}
\left(
\widetilde{\hbox{\boldmath ${\cal A}$}}^{0}_{~\widetilde{r}}
\!-\!
\widetilde{\hbox{\boldmath ${\cal B}$}}^\dagger _{0 \widetilde{r}}
\right) ,~
\hbox{\mathbf d}^\dagger_i
\!=\!
-\hbox{\mathbf d}^\star_i ,
\EA
\hbox{\boldmath ${\cal W}$}^\alpha_{\widetilde{p}}
\!\equiv\!
\left[ \!
\BA{c}
\hbox{\boldmath ${\cal B}$}_{\alpha \widetilde{p}}\\
\\
\hbox{\boldmath ${\cal A}$}^\alpha_{~\widetilde{p}}\\
\\
\hbox{\boldmath ${\cal Y}$}_{\widetilde{p}}
\EA
\right] .
\right\}
\label{bosonimagequasiLieop}
\ea\\[-12pt]

In the boson images of 
$SO(2N \!\!+\!\! 1)$ Lie operators,
particularly
the images $\hbox{\mathbf c}_\alpha$ and
$\hbox{\mathbf c}_\alpha^\dagger$
must satifsy the anticommutation relations
$\{\hbox{\mathbf c}^\dagger _\alpha, \hbox{\mathbf c}_\beta\}
\!=\! \delta_
{\alpha\beta }$, etc.,
when they
operate on the spinor subspace.
Therefore, we use in the regular representation space the following image of the Hamiltonian with the Lagrange multiplier terms
to select out the spinor subspace:
\ba
\left.
\BA{lll}
\hbox{\mathbf H}
\!\!\!&\!\!=\!\!&\!\!\!
h_{\alpha \beta } \!
\left( \!
\hbox{\mathbf E}^{\,\alpha }_{~\beta }
\!+\!
{\displaystyle \frac{1}{2}} \delta_{\alpha \beta } \!
\right)
\!+\!
{\displaystyle \frac{1}{4}} [\alpha\beta|\gamma\delta]
\left( \!
\{
\hbox{\mathbf E}^{\,\alpha }_{~\beta }
\!+\!
{\displaystyle \frac{1}{2}} \delta_{\alpha \beta },
\hbox{\mathbf E}^{\,\gamma }_{~\delta }
\!+\!
{\displaystyle \frac{1}{2}} \delta_{\gamma \delta }
\}
\!+\!
\{
\hbox{\mathbf E}^{\,\alpha \gamma },\hbox{\mathbf E}_{\delta \beta }
\} \!
\right)
\!+\!
\hbox{\mathbf H}^{~\!\prime } ,\\
\\[-10pt]
\hbox{\mathbf H}^{~\!\prime }
\!\!\!&\!=\!&\!\!\!
{\displaystyle \frac{1}{2}} k_{\alpha \beta } \!
\left( \!
\{\hbox{\mathbf c}^\dagger_\alpha,\hbox{\mathbf c}_\beta\}
\!-\!
\delta_{\alpha \beta } \!
\right)
\!+\!
{\displaystyle \frac{1}{4}}
l_{\alpha\beta }\{\hbox{\mathbf c}^\dagger_\alpha,
\hbox{\mathbf c}^\dagger_\beta\}
\!+\!
{\displaystyle \frac{1}{4}}
l^\star_{\alpha \beta }\{\hbox{\mathbf c}_\alpha,\hbox{\mathbf c}_
\beta\},~
(
k^\star_{\alpha \beta }
\!=\!
k_{\beta \alpha },~
l_{\alpha \beta }
\!=\!
l_{\beta \alpha }
) ,
\EA
\right\}
\label{Hamiltonianimage}
\ea
Using
(\ref{Hamiltonianimage}), (\ref{bosonops}) and (\ref{bosonrotation}),
the Heisenberg equations of motions for
$\hbox{\boldmath ${\cal W}$}^{\alpha }_{\widetilde{p}}~
(\widetilde{p} \!=\! 0, \ldots, N, N \!+\! 1, \ldots, 2 N \!+\! 1)$
are calculated as\\[-16pt]
\ba
\!\!\!\!
\BA{ll}
&i \hbar
\dot {\hbox{\boldmath ${\cal W}$}}^{\alpha } _{\widetilde{p}}
\!=\!
[
\hbox{\boldmath ${\cal W}$}^\alpha_{\widetilde{p}}
,~\hbox{\mathbf H}~
]
\!=\!
\left[ \!
\BA{l}
{\displaystyle \frac{1}{2}}
\{\hbox{\mathbf F}_{\alpha \beta },
\hbox{\boldmath ${\cal B}$}_{\beta \widetilde{p}}\}
\!+\!
{\displaystyle \frac{1}{2}}
\{\hbox{\mathbf D}_{\alpha \beta },
\hbox{\boldmath ${\cal A}$}^\beta _{~\widetilde{p}}\}
\!+\!
{\displaystyle \frac{1}{\sqrt 2}}
\{\hbox{\mathbf M}_{\alpha },
\hbox{\boldmath ${\cal Y}$}_{\widetilde{p}}\} \\
\\[-10pt]
\!\!\!\!
-{\displaystyle \frac{1}{2}}
\{\hbox{\mathbf F}^\dagger_{\alpha \beta },
\hbox{\boldmath ${\cal A}$}^\beta _{~\widetilde{p}}\}
\!-\!
{\displaystyle \frac{1}{2}}
\{\hbox{\mathbf D}^\dagger_{\alpha \beta },
\hbox{\boldmath ${\cal B}$}_{\beta \widetilde{p}}\}
\!+\!
{\displaystyle \frac{1}{\sqrt 2}}
\{\hbox{\mathbf M}^\dagger_{\alpha },
\hbox{\boldmath ${\cal Y}$}_{\widetilde{p}}\} \\
\\[-10pt]
{\displaystyle \frac{1}{\sqrt 2}}
\{\hbox{\mathbf M}_{\alpha },
\hbox{\boldmath ${\cal A}$}^\alpha _{~\widetilde{p}}\}
\!+\!
{\displaystyle \frac{1}{\sqrt 2}}
\{\hbox{\mathbf M}^\dagger_{\alpha },
\hbox{\boldmath ${\cal B}$}_{\alpha \widetilde{p}}\}
\EA \!
\right] \\
\\[-6pt]
&\!=\!
\left[ \!
\BA{l}
\left(
F_{\alpha \beta }{\cal B}_{\beta \widetilde{q}}
\!+\!
D_{\alpha \beta }{\cal A}^\beta _{~\widetilde{q}}
\!+\!
\sqrt{2}M_{\alpha }{\cal Y}_{\widetilde{q}}
\right) \!
\hbox{\boldmath ${\cal O}$}_{\widetilde{q}\widetilde{p}}
\!+\!
{\displaystyle \frac{1}{2}}
\{
\hbox{\mathbf f}_{\alpha \beta }{\cal B}_{\beta \widetilde{r}}
\!+\!
\hbox{\mathbf d}_{\alpha \beta }{\cal A}^\beta _{~\widetilde{r}}
\!+\!
\sqrt{2}\hbox{\mathbf m}_{\alpha } {\cal Y}_{\widetilde{r}},
\hbox{\boldmath ${\cal O}$}_{\widetilde{r}\widetilde{p}}
\} \\
\\[-8pt]
\left(
-
F^\star _{\alpha \beta }{\cal A}^\beta _{~\widetilde{q}}
\!-\!
D^\star _{\alpha \beta }{\cal B}_{\beta \widetilde{q}}
\!+\!
\sqrt{2}M^{\star }_{\alpha } {\cal Y}_{\widetilde{q}}
\right) \!
\hbox{\boldmath ${\cal O}$}_{\widetilde{q}\widetilde{p}}
\!+\!
{\displaystyle \frac{1}{2}}
\{
-
\hbox{\mathbf f}^{~\dagger }_{\alpha \beta }{\cal A}^\beta _{~\widetilde{r}}
\!-\!
\hbox{\mathbf d}^{~\dagger }_{\alpha \beta }{\cal B}_{\beta \widetilde{r}}
\!+\!
\sqrt{2}\hbox{\mathbf m}^{\dagger }_{\alpha } {\cal Y}_{\widetilde{r}},
\hbox{\boldmath ${\cal O}$}_{\widetilde{r}\widetilde{p}}
\} \\
\\[-8pt]
\left(
\sqrt{2}M_{\alpha }{\cal A}^\alpha _{~\widetilde{q}}
\!+\!
\sqrt{2}M^\star _{\alpha }{\cal B}_{\alpha \widetilde{q}}
\right) \!
\hbox{\mathbf O}_{\widetilde{q}\widetilde{p}}
\!+\!
{\displaystyle \frac{1}{2}}
\{
\sqrt{2}\hbox{\mathbf m}_{\alpha }{\cal A}^\alpha _{~\widetilde{r}}
\!+\!
\sqrt{2}\hbox{\mathbf m}^{\dagger }_{\alpha }{\cal B}_{\alpha \widetilde{r}},
\hbox{\boldmath ${\cal O}$}_{\widetilde{r}\widetilde{p}}
\}
\EA \!
\right]  \\
\\[-8pt]
&\!=
i \hbar
{\displaystyle \frac{\partial }{\partial t}}
{\cal W}^\alpha_{\widetilde{q}}
\!\cdot\!
\hbox{\boldmath ${\cal O}$}_{\widetilde{q} \widetilde{p}}
+
{\cal W}^\alpha_{\widetilde{q}}
\!\cdot\!
[\hbox{\boldmath ${\cal O}$}_{\widetilde{q} \widetilde{p}},
~\hbox{\mathbf H}~]
\!=
i \hbar
\dot {\cal W}^\alpha_{\widetilde{q}}
\!\cdot\!
\hbox{\boldmath ${\cal O}$}_{\widetilde{q} \widetilde{p}}
+
i \hbar
{\cal W}^\alpha_{\widetilde{q}}
\!\cdot\!
\dot {\hbox{\boldmath ${\cal O}$}}_{\widetilde{q} \widetilde{p}}.
\EA
\label{TDextendedHBeq2}
\ea
where the
${\cal W}^\alpha_{\widetilde{q}}$
denotes a classical part of the
$\hbox{\boldmath ${\cal W}$}^\alpha_{\widetilde{p}}$
and the
$\hbox{\mathbf F}_{\alpha \beta }$,
etc. are defined as follows:\\[-22pt]
\ba
\!\!\!\!
\left.
\BA{ll}
&\hbox{\mathbf F}_{\alpha \beta }
\!\equiv\!
h_{\alpha\beta }
\!+\!
[\alpha\beta|\gamma\delta]
\left( \!
\hbox{\mathbf E}^{\,\gamma }_{~\delta }
\!+\!
{\displaystyle \frac{1}{2}} \delta_{\gamma \delta } \!
\right)
\!=\!
F_{\alpha \beta }
\!+\!
\hbox{\mathbf f}_{\alpha \beta } ,~
\hbox{\mathbf f}_{\alpha \beta }
\!\equiv\!
[\alpha\beta|\gamma\delta]
\left(
\hbox{\mathbf E}^{\,\gamma }_{~\delta }
-\!
\erw{E^{\,\gamma }_{~\delta }}
\right)
\!=\!
\hbox{\mathbf f}^{~\dagger }_{\beta \alpha } ,\\
\\[-12pt]
&\hbox{\mathbf D}_{\alpha \beta }
\!\equiv\!
{\displaystyle \frac{1}{2}}
[\alpha\gamma|\beta\delta]\hbox{\mathbf E}_{\delta \gamma }
\!=\!
D_{\alpha \beta }
\!+\!
\hbox{\mathbf d}_{\alpha \beta } ,~
\hbox{\mathbf d}_{\alpha \beta }
\!\equiv\!
{\displaystyle \frac{1}{2}}
[\alpha\gamma|\beta\delta]
\left(
\hbox{\mathbf E}_{\delta \gamma }
-\!
\erw{E_{\delta \gamma }}
\right)
\!=\!
-
\hbox{\mathbf d}_{\beta \alpha } ,\\
\\[-6pt]
&\hbox{\mathbf M}_{\alpha }
\!\equiv\!
k_{\alpha\beta }\hbox{\mathbf c}_{\beta }
\!+\!
l_{\alpha\beta }\hbox{\mathbf c}_\beta ^{\dagger }
\!=\!
M_{\alpha }
\!+\!
\hbox{\mathbf m}_{\alpha } ,~
\hbox{\mathbf m}_{\alpha }
\!\equiv\!
k_{\alpha \beta }
\left(
\hbox{\mathbf c}_{\beta }
-\!
\erw{c_{\beta }}
\right)
\!+\!
l_{\alpha \beta }
\left(
\hbox{\mathbf c}_\beta ^{\dagger }
-\!
\erw{c_\beta ^{\dagger }}
\right) .
\EA \!\!
\right\}
\label{bosonizedSCFFandD}
\ea\\[-8pt]
The SCF parameters
$F\!=\!(F_{\alpha \beta })\!=\!F^\dagger,
D\!=\!(D_{\alpha \beta })\!=\!-D^{\mbox{\scriptsize T}}$
and
$M\!=\!(M_{\alpha })$
are defined by\\[-18pt]
\ba
\BA{ll}
F_{\alpha\beta }
\!=\!
h_{\alpha\beta }
\!+\!
[\alpha\beta|\gamma\delta]R_{\gamma\delta } ,~
D_{\alpha\beta }
\!=\!
{\displaystyle \frac{1}{2}}[\alpha\gamma|\beta\delta]\,(-K_{\delta\gamma }) ,
M_{\alpha }
\!=\!
k_{\alpha\beta }\erw{c_\beta }
\!+\!
l_{\alpha\beta }\erw{c_\beta^\dagger }~.
\EA
\label{SCFFandDandM}
\ea\\[-12pt]
which involve effects of unpaired-mode amplitudes as well as those of paired mode.

The Hamiltonian
(\ref{Hamiltonianimage})
is expressed in terms of
the boson operators, i.e., an assembly of oscillators, but
represents a $(2 N \!+\! 1)$-dimensional fermion top.
From
(\ref{TDextendedHBeq2})
the classical equation of motion for the
${\cal W}^\alpha_{\widetilde{p}}$
of the fermion top
is obtained as\\[-12pt]
\ba
i \hbar
\dot {\cal W}^\alpha_{\widetilde{p}}
\!=\!
{\cal F}_{\alpha \beta }
{\cal W}^\beta_{\widetilde{p}}
,~
{\cal F}_{\alpha \beta }
\!\equiv\!
\left[ \!
\BA{ccc}
F_{\alpha \beta }&D_{\alpha \beta }&\sqrt2M_{\alpha }\\
\\[-8pt]
\!\!\!\!\!\!-
D^\star_{\alpha \beta }&-F^\star_{\alpha \beta }&\sqrt2M^\star_{\alpha }\\
\\[-8pt]
\sqrt2M^\dagger _{\beta }&\sqrt2M^{\mbox{\scriptsize T}} _{\beta }&0
\EA \!\!
\right] \!.
\label{classicaleqofmotion}
\ea
Simultaneously
the direction of the axis of the rotation fluctuates quantum mechanically
around the classical axis.
This picture of the fermion dynamics was first given by
Fukutome, Yamamura and one of the present author (S.N.)
\cite{FYN.77}.
Equation
(\ref{classicaleqofmotion})
is transfomed to\\[-14pt]
\ba
i \hbar
\dot {\cal W}^\alpha_{p}
\!=\!
{\cal F}_{\alpha \beta }
{\cal W}^\beta_{p}
\Longrightarrow
i \hbar
\dot W^\alpha_{i}
\!=\!
{\cal F}_{\alpha \beta }
W^\beta_{i}
~\mbox{and}~
i \hbar \!
\left[ \!\!
\BA{c}
\dot {\displaystyle \frac{x_\alpha }{\sqrt 2}}\\
\\[-12pt]
- \dot {\displaystyle \frac{x^\star _\alpha }{\sqrt 2}}\\
\\[-12pt]
\dot z
\EA \!\!
\right]
\!=\!
{\cal F}_{\alpha \beta } \!
\left[ \!\!
\BA{c}
{\displaystyle \frac{x_\beta }{\sqrt 2}}\\
\\[-12pt]
-{\displaystyle \frac{x^\star _\beta }{\sqrt 2}}\\
\\[-12pt]
z
\EA \!\!
\right] \! ,
\label{classicaleqofmotion2}
\ea
using the relations in
(\ref{matAandB}),
from which
we obtain the equations of motion for
HB amplitudes $a$ and $b$ though including unpaired mode
amplitudes $x$ and $x^\star$.
\ba
\left.
\BA{ll}
&i\hbar\dot a
\!=\!
-\left[
F^\star \!-\! {\displaystyle \frac{1}{1\!+\!z}} \!
\left( \!
x^\star M^{\mbox{\scriptsize T}}
\!+\!
M^\star x^{\mbox{\scriptsize T}} \!
\right)
\right] \! a
\!-\!
\left[
D^\star -{\displaystyle \frac{1}{1\!+\!z}}
\left( \!
x^\star M^\dagger
\!-\!
M^\star x^\dagger \!
\right)
\right] \! b,\\
\\[-10pt]
&i\hbar\dot b
\!=\!
\left[
F \!-\! {\displaystyle \frac{1}{1\!+\!z}} \!
\left( \!
xM^\dagger
\!+\!
M x^\dagger \!
\right)
\right] \! b
\!+\!
\left[
D-{\displaystyle \frac{1}{1\!+\!z}} \!
\left( \!
xM^{\mbox{\scriptsize T}}
\!-\!
M x^{\mbox{\scriptsize T}} \! 
\right)
\right] \! a .
\EA
\right\}
\label{eqmotionaandbandx}
\ea
This is an extended TDHB (ETDHB) equation in which
the paired and the unpaired modes are treated in an equal manner
\cite{FYN.77}
and is applicable to both even and odd fermion systems.
Then it is the extended TDSCF theory having
the same level of a mean field approximation as the usual TDHB.
By putting $a$ and $b$ as $e^{\frac{iEt}{\hbar }}a$ and
$e^{\frac{iEt}{\hbar }}b$,
we obtain a static EHB equation
\ba
\!\!\!\!
\left[ \!\!
\BA{cc}
F^\star \!-\!{\displaystyle \frac{1}{1\!+\!z}} \!
\left(
x^\star M^{\mbox{\scriptsize T}} 
\!+\!
M^\star x^{\mbox{\scriptsize T}} 
\right)&\!\!\!
D^\star
\!-\!
{\displaystyle \frac{1}{1\!+\!z}} \!
\left( x^\star M^\dagger \!-\! M^\star x^\dagger \right)\\
\\
-D \!+\! {\displaystyle \frac{1}{1\!+\!z}} \!
\left(
xM^{\mbox{\scriptsize T}}
\!-\!
Mx^{\mbox{\scriptsize T}}
\right)&\!\!\!\!
-F \!+\! {\displaystyle \frac{1}{1\!+\!z}} \!
\left(xM^\dagger \!+\! Mx^\dagger \right)
\EA \!\!\!
\right] \!\!
\left[ \!\!
\BA{c}
a\\
\\ \\
b
\EA \!\!
\right]
\!=\!
E \!
\left[ \!\!
\BA{c}
a\\
\\ \\
b
\EA \!\!
\right] \! ,
{\cal F} \!
\left[ \!\!
\BA{c}
{\displaystyle \frac{x}{\sqrt 2}}\\
-{\displaystyle \frac{x^\star }{\sqrt 2}}\\
z
\EA \!\!
\right]
\!=\!
0 .
\label{staticextendedHBeq}
\ea
Though we use the same symbols, $a$, $b$, $x$ and $z$
in (\ref{staticextendedHBeq})
are time independent amplitudes.
The constraint term in (\ref{Hamiltonianimage}) should vanish in the physical fermion space but their classical part might not.
Using the third of
(\ref{SCFFandDandM})
we must determine the parameters $k$ and $l$ in $M$.

\newpage

%%%%%%%%%%%%%%%%%%%%%%%%%%%%%%%%%%%%%%%%%%%%%%%%%%%%%%
%                                                    %
%  4  A Solution of the extended Hartree-Bogoliubov  %
%                                                    %
%               eigenvalue equation                  %
%                                                    %
%%%%%%%%%%%%%%%%%%%%%%%%%%%%%%%%%%%%%%%%%%%%%%%%%%%%%%

\def\thesection{\arabic{section}}
\setcounter{equation}{0}
\renewcommand{\theequation}{\arabic{section}.\arabic{equation}}

\section{A Solution of the extended Hartee-Bogoliubov eigenvalue equation}

\vspace{-0.3cm}
~~~
We use a spherical symmetric single-particle state specified by
the set of quantum numbers $\{n_a,l_a,j_a,m_\alpha\}$
which is denoted as $\alpha$.
The time-reversed single-particle state
$\bar\alpha$ is obtained from $\alpha$ by changing the sign of
$m_\alpha$.
We use a phase factor $s_\alpha \!=\! (-1)^{j_a-m_\alpha }$ in the time reversed quantity.
The $SO(2N\!+\!1)$ density matrix has the time reversal symmetry
$
R^\star_{\alpha \beta }
\!=\!
s_\alpha s_\beta R_{\bar{\vphantom\beta \alpha }
\bar\beta }
$.
By introducing a new amplitude $\widetilde b$ defined by
$
\widetilde{b}_{\alpha i}
\!=\!
s_{\alpha }b_{\bar\alpha i}
$
\cite{FukuNishiFuku.92}
and using the time reversal symmetry
$F^\star_{\alpha\beta }
\!=\!
s_\alpha
s_\beta F_{\bar{\vphantom\beta \alpha }\bar\beta }
$,
the EHB eigenvalue equation is transformed into\\[-20pt]
\ba
\!\!\!\!\!\!\!\!
\left.
\BA{cc}
&\left\{ \!\!
F^\star_{\alpha\beta } \!-\! {\displaystyle \frac{1}{1\!+\!z}} \!
\left( \! x^\star_\alpha M_\beta \!\!+\!\! M^\star_\alpha x_\beta \!
\right) \!
\right\} \!
a^\beta_{~i}
\!\!+\!\!
\left\{ \!
s_\beta D^\star_{\alpha\bar \beta }
\!-\!
{\displaystyle \frac{1}{1\!\!+\!\!z}} \!
\left( \!
x^\star_\alpha s_\beta M^\star_{\bar\beta }
\!\!-\!\!
M^\star_\alpha s_\beta x^\star_{\bar\beta } \!
\right) \!\!
\right\} \!
\widetilde{b}_{\beta i}
\!\!=\!
E_i a^\alpha_{~i} ,\\
\\[-12pt]
&\!\!
\!-\!
\left\{ \!\!
F^\star_{\alpha\beta } \!-\! {\displaystyle \frac{1}{1\!\!+\!\!z}} \!
\left( \!
s_\alpha x_{\bar\alpha }s_\beta M^\star_{\bar\beta }
\!\!+\!\!
s_\alpha M_{\bar\alpha }s_\beta x^\star_{\bar\beta } \!
\right) \!\!
\right\} \!
\widetilde{b}_{\beta i}
\!\!-\!\!
\left\{ \!
s_\alpha D_{\bar\alpha\beta }\!-\!{\displaystyle \frac{1}{1\!\!+\!\!z}} \!
\left( \!
s_\alpha x_{\bar\alpha }M_\beta
\!\!-\!\!
s_\alpha M_{\bar\alpha }x_{\beta } \!
\right) \!
\right\} \!
a^\beta_{~i}
\!\!=\!
E_i \widetilde{b}_{\alpha i},
\EA \!\!\!
\right\}
\label{extendedHBeigenvalueeq1}
\ea
\vspace{-0.6cm}
\ba
-F^\star_{\alpha \beta } \!
\left( \!\! -\frac{x^\star_\beta }{\sqrt 2} \! \right)
\!\!-\!\!
D^\star_{\alpha\beta } \!
\left( \! \frac{x_\beta }{\sqrt 2} \! \right)
\!\!+\!\!
\sqrt 2 M^\star_\alpha z
\!=\!
0,
~
F^\star_{\alpha \beta } \!
\left( \! s_\beta \frac{x_{\bar\beta }}{\sqrt 2} \! \right)
\!\!+\!\!
s_\alpha D_{\bar\alpha \beta } \!
\left( \!\! -\frac{x^\star_\beta }{\sqrt 2} \! \right)
\!\!+\!\!
\sqrt 2 s_\alpha M_{\bar\alpha }z
\!=\!
0 .
\label{extendedHBeigenvalueeq2}
\ea\\[-14pt]
The SCF parameters $F_{\alpha \beta }$ and
$D_{\alpha \beta }$ are decomposed as\\[-24pt]
\ba
\!\!\!\!\!\!
\left.
\BA{ll}
&F_{\alpha \beta }
\!=\!
f_{\alpha \beta } \!+\!
{\displaystyle \frac{1}{2(1\!\!+\!\!z)}}
[\alpha\beta|\gamma\delta]
\,(x^\star_\gamma\erw{c_\delta }
\!+\!
\erw{c^\dagger_\gamma }x_\delta)
\!\equiv\!
f_{\alpha \beta } \!+\!
{\displaystyle \frac{f_{x, \alpha \beta }}{1\!\!+\!\!z}} ,~
f^\dagger _x \!=\! f_x , \\
\\[-14pt]
&D_{\alpha \beta }
\!=\!
d_{\alpha \beta } \!+\!
{\displaystyle \frac{1}{2(1\!\!+\!\!z)}\frac{1}{2}}
\,[\alpha\gamma|\beta\delta]
\,(x_\gamma\erw{c_\delta } \!-\! \erw{c_\gamma }x_\delta)
\!\equiv\!
d_{\alpha \beta } \!+\!
{\displaystyle \frac{d_{x, \alpha \beta }}{1\!\!+\!\!z}} ,~
d^{\mbox{\scriptsize T}} _x \!=\! -d_x ,
\EA
\right\}
\label{SCFparameters}
\ea\\[-14pt]
where $f_{\alpha\beta }$ and
$d_{\alpha\beta }$ are the usual $SO(2N)$ SCF parameters
defined as\\[-22pt]
\ba
\BA{l}
f_{\alpha\beta }
\!=\!
(\varepsilon_a \!-\! \lambda) \!\cdot\! \delta_{\alpha\beta }
\!+\!
[\alpha\beta|\gamma\delta]\,\rho_{\gamma\delta } ,~
(f^\dagger \!=\! f) , ~~
d_{\alpha\beta }
\!=\!
{\displaystyle \frac{1}{2}}
[\alpha\gamma|\beta\delta]\,(-\kappa_{\delta\gamma }) ,~
(d^{\mbox{\scriptsize T}} \!=\! -d) ,
\EA
\label{usualSCFparameters}
\ea\\[-16pt]
and the $SO(2N)$ density matrices
$\rho_{\alpha\beta }$ and
$\kappa_{\alpha\beta }$ are given by\\[-22pt]
\def\erw#1{{<\!#1\!>_g}}
\ba
\rho_{\alpha\beta }
\!=\!
\erw{E^\alpha_{~\beta } \!+\! \frac{1}{2} \delta_{\alpha\beta }} ,~
\kappa_{\alpha\beta }
\!=\! 
-\erw{E_{\alpha\beta }} ,~
\kappa^\star_{\alpha\beta }
\!=\! 
\erw{E^{\alpha\beta }} .
\label{rhokappa}
\ea\\[-18pt]
\def\erw#1{{<\!#1\!>_G}}
The expectation values of the annihilation and creation operators,
with respect to the state $|G \!>$, are given in terms of
the matrix entries $\rho_{\alpha\beta }$
and $\kappa_{\alpha\beta }$
and the amplitudes $x_{\alpha }$ and
$x^\star_{\alpha }$ as\\[-22pt]
\ba
\erw{c_\alpha }
\!=\!
\left( \!
\frac{1}{2} \delta_{\alpha \beta } \!-\! \rho^\star_{\alpha \beta } \!
\right) \!
x_\beta
\!+\!
\kappa_{\alpha \beta }x^\star_{\beta } ,~
\erw{c^\dagger_\alpha }
\!=\!
\erw{c_\alpha }^{\!\!\!\!\!\!\star } ~.
\label{expectGcandcdag}
\ea\\[-30pt]

We solve the EHB eigenvalue equation
by the two-step diagonalization method
\cite{Ba.61}.
First we diagonalize the separable particle-hole-type interaction and neglect its exchange effect.
Then, the Hartree approximation leads to the eigenvalue equation\\[-20pt]
\ba
f^\star_{\alpha \beta }w_{\beta i}
\!=\!
e_i w_{\alpha i} ,~~~~
w^\star_{\alpha i}w_{\alpha j}
\!=\!
\delta_{ij} ,~~~~
w^\star_{\alpha i}w_{\beta i}
\!=\!
\delta_{\alpha \beta } .
\label{Hartreeapprox}
\ea\\[-18pt]
The time-reversed Hartree state
$t_{\widetilde{\,i\,}}w_{\alpha \widetilde{\,i\,}}$ is degenerate
with the Hartree state $w_{\alpha i}$ in energy:\\[-20pt]
\ba
f^\star_{\alpha \beta }t_{\widetilde{\,i\,}}w_{\beta \widetilde{\,i\,}}
\!=\!
e_{\widetilde{\,i\,}}t_{\widetilde{\,i\,}}w_{\alpha \widetilde{\,i\,}} ,~~~~
e_{\widetilde{\,i\,}}
\!=\!
e_i .
\label{timerevHartreeapprox}
\ea\\[-18pt]
The phase $t_i$, the counterpart of $s_\alpha$,
satisfies the relation
$t_{\widetilde{\,i\,}} \!=\! -t_i$.
The Hartree state $w_{\alpha i}$ satisfies the relation
$w^\star_{\alpha i}
\!=\!
s_{\alpha }
t_{\widetilde{\,i\,}}
w_{\bar\alpha \widetilde{\,i\,}}$.
We put the HB amplitudes into product forms\\[-18pt]
\ba
a^\alpha_{~i}\!=\!u_i w_{\alpha i} ,~~~~
\widetilde{b}_{\alpha i}
\!=\!
v_i 
w_{\alpha i} ,~~\hbox{(not summed over $i$) ,}~~~~
|u_i|^2+|v_i|^2=1 ,
\label{HBamplitudes}
\ea\\[-18pt]
for which we impose also the time reversal ansatz
\\[-20pt]
\ba
s_\alpha t_i a^{\bar\alpha }_{~\widetilde{\,i\,}}
\!=\!
{a^\star }{}^\alpha_{~i} ,~
u_{\widetilde{\,i\,}}
\!=\!
u^\star _i ,~
s_\alpha t_i\widetilde{b}_{\bar\alpha \widetilde{\,i\,}}
\!=\!
\widetilde{b}^\star_{\alpha i} ,~
v_{\widetilde{\,i\,}}
\!=\!
v^\star_i .
\label{timerevHBamplitudes}
\ea\\[-18pt]
\vspace{-0.1cm}
We also introduce the following quantities associated with the unpaired mode amplitudes\\[-12pt]
\ba
M_i
\!\equiv\!
M_\alpha w_{\alpha i} ,~
{\overline M}_i
\!\equiv\!
s_\alpha M^\star _{\bar\alpha } w_{\alpha i} ,~
x_i
\!\equiv\!
x_\alpha w_{\alpha i} ,~
\bar x_i
\!\equiv\!
s_\alpha x^\star_{\bar\alpha } w_{\alpha i} ,
\label{unpaired}
\ea\\[-14pt]
which have the time reversal properties
$
t^\star_{\widetilde{\,i\,}}M^\star_{\widetilde{\,i\,}}
\!=\!
{\overline M}_i ,~
t^\star_{\widetilde{\,i\,}}{\overline M}^\star_{\widetilde{\,i\,}}
\!=\!
M_i ,~
t^\star_{\widetilde{\,i\,}}x^\star_{\widetilde{\,i\,}}
\!=\!
\bar x_i ,~
t^\star_{\widetilde{\,i\,}}\bar x^\star_{\widetilde{\,i\,}}
\!=\!
x_i
$.
\newpage
Further assume the pairing potentials
$d\!=\!(d_{\alpha \beta })$
and
$d_x\!=\!(d_{x,\alpha \beta })$
to be constant.
This makes the situation very simple as in the BCS theory does
\cite{BCS.80}.
Then, the pairing potentials have the following forms:\\[-24pt]
\ba
\!\!\!\!
\BA{l}
d_{\alpha \beta }
\!\!=\!\!
s_\alpha \delta_{\alpha \bar\beta }\Delta ,~
\Delta
\!\!\equiv\!\!
{\displaystyle \frac{1}{2}} g s_\gamma \kappa_{\gamma \bar\gamma }
~\mbox{and}~
d_{x,\alpha \beta }
\!\!=\!\!
s_\alpha \delta_{\alpha \bar\beta }\Delta_x,~
\Delta_x
\!\!\equiv\!\!
{\displaystyle \frac{1}{4}} g s_\gamma \!
\left(
x_\gamma \erw{c_{\bar \gamma }} \!-\! \erw{c_\gamma } x_{\bar\gamma }
\right) ,
\EA
\label{pairingpot}
\ea\\[-18pt]
where $g$ is the strength parameter for the pairing force.
By substituting (\ref{HBamplitudes}) into
(\ref{extendedHBeigenvalueeq1}) and using
the Hartree equation~
(\ref{Hartreeapprox})
and the definition
$
f^\star_{x,i}
\!\equiv\!
w^\star_{\alpha i} f^\star_{x,\alpha \beta } w_{\beta i}
( =\! f_{x,i} )
$,
the EHB eigenvalue equation~
(\ref{extendedHBeigenvalueeq1})
under the assumption~(\ref{pairingpot}) is converted into the following form:\\[-22pt]
\ba
\left.
\BA{ll}
&e_i u_i
\!-\!
{\displaystyle \frac{1}{1\!+\!z}}
\left(
x^\star_i M_i \!+\! M^\star_i x_i
\!-\!
f_{x,i}
\right) u_i
\!+\!
\Delta^\star 
v_i
\!-\!
{\displaystyle \frac{1}{1\!+\!z}}
\left(
x^\star_i {\overline M}_i \!-\! M^\star_i\bar x_i
\!-\!
\Delta^\star _x
\right) v_i
\!=\!
E_i u_i , \\
\\[-10pt]
&\!\!\!\!
-e_i v_i
\!+\!
{\displaystyle \frac{1}{1\!+\!z}}
\left(
\bar x^\star_i {\overline M}_i \!+\! {\overline M}^\star_i\bar x_i 
\!-\! f_{x,i}
\right) v_i
\!+\!
\Delta
u_i
\!+\!
{\displaystyle \frac{1}{1\!+\!z}}
\left(
\bar x^\star_i M_i \!-\! {\overline M}_i^\star x_i
\!+\!
\Delta _x
\right) u_i
\!=\!
E_i v_i .
\EA
\right\}
\label{assumptionHBeq1}
\ea\\[-16pt]
Here we have used the orthonormalization condition in 
(\ref{Hartreeapprox}) and the new quantities in (\ref{unpaired}).
Similarly, multiplying
(\ref{extendedHBeigenvalueeq2})
by $w^\star _{\alpha i}$ and summing over the index $\alpha$,
we obtain the equations\\[-20pt]
\ba
\!\!\!\!
\BA{l}
{\displaystyle \frac{1}{\sqrt 2}} \!\!
\left( \!\!
e_i \!\!+\!\! {\displaystyle \frac{f_{x,i}}{1\!\!+\!\!z}} \!\!
\right) \!
x_i 
\!\!-\!\!
{\displaystyle \frac{1}{\sqrt 2}} \!\!
\left( \!\!
\Delta \!\!+\!\! {\displaystyle \frac{\Delta _x}{1\!\!+\!\!z}} \!\!
\right) \!
\bar x_i
\!\!+\!\!
{\sqrt 2}z M_i
\!\!=\!\!
0 , 
{\displaystyle \frac{1}{\sqrt 2}} \!\!
\left( \!\!
e_i \!\!+\!\! {\displaystyle \frac{f_{x,i}}{1\!\!+\!\!z}} \!\!
\right) \!
\bar x_i
\!\!+\!\!
{\displaystyle \frac{1}{\sqrt 2}} \!\!
\left( \!\!
\Delta^\star \!\!+\!\!
{\displaystyle \frac{\Delta^\star _x }{1\!\!+\!\!z}} \!\!\right) \!
x_i 
\!\!+\!\!
{\sqrt 2}z {\overline M}_i
\!\!=\!\!
0 ,
\EA
\label{assumptionHBeq2}
\ea\\[-16pt]
from which we have the formal solutions for
$x_i$ and $\bar x_i$,\\[-18pt]
\ba
\!\!\!\!\!\!\!\!\!\!\!\!
\left.
\BA{ll}
&x_i
\!=\!
-
{\displaystyle \frac{2z}{{\widetilde E}_i^2}} \!
\left\{ \!\!
\left( \!
e_i \!\!+\!\! {\displaystyle \frac{f_{x,i}}{1\!+\!z}} \!
\right) \!
M_i
\!\!+\!\!
\left( \!
\Delta \!\!+\!\! {\displaystyle \frac{\Delta _x}{1\!+\!z}} \!
\right) \!
{\overline M}_i \!
\right\} \! ,
\bar x_i
\!=\!
-
{\displaystyle \frac{2z}{{\widetilde E}_i^2}} \!
\left\{ \!\!
\left( \!
e_i \!\!+\!\! {\displaystyle \frac{f_{x,i}}{1\!+\!z}} \!
\right) \!
{\overline M}_i
\!\!-\!\!
\left( \!
\Delta^\star
\!\!+\!\!
{\displaystyle \frac{\Delta^\star _x}{1\!+\!z}} \!
\right) \!
M_i \!
\right\} \! , \\
\\[-10pt]
&{\widetilde E}_i^2
\!\equiv\!
\left( \!
e_i \!+\! {\displaystyle \frac{f_{x,i}}{1\!+\!z}} \!
\right)^{\!2}
\!\!+\!
\left|\Delta
\!+\!
{\displaystyle \frac{\Delta _x}{1\!+\!z}}\right|^2 \! .
\EA \!\!
\right\}
\label{solutionforx}
\ea\\[-28pt]

In the second step we diagonalize the pairing interaction contribution only
to the SCF pairing potential $D \!=\! (D_{\alpha \beta })$.
In order to diagonalize the EHB eigenvalue equation
(\ref{assumptionHBeq1}),
it is very useful to notice some important
relations satisfied with $x_i$, $\bar x_i$, $M_i$ and ${\overline M}_i$.
Using the above formal solutions
(\ref{solutionforx}),
we obtain the relations\\[-18pt]
\ba
\!\!\!\!\!\!\!\!\!\!\!\!\!
\left.
\BA{ll}
&x^\star_iM_{\!i} \!\!+\!\! M^\star_{\!i} x_i
\!\!=\!\!
-{\displaystyle \frac{2z}{{\widetilde E}_i^2}} \!
\left\{ \!
2 \! \left( \!
e_i \!\!+\!\! {\displaystyle \frac{f_{x,i}}{1\!\!+\!\!z}} \!
\right) \! |M_{\!i}|^2
\!\!+\!\!
\left( \!
\Delta \!\!+\!\! {\displaystyle \frac{\Delta _x}{1\!\!+\!\!z}} \!
\right) \! M^\star_{\!i} \! {\overline M}_{\!i}
\!\!+\!\!
\left( \!
\Delta^\star
\!\!+\!\!
{\displaystyle \frac{\Delta^\star _x}{1\!\!+\!\!z}} \!
\right) \! {\overline M}^\star_{\!i} \! M_{\!i} \!
\right\} \!\! , \\
\\[-10pt]
&\bar x^\star_i {\overline M}_{\!i}
\!\!+\!\!
{\overline M}^\star_{\!i} \bar x_i
\!\!=\!\!
-{\displaystyle \frac{2z}{{\widetilde E}_i^2}} \!
\left\{ \!
2 \! \left( \!
e_i \!\!+\!\! {\displaystyle \frac{f_{x,i}}{1\!\!+\!\!z}} \!
\right) \!\! |{\overline M}_{\!i}|^2
\!\!-\!\!
\left( \!
\Delta \!\!+\!\! {\displaystyle \frac{\Delta _x}{1\!\!+\!\!z}} \!
\right) \! M^\star_{\!i} \! {\overline M}_{\!i}
\!\!-\!\!
\left( \!
\Delta^\star
\!\!+\!\!
{\displaystyle \frac{\Delta^\star _x}{1\!\!+\!\!z}} \!
\right) \! {\overline M}_{\!i}^\star \! M_{\!i} \!
\right\} \!\! , \\
\\[-12pt]
&x^\star_i {\overline M}_{\!i} \!\!-\!\! M^\star_{\!i} \bar x_i
\!\!=\!\!
-{\displaystyle \frac{2z}{{\widetilde E}_i^2}} \!\!
\left( \!
\Delta^\star
\!\!+\!\!
{\displaystyle \frac{\Delta^\star _x}{1\!\!+\!\!z}} \!
\right) \!\!
\left( |M_{\!i}|^2 \!\!+\!\! |{\overline M}_{\!i}|^2\right) \! ,
\bar x^\star_i M_{\!i}
\!\!-\!\!
{\overline M}^\star_{\!i} x_i
\!\!=\!\!
{\displaystyle \frac{2z}{{\widetilde E}_i^2}} \!\!
\left( \!
\Delta \!\!+\!\! {\displaystyle \frac{\Delta _x}{1\!\!+\!\!z}} \!
\right) \!\!
\left( |M_{\!i}|^2 \!\!+\!\! |{\overline M}_{\!i}|^2\right) \! .
\EA \!\!\!\!
\right\}
\label{relations}
\ea\\[-14pt]
Substituting (\ref{relations}) into (\ref{assumptionHBeq1}),
we have a secular equation\\[-20pt]
\ba
\!\!\!\!
\left| \!\!
\BA{cc}
e_i
\!+\!
F^{(+)}_M
\!-\!
E_i
&~
D^\star _M\\
\\[-10pt]
D_M&~
-e_i
\!-\!
F^{(-)}_M
\!-\!
E_i
\EA \!\!
\right| 
\!=\!
0 ,
\label{seculareq}
\ea\\[-10pt]
where the quantities
$F^{(\pm)}_M$ and $D_M$ are defined as\\[-24pt]
\ba
\!\!\!\!\!
\left.
\BA{ll}
&F^{(+)}_M
\!\equiv\!
{\displaystyle \frac{1}{1\!\!+\!\!z}} \!
\left[ \!
{\displaystyle \frac{2z}{{\widetilde E}_i^2}}
\left\{ \!
2 \!
\left( \!
e_i \!\!+\!\! {\displaystyle \frac{f_{x,i}}{1\!\!+\!\!z}} \!
\right) \! |M_i|^2
\!+\!
\left( \!
\Delta \!\!+\!\! {\displaystyle \frac{\Delta _x}{1\!\!+\!\!z}} \!
\right) \! 
M^\star_i {\overline M}_i
\!+\!
\left( \!
\Delta^\star
\!\!+\!\!
{\displaystyle \frac{\Delta^\star _x}{1\!\!+\!\!z}} \!
\right) \! 
{\overline M}^\star_i M_i \!
\right\}
\!+\!
f_{x,i} \!
\right], \!\! \\
\\[-10pt]
&F^{(-)}_M
\!\equiv\!
{\displaystyle \frac{1}{1\!\!+\!\!z}} \!
\left[ \!
{\displaystyle \frac{2z}{{\widetilde E}_i^2}}
\left\{ \!
2 \!
\left( \!
e_i \!\!+\!\! {\displaystyle \frac{f_{x,i}}{1\!\!+\!\!z}} \!
\right) \! 
|{\overline M}_{\!i}|^2
\!-\!
\left( \!
\Delta \!\!+\!\! {\displaystyle \frac{\Delta _x}{1\!\!+\!\!z}} \!
\right) \!
M^\star_i {\overline M}_i
\!-\!
\left( \!
\Delta^\star
\!\!+\!\!
{\displaystyle \frac{\Delta^\star _x}{1\!\!+\!\!z}} \!
\right) \!
{\overline M}^\star_i M_i \!
\right\}
\!+\!
f_{x,i} \!
\right] , \!\! \\
\\[-12pt]
&D_M ~
\!\equiv\!
\left( \!
\Delta \!\!+\!\! {\displaystyle \frac{\Delta _x}{1\!\!+\!\!z}} \!
\right) \! \!
\left( \!
{\displaystyle 1 \!\!+\!\! \frac{2z}{1\!\!+\!\!z}}
{\displaystyle 
\frac{|M_i|^2 \!\!+\!\! |{\overline M}_i|^2}{{\widetilde E}_i^2}
}
\right) \! .
\EA
\right\}
\label{seculareq2}
\ea\\[-14pt]
From
(\ref{seculareq}) and (\ref{seculareq2})
we obtain an eigenvalue

\ba
\!\!\!\!
\BA{l}
E_i
\!\!=\!\!
{\widetilde E}_i
\!+\!
{\displaystyle \frac{2z}{1\!\!+\!\!z}}
{\displaystyle 
\frac{|M_{\!i}|^2 \!\!+\!\! |{\overline M}_{\!i}|^2}
{{\widetilde E}_i}
}
\!+\!
{\displaystyle \frac{2z}{1\!\!+\!\!z}}
{\displaystyle 
\frac{
\left( \!\!
e_i \!\!+\!\! {\displaystyle \frac{f_{x,i}}{1\!\!+\!\!z}} \!\!
\right) \!\!
\left( \!
|M_{\!i}|^{\!2} \!\!-\!\! |{\overline M}_{\!i}|^{\!2} \right)
\!\!+\!\!
\left( \!\!
\Delta \!\!+\!\! {\displaystyle \frac{\Delta _x}{1\!\!+\!\!z}} \!\!
\right) \!\!
M^\star_{\!i} \! {\overline M}_{\!i}
\!\!+\!\!
\left( \!\!
\Delta^\star
\!\!+\!\!
{\displaystyle \frac{\Delta^\star _x}{1\!\!+\!\!z}} \!\!
\right) \!\!
{\overline M}^\star_{\!i}\!  M_{\!i}}
{{\widetilde E}_i^2}
} ,
\EA
\label{qpenergy}
\ea\\[-12pt]
which reduces to the usual HB quasiparticle energy if $z=1$
\cite{RS.80}.
It involves unpaired-mode effects in contrast to
the usual HB theory, which is unable to describe the unpaired modes.

With the help of the EHB eigenvalue equation
(\ref{assumptionHBeq1}),
the eigenvectors are obtained as\\[-16pt]
\ba
\!\!\!\!\!\!\!\!\!\!\!\!
\left.
\BA{ll}
&|u_i|^2
\!=\!
{\displaystyle \frac{1}{2}}
\frac{
{\displaystyle E_i \!\!+\!\! e_i
\!\!+\!\!
\frac{f_{x,i}}{1 \!\!+\!\! z}
+
}
{\displaystyle
\frac{2z}{1 \!\!+\!\! z}
\frac{1}{{\widetilde E}_i^2}}
\left\{ \!
{\displaystyle
2 \! \left( \!
e_i \!\!+\!\! {\displaystyle \frac{f_{x,i}}{1 \!\!+\!\! z}} \!
\right) \! |{\overline M}_{\!i}|^2
\!\!-\!\!
\left( \!
\Delta \!\!+\!\! {\displaystyle \frac{\Delta _x}{1 \!\!+\!\! z}} \!
\right) \!\!
M^\star_{\!i} \! {\overline M}_{\!i}
\!\!-\!\!
\left( \!
\Delta^\star
\!\!+\!\!
{\displaystyle \frac{\Delta^\star _x}{1 \!\!+\!\! z}} \!
\right) \!\!
{\overline M}^\star_{\!i} \! M_{\!i}
} \!
\right\}
}{
{\displaystyle E_i
-
}
{\displaystyle
\frac{2z}{1\!+\!z}
\frac{1}{{\widetilde E}_i^2}}
\left\{ \!
{\displaystyle
\left( \!
e_i \!\!+\!\! {\displaystyle \frac{f_{x,i}}{1 \!\!+\!\! z}} \!
\right) \!
( |M_{\!i}|^2 \!-\! |{\overline M}_{\!i}|^2 )
\!\!+\!\!
\left( \!
\Delta \!\!+\!\! {\displaystyle \frac{\Delta _x}{1 \!\!+\!\! z}} \!
\right) \!\!
M^\star_{\!i} \! {\overline M}_{\!i}
\!\!+\!\!
\left( \!
\Delta^\star
\!\!+\!\!
{\displaystyle \frac{\Delta^\star _x}{1 \!\!+\!\! z}} \!
\right) \!\!
{\overline M}^\star_{\!i} \! M_{\!i}
} \!
\right\}
} ,\\
\\[-10pt]
&|v_i|^2
\!=\!
{\displaystyle \frac{1}{2}}
\frac{
{\displaystyle E_i \!\!-\!\! e_i
\!\!-\!\!
\frac{f_{x,i}}{1 \!\!+\!\! z}
-
}
{\displaystyle
\frac{2z}{1\!+\!z}
\frac{1}{{\widetilde E}_i^2}}
\left\{ \!
{\displaystyle
2 \! \left( \!
e_i \!\!+\!\! {\displaystyle \frac{f_{x,i}}{1 \!\!+\!\! z}} \!
\right) \! |M_{\!i}|^2
\!\!+\!\!
\left( \!
\Delta \!\!+\!\! {\displaystyle \frac{\Delta _x}{1 \!\!+\!\! z}} \!
\right) \!\!
M^\star_{\!i} \! {\overline M}_{\!i}
\!\!+\!\!
\left( \!
\Delta^\star
\!\!+\!\!
{\displaystyle \frac{\Delta^\star _x}{1 \!\!+\!\! z}} \!
\right) \!\!
{\overline M}^\star_{\!i} \! M_{\!i}
} \!
\right\}
}{
{\displaystyle E_i
-
}
{\displaystyle
\frac{2z}{1\!+\!z}
\frac{1}{{\widetilde E}_i^2}}
\left\{ \!
{\displaystyle
\left( \!
e_i \!\!+\!\! {\displaystyle \frac{f_{x,i}}{1 \!\!+\!\! z}} \!
\right) \!
( |M_{\!i}|^2 \!-\! |{\overline M}_{\!i}|^2 )
\!\!+\!\!
\left( \!
\Delta \!\!+\!\! {\displaystyle \frac{\Delta _x}{1 \!\!+\!\! z}} \!
\right) \!\!
M^\star_{\!i} \! {\overline M}_{\!i}
\!\!+\!\!
\left( \!
\Delta^\star
\!\!+\!\!
{\displaystyle \frac{\Delta^\star _x}{1 \!\!+\!\! z}} \!
\right) \!\!
{\overline M}^\star_{\!i} \! M_{\!i}
} \!
\right\}
} .
\EA \!\!
\right\}
\label{eigenvectors}
\ea\\[-10pt]
From
(\ref{solutionforx}),
we can get the
solutions for the unpaired-mode amplitudes and for $z$ as\\[-18pt]
\ba
\left.
\BA{ll}
&|x_i|^2
\!=\!
{\displaystyle
\frac{4z^2}{{\widetilde E}_i^4}} \!
\left\{ \!
\left( \!
e_i \!\!+\!\! {\displaystyle \frac{f_{x,i}}{1 \!\!+\!\! z}} \!
\right) \! |M_i|^2
\!+\!
|\Delta \!\!+\!\! {\displaystyle \frac{\Delta _x}{1 \!\!+\!\! z}} |^2
|{\overline M}_i|^2
\right. \\
\\[-12pt]
&~~~~~~~~
\left.
\!+\!
\left( \!
e_i \!\!+\!\! {\displaystyle \frac{f_{x,i}}{1 \!\!+\!\! z}} \!
\right) \!\!
\left( \!
\Delta \!\!+\!\! {\displaystyle \frac{\Delta _x}{1 \!\!+\!\! z}} \!
\right) \! M^\star_i {\overline M}_i
\!+\!
\left( \!
e_i \!\!+\!\! {\displaystyle \frac{f_{x,i}}{1 \!\!+\!\! z}} \!
\right) \!\!
\left( \!
\Delta^\star
\!\!+\!\!
{\displaystyle \frac{\Delta^\star _x}{1 \!\!+\!\! z}} \!
\right) \! {\overline M}^\star_i M_i \!
\right\} , \\
\\ \\[-20pt]
&|\bar x_i|^2
\!=\!
{\displaystyle
\frac{4z^2}{{\widetilde E}_i^4}} \!
\left\{ \!
\left( \!
e_i \!\!+\!\! {\displaystyle \frac{f_{x,i}}{1 \!\!+\!\! z}} \!
\right) \! |{\overline M}_i|^2
\!+\!
|\Delta \!\!+\!\! {\displaystyle \frac{\Delta _x}{1 \!\!+\!\! z}}|^2
|M_i|^2
\right. \\
\\[-12pt]
&~~~~~~~~
\left.
\!-\!
\left( \!
e_i \!\!+\!\! {\displaystyle \frac{f_{x,i}}{1 \!\!+\!\! z}} \!
\right) \!\!
\left( \!
\Delta \!\!+\!\! {\displaystyle \frac{\Delta _x}{1 \!\!+\!\! z}} \!
\right) \! M^\star_i {\overline M}_i
\!\!-\!\!
\left( \!
e_i \!\!+\!\! {\displaystyle \frac{f_{x,i}}{1 \!\!+\!\! z}} \!
\right) \!\!
\left( \!
\Delta^\star
\!\!+\!\!
{\displaystyle \frac{\Delta^\star _x}{1 \!\!+\!\! z}} \!
\right) \! {\overline M}^\star_i M_i \!
\right\} , \\
\\ \\[-26pt]
&z^2
\!=\!
\left[
1
\!+\!
2\sum_i
{\displaystyle
\frac{1}{{\widetilde E}_i^2}
}
\left(
|M_i|^2 \!+\! |{\overline M}_i|^2
\right)
\right]^{-1} ,
\EA \!
\right\}
\label{eigenvectorsxandz}
\ea\\[-8pt]
where we have used the normalization condition
$\sum_i|x_i|^2 \!=\! \sum_i|\bar x_i|^2 \!=\! 1 \!-\! z^2$.

Up to the present stage, the SCF Hartree potential,
the chemical potential $\lambda$
and the pairing potential
remain
undetermined. With the use of
(\ref{SCFparameters})$\sim$(\ref{expectGcandcdag})
and
the product form of the HB amplitude
(\ref{HBamplitudes}),
the SCF Hartree potential given through 
(\ref{SCFparameters})
is determined as\\[-16pt]
\ba
\!\!\!\!\!\!\!\!\!\!\!\!
\left.
\BA{ll}
&f_{\alpha \beta }
\!\!=\!\!
(\varepsilon_a \!-\! \lambda) \!\cdot\! \delta_{\alpha \beta } \!
\sum_i|v_i|^2 w_{\alpha i} w^\star_{\beta i}
\!+\!
[\alpha\beta|\gamma\delta]
\sum_i |v_i|^2
w_{\gamma i} w^\star_{\delta i} , \\
\\[-10pt]
&f_{x,\alpha \beta }
\!\!=\!\!
{\displaystyle \frac{1}{2}}
[\alpha\beta|\gamma\delta] \!
\sum_{ij} \!\!
\left\{ \!
\left( 1 \!\!-\!\! |v_i|^2 \!\!-\!\! |v_j|^2 \right) \! x^\star_i x_j
\!\!+\!\!
{\displaystyle \frac{1}{2}} \!
\left( u^\star_j v_j \!\!+\!\! u_j v^\star_j \right) \!
x^\star_i \bar x_j
\!\!+\!\!
{\displaystyle \frac{1}{2}} \!
\left( u^\star_i v_i \!\!+\!\! u_i v^\star_i \right) \!
\bar x^\star_i x_j \!
\right\} \!
w_{\gamma i}
w^\star_{\delta j} ,
\EA \!\!\!
\right\}
\label{SCFHartreepot}
\ea\\[-16pt]
and the pairing potentials
and the chemical potential are determined
by the conditions\\[-18pt]
\ba
\BA{l}
\Delta
\!=\!
{\displaystyle \frac{1}{2}} g
\sum_i \!
\left(
u^\star_i v_i \!+\! u_i v^\star_i
\right) , ~
\Delta_x
\!=\!
-{\displaystyle \frac{1}{8}} g
\sum_{i} \!
\left( u_i^\star v_i \!+\! v_i^\star u_i \right) \!
\left( \!
|x_i|^2
\!+\!
|\bar x_i|^2 \!
\right) ,~\mbox{and}
\EA
\label{gap}
\ea\\[-36pt]
\ba
\BA{c}
{\cal N}
\!\!=\!\!
R_{\alpha \alpha }
\!\!=\!\!
\sum_i \!
\left[
1 \!-\! {\displaystyle \frac{1}{1\!+\!z}} \!
\left\{
|v_i|^2
|x_i|^2
\!-\!
{\displaystyle \frac{1}{4}} \!
\left(
u^\star_i v_i \!+\! u_i v^\star_i
\right) \!
\left( x^\star_i \bar x_i \!+\! \bar x_i^\star x_i \right) \!
\right\} \!
\right]
\!\!+\!\!
{\displaystyle \frac{1\!-\!z}{2}} ,
\EA
\label{number}
\ea\\[-14pt]
where ${\cal N}$ is a number of fermions in the system.
They contain unpaired-mode effects and reduce to
the equations for the gap energy
and the chemical potential of the usual HB theory if we put $z \!\!=\!\! 1$
\cite{RS.80}.
The magnitude of the gap energy evidently
decreases due to the presence of the unpaired-mode effects (Pauli blocking effects),
which is also seen in the solution of the improper-HB eigenvalue equation presented by one of the present authors (S.N.)
et al.
\cite{NIFY.80}-\cite{NIFY2.80}.

\newpage

%%%%%%%%%%%%%%%%%%%%%%%%%%%%%%%%%%%%%%%%%%%%%%%%%%%%%%%
%                                                     %
% 5     Energies of classical motion and quantum      %
%                                                     %
%  mechanical fluctuation of fermion SO(2N+1) rotator %
%                                                     %
%%%%%%%%%%%%%%%%%%%%%%%%%%%%%%%%%%%%%%%%%%%%%%%%%%%%%%%

\def\thesection{\arabic{section}}
\setcounter{equation}{0}
\renewcommand{\theequation}{\arabic{section}.\arabic{equation}}

\section{Energies of classical motion and quantum mechanical fluctuation of fermion SO(2N+1) top}

~~~Through the coordinate transformations
(\ref{bosonrotation}) and (\ref{bosonrotationmat}),
the fermion $SO(2N \!+\! 1)$ Lie operators are expressed
in terms of their quasiparticle expectation values ($c$-number) and of the quasiparticle $SO(2N \!+\! 1)$ Lie operators
(quantum mechanical fluctuations) as follows:\\[-20pt]
\ba
\!\!\!\!
\left.
\BA{ll}
\hbox{\mathbf E}^{\alpha }_{~\beta }
\!\!\!&\!=\!
\erw{E^{\alpha }_{~\beta } }
\!+\!\!
\widetilde{\hbox{\mathbf E}}^{\alpha }_{~\beta } 
\!=\!
\erw{E^{\alpha }_{~\beta } }
\!+\!\!
\left( \!\!
{\mathbf A}^{\alpha }_{~i}
{\mathbf A}^{\beta \star }_{~j}
\!\!-\!\!
{\mathbf B}_{\beta i}
{\mathbf B}^\star_{\alpha j} \!\!
\right) \!\!
\left( \!\!
\hbox{{\mathbf E}}^{~\!i}_{~j}
\!\!+\!\!
{\displaystyle \frac{1}{2} \delta_{ij}} \!\!
\right)
\!\!+\!\!
{\mathbf B}^\star_{\alpha i}{\mathbf A}^{\beta \star }_{~j} \!
\hbox{{\mathbf E}}_{~\!ij}
\!\!+\!\!
{\mathbf A}^{\alpha }_{~i}{\mathbf B}_{\beta j} \!
\hbox{{\mathbf E}}^{~\!ij} \\
\\[-16pt]
&~~~~~~~~~~~~~~
\!+\!
{\displaystyle \frac{1}{2}} \!
\left( \!
{\mathbf B}^\star_{\alpha i} x_\beta
\!+\!
{\mathbf A}^{\beta \star }_{~i} x^\star_\alpha \!
\right) \!
\hbox{{\mathbf d}}_{~\!\!i}
\!+\!
{\displaystyle \frac{1}{2}} \!
\left( \!\!
{\mathbf A}^{\alpha }_{~i} x_\beta
\!+\!
{\mathbf B}_{\beta i} x^\star_\alpha \!\!
\right) \!
\hbox{{\mathbf d}}^\dagger_{~\!\!i} , \\
\\[-12pt]
\hbox{\mathbf E}_{\alpha \beta }
\!\!\!&\!=\!
\erw{E_{\alpha \beta } }
\!+\!\!
\widetilde{\hbox{\mathbf E}}_{\alpha \beta }
\!=\!
\erw{E_{\alpha \beta } }
\!+\!\!
\left( \!
{\mathbf B}_{\alpha i}
{\mathbf A}^{\beta \star }_{~j}
\!\!-\!\!
{\mathbf B}_{\beta i}
{\mathbf A}^{\alpha \star }_{~j} \!
\right) \!\!
\left( \!\!
\hbox{{\mathbf E}}^{~\!i}_{~j}
\!\!+\!\!
{\displaystyle \frac{1}{2} \delta_{ij}} \!\!
\right)
\!\!+\!\!
{\mathbf A}^{\alpha \star }_{~i} \! {\mathbf A}^{\beta \star }_{~j} \!
\hbox{{\mathbf E}}_{~\!ij}
\!\!+\!\!
{\mathbf B}_{\alpha i} \! {\mathbf B}_{\beta j} \!
\hbox{{\mathbf E}}^{~\!ij} \\
\\[-16pt]
&~~~~~~~~~~~~~~~\!
\!+\!
{\displaystyle \frac{1}{2}} \!
\left( \!
{\mathbf A}^{\alpha \star }_{~i} x_\beta
\!-\!
{\mathbf A}^{\beta \star }_{~i} x_\alpha \!
\right) \!
\hbox{{\mathbf d}}_{~\!\!i}
\!+\!
{\displaystyle \frac{1}{2}} \!
\left( \!
{\mathbf B}_{\alpha i} x_\beta
\!-\!
{\mathbf B}_{\beta i} x_\alpha \!
\right) \!
\hbox{{\mathbf d}}^\dagger_{~\!\!i} , \\
\\[-12pt]
\hbox{\mathbf c}_\alpha
&\!\!\!\!\!=\!
\erw{c_\alpha }
\!+\!
\widetilde{\hbox{\mathbf c}}_\alpha
\!=\!
\erw{c_\alpha }
\!-\!
\left( \!
{\mathbf B}_{\alpha i} y^\star_{j} \!
\!+\!
{\mathbf A}^{\alpha \star }_{~j}y_{i} \!
\right) \!\!
\left( \!\!
\hbox{{\mathbf E}}^{~\!i}_{~j}
\!\!+\!\!
{\displaystyle \frac{1}{2} \delta_{ij}} \!\!
\right)
\!\!-\!\!
{\mathbf A}^{\alpha \star }_{~i} y^\star_{j} \!
\hbox{{\mathbf E}}_{~\!ij}
\!\!+\!\!
{\mathbf B}_{\alpha i} y_{j} \!
\hbox{{\mathbf E}}^{~\!ij} \\
\\[-16pt]
&~~~~~~~~~~~~~~~\!
\!+\!
\left( \!\!
z {\mathbf A}^{\alpha \star }_{~i}
\!+\!
{\displaystyle \frac{1}{2}} x_\alpha y^\star_{i} \!
\right) \!\!
\hbox{{\mathbf d}}_{~\!\!i}
\!+\!
\left( \!\!
z {\mathbf B}_{\alpha i}
\!-\!
{\displaystyle \frac{1}{2}} x_\alpha y_{i} \!
\right) \!\!
\hbox{{\mathbf d}}^\dagger_{~\!\!i} .
\EA \!\!\!\!
\right\}
\label{bosonrotation2}
\ea\\[-16pt]
The combination
$
\hbox{{\mathbf E}}^{~\!i}_{~j}
\!+\!
{\displaystyle \frac{1}{2} \delta_{ij}}
$,
in which the constant term
$- {\displaystyle \frac{1}{2} \delta_{ij}}$ in
$\hbox{{\mathbf E}}^{~\!i}_{~j}$
is eliminated,
is relevant to describe the quantum mechanical fluctuations.

Using
(\ref{bosonrotation2}),
the boson image of the Hamiltonian with the Lagrange multipliers 
(\ref{Hamiltonianimage})
is transformed to the following form:\\[-18pt]
\ba
\!\!\!\!\!\!\!\!
\left.
\BA{ll}
&\hbox{\mathbf H}
\!=\!
E_0
\!+\!
\widetilde{\hbox{\mathbf H}} ,\\
\\[-14pt]
&E_0
\!=\!
\erw{H}
\!+\!
k_{\alpha \beta } \!
\left( \!\!
\erw{c^\dagger_\alpha } \erw{c_\beta }
\!-\!\!
{\displaystyle \frac{1}{2}} \delta_{\alpha \beta } \!\!
\right) \!
\!+\!\!
{\displaystyle \frac{1}{2}} l_{\alpha \beta }
\erw{c^\dagger_\alpha } \erw{c^\dagger_\beta }
\!+\!\!
{\displaystyle \frac{1}{2}} l^\star_{\alpha \beta }
\erw{c_\alpha } \erw{c_\beta } , \\
\\[-14pt]
&\widetilde{\hbox{\mathbf H}}
\!=\!
{\displaystyle \frac{1}{2}}
F_{\alpha \beta } \widetilde{\hbox{\mathbf E}}^{\,\alpha }_{~\beta }
\!+\!
{\displaystyle \frac{1}{2}}
F^\star_{\alpha \beta } \widetilde{\hbox{\mathbf E}}^{\alpha \dagger }_{~\beta }
\!+\!
{\displaystyle \frac{1}{2}}
D_{\alpha \beta } \widetilde{\hbox{\mathbf E}}^{\alpha \beta }
\!-\!
{\displaystyle \frac{1}{2}}
D^\star_{\alpha \beta } \widetilde{\hbox{\mathbf E}}_{\alpha \beta }
\!+\!
M_\alpha \widetilde{\hbox{\mathbf c}}^\dagger_\alpha
\!+\!
M^\star_\alpha \widetilde{\hbox{\mathbf c}}_\alpha \\
\\[-12pt]
\!\!\!&
\!+\!
{\displaystyle \frac{1}{4}} \!
\left[
[\alpha \beta|\gamma \delta] \!
\left( \!\!
\{ \!
\widetilde{\hbox{\mathbf E}}^{\,\alpha }_{~\beta },
\widetilde{\hbox{\mathbf E}}^{\,\gamma }_{~\delta } \!
\}
\!\!+\!\!
{\displaystyle \frac{1}{2}} \!
\{ \!
\widetilde{\hbox{\mathbf E}}^{\,\alpha \gamma },
\widetilde{\hbox{\mathbf E}}_{\delta \beta } \!
\} \!\!
\right)
\!\!+\!\!
k_{\alpha \beta } \!
\{ \!
\widetilde{\hbox{\mathbf c}}^\dagger_\alpha,
\widetilde{\hbox{\mathbf c}}_\beta \!
\}
\!\!+\!\!
k^\star_{\alpha \beta } \!
\{ \!
\widetilde{\hbox{\mathbf c}}^\dagger_\beta,
\widetilde{\hbox{\mathbf c}}_\alpha \!
\}
\!\!+\!\!
l_{\alpha\beta }
\{ \!
\widetilde{\hbox{\mathbf c}}^\dagger_\alpha,
\widetilde{\hbox{\mathbf c}}^\dagger_\beta \!
\}
\!\!+\!\!
l^\star_{\alpha \beta }
\{ \!
\widetilde{\hbox{\mathbf c}}_\alpha,
\widetilde{\hbox{\mathbf c}}_\beta \!
\}
\right] ,
\EA \!\!\!
\right\}
\label{Hamiltonianimage2}
\ea\\[-14pt]
where $E_0$ is the energy of the classical motion of the top
and deviates from the quasiparticle expectation value
of the Hamiltonian $\erw{H}$ owing to the presence of
the constraint term $\hbox{\mathbf H}^{~\!\prime }$ in 
(\ref{Hamiltonianimage}).
The coordinate transformation method contrasts with
the coherent-state method
\cite{MNY.77}.
The Hamiltonian $\widetilde{\hbox{\mathbf H}}$ is the boson image of Hamiltonian for the quantum mechanical fluctuation.
To treat such a quantum mechanical fluctuation,
we make the successive coordinate transformation
$\widetilde{\cal G}$
similar to
(\ref{bosonrotationmat}).
We may decompose the tranformation matrix
$\widetilde{\cal G}$
approximately into a product form of two matrices in the following form:
\\[-18pt]
\ba
\!\!
\left[ \!\!
\BA{cc}
\hbox{\boldmath $\widetilde{{\cal A}}$}^p_{~q}&
\hbox{\boldmath $\widetilde{{\cal B}}$}^\star _{p q}\\
\\[-4pt]
\hbox{\boldmath $\widetilde{{\cal B}}$}_{p q}&
\hbox{\boldmath $\widetilde{{\cal A}}$}^{p\star }_{~q}
\EA \!\!
\right] \!
\!=\!
\left[ \!\!
\BA{cc}
\widetilde{{\cal A}}^p_{~r}&\widetilde{{\cal B}}^\star _{p r}\\
\\[-4pt]
\widetilde{{\cal B}}_{p r}&\widetilde{{\cal A}}^{p\star }_{~r}
\EA \!\!
\right] \!\!
\left[ \!\!
\BA{cc}
{\hbox{\boldmath $\widetilde{\widetilde{{\cal A}}}$}}{}^r_{~q}&
{\hbox{\boldmath $\widetilde{\widetilde{{\cal B}}}$}}{}^\star_{r q}\\
\\[-6pt]
{\hbox{\boldmath $\widetilde{\widetilde{{\cal B}}}$}}_{r q}&
{\hbox{\boldmath $\widetilde{\widetilde{{\cal A}}}$}}{}^{r \star }_{~q}
\EA \!\!
\right] \! ,~~
\left[ \!\!
\BA{cc}
\widetilde{{\cal A}}^p_{~q}&\widetilde{{\cal B}}^\star _{p q}\\
\\[-6pt]
\widetilde{{\cal B}}_{p q}&\widetilde{{\cal A}}^{p\star }_{~q}
\EA \!\!
\right] \!
\!\approx\!
\left[ \!\!
\BA{cc}
{\cal A}^p_{~r}&{\cal B}^\star _{p r}\\
\\[-2pt]
{\cal B}_{p r}&{\cal A}^{p\star }_{~r}
\EA \!\!
\right] \!\!
\left[ \!\!
\BA{cc}
{\mathfrak A}^r_{~q}&{\mathfrak B}^\star _{r q}\\
\\[-2pt]
{\mathfrak B}_{r q}&{\mathfrak A}^{r \star }_{~q}
\EA \!\!
\right] \! ,
\label{2ndbosonrotationmat}
\ea\\[-12pt]
in the second equation of which,
the matrix
${\mathfrak G}$
stands for the fluctuation matrix deviated from the statioary matrix
${\cal G}$.
The decomposition of the matrix $\widetilde{\cal G}$
(\ref{2ndbosonrotationmat})
first proposed in Ref.
\cite{FukuNishi.84}
is equivalent to making the Bogoliubov transformation by the stationary amplitudes ${\cal A}$ and ${\cal B}$ that brings state vectors and operators into the quasiparticle frame determined by the stationary solution.
From the decomposition we have approximate relations between
the amplitude in the $\widetilde{\cal G}$ quasiparticle frame
and
the amplitude in the ${\mathfrak G}$ quasiparticle frame as 
$
\widetilde{{\cal A}}^p_{~q}
\!\approx\!
{\cal A}^p_{~r}{\mathfrak A}^r_{~q}
\!\!+\!\!
{\cal B}^\star _{p r}{\mathfrak B}_{r q} 
$
and
$
\widetilde{{\cal B}}_{p q}
\!\approx\!
{\cal B}_{p r}{\mathfrak A}^r_{~q}
\!\!+\!\!
{\cal A}^{p\star }_{~r}{\mathfrak B}_{r q}
$.

The bosonized quasiparticle
$SO(2N \!+\! 1)$ Lie operators in the
$\widetilde{\cal G}$ quasiparticle frame
with indices $i$ and $j$
$(i,j \!=\! 1, \ldots, N)$
are also constructed from operators similar to the operators in
(\ref{bosonrotation})
along the same way as
(\ref{bosonimagequasiLieop})
in the following forms:\\[-20pt]
\ba
\left.
\BA{ll}
&
\widetilde{\hbox{\mathbf E}}{}^{\!~i}_{~j}
\!=\!
{\displaystyle \frac{1}{2}}
[\widetilde{\hbox{\mathbf d}}^\dagger_i,\widetilde{\hbox{\mathbf d}}_j]
\!=\!
{\hbox{\boldmath $\widetilde{\widetilde{{\cal B}}}$}}{}^\dagger_{i \widetilde{r}}
{\hbox{\boldmath $\widetilde{\widetilde{{\cal B}}}$}}_{j \widetilde{r}}
\!-\!
{\hbox{\boldmath $\widetilde{\widetilde{{\cal A}}}$}}{}^{j~\dagger }_{~\widetilde{r}}
{\hbox{\boldmath $\widetilde{\widetilde{{\cal A}}}$}}{}^i_{~\widetilde{r}},~\\
\\[-12pt]
&
\widetilde{\hbox{\mathbf E}}_{i j}
\!=\!
{\displaystyle \frac{1}{2}}
[\widetilde{\hbox{\mathbf d}}_i,\widetilde{\hbox{\mathbf d}}_j]
\!=\!
{\hbox{\boldmath $\widetilde{\widetilde{{\cal A}}}$}}{}^{i~\dagger }_{~\widetilde{r}}
{\hbox{\boldmath $\widetilde{\widetilde{{\cal B}}}$}}_{j \widetilde{r}}
\!-\!
{\hbox{\boldmath $\widetilde{\widetilde{{\cal A}}}$}}{}^{j~\dagger }_{~\widetilde{r}}
{\hbox{\boldmath $\widetilde{\widetilde{{\cal B}}}$}}_{i \widetilde{r}},~
\widetilde{\hbox{\mathbf E}}^{i j}
\!=\!
-\widetilde{\hbox{\mathbf E}}_{i j}^\dagger
\!=\!
\widetilde{\hbox{\mathbf E}}_{i j}^\star, \\
\\[-12pt]
&
\widetilde{\hbox{\mathbf d}}_i
\!=\!
\sqrt{2}
\left( \!
{\hbox{\boldmath $\widetilde{\widetilde{{\cal A}}}$}}{}^{i\dagger }_{~~\widetilde{r}}
{\hbox{\boldmath $\widetilde{\widetilde{{\cal Y}}}$}}_{\widetilde{r}}
\!+\!
{\hbox{\boldmath $\widetilde{\widetilde{{\cal Y}}}$}}{}^\dagger_{\widetilde{r}}
{\hbox{\boldmath $\widetilde{\widetilde{{\cal B}}}$}}_{i \widetilde{r}} \!
\right) ,~
{\hbox{\boldmath $\widetilde{\widetilde{{\cal Y}}}$}}_{\widetilde{r}}
\!\equiv\!
{\displaystyle \frac{1}{\sqrt 2}}
\left( \!
{\hbox{\boldmath $\widetilde{\widetilde{{\cal A}}}$}}{}^{0}_{~\widetilde{r}}
\!-\!
{\hbox{\boldmath $\widetilde{\widetilde{{\cal B}}}$}}{}^\dagger _{0 \widetilde{r}} \!
\right) ,~
\widetilde{\hbox{\mathbf d}}^\dagger_i
\!=\!
-\widetilde{\hbox{\mathbf d}}^\star_i .
\EA
\right\}
\label{2ndbosonimagequasiLieop}
\ea\\[-12pt]
Using
(\ref{2ndbosonrotationmat}),
the fermion $SO(2N \!+\! 1)$ Lie operators describing
the quantum mechanical fluctuation are expressed
in terms of the quasiparticle expectation values ($c$-number) of them in the $\widetilde{\cal G}$ quasiparticle frame and
the new quasiparticle $SO(2N \!+\! 1)$ Lie operators as follows:\\[-18pt]
\ba
\!\!\!\!
\left.
\BA{ll}
\widetilde{\hbox{\mathbf E}}{}^{\!~\alpha }_{~\beta }
&\!\!\!\!
=
< \! \widetilde{E}^{\alpha }_{~\beta } \! >_{\widetilde G}
+\!
\widetilde{\widetilde{\hbox{\mathbf E}}}{}^{\!~\alpha }_{~\beta }
\!=
< \! \widetilde{E}^{\alpha }_{~\beta } \! >_{\widetilde G}
\!\!+\!\!
\left( \!
\widetilde{{\mathbf A}}^{\alpha }_{~i}
\widetilde{{\mathbf A}}^{\beta \star }_{~j}
\!\!-\!\!
\widetilde{{\mathbf B}}_{\beta i}
\widetilde{{\mathbf B}}^\star_{\alpha j} \!
\right) \!\!
\left( \!\!
\widetilde{\hbox{{\mathbf E}}}{}^{~\!i}_{~j}
\!\!+\!\!
{\displaystyle \frac{1}{2} \delta_{ij}} \!\!
\right)
\!\!+\!\!
\widetilde{{\mathbf B}}^\star_{\alpha i}
\widetilde{{\mathbf A}}^{\beta \star }_{~j} \!
\widetilde{\hbox{{\mathbf E}}}_{~\!\!ij}
\!\!+\!\!
\widetilde{{\mathbf A}}^{\alpha }_{~i}
\widetilde{{\mathbf B}}_{\beta j} \!
\widetilde{\hbox{{\mathbf E}}}{}^{~\!ij} \\
\\[-16pt]
&~~~~~~~~~~~~~
\!+\!
{\displaystyle \frac{1}{2}} \!
\left( \!
\widetilde{{\mathbf B}}^\star_{\alpha i} \widetilde{x}_\beta
\!+\!
\widetilde{{\mathbf A}}^{\beta \star }_{~i}
\widetilde{x}^\star_\alpha \!
\right) \!
\widetilde{\hbox{{\mathbf d}}}_{~\!\!i}
\!\!+\!\!
{\displaystyle \frac{1}{2}} \!
\left( \!
\widetilde{{\mathbf A}}^{\alpha }_{~i} \widetilde{x}_\beta
\!+\!
\widetilde{{\mathbf B}}_{\beta i} \widetilde{x}^\star_\alpha \!
\right) \!
\widetilde{\hbox{{\mathbf d}}}^\dagger_{~\!\!i} , \\
\\[-14pt]
\widetilde{\hbox{\mathbf E}}_{\alpha \beta }
&\!\!\!\!
=
< \! \widetilde{E}_{\alpha \beta } \! >_{\widetilde G}
\!+\!
\widetilde{\widetilde{\hbox{\mathbf E}}}_{\alpha \beta } 
\!=
< \! \widetilde{E}_{\alpha \beta } \! >_{\widetilde G}
\!\!+\!\!
\left( \!
\widetilde{{\mathbf B}}_{\alpha i}
\widetilde{{\mathbf A}}^{\beta \star }_{~j}
\!\!-\!\!
\widetilde{{\mathbf B}}_{\beta i}
\widetilde{{\mathbf A}}^{\alpha \star }_{~j} \!
\right) \!\!
\left( \!\!
\widetilde{\hbox{{\mathbf E}}}{}^{~\!i}_{~j}
\!\!+\!\!
{\displaystyle \frac{1}{2} \delta_{ij}} \!
\right)
\!\!+\!\!
\widetilde{{\mathbf A}}^{\alpha \star }_{~i} \!
\widetilde{{\mathbf A}}^{\beta \star }_{~j} \!
\widetilde{\hbox{{\mathbf E}}}_{~\!\!ij}
\!\!+\!\!
\widetilde{{\mathbf B}}_{\alpha i} \!
\widetilde{{\mathbf B}}_{\beta j} \!
\widetilde{\hbox{{\mathbf E}}}{}^{~\!ij} \\
\\[-16pt]
&~~~~~~~~~~~~~
\!+\!
{\displaystyle \frac{1}{2}} \!
\left( \!
\widetilde{{\mathbf A}}^{\alpha \star }_{~i} \widetilde{x}_\beta
\!-\!
\widetilde{{\mathbf A}}^{\beta \star }_{~i} \widetilde{x}_\alpha \!
\right) \!
\widetilde{\hbox{{\mathbf d}}}_{~\!\!i}
\!\!+\!\!
{\displaystyle \frac{1}{2}} \!
\left( \!
\widetilde{{\mathbf B}}_{\alpha i} \widetilde{x}_\beta
\!-\!
\widetilde{{\mathbf B}}_{\beta i} \widetilde{x}_\alpha \!
\right) \!
\widetilde{\hbox{{\mathbf d}}}^\dagger_{~\!\!i} , \\
\\[-14pt]
\widetilde{\hbox{\mathbf c}}_\alpha
&\!\!\!\!\!\!\!\!=
< \! \widetilde{c}_\alpha \! >_{\widetilde G}
\!+\!
\widetilde{\widetilde{\hbox{\mathbf c}}}_\alpha
\!
=
< \! \widetilde{c}_\alpha \! >_{\widetilde G}
\!-\!
\left( \!
\widetilde{{\mathbf B}}_{\alpha i} \widetilde{y}^\star_{j} \!
\!\!+\!\!
\widetilde{{\mathbf A}}^{\alpha \star }_{~j} \widetilde{y}_{i} \!
\right) \!\!
\left( \!\!
\widetilde{\hbox{{\mathbf E}}}{}^{~\!i}_{~j}
\!\!+\!\!
{\displaystyle \frac{1}{2} \delta_{ij}} \!\!
\right)
\!\!-\!\!
\widetilde{{\mathbf A}}^{\alpha \star }_{~i} \widetilde{y}^\star_{j} \!
\widetilde{\hbox{{\mathbf E}}}_{~\!\!ij}
\!\!+\!\!
\widetilde{{\mathbf B}}_{\alpha i} \widetilde{y}_{j} \!
\widetilde{\hbox{{\mathbf E}}}{}^{~\!ij} \\
\\[-16pt]
&~~~~~~~~~~~~~
\!+\!
\left( \!\!
\widetilde{z} \widetilde{{\mathbf A}}^{\alpha \star }_{~i}
\!\!+\!\!
{\displaystyle \frac{1}{2}} \widetilde{x}_\alpha
\widetilde{y}^\star_{i} \!
\right) \!
\widetilde{\hbox{{\mathbf d}}}_{~\!\!i}
\!\!+\!\!
\left( \!
\widetilde{z} \widetilde{{\mathbf B}}_{\alpha i}
\!\!-\!\!
{\displaystyle \frac{1}{2}} \widetilde{x}_\alpha
\widetilde{y}_{i} \!
\right) \!
\widetilde{\hbox{{\mathbf d}}}^\dagger_{~\!\!i} ,
\EA \!\!\!\!
\right\}
\label{2ndbosonrotation}
\ea\\[-12pt]
where the quasiparticle expectation values in the $\widetilde{\cal G}$
quasiparticle frame are given as
\\[-18pt]
\ba
\!\!\!\!
\left.
\BA{ll}
&
< \! \widetilde{E}^{\alpha }_{~\beta }
\!+\!
{\displaystyle \frac{1}{2}}\delta_{\alpha \beta } \! >_{\widetilde G}
~\!\!\!=
\widetilde{R}^{\alpha }_{~\beta }
\!=\!
{\displaystyle \frac{1}{2}} \!
\left( \!
\widetilde{B}^\star_{\alpha i^\prime } \widetilde{B}_{\beta i^\prime }
\!-\!
\widetilde{A}^{\alpha } _{~i^\prime } \widetilde{A}^{\beta \star }_{~i^\prime } \!
\right)
\!+\!
{\displaystyle \frac{1}{2}}\delta_{\alpha \beta } ,\\
\\[-12pt]
&
< \! \widetilde{E}_{\alpha \beta } \! >_{\widetilde G}
~\!\!\!=
-
\widetilde{K}_{\alpha \beta }
\!=\!
{\displaystyle \frac{1}{2}} \!
\left( \!
\widetilde{A}^{\alpha \star }_{~i^\prime }
\widetilde{B}_{\beta i^\prime }
\!-\!
\widetilde{B}_{\alpha i^\prime }
\widetilde{A}^{\beta \star }_{~i^\prime } \!
\right),~
< \! \widetilde{E}^{\alpha \beta } \! >_{\widetilde G}
~\!\!\!=
-
< \! \widetilde{E}_{\alpha \beta } \! >^\star_{\widetilde G} ,\\
\\[-14pt]
&
< \! \widetilde{c}_{\alpha } \! >_{\widetilde G}
~\!\!\!=
{\displaystyle \frac{1}{2}} \!
\left( \!
\widetilde{A}^{\alpha \star }_{~i^\prime } \widetilde{y}_{i^\prime }
\!+\!
\widetilde{B}_{\alpha i^\prime } \widetilde{y}^{\star }_{i^\prime } \!
\right),~
< \! \widetilde{c}^\dagger_{\alpha } \! >_{\widetilde G}
~\!\!\!=
< \! \widetilde{c}_{\alpha } \! >^\star_{\widetilde G} .
\EA \!\!
\right\}
\label{expectGtilde}
\ea\\[-14pt]
whose forms are the same as those of
(\ref{expectG}).
Substituting
(\ref{2ndbosonrotation}) and (\ref{expectGtilde})
into
(\ref{Hamiltonianimage2}),
the fluctuating Hamiltonian
$\widetilde{\hbox{\mathbf H}}$
with the Lagrange multipliers
is converted into\\[-16pt]
\ba
\!\!\!\!\!\!\!\!
\left.
\BA{ll}
&
\widetilde{\hbox{\mathbf H}}
\!=\!
\widetilde{E_0}
\!+\!
\widetilde{\widetilde{\hbox{\mathbf H}}} \\
\\[-12pt]
&
\!\!+\!\!
{\displaystyle \frac{1}{4}} [\alpha \beta|\gamma \delta] \!\!
\left( \!\!
\{ \!
\widetilde{\widetilde{\hbox{\mathbf E}}}{}^{\alpha }_{~\beta },
\widetilde{\widetilde{\hbox{\mathbf E}}}{}^{\gamma }_{~\delta } \!
\}
\!\!+\!\!
{\displaystyle \frac{1}{2}} \!
\{ \!
\widetilde{\widetilde{\hbox{\mathbf E}}}{}^{\alpha \gamma },
\widetilde{\widetilde{\hbox{\mathbf E}}}_{\delta \beta } \!
\} \!\!
\right)
\!\!\!+\!\!\!
{\displaystyle \frac{1}{4}} k_{\alpha \beta }
\{ \!
\widetilde{\widetilde{\hbox{\mathbf c}}}{}^\dagger_\alpha,
\widetilde{\widetilde{\hbox{\mathbf c}}}_\beta \!
\}
\!\!+\!\!\!
{\displaystyle \frac{1}{4}} k^\star_{\alpha \beta }
\{ \!
\widetilde{\widetilde{\hbox{\mathbf c}}}{}^\dagger_\beta,
\widetilde{\widetilde{\hbox{\mathbf c}}}_\alpha \!
\}
\!\!+\!\!\!
{\displaystyle \frac{1}{4}}
l_{\alpha \beta }
\{ \!
\widetilde{\widetilde{\hbox{\mathbf c}}}{}^\dagger_\alpha,
\widetilde{\widetilde{\hbox{\mathbf c}}}{}^\dagger_\beta \!
\}
\!\!+\!\!\!
{\displaystyle \frac{1}{4}}
l^\star_{\alpha \beta }
\{ \!
\widetilde{\widetilde{\hbox{\mathbf c}}}_\alpha,
\widetilde{\widetilde{\hbox{\mathbf c}}}_\beta \!
\} , \\
\\[-12pt]
&
\widetilde{E_0}
\!\equiv\!\!
{\displaystyle \frac{1}{2}} \!
F_{\alpha \beta } \!\!
< \!\! \widetilde{E}^{\alpha }_{~\beta } \!\! >_{\widetilde G}
\!\!+\!
{\displaystyle \frac{1}{2}} \!
F^\star_{\alpha \beta } \!\!
< \!\! \widetilde{E}^{\alpha \dagger }_{~\beta } \!\! >_{\widetilde G}
\!\!+\!
{\displaystyle \frac{1}{2}} \!
D_{\alpha \beta } \!\!
< \!\! \widetilde{E}^{\alpha \beta } \!\! >_{\widetilde G}
\!\!-\!
{\displaystyle \frac{1}{2}} \!
D^\star_{\alpha \beta } \!\!
< \!\! \widetilde{E}_{\alpha \beta } \!\! >_{\widetilde G}
\!\!+\!
M_\alpha \!\!
< \!\! \widetilde{c}^\dagger_\alpha \!\! >_{\widetilde G}
\!\!+\!
M^\star_\alpha \!\!
< \!\! \widetilde{c}_\alpha \!\! >_{\widetilde G} \\
\\[-12pt]
&
\!+\!
{\displaystyle \frac{1}{4}} [\alpha \beta|\gamma \delta] \!
\left( \!
2 \!
< \!\! \widetilde{E}^{\alpha }_{~\beta } \!\! >_{\widetilde G}
< \!\! \widetilde{E}^{\gamma }_{~\delta } \!\! >_{\widetilde G}
\!+\!
< \!\! \widetilde{E}^{\alpha \gamma } \!\! >_{\widetilde G}
< \!\! \widetilde{E}_{\delta \beta } \!\! >_{\widetilde G}
\right) \\
\\[-12pt]
&
+
k_{\alpha \beta } \!
< \!\! \widetilde{c}^\dagger_\alpha \!\! >_{\widetilde G}
< \!\! \widetilde{c}_\beta \!\! >_{\widetilde G}
+
{\displaystyle \frac{1}{2}}
l_{\alpha \beta } \!
< \!\! \widetilde{c}^\dagger_\alpha \!\! >_{\widetilde G}
< \!\! \widetilde{c}^\dagger_\beta \!\! >_{\widetilde G}
+
{\displaystyle \frac{1}{2}}
l^\star_{\alpha \beta } \!
< \!\! \widetilde{c}_\alpha \!\! >_{\widetilde G}
< \!\! \widetilde{c}_\beta \!\! >_{\widetilde G} , \\
\\[-12pt]
&
\widetilde{\widetilde{\hbox{\mathbf H}}}
\equiv\!\!
{\displaystyle \frac{1}{2}}
F_{\alpha \beta }
\widetilde{\widetilde{\hbox{\mathbf E}}}{}^{\alpha }_{~\beta }
\!+\!
{\displaystyle \frac{1}{2}}
F^\star_{\alpha \beta }
\widetilde{\widetilde{\hbox{\mathbf E}}}{}^{\alpha \dagger }_{~\beta }
\!+\!
{\displaystyle \frac{1}{2}}
D_{\alpha \beta }
\widetilde{\widetilde{\hbox{\mathbf E}}}{}^{\alpha \beta }
\!-\!
{\displaystyle \frac{1}{2}}
D^\star_{\alpha \beta }
\widetilde{\widetilde{\hbox{\mathbf E}}}_{\alpha \beta }
\!+\!
M_\alpha
\widetilde{\widetilde{\hbox{\mathbf c}}}{}^\dagger_\alpha
\!+\!
M^\star_\alpha
\widetilde{\widetilde{\hbox{\mathbf c}}}_\alpha \\
\\[-12pt]
&~~~~~
+\!\!
{\displaystyle \frac{1}{4}} [\alpha \beta|\gamma \delta] \!
\left( \!\!
< \!\! \widetilde{E}^{\gamma }_{~\delta } \!\! >_{\widetilde G} \!
\widetilde{\widetilde{\hbox{\mathbf E}}}{}^{\alpha }_{~\beta }
\!+\!
< \!\! \widetilde{E}^{\alpha }_{~\beta } \!\! >_{\widetilde G} \!
\widetilde{\widetilde{\hbox{\mathbf E}}}{}^{\gamma }_{~\delta }
\!\!+\!\!
{\displaystyle \frac{1}{2}} \!\!
< \!\! \widetilde{E}_{\delta \beta } \!\! >_{\widetilde G} \!
\widetilde{\widetilde{\hbox{\mathbf E}}}{}^{\alpha \gamma }
\!\!\!+\!\!
{\displaystyle \frac{1}{2}} \!\!
< \!\! \widetilde{E}^{\alpha \gamma } \!\! >_{\widetilde G} \!
\widetilde{\widetilde{\hbox{\mathbf E}}}_{\delta \beta } \!\!
\right) \\
\\[-14pt]
&
~~~~~~
\!\!+\!\!
{\displaystyle \frac{1}{4}} k_{\alpha \beta } \!
\left( \!
< \!\! \widetilde{c}_\beta \!\! >_{\widetilde G} \!
\widetilde{\widetilde{\hbox{\mathbf c}}}{}^\dagger_\alpha
\!+\!
< \!\! \widetilde{c}^\dagger_\alpha \!\! >_{\widetilde G} \!
\widetilde{\widetilde{\hbox{\mathbf c}}}_\beta
\right)
\!+\!
\mbox{h.c.}
\!+\!
{\displaystyle \frac{1}{4}}~
l_{\alpha \beta } \!
\left( \!
< \!\! \widetilde{c}^\dagger_\beta \!\! >_{\widetilde G} \!
\widetilde{\widetilde{\hbox{\mathbf c}}}{}^\dagger_\alpha
+\!
< \!\! \widetilde{c}^\dagger_\alpha \!\! >_{\widetilde G} \!
\widetilde{\widetilde{\hbox{\mathbf c}}}{}^\dagger_\beta \!
\right)
\!+\!
\mbox{h.c.} ,
\EA \!\!\!\!
\right\}
\label{Hamiltonianimage2p}
\ea\\[-12pt]
where h.c.
denotes the Hermtian conjugate.
A part of the fluctuating Hamiltonian
$\widetilde{\widetilde{\hbox{\mathbf H}}}$
is given
up to the first order in the boson images
$\widetilde{\widetilde{\hbox{\mathbf E}}}$
and
$\widetilde{\widetilde{\hbox{\mathbf c}}}$.

\newpage

$\!\!\!\!\!\!\!\!\!\!\!$
The part of the fluctuating Hamiltonian
$\widetilde{\widetilde{\hbox{\mathbf H}}}$
is devided into two parts as
$\widetilde{\widetilde{\hbox{\mathbf H}}}
\!=\! 
\widetilde{\widetilde{\hbox{\mathbf H}}}_{\widetilde{\hbox{\mathbf d}}}
\!+\!
\widetilde{\widetilde{\hbox{\mathbf H}}}_{\widetilde{\hbox{\mathbf E}}}
$.
Above all the
$\widetilde{\widetilde{\hbox{\mathbf H}}}_{\widetilde{\hbox{\mathbf d}}}$
is important.
$\widetilde{\widetilde{\hbox{\mathbf H}}}_{\widetilde{\hbox{\mathbf d}}}$
is given as\\[-14pt]
\ba
\!\!\!\!\!\!
\BA{ll}
&
\widetilde{\widetilde{\hbox{\mathbf H}}}_{\widetilde{\hbox{\mathbf d}}} \\
\\[-12pt]
&
\!=\!
{\displaystyle \frac{1}{4}} \!\!
\left[ \!
F_{\alpha \beta } \!\!
\left( \!\!
\widetilde{{\mathbf B}}^{\star }_{\alpha i} \widetilde{x}_\beta
\!\!+\!\!
\widetilde{{\mathbf A}}^{\beta \star }_{~i}
\widetilde{x}^\star_\alpha \!\!
\right)
\!\!+\!\!\!
F^\star_{\alpha \beta } \!\!
\left( \!\!
\widetilde{{\mathbf A}}^{\alpha \star }_{~i}
\widetilde{x}^\star_\beta \!
\!\!+\!\!
\widetilde{{\mathbf B}}^\star_{\beta i} \widetilde{x}_\alpha \!\!
\right)
\!\!-\!\!
D_{\alpha \beta } \!\!
\left( \!\!
\widetilde{{\mathbf B}}^{\star }_{\alpha i} \widetilde{x}^\star_\beta
\!\!-\!\!
\widetilde{{\mathbf B}}^{\star }_{\beta i} \widetilde{x}^\star_\alpha \!\!
\right)
\!\!-\!\!
D^\star_{\alpha \beta } \!\!
\left( \!\!
\widetilde{{\mathbf A}}^{\alpha \star }_{~i} \widetilde{x}_\beta
\!\!-\!\!
\widetilde{{\mathbf A}}^{\beta \star }_{~i} \widetilde{x}_\alpha \!\!
\right) \!
\right] \!\!
\widetilde{\hbox{{\mathbf d}}}_{~\!\!i} \!
\!+\!
\mbox{h.c.} \\
\\[-12pt]
&
\!+
\left[ \!
M_\alpha \!\!
\left( \!
\widetilde{z} \widetilde{{\mathbf B}}^{\star }_{\alpha i}
\!\!-\!\!
{\displaystyle \frac{1}{2}} \widetilde{x}^\star_\alpha
\widetilde{y}^\star_{i} \!
\right)
\!+\!
M^\star_\alpha \!
\left( \!
\widetilde{z} \widetilde{{\mathbf A}}^{\alpha \star }_{~i}
\!\!+\!\!
{\displaystyle \frac{1}{2}} \widetilde{x}_\alpha
\widetilde{y}^\star_{i} \!
\right) \!
\right] \!
\widetilde{\hbox{{\mathbf d}}}_{~\!\!i} \!
\!+\!
\mbox{h.c.} \\
\\[-12pt]
&
\!+
{\displaystyle \frac{1}{4}} [\alpha \beta|\gamma \delta] \!
\left[
< \!\! \widetilde{E}^{\alpha }_{~\beta } \!\! >_{\widetilde G} \!
\left( \!
\widetilde{{\mathbf B}}^\star_{\gamma i} \widetilde{x}_\delta
\!\!+\!\!
\widetilde{{\mathbf A}}^{\delta \star }_{~i}
\widetilde{x}^\star_\gamma \!
\right)
\right] \!
\widetilde{\hbox{{\mathbf d}}}_{~\!\!i} \!
\!+\!
\mbox{h.c.} \\
\\[-12pt]
&
\!-
{\displaystyle \frac{1}{16}} [\alpha \beta|\gamma \delta] \!
\left[
< \!\! \widetilde{E}_{\delta \beta } \!\! >_{\widetilde G} \!
\left( \!
\widetilde{{\mathbf B}}^{\star }_{\alpha i} \widetilde{x}^\star_\gamma
\!\!-\!\!
\widetilde{{\mathbf B}}^{\star }_{\gamma i} \widetilde{x}^\star_\alpha \!
\right)
\!-\!
< \!\! \widetilde{E}^{\alpha \gamma } \!\! >_{\widetilde G} \!
\left( \!
\widetilde{{\mathbf A}}^{\delta \star }_{~i} \widetilde{x}_\beta
\!\!-\!\!
\widetilde{{\mathbf A}}^{\beta \star }_{~i} \widetilde{x}_\delta \!
\right)
\right] \!
\widetilde{\hbox{{\mathbf d}}}_{~\!\!i} \!
\!-\!
\mbox{h.c.} \\
\\[-10pt]
&
\!+\!
{\displaystyle \frac{1}{4}} \! k_{\alpha \beta } \!\!
\left\{ \!
\left[ \!
< \!\! \widetilde{c}_\beta \!\! >_{\widetilde G} \!\!
\left( \!\!
\widetilde{z} \widetilde{{\mathbf B}}^{\star }_{\alpha i}
\!\!-\!\!
{\displaystyle \frac{1}{2}} \! \widetilde{x}^\star_\alpha
\widetilde{y}^\star_{i} \!\!
\right) \!
\!\!+\!\!\!
< \!\! \widetilde{c}^\dagger_\alpha \!\! >_{\widetilde G} \!\!
\left( \!\!
\widetilde{z} \widetilde{{\mathbf A}}^{\beta \star }_{~i}
\!\!+\!\!
{\displaystyle \frac{1}{2}} \! \widetilde{x}_\beta
\widetilde{y}^\star_{i} \!\!
\right) \!
\right] \!\!
\widetilde{\hbox{{\mathbf d}}}_{~\!\!i} 
\right. \\
\\[-12pt]
&
\left.
~~~~~~~~~~~~~~~~~~~~~~~~~~~~~~~~~~~~~~~~~~~~
\!\!+\!\!
\left[ \!
< \!\! \widetilde{c}_\beta \!\! >_{\widetilde G} \!\!
\left( \!\!
\widetilde{z} \widetilde{{\mathbf A}}^{\alpha }_{~i}
\!\!+\!\!
{\displaystyle \frac{1}{2}} \! \widetilde{x}^\star_\alpha
\widetilde{y}_{i} \!\!
\right) \!
\!+\!\!\!
< \!\! \widetilde{c}^\dagger_\alpha \!\! >_{\widetilde G} \!\!
\left( \!\!
\widetilde{z} \widetilde{{\mathbf B}}_{\beta i}
\!\!-\!\!
{\displaystyle \frac{1}{2}} \! \widetilde{x}_\beta
\widetilde{y}_{i} \!\!
\right) \!
\right] \!\!
\widetilde{\hbox{{\mathbf d}}}^\dagger_{~\!\!i} \!
\right\} \!
\!+\!
\mbox{h.c.} \\
\\[-12pt]
&
\!+
{\displaystyle \frac{1}{4}} \! l_{\alpha \beta } \!\!
\left\{ \!
\left[ \!
< \!\! \widetilde{c}^\dagger_\beta \!\! >_{\widetilde G} \!\!
\left( \!\!
\widetilde{z} \widetilde{{\mathbf B}}^{\star }_{\alpha i}
\!\!-\!\!
{\displaystyle \frac{1}{2}} \! \widetilde{x}^\star_\alpha
\widetilde{y}^\star_{i} \!\!
\right) \!
\!\!+\!\!
< \!\! \widetilde{c}^\dagger_\alpha \!\! >_{\widetilde G} \!\!
\left( \!\!
\widetilde{z} \widetilde{{\mathbf B}}^{\star }_{\beta i}
\!\!-\!\!
{\displaystyle \frac{1}{2}} \! \widetilde{x}^\star_\beta
\widetilde{y}^\star_{i} \!\!
\right) \!
\right] \!\!
\widetilde{\hbox{{\mathbf d}}}_{~\!\!i}
\right. \\
\\[-12pt]
&
\left.
~~~~~~~~~~~~~~~~~~~~~~~~~~~~~~~~~~~~~~~~~~~~
\!\!+\!\!
\left[ \!
< \!\! \widetilde{c}^\dagger_\beta \!\! >_{\widetilde G} \!\!
\left( \!\!
\widetilde{z} \widetilde{{\mathbf A}}^{\alpha }_{~i}
\!\!+\!\!
{\displaystyle \frac{1}{2}} \! \widetilde{x}^\star_\alpha
\widetilde{y}_{i} \!\!
\right) \!
\!+\!
< \!\! \widetilde{c}^\dagger_\alpha \!\! >_{\widetilde G} \!\!
\left( \!\!
\widetilde{z} \widetilde{{\mathbf A}}^{\beta }_{~i}
\!\!+\!\!\!
{\displaystyle \frac{1}{2}} \! \widetilde{x}^\star_\beta
\widetilde{y}_{i} \!\!
\right) \!
\right] \!\!
\widetilde{\hbox{{\mathbf d}}}^\dagger_{~\!\!i} \!
\right\} \!
\!+\!
\mbox{h.c.} ,
\EA
\label{Hamiltonianimage4p}
\ea\\[-10pt]
which is exactly the so-called dangerous term
and should vanish.
As for the annihilation opertor
$\widetilde{\hbox{{\mathbf d}}}_{~\!\!i^\prime }$
of the quasiparticle state $i^\prime $,
the right-hand-side of
(\ref{Hamiltonianimage4p})
is calculated as follows:\\[-12pt]
\ba
\!\!\!\!\!\!\!\!
\BA{ll}
&
F_{\alpha \beta } \!
\left( \!
\widetilde{{\mathbf B}}^{\star }_{\alpha i^\prime } \widetilde{x}_\beta
\!\!+\!\!
\widetilde{{\mathbf A}}^{\beta \star }_{~i^\prime }
\widetilde{x}^\star_\alpha \!
\right)
\!\!+\!\!
F^\star_{\alpha \beta } \!
\left( \!
\widetilde{{\mathbf A}}^{\alpha \star }_{~i^\prime }
\widetilde{x}^\star_\beta \!
\!\!+\!\!
\widetilde{{\mathbf B}}^\star_{\beta i^\prime } \widetilde{x}_\alpha \!
\right)
\!\!-\!\!
D_{\alpha \beta } \!
\left( \!
\widetilde{{\mathbf B}}^{\star }_{\alpha i^\prime }
\widetilde{x}^\star_\beta
\!\!-\!\!
\widetilde{{\mathbf B}}^{\star }_{\beta i^\prime }
\widetilde{x}^\star_\alpha \!
\right)
\!\!-\!\!
D^\star_{\alpha \beta } \!
\left( \!
\widetilde{{\mathbf A}}^{\alpha \star }_{~i^\prime }
\widetilde{x}_{\!\beta }
\!\!-\!\!
\widetilde{{\mathbf A}}^{\beta \star }_{~i^\prime }
\widetilde{x}_\alpha \!
\right) \\
\\[-12pt]
&
\!+\!
4
M_\alpha \!
\left( \!
\widetilde{z} \widetilde{{\mathbf B}}^{\star }_{\alpha i^\prime }
\!\!-\!\!
{\displaystyle \frac{1}{2}} \widetilde{x}^\star_\alpha
\widetilde{y}^\star_{i^\prime } \!
\right)
\!\!+\!
4
M^\star_\alpha \!
\left( \!
\widetilde{z} \widetilde{{\mathbf A}}^{\alpha \star }_{~i^\prime }
\!\!+\!\!
{\displaystyle \frac{1}{2}} \widetilde{x}_\alpha
\widetilde{y}^\star_{i^\prime } \!
\right) \\
\\[-12pt]
&
\!+
[\alpha \beta|\gamma \delta] \!
\left[ \!
< \!\! \widetilde{E}^{\alpha }_{~\beta } \!\! >_{\!\widetilde G} \!
\left( \!
\widetilde{{\mathbf B}}^\star_{\gamma i^\prime } \widetilde{x}_\delta
\!\!+\!\!
\widetilde{{\mathbf A}}^{\delta \star }_{~i^\prime }
\widetilde{x}^\star_\gamma \!
\right)
\!-\!\!
{\displaystyle \frac{1}{4}} \!
\left\{ \!
< \!\! \widetilde{E}_{\delta \beta } \!\! >_{\!\widetilde G} \!
\left( \!
\widetilde{{\mathbf B}}^{\star }_{\alpha i^\prime }
\widetilde{x}^\star_\gamma
\!\!-\!\!
\widetilde{{\mathbf B}}^{\star }_{\gamma i^\prime }
\widetilde{x}^\star_\alpha \!
\right)
\!\!-\!\!
< \!\! \widetilde{E}^{\alpha \gamma } \!\! >_{\!\widetilde G} \!
\left( \!
\widetilde{{\mathbf A}}^{\delta \star }_{~i^\prime }
\widetilde{x}_\beta
\!\!-\!\!
\widetilde{{\mathbf A}}^{\beta \star }_{~i^\prime }
\widetilde{x}_{\!\delta } \!
\right) \!
\right\} \!
\right] \\
\\[-12pt]
&
\!+\!
2 k_{\alpha \beta } \!\!
\left[ \!
< \!\! \widetilde{c}_\beta \!\! >_{\widetilde G} \!\!
\left( \!\!
\widetilde{z} \widetilde{{\mathbf B}}^{\star }_{\alpha i^\prime }
\!\!-\!\!
{\displaystyle \frac{1}{2}} \! \widetilde{x}^\star_\alpha
\widetilde{y}^\star_{i^\prime } \!\!
\right) \!
\!\!+\!\!\!
< \!\! \widetilde{c}^\dagger_\alpha \!\! >_{\widetilde G} \!\!
\left( \!\!
\widetilde{z} \widetilde{{\mathbf A}}^{\beta \star }_{~i^\prime }
\!\!+\!\!
{\displaystyle \frac{1}{2}} \! \widetilde{x}_{\!\beta }
\widetilde{y}^\star_{i^\prime } \!\!
\right) \!
\right] \\
\\[-12pt]
&
\!+\!
l_{\!\alpha \beta } \!\!
\left[ \!
< \!\! \widetilde{c}^\dagger_{\!\beta } \!\! >_{\!\widetilde G} \!\!
\left( \!\!
\widetilde{z} \! \widetilde{{\mathbf B}}^{\star }_{\alpha i^\prime }
\!\!-\!\!
{\displaystyle \frac{1}{2}} \! \widetilde{x}^\star_{\!\alpha }
\widetilde{y}^\star_{i^\prime } \!\!
\right) \!
\!\!+\!\!
< \!\! \widetilde{c}^\dagger_\alpha \!\! >_{\!\widetilde G} \!\!
\left( \!\!
\widetilde{z} \! \widetilde{{\mathbf B}}^{\star }_{\beta i^\prime }
\!\!-\!\!
{\displaystyle \frac{1}{2}} \! \widetilde{x}^\star_{\!\beta }
\widetilde{y}^\star_{i^\prime } \!\!
\right) \!\!
\right]
\!\!\!+\!\!
l^\star_{\!\alpha \beta } \!\!
\left[ \!
< \!\! \widetilde{c}_{\!\beta } \!\! >_{\!\widetilde G} \!\!
\left( \!\!
\widetilde{z} \! \widetilde{{\mathbf A}}^{\alpha \star }_{~i^\prime }
\!\!+\!\!\!
{\displaystyle \frac{1}{2}} \! \widetilde{x}_{\!\alpha }
\widetilde{y}^\star_{i^\prime } \!\!
\right) \!
\!\!+\!\!
< \!\! \widetilde{c}_\alpha \!\! >_{\!\widetilde G} \!\!
\left( \!\!
\widetilde{z} \! \widetilde{{\mathbf A}}^{\beta \star }_{~i^\prime }
\!\!+\!\!\!
{\displaystyle \frac{1}{2}} \! \widetilde{x}_{\!\beta }
\widetilde{y}^\star_{i^\prime } \!\!
\right) \!\!
\right] \\
\\[-12pt]
&
\!=\!
2 \!
\left[ \!
F_{\alpha \beta }
\widetilde{{\mathbf A}}^{\beta \star }_{~i^\prime }
\!+\!
D_{\alpha \beta }
\widetilde{{\mathbf B}}^{\star }_{\beta i^\prime }
\!+\!
\sqrt{2}
M_\alpha \!
\left( \!
-{\displaystyle \frac{\widetilde{y}^\star_{i^\prime }}{\sqrt{2}}} \!
\right) \!
\right] \!
\widetilde{x}^\star_\alpha \!
\!+\!
2 \!
\left[ \!
F^\star_{\alpha \beta }
\widetilde{{\mathbf B}}^{\star }_{\beta i^\prime }
\!+\!
D^\star_{\alpha \beta }
\widetilde{{\mathbf A}}^{\beta \star }_{~i^\prime }
\!+\!
\sqrt{2}
M^\star_\alpha \!
\left( \!
{\displaystyle \frac{\widetilde{y}^\star_{i^\prime }}{\sqrt{2}}} \!
\right) \!
\right] \!
\widetilde{x}_\alpha \\
\\[-12pt]
&
\!+
2
\left[
\sqrt{2} M_\alpha
\widetilde{{\mathbf B}}^{\star }_{\alpha i^\prime }
\!+\!
\sqrt{2} M^\star_\alpha
\widetilde{{\mathbf A}}^{\alpha \star }_{~i^\prime }
\right] \!
\sqrt{2} \widetilde{z} \\
\\[-12pt]
&
\!+\!
\left( \!
\widetilde{F}_{\alpha \beta }
\!-\!
h_{\alpha \beta }
\!-\!
{\displaystyle \frac{1}{2}}
[\alpha \beta|\gamma \gamma] \!
\right) \!\!
\left( \!
\widetilde{{\mathbf B}}^\star_{\alpha i^\prime } \widetilde{x}_\beta
\!+\!
\widetilde{{\mathbf A}}^{\beta \star }_{~i^\prime }
\widetilde{x}^\star_\alpha \!
\right)
\!\!-\!\!
{\displaystyle \frac{1}{2}} \!
\left\{ \!
\widetilde{D}_{\alpha \beta } \!
\left( \!
\widetilde{{\mathbf B}}^{\star }_{\alpha i^\prime }
\widetilde{x}^\star_\beta
\!-\!
\widetilde{{\mathbf B}}^{\star }_{\beta i^\prime }
\widetilde{x}^\star_\alpha \!
\right) \!
\!+\!
\widetilde{D}^\star_{\alpha \beta } \!
\left( \!
\widetilde{{\mathbf A}}^{\alpha \star }_{~i^\prime }
\widetilde{x}_\beta
\!-\!
\widetilde{{\mathbf A}}^{\beta \star }_{~i^\prime }
\widetilde{x}_{\!\alpha } \!
\right) \!
\right\} \\
\\[-12pt]
&
\!+\!
2
\widetilde{M}_\alpha \!
\left( \!
\widetilde{z} \widetilde{{\mathbf B}}^{\star }_{\alpha i^\prime }
\!\!-\!\!
{\displaystyle \frac{1}{2}} \widetilde{x}^\star_\alpha
\widetilde{y}^\star_{i^\prime } \!
\right)
\!\!+\!
2
\widetilde{M}^\star_\alpha \!
\left( \!
\widetilde{z} \widetilde{{\mathbf A}}^{\alpha \star }_{~i^\prime }
\!\!+\!\!
{\displaystyle \frac{1}{2}} \widetilde{x}_\alpha
\widetilde{y}^\star_{i^\prime } \!
\right) \\
\\[-12pt]
&
\!=\!
2
E_{i^\prime } \!
\left( \!
\widetilde{{\mathbf A}}^{\alpha \star }_{~i^\prime }
\widetilde{x}^\star_\alpha \!
\!-\!
\widetilde{{\mathbf B}}^{\star }_{\alpha i^\prime }
\widetilde{x}_\alpha
\!-\!
\widetilde{z}
\widetilde{y}^\star_{i^\prime } \!
\right)
\!+\!
\sqrt{2} \widetilde{z}
\left( \!
\sqrt{2}
\widetilde{M}_\alpha
\widetilde{{\mathbf B}}^{\star }_{\alpha i^\prime } \!
\!+\!
\sqrt{2}
\widetilde{M}^\star_\alpha
\widetilde{{\mathbf A}}^{\alpha \star }_{~i^\prime } \!
\right) \\
\\[-12pt]
&
\!+\!
\left[ \!
\widetilde{F}_{\alpha \beta }
\widetilde{{\mathbf A}}^{\beta \star }_{~i^\prime }
\!+\!
\widetilde{D}_{\alpha \beta }
\widetilde{{\mathbf B}}^{\star }_{\beta i^\prime }
\!+\!
\sqrt{2}
\widetilde{M}_\alpha \!
\left( \!
-
{\displaystyle \frac{\widetilde{y}^\star_{i^\prime } }{\sqrt{2}}} \!
\right) \!
\right] \!
\widetilde{x}^\star_\alpha
\!+\!
\left[ \!
\widetilde{F}^\star_{\alpha \beta }
\widetilde{{\mathbf B}}^\star_{\beta i^\prime }
\!+\!
\widetilde{D}^\star_{\alpha \beta }
\widetilde{{\mathbf A}}^{\beta \star }_{~i^\prime }
\!+\!
\sqrt{2}
\widetilde{M}^\star_\alpha \!
\left( \!
{\displaystyle \frac{\widetilde{y}^\star_{i^\prime } }{\sqrt{2}}} \!
\right) \!
\right] \!
\widetilde{x}_\alpha \\
\\[-12pt]
&
\!=\!
\left( \!
2
E_{i^\prime }
\!+\!
\widetilde{E}_{i^\prime } \!
\right) \!
\left( \!
\widetilde{{\mathbf A}}^{\alpha \star }_{~i^\prime }
\widetilde{x}^\star_\alpha \!
\!-\!
\widetilde{{\mathbf B}}^{\star }_{\alpha i^\prime }
\widetilde{x}_\alpha
\!-\!
\widetilde{z}
\widetilde{y}^\star_{i^\prime } \!
\right) .
\EA
\label{Hamiltonianimage4dahgerous}
\ea
The
$\widetilde{\widetilde{\hbox{\mathbf H}}}_{\widetilde{\hbox{\mathbf d}}}$
vanishes if we use the normalization condition
$
\widetilde{{\mathbf A}}^{\alpha \star }_{~i^\prime }
\widetilde{x}^\star_\alpha \!
\!-\!
\widetilde{{\mathbf B}}^{\star }_{\alpha i^\prime }
\widetilde{x}_\alpha
\!-\!
\widetilde{z}
\widetilde{y}^\star_{i^\prime }
\!=\!
0
$
which is derived from
(\ref{matG}).
We define the tilded SCF parameters in the $\widetilde{G}$ quasiparticle frame as\\[-18pt]
\ba
\BA{c}
\widetilde{F}_{\alpha\beta }
\!\equiv\!
h_{\alpha\beta }
\!+\!
[\alpha\beta|\gamma\delta]\widetilde{R}_{\gamma\delta } , ~
\widetilde{D}_{\alpha\beta }
\!\equiv\!
{\displaystyle \frac{1}{2}}[\alpha\gamma|\beta\delta] 
(-\widetilde{K}_{\delta \gamma }) , ~
\widetilde{M}_{\alpha }
\!\equiv\!
k_{\alpha\beta } \! < \! \widetilde{c}_{\beta } \! >_{\widetilde G}
+
l_{\alpha\beta } \!
< \! \widetilde{c}{}^{\!~\dagger }_{\beta } \! >_{\widetilde G} .
\EA
\label{SCFFandDandMtilde}
\ea\\[-16pt]
In
(\ref{Hamiltonianimage4dahgerous}) 
we also have used another eigenvalue
$\widetilde{E}_{i^\prime }$
related to a quantum fluctuation
given by the same type of the eigenvalue equation as
(\ref{staticextendedHBeq}).
Namely,
the eigenvalue equation for the fluctuating Hamiltonian
with the eigenvalue
$\widetilde{E}_{i^\prime }$
is given as\\[-16pt]
\ba
\widetilde{{\cal F}}_{\alpha \beta } \!
\left[ \!\!
\BA{c}
\widetilde{A}^{\beta \star }_{~i^\prime }\\
\\[-8pt]
\widetilde{B}^\star_{\beta i^\prime }\\
\\[-8pt]
-{\displaystyle \frac{\widetilde{y}^\star_{i^\prime }}{\sqrt 2}}
\EA \!\!
\right]
\!=\!
\widetilde{E}_{i^\prime }
\left[ \!\!
\BA{c}
\widetilde{A}^{\beta \star }_{~i^\prime }\\
\\[-8pt]
\widetilde{B}^\star_{\beta i^\prime }\\
\\[-8pt]
-{\displaystyle \frac{\widetilde{y}^\star_{i^\prime }}{\sqrt 2}}
\EA \!\!
\right] , ~
\widetilde{{\cal F}}_{\alpha \beta } \!
\left[ \!\!
\BA{c}
{\displaystyle \frac{\widetilde{x}_\beta }{\sqrt 2}}\\
\\[-12pt]
-{\displaystyle \frac{\widetilde{x}^\star _\beta }{\sqrt 2}}\\
\\[-12pt]
\widetilde{z}
\EA \!\!
\right]
\!=\!
0 ,~~
\widetilde{{\cal F}}_{\alpha \beta }
\!\equiv\!
\left[ \!
\BA{ccc}
\widetilde{F}_{\alpha \beta }&\widetilde{D}_{\alpha \beta }&
\sqrt2\widetilde{M}_{\alpha }\\
\\[-8pt]
\!\!\!\!\!\!-
\widetilde{D}^\star_{\alpha \beta }&-\widetilde{F}^\star_{\alpha \beta }&\sqrt2\widetilde{M}^\star_{\alpha }\\
\\[-8pt]
\sqrt2\widetilde{M}^\dagger _{\beta }&
\sqrt2\widetilde{M}^{\mbox{\scriptsize T}} _{\beta }&0
\EA \!\!
\right] .
\label{eqofmotiontilde}
\ea\\[-14pt]
The Hamiltonian
$\widetilde{\widetilde{\hbox{\mathbf H}}}_{\widetilde{\hbox{\mathbf E}}}$
describes the quantum mechanical fluctuation and is given by\\[-20pt]
\ba
\!\!\!\!\!\!
\BA{ll}
&
\widetilde{\widetilde{\hbox{\mathbf H}}}_{\widetilde{\hbox{\mathbf E}}} \\
\\[-14pt]
&
\!=\!\!
\left[ \!
\widetilde{{\mathbf A}}^{\alpha }_{~i} \!\!
\left\{ \!
F_{\alpha \beta } \!
\widetilde{{\mathbf A}}^{\beta \star }_{~j}
\!\!+\!\!
D_{\alpha \beta } \!
\widetilde{{\mathbf B}}^\star_{\beta j} \!
\!+\!\!
\sqrt{2}
M_\alpha \!
\left( \!\!
-{\displaystyle \frac{\widetilde{y}^\star_{j}}{\sqrt{2}}} \!
\right) \!\!
\right\}
\!\!-\!\!
\widetilde{{\mathbf B}}_{\alpha i} \!\!
\left\{ \!
F^\star_{\alpha \beta } \!
\widetilde{{\mathbf B}}^\star_{\beta j}
\!\!+\!\!
D^\star_{\alpha \beta } \!
\widetilde{{\mathbf A}}^{\beta \star }_{~j} \!
\!+\!\!
\sqrt{2}
M^\star_\alpha \!
\left( \!
{\displaystyle \frac{\widetilde{y}^\star_{j}}{\sqrt{2}}} \!
\right) \!\!
\right\} \!
\right] \!\!
\left( \!\!
\widetilde{\hbox{{\mathbf E}}}{}^{~\!i}_{~j}
\!\!+\!\!
{\displaystyle \frac{1}{2} \delta_{ij}} \!\!
\right) \\
\\[-12pt]
&
\!+
{\displaystyle \frac{1}{2}} \!
\left[ \!
\widetilde{{\mathbf B}}^{\star }_{\alpha i} \!
\left\{ \!
F_{\alpha \beta }
\widetilde{{\mathbf A}}^{\beta \star }_{~j}
\!+\!
D_{\alpha \beta }
\widetilde{{\mathbf B}}^\star_{\beta j} \!
\!+\!\!
\sqrt{2}
M_\alpha \!
\left( \!\!
-{\displaystyle \frac{\widetilde{y}^\star_{j}}{\sqrt{2}}} \!
\right) \!\!
\right\}
\!\!+\!\!
\widetilde{{\mathbf A}}^{\alpha \star }_{~j} \!
\left\{ \!
F^\star_{\alpha \beta }
\widetilde{{\mathbf B}}^\star_{\beta i}
\!+\!
D^\star_{\alpha \beta }
\widetilde{{\mathbf A}}^{\beta \star }_{~i} \!
\!+\!\!
\sqrt{2}
M^\star_\alpha \!
\left( \!
{\displaystyle \frac{\widetilde{y}^\star_{i}}{\sqrt{2}}} \!
\right) \!\!
\right\} \!
\right] \!
\widetilde{\hbox{{\mathbf E}}}_{~\!\!ij} \\
\\[-12pt]
&
\!+
{\displaystyle \frac{1}{2}} \!
\left[ \!
\widetilde{{\mathbf B}}_{\alpha j} \!
\left\{ \!
F^\star_{\alpha \beta }
\widetilde{{\mathbf A}}^{\beta }_{~i}
\!+\!
D^\star_{\alpha \beta }
\widetilde{{\mathbf B}}_{\beta i} \!
\!+\!\!
\sqrt{2}
M^\star_\alpha \!
\left( \!\!
-{\displaystyle \frac{\widetilde{y}_{i}}{\sqrt{2}}} \!
\right) \!\!
\right\}
\!+\!
\widetilde{{\mathbf A}}^{\alpha }_{~i}
\left\{ \!
F_{\alpha \beta }
\widetilde{{\mathbf B}}_{\beta j}
\!+\!
D_{\alpha \beta }
\widetilde{{\mathbf A}}^{\beta }_{~j} \!
\!+\!\!
\sqrt{2}
M_\alpha \!
\left( \!
{\displaystyle \frac{\widetilde{y}_{j}}{\sqrt{2}}} \!
\right) \!\!
\right\} \!
\right] \!
\widetilde{\hbox{{\mathbf E}}}{}^{~\!ij} \\
\\[-12pt]
&
\!+
{\displaystyle \frac{1}{2}} \!
\left[ \!
\widetilde{{\mathbf A}}^{\alpha }_{~i} \!
\left\{ \!
\widetilde{F}_{\alpha \beta }
\widetilde{{\mathbf A}}^{\beta \star }_{~j}
\!+\!
\widetilde{D}_{\alpha \beta }
\widetilde{{\mathbf B}}^\star_{\beta j}
\!+\!\!
\sqrt{2}
\widetilde{M}_\alpha \!
\left( \!\!
-{\displaystyle \frac{\widetilde{y}^\star_{j}}{\sqrt{2}}} \!
\right) \!\!
\right\}
\!\!-\!\!
\widetilde{{\mathbf B}}_{\alpha i} \!
\left\{ \!
\widetilde{F}^\star_{\alpha \beta }
\widetilde{{\mathbf B}}^\star_{\beta j}
\!+\!
\widetilde{D}^\star_{\alpha \beta }
\widetilde{{\mathbf A}}^{\beta \star }_{~j}
\!+\!\!
\sqrt{2}
\widetilde{M}^\star_\alpha \!
\left( \!
{\displaystyle \frac{\widetilde{y}^\star_{j}}{\sqrt{2}}} \!
\right) \!\!
\right\}
\right. \\
\\[-12pt]
&
\left.
\!-
\widetilde{y}_{i} \!
\left( \!
\widetilde{M}_{\alpha }
\widetilde{{\mathbf B}}^{\star }_{\alpha j}
\!+\!
\widetilde{M}^\star_{\alpha }
\widetilde{{\mathbf A}}^{\alpha \star }_{~j}
\right) \!\!\!\!
^{^{^{^{^{^{.}}}}}}
\right] \!\!
\left( \!
\widetilde{\hbox{{\mathbf E}}}{}^{~\!i}_{~j}
\!\!+\!\!
{\displaystyle \frac{1}{2} \delta_{ij}} \!\!
\right) \\
\\[-12pt]
&
\!+\!
{\displaystyle \frac{1}{4}} \!
\left[ \!
\widetilde{{\mathbf B}}^\star_{\alpha i} \!
\left\{ \!
\widetilde{F}_{\alpha \beta }
\widetilde{{\mathbf A}}^{\beta \star }_{~j}
\!+\!
\widetilde{D}_{\alpha \beta }
\widetilde{{\mathbf B}}^\star_{\beta j} \!
\!+\!
\sqrt{2}
\widetilde{M}_\alpha \!
\left( \!\!
-{\displaystyle \frac{\widetilde{y}^\star_{j}}{\sqrt{2}}} \!
\right) \!\!
\right\}
\!+\!
\widetilde{{\mathbf A}}^{\alpha \star }_{~j} \!
\left\{ \!
\widetilde{F}^\star_{\alpha \beta }
\widetilde{{\mathbf B}}^\star_{\beta i}
\!+\!
\widetilde{D}^\star_{\alpha \beta }
\widetilde{{\mathbf A}}^{\beta \star }_{~i}
\!\!+\!\!
\sqrt{2}
\widetilde{M}^\star_\alpha \!
\left( \!
{\displaystyle \frac{\widetilde{y}^\star_{i}}{\sqrt{2}}} \!
\right) \!\!
\right\}
\right. \\
\\[-12pt]
&
\left.
\!-
\widetilde{y}^\star_{j} \!
\left( \!
\widetilde{M}_{\alpha }
\widetilde{{\mathbf B}}^{\star }_{\alpha i}
\!+\!
\widetilde{M}^\star_{\alpha }
\widetilde{{\mathbf A}}^{\alpha \star }_{~i}
\right) \!\!\!
^{^{^{^{^{^{.}}}}}} \!
\right] \!
\widetilde{\hbox{{\mathbf E}}}_{~\!\!ij} \\
\\[-14pt]
&
\!+\!
{\displaystyle \frac{1}{4}} \!
\left[ \!
\widetilde{{\mathbf B}}_{\alpha j} \!
\left\{ \!
\widetilde{F}^\star _{\alpha \beta }
\widetilde{{\mathbf A}}^{\beta }_{~i}
\!+\!
\widetilde{D}^\star_{\alpha \beta }
\widetilde{{\mathbf B}}_{\beta i}
\!+\!
\sqrt{2}
\widetilde{M}^\star_\alpha \!
\left( \!\!
-{\displaystyle \frac{\widetilde{y}_{i}}{\sqrt{2}}} \!
\right) \!\!
\right\}
\!+\!
\widetilde{{\mathbf A}}^{\alpha }_{~i} \!
\left\{ \!
\widetilde{F}_{\alpha \beta }
\widetilde{{\mathbf B}}_{\beta j}
\!+\!
\widetilde{D}_{\alpha \beta }
\widetilde{{\mathbf A}}^{\beta }_{~j}
\!+\!
\sqrt{2}
\widetilde{M}_\alpha \!
\left( \!
{\displaystyle \frac{\widetilde{y}_{j}}{\sqrt{2}}} \!
\right) \!\!
\right\}
\right. \\
\\[-12pt]
&
\left.
\!+
\widetilde{y}_{j} \!
\left( \!
\widetilde{M}_{\alpha }
\widetilde{{\mathbf A}}^{\alpha }_{~i}
\!+
\widetilde{M}^{\star }_{\alpha }
\widetilde{{\mathbf B}}_{\alpha i}
\right) \!\!\!
^{^{^{^{^{^{.}}}}}} \!
\right] \!
\widetilde{\hbox{{\mathbf E}}}{}^{~\!ij} .
\EA 
\label{Hamiltonianimage2pEterm}
\ea\\[-16pt]
Using eigenvalue equations
(\ref{staticextendedHBeq})
and
(\ref{eqofmotiontilde}) with eigenvalues
$E_i$
and
$\widetilde{E}_{i}$ and
normalization condition
$\widetilde{G}^\dagger \widetilde{G}
\!=\!
\widetilde{G} \widetilde{G}^\dagger
\!\!=\!\!
1_{\!2N\!+\!1}$,
the Hamiltonian
$\widetilde{\widetilde{\hbox{\mathbf H}}}_{\widetilde{\hbox{\mathbf E}}}$
(\ref{Hamiltonianimage2pEterm})
is simply rewritten as follows:\\[-18pt]
\ba
\!\!\!\!\!\!\!\!
\BA{ll}
&
\widetilde{\widetilde{\hbox{\mathbf H}}}_{\widetilde{\hbox{\mathbf E}}}
\!=\!
E_j \!
\left( \!
\widetilde{{\mathbf A}}^{\alpha }_{~i}
\widetilde{{\mathbf A}}^{\alpha \star }_{~j}
\!+\!
\widetilde{{\mathbf B}}_{\alpha i}
\widetilde{{\mathbf B}}^\star_{\alpha j} \!
\right) \!\!
\left( \!\!
\widetilde{\hbox{{\mathbf E}}}{}^{~\!i}_{~j}
\!\!+\!\!
{\displaystyle \frac{1}{2} \delta_{ij}} \!\!
\right)
\!\!+\!\!
{\displaystyle \frac{1}{2}}
\widetilde{E}_j \!
\left( \!\!
\widetilde{{\mathbf A}}^{\alpha }_{~i}
\widetilde{{\mathbf A}}^{\alpha \star }_{~j}
\!+\!
\widetilde{{\mathbf B}}_{\alpha i}
\widetilde{{\mathbf B}}^\star_{\alpha j}
\!+\!
{\displaystyle \frac{1}{2}}
\widetilde{y}_{i}
\widetilde{y}^\star_{j} \!\!
\right) \!\!
\left( \!\!
\widetilde{\hbox{{\mathbf E}}}^{~\!i}_{~j}
\!\!+\!\!
{\displaystyle \frac{1}{2} \delta_{ij}} \!\!
\right) \\
\\[-8pt]
&
\!+
{\displaystyle \frac{1}{2}}
E_j \!
\left( \!
\widetilde{{\mathbf B}}^\star_{\alpha i}
\widetilde{{\mathbf A}}^{\alpha \star }_{~j}
\!+\!
\widetilde{{\mathbf A}}^{\alpha \star }_{~i}
\widetilde{{\mathbf B}}^\star_{\alpha j} \!
\right) \!
\widetilde{\hbox{{\mathbf E}}}_{~\!\!ij}
\!-\!
{\displaystyle \frac{1}{2}}
E_j \!
\left( \!
\widetilde{{\mathbf B}}_{\alpha i}
\widetilde{{\mathbf A}}^{\alpha }_{~j}
\!+\!
\widetilde{{\mathbf A}}^{\alpha }_{~i}
\widetilde{{\mathbf B}}_{\alpha j} \!
\right) \!
\widetilde{\hbox{{\mathbf E}}}{}^{~\!ij} \\
\\[-8pt]
&
\!+
{\displaystyle \frac{1}{4}}
\widetilde{E}_j \!
\left( \!\!
\widetilde{{\mathbf B}}^\star_{\alpha i}
\widetilde{{\mathbf A}}^{\alpha \star }_{~j}
\!+\!
\widetilde{{\mathbf A}}^{\alpha \star }_{~i}
\widetilde{{\mathbf B}}^\star_{\alpha j}
\!-\!
{\displaystyle \frac{1}{2}}
\widetilde{y}^\star_{i}
\widetilde{y}^\star_{j} \!\!
\right) \!
\widetilde{\hbox{{\mathbf E}}}_{~\!\!ij}
\!-\!
{\displaystyle \frac{1}{4}}
\widetilde{E}_j \!
\left( \!\!
\widetilde{{\mathbf B}}_{\alpha i}
\widetilde{{\mathbf A}}^{\alpha }_{~j}
\!+\!
\widetilde{{\mathbf A}}^{\alpha }_{~i}
\widetilde{{\mathbf B}}_{\alpha j}
\!-\!
{\displaystyle \frac{1}{2}}
\widetilde{y}_{i}
\widetilde{y}_{j} \!\!
\right) \!
\widetilde{\hbox{{\mathbf E}}}^{~\!ij} \\
\\[-8pt]
&
\!=\!\!
\sum_i \!\!
\left( \!\!
E_i
\!\!+\!\!
{\displaystyle \frac{1}{2}}
\widetilde{E}_i \!\!
\right) \!\!
\left( \!
\widetilde{\hbox{{\mathbf E}}}{}^{~\!i}_{~i}
\!\!+\!\!
{\displaystyle \frac{1}{2}} \!
\right)
\!\!-\!\!
{\displaystyle \frac{1}{4}}
E_j
\widetilde{y}^\star_{j}
\widetilde{y}_{i} \!\!
\left( \!\!
\widetilde{\hbox{{\mathbf E}}}{}^{~\!i}_{~j}
\!\!+\!\!
{\displaystyle \frac{1}{2} \delta_{ij}} \!\!
\right)
\!\!-\!\!
{\displaystyle \frac{1}{4}} \!
E_i
\widetilde{y}^\star_{i}
\widetilde{y}_{j} \!\!
\left( \!\!
\widetilde{\hbox{{\mathbf E}}}{}^{~\!i \dagger }_{~j}
\!\!+\!\!
{\displaystyle \frac{1}{2} \delta_{ij}} \!\!
\right)
\!\!+\!\!
{\displaystyle \frac{1}{4}} \!
E_j
\widetilde{y}^\star_{j}
\widetilde{y}^\star_{i} \!
\widetilde{\hbox{{\mathbf E}}}_{~\!\!ij}
\!\!-\!\!
{\displaystyle \frac{1}{4}} \!
E_j
\widetilde{y}_{j}
\widetilde{y}_{i} \!
\widetilde{\hbox{{\mathbf E}}}{}^{~\!ij} \! .
\EA
\label{Hamiltonianimage2pEterm2}
\ea\\[-10pt]
Now we consider an excitation energy due to
$
\widetilde{\widetilde{\hbox{\mathbf H}}}_{\widetilde{\hbox{\mathbf E}}}
~
(
=\!
\widetilde{\widetilde{\hbox{\mathbf H}}}
)
$
(\ref{Hamiltonianimage2pEterm2})
in which $\widetilde{y}$ and $\widetilde{y}^\star$
are relevant variables to treat the excitation energy.
This contrasts with the role played by the variables $x$ and $x^\star$ in
(\ref{staticextendedHBeq}).
We attempt to derive the excitation energy
by the Sawada's eigenmode method
\cite{Sawada.57}-\cite{Sawada.60}.
The method, however, does not work well because
the Hamiltonian
$\widetilde{\widetilde{\hbox{\mathbf H}}}_{\widetilde{\hbox{\mathbf E}}}$
contains the unpaired-mode amplitudes.
We return to the traditional Bogoliubov's procedure
\cite{Bogo.59}.
Let us introduce the following annihilation and creation operators
$\widetilde{\widetilde{\hbox{\mathbf d}}}_i$
and
$\widetilde{\widetilde{\hbox{\mathbf d}}}{}^\dagger_i$:\\[-16pt]
\ba
\left[
\widetilde{\widetilde{\hbox{\mathbf d}}}_i,~
\widetilde{\widetilde{\hbox{\mathbf d}}}{}^\dagger_i
\right]
\!=\!
\left[
\widetilde{\hbox{\mathbf d}}_i,~
\widetilde{\hbox{\mathbf d}}{}^\dagger_i
\right] \!
\left[ \!
\begin{array}{cc}
\widetilde{\widetilde{u}}_i & \widetilde{\widetilde{v}}_i {}^{\ast } \\
\\[-12pt]
-\widetilde{\widetilde{v}}_i & \widetilde{\widetilde{u}}_i {}^{\ast }
\end{array} \!
\right]
\!\equiv\!
\left[
\widetilde{\hbox{\mathbf d}}_i,~
\widetilde{\hbox{\mathbf d}}{}^\dagger_i
\right] \!
\widetilde{\widetilde{g}}_i , ~~
\widetilde{\widetilde{g}}^\dag_i
\widetilde{\widetilde{g}}_i 
\!=\!
\widetilde{\widetilde{g}}_i
\widetilde{\widetilde{g}}^\dag_i
\!=\!
1_2 .
\label{TransBogo}
\ea\\[-12pt]
We also define a new bosonized quasi-particle $SO(2N)$ Lie operators as follows:\\[-16pt]
\ba
\widetilde{\widetilde{\hbox{\mathbf E}}}{}^{\!~i}_{~j}
\!=\!
{\displaystyle \frac{1}{2}}
[\widetilde{\widetilde{\hbox{\mathbf d}}}{}^\dagger_i,
\widetilde{\widetilde{\hbox{\mathbf d}}}_j] ,~
\widetilde{\widetilde{\hbox{\mathbf E}}}{}^{\!~j}_{~i}
\!=\!
\widetilde{\widetilde{\hbox{\mathbf E}}}{}^{\!~i \dagger }_{~j} , ~~
\widetilde{\widetilde{\hbox{\mathbf E}}}_{i j}
\!=\!
{\displaystyle \frac{1}{2}}
[\widetilde{\widetilde{\hbox{\mathbf d}}}_i,
\widetilde{\widetilde{\hbox{\mathbf d}}}_j], ~
\widetilde{\widetilde{\hbox{\mathbf E}}}{}^{\!~i j}
\!=\!
-\widetilde{\widetilde{\hbox{\mathbf E}}}{}^{\!~\dagger }_{i j}
\!=\!
\widetilde{\widetilde{\hbox{\mathbf E}}}{}^{\!~\star }_{i j} .
\label{3rdbosonimagequasiLieop}
\ea\\[-16pt]
Through the transformation
(\ref{TransBogo}),
the old bosonized quasiparticle $SO(2N)$ Lie operators
(\ref{2ndbosonimagequasiLieop})
are converted into\\[-14pt]
\ba
\!\!\!\!\!\!\!\!\!\!
\left.
\BA{ll}
&
\widetilde{\hbox{\mathbf E}}{}^{\!~i}_{~j}
\!=\!
{\displaystyle \frac{1}{2}}
[
- \widetilde{\widetilde{v}}{}^{\ast } _i
\widetilde{\widetilde{\hbox{\mathbf d}}}_i
\!+\!
\widetilde{\widetilde{u}}_i
\widetilde{\widetilde{\hbox{\mathbf d}}}{}^\dagger_i
,
\widetilde{\widetilde{u}}{}^{\ast } _j
\widetilde{\widetilde{\hbox{\mathbf d}}}_j
\!+\!
\widetilde{\widetilde{v}}_j
\widetilde{\widetilde{\hbox{\mathbf d}}}{}^\dagger_j
]
\!=\!
\widetilde{\widetilde{u}}_i
\widetilde{\widetilde{u}} {}^{\ast } _j
\widetilde{\widetilde{\hbox{\mathbf E}}}{}^{\!~i}_{~j}
\!+\!
\widetilde{\widetilde{v}} {}^{\ast } _i
\widetilde{\widetilde{v}}_j
\widetilde{\widetilde{\hbox{\mathbf E}}}{}^{\!~i \dagger }_{~j}
\!-\!
\widetilde{\widetilde{v}}{}^{\ast } _i
\widetilde{\widetilde{u}} {}^{\ast } _j
\widetilde{\widetilde{\hbox{\mathbf E}}}_{i j}
\!+\!
\widetilde{\widetilde{u}}_i
\widetilde{\widetilde{v}}_j
\widetilde{\widetilde{\hbox{\mathbf E}}}{}^{\!~i j} ,~
\widetilde{\hbox{\mathbf E}}{}^{\!~i}_{~~\!\!i}
\!=\!
\widetilde{\widetilde{\hbox{\mathbf E}}}{}^{\!~i}_{~i} , \\
\\[-6pt]
&
\widetilde{\hbox{\mathbf E}}{}^{\!~j \dagger }_{~i}
\!=\!
\widetilde{\widetilde{v}}_i
\widetilde{\widetilde{v}} {}^{\ast } _j
\widetilde{\widetilde{\hbox{\mathbf E}}}{}^{\!~i}_{~j}
\!+\!
\widetilde{\widetilde{u}} {}^{\ast } _i
\widetilde{\widetilde{u}}_j
\widetilde{\widetilde{\hbox{\mathbf E}}}{}^{\!~i \dagger }_{~j}
\!-\!
\widetilde{\widetilde{u}}{}^{\ast } _i
\widetilde{\widetilde{v}} {}^{\ast } _j
\widetilde{\widetilde{\hbox{\mathbf E}}}_{i j}
\!+\!
\widetilde{\widetilde{v}}_i
\widetilde{\widetilde{u}}_j
\widetilde{\widetilde{\hbox{\mathbf E}}}{}^{\!~i j} , \\
\\[-6pt]
&
\widetilde{\hbox{\mathbf E}}_{i j}
\!=\!
{\displaystyle \frac{1}{2}}
[
\widetilde{\widetilde{u}}{}^{\ast } _i
\widetilde{\widetilde{\hbox{\mathbf d}}}_i
\!+\!
\widetilde{\widetilde{v}}_i
\widetilde{\widetilde{\hbox{\mathbf d}}}{}^\dagger_i
,
\widetilde{\widetilde{u}}{}^{\ast } _j
\widetilde{\widetilde{\hbox{\mathbf d}}}_j
\!+\!
\widetilde{\widetilde{v}}_j
\widetilde{\widetilde{\hbox{\mathbf d}}}{}^\dagger_j
]
\!=\!
\widetilde{\widetilde{v}}_i
\widetilde{\widetilde{u}} {}^{\ast } _j
\widetilde{\widetilde{\hbox{\mathbf E}}}{}^{\!~i}_{~j}
\!-\!
\widetilde{\widetilde{u}} {}^{\ast } _i
\widetilde{\widetilde{v}}_j
\widetilde{\widetilde{\hbox{\mathbf E}}}{}^{\!~i \dagger }_{~j}
\!+\!
\widetilde{\widetilde{u}}{}^{\ast } _i
\widetilde{\widetilde{u}} {}^{\ast } _j
\widetilde{\widetilde{\hbox{\mathbf E}}}_{i j}
\!+\!
\widetilde{\widetilde{v}}_i
\widetilde{\widetilde{v}}_j
\widetilde{\widetilde{\hbox{\mathbf E}}}{}^{\!~i j} , \\
\\[-6pt]
&
\widetilde{\hbox{\mathbf E}}{}^{\!~i j}
\!=\!
-
\widetilde{\widetilde{u}}_i
\widetilde{\widetilde{v}} {}^{\ast } _j
\widetilde{\widetilde{\hbox{\mathbf E}}}{}^{\!~i}_{~j}
\!+\!
\widetilde{\widetilde{v}} {}^{\ast } _i
\widetilde{\widetilde{u}}_j
\widetilde{\widetilde{\hbox{\mathbf E}}}{}^{\!~i \dagger }_{~j}
\!+\!
\widetilde{\widetilde{v}}{}^{\ast } _i
\widetilde{\widetilde{v}} {}^{\ast } _j
\widetilde{\widetilde{\hbox{\mathbf E}}}_{i j}
\!+\!
\widetilde{\widetilde{u}}_i
\widetilde{\widetilde{u}}_j
\widetilde{\widetilde{\hbox{\mathbf E}}}{}^{\!~i j} .
\EA \!\!
\right\}
\label{4thdbosonimagequasiLieop}
\ea\\[-12pt]
Substituting
(\ref{4thdbosonimagequasiLieop}) 
into
each term in the last line of
(\ref{Hamiltonianimage2pEterm2}),
we obtain the Hamiltonian
$\widetilde{\widetilde{\hbox{\mathbf H}}}_{\widetilde{\hbox{\mathbf E}}}$
in terms of the new bosonized quasiparticle $SO(2N)$ Lie operators as follows:
\ba
\!\!\!\!\!\!\!\!
\BA{ll}
&
\widetilde{\widetilde{\hbox{\mathbf H}}}_{\widetilde{\hbox{\mathbf E}}}
\!=\!
\sum_i \!
\left( \!\!
E_i
\!+\!
{\displaystyle \frac{1}{2}}
\widetilde{E}_i \!\!
\right) \!\!
\left( \!\!
\widetilde{\widetilde{\hbox{\mathbf E}}}{}^{\!~i}_{~i}
\!+\!
{\displaystyle \frac{1}{2}} \!\!
\right) \\
\\[-10pt]
&
\!-\!
{\displaystyle \frac{1}{4}}
E_j
\widetilde{y}^\star_{j}
\widetilde{y}_{i} \!\!
\left( \!
\widetilde{\widetilde{u}}_i
\widetilde{\widetilde{u}} {}^{\ast } _j
\widetilde{\widetilde{\hbox{\mathbf E}}}{}^{\!~i}_{~j}
\!+\!
\widetilde{\widetilde{v}} {}^{\ast } _i
\widetilde{\widetilde{v}}_j
\widetilde{\widetilde{\hbox{\mathbf E}}}{}^{\!~i \dagger }_{~j}
\!-\!
\widetilde{\widetilde{v}}{}^{\ast } _i
\widetilde{\widetilde{u}} {}^{\ast } _j
\widetilde{\widetilde{\hbox{\mathbf E}}}_{i j}
\!+\!
\widetilde{\widetilde{u}}_i
\widetilde{\widetilde{v}}_j
\widetilde{\widetilde{\hbox{\mathbf E}}}{}^{\!~i j}
\!+\!
{\displaystyle \frac{1}{2} \delta_{ij}} \!
\right) \\
\\[-10pt]
&
\!-\!
{\displaystyle \frac{1}{4}}
E_i
\widetilde{y}^\star_{i}
\widetilde{y}_{j} \!\!
\left( \!
\widetilde{\widetilde{v}}_i
\widetilde{\widetilde{v}} {}^{\ast } _j
\widetilde{\widetilde{\hbox{\mathbf E}}}{}^{\!~i}_{~j}
\!+\!
\widetilde{\widetilde{u}} {}^{\ast } _i
\widetilde{\widetilde{u}}_j
\widetilde{\widetilde{\hbox{\mathbf E}}}{}^{\!~i \dagger }_{~j}
\!-\!
\widetilde{\widetilde{u}}{}^{\ast } _i
\widetilde{\widetilde{v}} {}^{\ast } _j
\widetilde{\widetilde{\hbox{\mathbf E}}}_{i j}
\!+\!
\widetilde{\widetilde{v}}_i
\widetilde{\widetilde{u}}_j
\widetilde{\widetilde{\hbox{\mathbf E}}}{}^{\!~i j}
\!+\!
{\displaystyle \frac{1}{2} \delta_{ij}} \!
\right) \\
\\[-10pt]
&
\!+\!
{\displaystyle \frac{1}{4}}
E_j
\widetilde{y}^\star_{j}
\widetilde{y}^\star_{i} \!\!
\left( \!
\widetilde{\widetilde{v}}_i
\widetilde{\widetilde{u}} {}^{\ast } _j
\widetilde{\widetilde{\hbox{\mathbf E}}}{}^{\!~i}_{~j}
\!-\!
\widetilde{\widetilde{u}} {}^{\ast } _i
\widetilde{\widetilde{v}}_j
\widetilde{\widetilde{\hbox{\mathbf E}}}{}^{\!~i \dagger }_{~j}
\!+\!
\widetilde{\widetilde{u}}{}^{\ast } _i
\widetilde{\widetilde{u}} {}^{\ast } _j
\widetilde{\widetilde{\hbox{\mathbf E}}}_{i j}
\!+\!
\widetilde{\widetilde{v}}_i
\widetilde{\widetilde{v}}_j
\widetilde{\widetilde{\hbox{\mathbf E}}}{}^{\!~i j} \!
\right) \\
\\[-10pt]
&
\!-\!
{\displaystyle \frac{1}{4}}
E_j
\widetilde{y}_{j}
\widetilde{y}_{i} \!\!
\left( \!
-
\widetilde{\widetilde{u}}_i
\widetilde{\widetilde{v}} {}^{\ast } _j
\widetilde{\widetilde{\hbox{\mathbf E}}}{}^{\!~i}_{~j}
\!+\!
\widetilde{\widetilde{v}} {}^{\ast } _i
\widetilde{\widetilde{u}}_j
\widetilde{\widetilde{\hbox{\mathbf E}}}{}^{\!~i \dagger }_{~j}
\!+\!
\widetilde{\widetilde{v}}{}^{\ast } _i
\widetilde{\widetilde{v}} {}^{\ast } _j
\widetilde{\widetilde{\hbox{\mathbf E}}}_{i j}
\!+\!
\widetilde{\widetilde{u}}_i
\widetilde{\widetilde{u}}_j
\widetilde{\widetilde{\hbox{\mathbf E}}}{}^{\!~i j} \!
\right) \\
\\[-10pt]
&
\!=\!
\sum_i \!
\left( \!\!
E_i
\!\!+\!\!
{\displaystyle \frac{1}{2}}
\widetilde{E}_i \!\!
\right) \!\!
\left( \!\!
\widetilde{\widetilde{\hbox{\mathbf E}}}{}^{\!~i}_{~i}
\!\!+\!\!
{\displaystyle \frac{1}{2}} \!\!
\right)
\!-\!
{\displaystyle \frac{1}{4}} \!
\left( \!
E_j
\widetilde{y}^\star_{j}
\widetilde{y}_{i}
\widetilde{\widetilde{u}}_i
\widetilde{\widetilde{u}} {}^{\ast } _j
\!\!+\!\!
E_i
\widetilde{y}^\star_{i}
\widetilde{y}_{j}
\widetilde{\widetilde{v}}_i
\widetilde{\widetilde{v}} {}^{\ast } _j
\!\!-\!\!
E_j
\widetilde{y}^\star_{j}
\widetilde{y}^\star_{i}
\widetilde{\widetilde{v}}_i
\widetilde{\widetilde{u}} {}^{\ast } _j
\!\!-\!\!
E_j
\widetilde{y}_{j}
\widetilde{y}_{i}
\widetilde{\widetilde{u}}_i
\widetilde{\widetilde{v}} {}^{\ast } _j
\right) \!\!
\left( \!\!
\widetilde{\widetilde{\hbox{\mathbf E}}}{}^{\!~i}_{~j}
\!\!+\!\!
{\displaystyle \frac{1}{2} \delta_{ij}}  \!\!
\right) \\
\\[-10pt]
&
\!-
{\displaystyle \frac{1}{4}} \!
\left( \!
E_j
\widetilde{y}^\star_{j}
\widetilde{y}_{i}
\widetilde{\widetilde{v}} {}^{\ast } _i
\widetilde{\widetilde{v}}_j
\!\!+\!\!
E_i
\widetilde{y}^\star_{i}
\widetilde{y}_{j}
\widetilde{\widetilde{u}} {}^{\ast } _i
\widetilde{\widetilde{u}}_j
\!\!+\!\!
E_j
\widetilde{y}^\star_{j}
\widetilde{y}^\star_{i}
\widetilde{\widetilde{u}} {}^{\ast } _i
\widetilde{\widetilde{v}}_j
\!\!+\!\!
E_j
\widetilde{y}_{j}
\widetilde{y}_{i}
\widetilde{\widetilde{v}} {}^{\ast } _i
\widetilde{\widetilde{u}}_j
\right) \!\!
\left( \!\!
\widetilde{\widetilde{\hbox{\mathbf E}}}{}^{\!~i \dagger }_{~j}
\!\!+\!\!
{\displaystyle \frac{1}{2} \delta_{ij}} \!\!
\right) \\
\\[-10pt]
&
\!+
{\displaystyle \frac{1}{4}} \!
\left( \!
E_j
\widetilde{y}_{i}
\widetilde{\widetilde{v}}{}^{\ast } _i
\!\!-\!\!
E_i
\widetilde{y}^\star_{i}
\widetilde{\widetilde{u}}{}^{\ast } _i
\right) \!\!
\left( \!
\widetilde{y}^\star_{j}
\widetilde{\widetilde{u}} {}^{\ast } _j
\!\!-\!\!
\widetilde{y}_{j}
\widetilde{\widetilde{v}} {}^{\ast } _j
\right) \!\!
\widetilde{\widetilde{\hbox{\mathbf E}}}_{i j}
\!-\!
{\displaystyle \frac{1}{4}} \!
\left( \!
E_j
\widetilde{y}_{i}
\widetilde{\widetilde{u}}_i
\!\!-\!\!
E_i
\widetilde{y}^\star_{i}
\widetilde{\widetilde{v}}_i
\right) \!\!
\left( \!
\widetilde{y}_{j}
\widetilde{\widetilde{u}}_j
\!\!-\!\!
\widetilde{y}{}^{\ast }_{j}
\widetilde{\widetilde{v}}_j
\right) \!\!
\widetilde{\widetilde{\hbox{\mathbf E}}}{}^{\!~i j} .
\EA
\label{Hamiltonianimage2pEterm3}
\ea
The last two terms
correspond to the so-called dangerous terms and should vanish.
Then we can get the relation between
$\widetilde{\widetilde{u}}{}^{\ast } _j$
and
$\widetilde{\widetilde{v}}{}^{\ast } _j$
and
consequently have a simple and important result for the state $j$
and also the same for the state $i$\\[-14pt]
\ba
\widetilde{\widetilde{u}}{}^{\ast } _j
\!=\!
{\displaystyle
\frac{\widetilde{y}_{j}}{\widetilde{y}{}^{\ast }_{j}}
}
\widetilde{\widetilde{v}}{}^{\ast } _j ,~
\left( \!
\widetilde{\widetilde{u}}{}^{\ast } _j
\widetilde{\widetilde{u}}_j
\!+\!
\widetilde{\widetilde{v}}{}^{\ast } _j
\widetilde{\widetilde{v}}_j
\!=\!
1 \!
\right)
\Longrightarrow 
\widetilde{\widetilde{u}}{}^{\ast } _j
\widetilde{\widetilde{u}}_j
\!=\!
\widetilde{\widetilde{v}}{}^{\ast } _j
\widetilde{\widetilde{v}}_j
\!=\!
{\displaystyle \frac{1}{2}} ,~~
\widetilde{\widetilde{u}}{}^{\ast } _i
\widetilde{\widetilde{u}}_i
\!=\!
\widetilde{\widetilde{v}}{}^{\ast } _i
\widetilde{\widetilde{v}}_i
\!=\!
{\displaystyle \frac{1}{2}} .
\label{uvrelationsandsimpleresult}
\ea\\[-12pt]
Substituting
(\ref{uvrelationsandsimpleresult})
into the second and third lines from the bottom of
(\ref{Hamiltonianimage2pEterm3}),
and putting
$
\widetilde{\widetilde{v}} _j
\!=\!
\widetilde{\widetilde{v}} _i
e^{i \theta }
$
which is due to
(\ref{uvrelationsandsimpleresult}),
then we have the Hamiltonian
$\widetilde{\widetilde{\hbox{\mathbf H}}}_{\widetilde{\hbox{\mathbf E}}}$
in the following form:\\[-12pt]
\ba
\BA{c}
\widetilde{\widetilde{\hbox{\mathbf H}}}_{\widetilde{\hbox{\mathbf E}}}
\!=\!
\sum_i \!
\left( \!\!
E_i
\!\!+\!\!
{\displaystyle \frac{1}{2}}
\widetilde{E}_i \!\!
\right) \!\!
\left( \!\!
\widetilde{\widetilde{\hbox{\mathbf E}}}{}^{~\!i}_{~i}
\!\!+\!\!
{\displaystyle \frac{1}{2}} \!\!
\right)
\!-\!
{\displaystyle \frac{1}{8}} \!
\left( \!
E_i
\!-\!
E_j
\right) \!
\left\{ \!
\widetilde{y}^\star_{i}
\widetilde{y}_{j}
e^{-i \theta } \!
\left( \!\!
\widetilde{\widetilde{\hbox{\mathbf E}}}{}^{~\!i}_{~j}
\!\!+\!\!
{\displaystyle \frac{1}{2} \delta_{ij}} \!\!
\right)
\!\!+\!
\widetilde{y}_{i}
\widetilde{y}^\star_{j}
e^{i \theta } \!
\left( \!\!
\widetilde{\widetilde{\hbox{\mathbf E}}}{}^{~\!i \dagger }_{~j}
\!\!+\!\!
{\displaystyle \frac{1}{2} \delta_{ij}} \!\!
\right) \!\!
\right\} .
\EA
\label{Hamiltonianimage2pEterm5}
\ea\\[-10pt]
The Hamiltonian
$\widetilde{\widetilde{\hbox{\mathbf H}}}_{\widetilde{\hbox{\mathbf E}}}$
can be diagonalized as
$
\widetilde{\widetilde{\hbox{\mathbf H}}}_{\widetilde{\hbox{\mathbf E}}}
\!=\!
\sum_i \!
E_{\widetilde{y}i} \!\!
\left( \!\!
\widetilde{\widetilde{\hbox{\mathbf E}}}{}^{~\!i}_{~i}
\!\!+\!\!
{\displaystyle \frac{1}{2}} \!\!
\right)
$
where
$E_{\widetilde{y}i}$
is a diagonalized energy which is given,
e.g., for $i,~j \!=\! 1 \!\sim\! 2$ as\\[-16pt]
\ba
\!\!\!\!
\BA{c}
E_{\widetilde{y}i}
\!=\!
{\displaystyle \frac{1}{2}} \!\!
\left[ \!
\left( \!\!
E_1
\!\!+\!\!
{\displaystyle \frac{1}{2}}
\widetilde{E}_1 \!\!
\right) \!
\!+\!
\left( \!\!
E_2
\!\!+\!\!
{\displaystyle \frac{1}{2}}
\widetilde{E}_2 \!\!
\right) \!
\!+\!
\sqrt{ \!
\left\{ \!\!
\left( \!\!
E_1
\!\!+\!\!
{\displaystyle \frac{1}{2}}
\widetilde{E}_1 \!\!
\right) \!
\!-\!
\left( \!\!
E_2
\!\!+\!\!
{\displaystyle \frac{1}{2}}
\widetilde{E}_2 \!\!
\right) \!\!
\right\}^{\!2}
\!\!+\!
\left\{ \!
{\displaystyle \frac{1}{2}} \!
\left( \!
E_1
\!\!-\!\!
E_2
\right) \!
|\widetilde{y}_{1}|
|\widetilde{y}_{2}| \!
\sin \theta \!
\right\}^{\!2}
}
\right] \!\! .
\EA
\label{Hamiltoniandiagtildey}
\ea\\[-12pt]
The diagonalized energy
$\!E_{\widetilde{y}i}$
is given also in the case,
$i,j \!\!=\!\! 1 \!\!\sim\!\! N$
but then it becomes complex.

Using
(\ref{Hamiltonianimage2}), (\ref{Hamiltonianimage2p})
and
(\ref{Hamiltonianimage2pEterm5}),
finally we obtain the boson image of the Hamiltonian $\hbox{\mathbf H}$
with the Lagrange multipliers
in terms of $E_0$ (energy of classical motion of the top), $\widetilde{E_0}$ (energy of fluctuating motion) and quadratic terms of fluctuating operators
$\widetilde{\widetilde{\hbox{\mathbf E}}}$
and
$\widetilde{\widetilde{\hbox{\mathbf c}}}$
as
\ba
\!\!\!\!\!\!\!\!\!\!\!\!
\left.
\BA{ll}
&
\hbox{\mathbf H}
\!=\!
E_0
\!+\!
\widetilde{\hbox{\mathbf H}} \\
\\[-14pt]
&~~~
\!=\!
E_0
\!+\!
\sum_i \!
E_i \!
\left( \!\!
\hbox{\mathbf E}^{\,i }_{~i }
\!+\!
{\displaystyle \frac{1}{2}} \!\!
\right)
\!+\!
{\displaystyle \frac{1}{4}} [\alpha \beta|\gamma \delta]
\!
\left( \!
\{
\widetilde{\hbox{\mathbf E}}{}^{~\!\alpha }_{~\beta },
\widetilde{\hbox{\mathbf E}}{}^{~\!\gamma }_{~\delta }
\}
\!+\!
{\displaystyle \frac{1}{2}}
\{\widetilde{\hbox{\mathbf E}}{}^{~\!\alpha \gamma },
\widetilde{\hbox{\mathbf E}}_{\delta \beta }
\} \!
\right) \\
\\[-12pt]
&~~~~~~~~~~~~~~~~~~~~~~~~~~~~~~~\!
\!+\!
{\displaystyle \frac{1}{4}} k_{\alpha \beta } \!
\{
\widetilde{\hbox{\mathbf c}}^\dagger_\alpha,
\widetilde{\hbox{\mathbf c}}_\beta
\}
\!+\!
{\displaystyle \frac{1}{4}} k^\star_{\alpha \beta } \!
\{
\widetilde{\hbox{\mathbf c}}^\dagger_\beta,
\widetilde{\hbox{\mathbf c}}_\alpha
\}
\!+\!
{\displaystyle \frac{1}{4}}
l_{\alpha\beta }
\{
\widetilde{\hbox{\mathbf c}}^\dagger_\alpha,
\widetilde{\hbox{\mathbf c}}^\dagger_\beta
\}
\!+\!
{\displaystyle \frac{1}{4}}
l^\star_{\alpha \beta }
\{
\widetilde{\hbox{\mathbf c}}_\alpha,
\widetilde{\hbox{\mathbf c}}_\beta
\} , \\
\\[-10pt]
&
\!=\!
E_0
\!+\!
\widetilde{E_0}
\!+\!
\sum_i \!
E_{\widetilde{y}i} \!
\left( \!\!
\widetilde{\widetilde{\hbox{\mathbf E}}}{}^{~\!i}_{~i}
\!\!+\!\!
{\displaystyle \frac{1}{2}} \!\!
\right)
+
{\displaystyle \frac{1}{4}} [\alpha \beta|\gamma \delta]
\!
\left( \!
\{
\widetilde{\widetilde{\hbox{\mathbf E}}}{}^{\alpha }_{~\beta },
\widetilde{\widetilde{\hbox{\mathbf E}}}{}^{\gamma }_{~\delta }
\}
\!+\!
{\displaystyle \frac{1}{2}}
\{
\widetilde{\widetilde{\hbox{\mathbf E}}}{}^{\alpha \gamma },
\widetilde{\widetilde{\hbox{\mathbf E}}}_{\delta \beta }
\} \!
\right) \\
\\[-12pt]
&
+
{\displaystyle \frac{1}{4}} k_{\alpha \beta }
\{
\widetilde{\widetilde{\hbox{\mathbf c}}}{}^\dagger_\alpha,
\widetilde{\widetilde{\hbox{\mathbf c}}}_\beta 
\}
\!+\!
{\displaystyle \frac{1}{4}} k^\star_{\alpha \beta }
\{
\widetilde{\widetilde{\hbox{\mathbf c}}}{}^\dagger_\beta,
\widetilde{\widetilde{\hbox{\mathbf c}}}_\alpha
\}
\!+\!
{\displaystyle \frac{1}{4}}
l_{\alpha \beta }
\{
\widetilde{\widetilde{\hbox{\mathbf c}}}{}^\dagger_\alpha,
\widetilde{\widetilde{\hbox{\mathbf c}}}{}^\dagger_\beta
\}
\!+\!
{\displaystyle \frac{1}{4}}
l^\star_{\alpha \beta }
\{
\widetilde{\widetilde{\hbox{\mathbf c}}}_\alpha,
\widetilde{\widetilde{\hbox{\mathbf c}}}_\beta
\} ,
\EA \!\!
\right\}
\label{Hamiltonianimage2pEtermfinal}
\ea
\vspace{-0.3cm}
\ba
\!\!\!\!\!\!\!\!\!\!\!\!
\left.
\BA{ll}
&
E_0
\!\equiv\!
\erw{H}
\!+\!
{\displaystyle \frac{1}{2}}
\left( \!
M_{\alpha \beta }
\erw{c^\dagger_\alpha }
\!+\!
M^\star_{\alpha }
\erw{c_\alpha } \!
\right)
\!-\!
{\displaystyle \frac{1}{2}}
k_{\alpha \alpha } , \\
\\[-8pt]
&
\erw H
\!=\!
{\displaystyle \frac{1}{4}}
\left\{ \!
F_{\alpha \beta } \!
< \!\! E^{\alpha }_{~\beta } \!\! >_G
\!+
F^\star_{\alpha \beta } \!
< \!\! E^{\alpha \dagger }_{~\beta } \!\! >_G
\!+
D_{\alpha \beta } \!
< \!\! E^{\alpha \beta } \!\! >_G
\!-
D^\star_{\alpha \beta } \!
< \!\! E_{\alpha \beta } \!\! >_G
\right. \\
\\[-10pt]
&
\left.
~~~~~~~~~~~~~
\!+
h_{\alpha \beta } \!
< \!\! E^{\alpha }_{~\beta } \!\! >_G
\!+
h^\star_{\alpha \beta }< \!\! E^{\alpha \dag }_{~\beta } \!\! >_G
\!+
\left(
F_{\alpha \alpha }
\!-\!
h_{\alpha \alpha }
\right)
\right\} , \\
\\[-8pt]
&
\widetilde{E_0}
\!\!\equiv\!\!
{\displaystyle \frac{1}{2}} \!\!
\left( \!
F_{\alpha \beta } \!\!
< \!\! \widetilde{E}^{\alpha }_{~\beta } \!\! >_{\widetilde G}
\!\!+\!
F^\star_{\alpha \beta } \!\!
< \!\! \widetilde{E}^{\alpha \dagger }_{~\beta } \!\! >_{\widetilde G}
\!\!+\!
D_{\alpha \beta } \!\!
< \!\! \widetilde{E}^{\alpha \beta } \!\! >_{\widetilde G}
\!\!-\!
D^\star_{\alpha \beta } \!\!
< \!\! \widetilde{E}_{\alpha \beta } \!\! >_{\widetilde G} \!
\right)
\!\!+\!\!
M_\alpha \!\!
< \!\! \widetilde{c}^\dagger_\alpha \!\! >_{\widetilde G}
\!\!+\!
M^\star_\alpha \!\!
< \!\! \widetilde{c}_\alpha \!\! >_{\widetilde G} \\
\\[-8pt]
&
\!+\!
{\displaystyle \frac{1}{4}} \!\!
\left( \!
\widetilde{F}_{\alpha \beta } \!\!
< \!\! \widetilde{E}^{\alpha }_{~\beta } \!\! >_{\widetilde G}
\!\!+\!
\widetilde{F}^\star_{\alpha \beta } \!\!
< \!\! \widetilde{E}^{\alpha \dagger }_{~\beta } \!\! >_{\widetilde G}
\!\!+\!
\widetilde{D}_{\alpha \beta } \!\!
< \!\! \widetilde{E}^{\alpha \beta } \!\! >_{\widetilde G}
\!\!-\!
\widetilde{D}^\star_{\alpha \beta } \!\!
< \!\! \widetilde{E}_{\alpha \beta } \!\! >_{\widetilde G}
\!\!-\!
h_{\alpha \beta } \!\!
< \!\! \widetilde{E}^{\alpha }_{~\beta } \!\! >_{\widetilde G}
\!\!-\!
h^\star_{\alpha \beta } \!\!
< \!\! \widetilde{E}^{\alpha \dagger }_{~\beta } \!\! >_{\widetilde G} \!
\right) \\
\\[-12pt]
&
~~~~~~~~~~~~~~~~~~~~~~~~~~~~~~~~~~~~~~~~~~~~~~~~~~~~~~~~~~~~~~~~~~~~~~~~
\!+\!
{\displaystyle \frac{1}{2}} \!
\left( \!
\widetilde{M}_\alpha \!\!
< \!\! \widetilde{c}^\dagger_\alpha \!\! >_{\widetilde G}
\!+\!
\widetilde{M}^\star_\alpha \!\!
< \!\! \widetilde{c}_\alpha \!\! >_{\widetilde G} \!
\right) \! . \!\!\!\!
\EA \!
\right\}
\label{E0andE0tilde}
\ea
The components of the generalized angular momentum,
i.e.,
the fermion $SO(2N\!+\!1)$ Lie operators, are not in general commutable.
Therefore,
owing to the presence of the fluctuating Hamiltonian
$\widetilde{\hbox{\mathbf H}}$,
the direction of the axis of rotation fluctuates quantum mechanically around the classical axis determined by
(\ref{classicaleqofmotion}).
If we consider the quadratic terms of the fluctuating operators in
(\ref{Hamiltonianimage2pEtermfinal}),
then the Sawada's eigenmode method
\cite{Sawada.57,Sawada.60}
is useful.

\newpage

%%%%%%%%%%%%%%%%%%%%%%%%%%%%%%%%%%%%%%%%%%%%%%%%%%%%
%                                                  %
%  6  Basic equations for parameters $k$ and $l$   %
%                                                  %
%           in Lagrange multiplier terms           %
%                                                  %
%%%%%%%%%%%%%%%%%%%%%%%%%%%%%%%%%%%%%%%%%%%%%%%%%%%%

%%%%%%%%%%%%%%%%%%%%%%%%%%%%%%%%%%%%%%%%%%%%%%%%%%%%

\def\thesection{\arabic{section}}
\setcounter{equation}{0}
\renewcommand{\theequation}{\arabic{section}.\arabic{equation}}
\section{Basic equations for determination of parameters $k$ and $l$ in Lagrange multiplier terms}

%%%%%%%%%%%%%%%%%%%%%%%%%%%%%%%%%%%%%%%%%%%%%%%%%%%%

~~~The constraint term $\hbox{\mathbf H}^{~\!\prime }$
(\ref{Hamiltonianimage})
should vanish in the physical fermion space
but its classical part in $E_0$ in 
(\ref{Hamiltonianimage2})
might not.
Thus we have an unsolved problem
to determine the unknown parameters
$k_{\alpha\beta }$ and $l_{\alpha\beta }$
in $M_{\alpha }$.
They cannot be determined in the classical limit only.
The determination of them requires taking the quantum mechanical fluctuations into account
but has a difficult prescription.
Instead, here we attempt another approach
which has been proposed by one of the present author (S.N.)
Ref.
\cite{Nishi.98}.
Following Eq. (6.1) in Ref.
\cite{Nishi.98},
using the relations\\[-18pt]
\ba
\BA{c}
\hbox{\mathbf c}_\alpha \Phi_{00}(G)
\!=\!
0 ,~
\hbox{\mathbf c}^\dagger_\alpha \Phi_{00}(G)
\!=\!
r^\star_\alpha
\Phi_{00}(G) , ~~~
\Phi^\star_{00}(G)
\!\equiv\!
\bra0 U(G) \ket0 ,~~~
\hbox{\mathbf c}^\dagger_\alpha
\!=\!
-\hbox{\mathbf c}^\star_\alpha ,
\EA
\label{vacfunc}
\ea\\[-16pt]
we obtain the identities\\[-18pt]
\ba
\BA{c}
\hbox{\mathbf c}_\alpha U({\cal G}) \ket0
\!=\!
(-r_\alpha \!+\! r_\alpha r_\xi c_\xi^\dagger
\!-\!
q_{\alpha \xi }c_\xi^\dagger) \!\cdot\! U(G) \ket0 ,~~~
\hbox{\mathbf c}_\alpha^\dagger U({\cal G}) \ket0
\!=\!
-c_\alpha^\dagger \!\cdot\! U(G) \ket0 ,
\EA
\label{identities}
\ea
which are just the identities given by Eq. (6.2) in Ref.
\cite{Nishi.98}
and whose proofs are given in App. A
where we have used
$U({\cal G}) \ket0 \!\!=\! U(G)\ket0$ and
(\ref{ketG}),
repeatedly.
It is shown that
on the $U({\cal G})\ket0$ the fermion operators
$\hbox{\mathbf c}_\alpha$ and
$\hbox{\mathbf c}_\alpha^\dagger$ satisfy exactly the
anticommutation relations\\[-14pt]
\ba
\left.
\BA{ll}
&\left(
\hbox{\mathbf c}_\alpha^\dagger \hbox{\mathbf c}_\beta
\!+\!
\hbox{\mathbf c}_\beta \hbox{\mathbf c}_\alpha^\dagger
\right) \!
U({\cal G}) \ket0
\!=\!
\delta_{\alpha \beta }
\!\cdot\! U(G) \ket0 , \\
\\[-6pt]
&\left(
\hbox{\mathbf c}_\alpha^\dagger \hbox{\mathbf c}_\beta^\dagger
\!+\!
\hbox{\mathbf c}^\dagger_\beta\hbox{\mathbf c}_\alpha^\dagger
\right) \!
U({\cal G}) \ket0
\!=\!
\left(
\hbox{\mathbf c}_\alpha \hbox{\mathbf c}_\beta
\!+\!
\hbox{\mathbf c}_\beta \hbox{\mathbf c}_\alpha
\right) \!
U({\cal G}) \ket0
\!=\!
0 ,
\EA
\right\}
\label{anticommurel}
\ea
for details of the corresponding proofs
see App. A.
The relations
(\ref{anticommurel}),
however, play no role to determine the parameters
$k_{\alpha \beta }$
and
$l_{\alpha \beta }$.
At a region very near $z \!\!=\!\! 1$
where we have no effects of the unpaired modes
discussed in the previous sections,
$
{\displaystyle \sqrt{\frac{1\!\!+\!\!z}{2}}}
\left( \!\equiv\! \mbox{$\slashed{z}$} \right)
$
is approximated to
$
{\displaystyle \frac{1}{2}} \!
\left( \!
1
\!\!+\!\!
{\displaystyle \frac{1\!\!+\!\!z}{2}} \!
\right)
(0 \!\ll\! z \!\approx\! 1)
$.
Then introducing the approximate function for
the vacuum function which has first been proposed
in Ref.
\cite{Nishi.98},\\[-18pt]
\ba
\widetilde{\Phi }_{00}^\star(G)
\!=\!
\frac{1}{2}
\left\{
\Phi^\star_{00}(g)
\!+\!
\mbox{$\slashed{z}$}
\Phi^\star_{00}(G)
\right\}
\!=\!
{\displaystyle \frac{1}{2}} \!
\left( \!
\mbox{$\slashed{z}$}
\!+\!
{\displaystyle \frac{1}{\mbox{$\slashed{z}$}}} \!
\right) \!
\Phi^\star_{00}(G) ,~
\Phi^\star_{00}(g)
\!\equiv\!
\bra0 U(g) \ket0
\!=\!
{\displaystyle \frac{1}{\mbox{$\slashed{z}$}}} \!
\Phi^\star_{00}(G) ,
\ea\\[-12pt]
we adopt an approximate
$SO(2N\!+\!1)$ WF expressed as\\[-18pt]
\ba
\left.
\BA{ll}
&U({\cal G}) \ket0
\!\cong\!
\widetilde{\Phi }^\star_{00}(G)
(1 \!+\! r_\alpha c_\alpha^\dagger)
\exp \!
\left( \! {\displaystyle \frac{1}{2}}
q_{\alpha \beta }
c^\dagger_\alpha c^\dagger_\beta \!
\right) \!
\ket0
\!=\!
\widetilde{U}(G) \ket0 , \\
\\[-14pt]
&\widetilde{U}(G) \ket0
\!=\!
{\displaystyle
\frac{\widetilde{\Phi }^\star_{00}(G)}{\Phi^\star_{00}(G)}
}
U(G) \ket0
\!=\!
{\displaystyle \frac{1}{2}} \!
\left( \!
\mbox{$\slashed{z}$}
\!+\!
{\displaystyle \frac{1}{\mbox{$\slashed{z}$}}} \!
\right) \!
U(G) \ket0 .~
\mbox{(equivalent to Eq. (6.6) in Ref.
\cite{Nishi.98})}
\EA \!
\right\}
\label{approxwf}
\ea\\[-6pt]
We summarize some results in Ref.
\cite{Nishi.98}.
Using
(\ref{approxwf}) and (\ref{calphadiffrential}),
we get approximate identities\\[-12pt]
\ba
\!\!\!\!
\left.
\BA{lll}
&\hbox{\mathbf c}_\alpha U({\cal G}) \ket0
\!\approx\!
\hbox{\mathbf c}_\alpha \widetilde{U}(G) \ket0
\!\!&\!=\!
{\displaystyle \frac{1}{2}} \!
\left( \!
\mbox{$\slashed{z}$}
\!+\!
{\displaystyle \frac{1}{\mbox{$\slashed{z}$}}} \!
\right) \!
\hbox{\mathbf c}_\alpha
U(G) \ket0 \\
\\[-8pt]
&\!\!&
-
{\displaystyle \frac{1}{4}} \!
\left( \!
\mbox{$\slashed{z}$}
\!-\!
{\displaystyle \frac{1}{\mbox{$\slashed{z}$}}} \!
\right) \!
\left\{
r^{\mbox{\scriptsize T}} (1\!+\!q^\dagger q)^{-1}
\!+\!
r^\dagger q(1\!+\!q^\dagger q)^{-1}
\right\}_\alpha \!
U(G) \ket0 , \\
\\[-8pt]
&\hbox{\mathbf c}^\dagger_\alpha U({\cal G}) \ket0
\!\approx\!
\hbox{\mathbf c}^\dagger_\alpha \widetilde{U}(G) \ket0
\!\!&\!=\!
{\displaystyle \frac{1}{2}} \!
\left( \!
\mbox{$\slashed{z}$}
\!+\!
{\displaystyle \frac{1}{\mbox{$\slashed{z}$}}} \!
\right) \!
\hbox{\mathbf c}^\dagger_\alpha
U(G) \ket0 \\
\\[-8pt]
&\!\!&
+
{\displaystyle \frac{1}{4}} \!
\left( \!
\mbox{$\slashed{z}$}
\!-\!
{\displaystyle \frac{1}{\mbox{$\slashed{z}$}}} \!
\right) \!
\left\{
r^\dagger (1\!+\!qq^\dagger )^{-1}
\!-\!
r^{\mbox{\scriptsize T}} q^\dagger (1\!+\!qq^\dagger )^{-1}
\right\}_\alpha \!
U(G) \ket0 .
\EA
\right\}
\label{approxidentities}
\ea
With the aid of
(\ref{calphadiffrentialproof2}),
successive operations of the differential form
(\ref{Lieopratorc2})
for the fermion operators
$\hbox{\mathbf c}_\alpha$ and
$\hbox{\mathbf c}_\alpha^\dagger$ on the approximate
identities,
though the formulas for such successive operations
in Ref.
\cite{Nishi.98}
have been incorrect,
leads to the correct formulas
\\[-16pt]
\ba
\!\!\!\!
\left.
\BA{lll}
&\hbox{\mathbf c}^\dagger_\alpha
\hbox{\mathbf c}_\beta \widetilde{U}(G) \ket0
\!\!
&\!\!=\!
\left[
{\displaystyle \frac{1}{2}} \!
\left( \!
\mbox{$\slashed{z}$}
\!+\!
{\displaystyle \frac{1}{\mbox{$\slashed{z}$}}} \!
\right) \!
\hbox{\mathbf c}^\dagger_\alpha \hbox{\mathbf c}_\beta
\!\!+\!\!
{\displaystyle \frac{1}{2}} \!
\left( \!
\mbox{$\slashed{z}$}
\!-\!
{\displaystyle \frac{1}{\mbox{$\slashed{z}$}}} \!
\right) \!\!
\left( \!\!
1 \!\!-\!\! {\displaystyle \frac{1}{2 \mbox{$\slashed{z}$}^2}} \!\!
\right) \!
\delta_{\alpha \beta }
\!\!+\!\!
{\displaystyle \frac{1}
{32 \mbox{$\slashed{z}$}^4}} \!\!
\left( \!
\mbox{$\slashed{z}$}
\!-\!
{\displaystyle \frac{3}{\mbox{$\slashed{z}$}}} \!
\right) \!
x^\star_\alpha x^{\mbox{\scriptsize T}}_\beta
\right. \\
\\[-12pt]
&\!\!&
\left.
~~~~~~~~~~~~~~~~~~~~~
-
{\displaystyle \frac{1}{8 \mbox{$\slashed{z}$}^2}} \!
\left( \!
\mbox{$\slashed{z}$}
\!-\!
{\displaystyle \frac{1}{\mbox{$\slashed{z}$}}} \!
\right) \!
\left(
x^{\mbox{\scriptsize T}}_\beta
\hbox{\mathbf c}^\dagger_\alpha
\!-\!
x^\star_\alpha
\hbox{\mathbf c}_\beta
\right)
\right] \!
U(G) \ket0 , \\
\\[-12pt]
&\hbox{\mathbf c}_\beta
\hbox{\mathbf c}^\dagger_\alpha \widetilde{U}(G) \ket0
\!\!
&\!\!=\!
\left[
{\displaystyle \frac{1}{2}} \!
\left( \!
\mbox{$\slashed{z}$}
\!+\!
{\displaystyle \frac{1}{\mbox{$\slashed{z}$}}} \!
\right) \!
\hbox{\mathbf c}_\beta \hbox{\mathbf c}^\dagger_\alpha
\!\!+\!\!
{\displaystyle \frac{1}{2}} \!
\left( \!
\mbox{$\slashed{z}$}
\!-\!
{\displaystyle \frac{1}{\mbox{$\slashed{z}$}}} \!
\right) \!\!
\left( \!\!
1 \!\!-\!\! {\displaystyle \frac{1}{2 \mbox{$\slashed{z}$}^2}} \!\!
\right) \!
\delta_{\alpha \beta }
\!\!+\!\!
{\displaystyle \frac{1}
{32 \mbox{$\slashed{z}$}^4}} \!\!
\left( \!
\mbox{$\slashed{z}$}
\!-\!
{\displaystyle \frac{3}{\mbox{$\slashed{z}$}}} \!
\right) \!
x^\star_\alpha x^{\mbox{\scriptsize T}}_\beta
\right. \\
\\[-12pt]
&\!\!&
\left.
~~~~~~~~~~~~~~~~~~~~~
-
{\displaystyle \frac{1}{8 \mbox{$\slashed{z}$}^2}} \!
\left( \!
\mbox{$\slashed{z}$}
\!-\!
{\displaystyle \frac{1}{\mbox{$\slashed{z}$}}} \!
\right) \!
\left(
x^{\mbox{\scriptsize T}}_\beta
\hbox{\mathbf c}^\dagger_\alpha
\!-\!
x^\star_\alpha
\hbox{\mathbf c}_\beta
\right)
\right] \!
U(G) \ket0 ,
\EA
\right\}
\label{approxidentities2}
\ea
\vspace{-0.7cm}
\ba
\!\!\!\!
\left.
\BA{lll}
&\hbox{\mathbf c}^\dagger_\alpha
\hbox{\mathbf c}^\dagger_\beta \widetilde{U}(G) \ket0
\!\!
&\!\!=\!
\left[
{\displaystyle \frac{1}{2}} \!
\left( \!
\mbox{$\slashed{z}$}
\!+\!
{\displaystyle \frac{1}{\mbox{$\slashed{z}$}}} \!
\right) \!
\hbox{\mathbf c}^\dagger_\alpha \hbox{\mathbf c}^\dagger_\beta
\!-\!
{\displaystyle \frac{1}{32 \mbox{$\slashed{z}$}^4}} \!
\left( \!
\mbox{$\slashed{z}$}
\!-\!
{\displaystyle \frac{3}{\mbox{$\slashed{z}$}}} \!
\right) \!
x^\star_\alpha x^\dagger_\beta
\right. \\
\\[-10pt]
&\!\!&
\left.
~~~~~~~~~~~~~~~~~~~~~
+
{\displaystyle \frac{1}{8 \mbox{$\slashed{z}$}^2}} \!
\left( \!
\mbox{$\slashed{z}$}
\!-\!
{\displaystyle \frac{1}{\mbox{$\slashed{z}$}}} \!
\right) \!
\left(
x^\star_\alpha
\hbox{\mathbf c}^\dagger_\beta
\!+\!
x^\dagger_\beta
\hbox{\mathbf c}^\dagger_\alpha
\right)
\right] \!
U(G) \ket0 , \\
\\[-10pt]
&\hbox{\mathbf c}_\alpha
\hbox{\mathbf c}_\beta \widetilde{U}(G) \ket0
\!\!
&\!\!=\!
\left[
{\displaystyle \frac{1}{2}} \!
\left( \!
\mbox{$\slashed{z}$}
\!+\!
{\displaystyle \frac{1}{\mbox{$\slashed{z}$}}} \!
\right) \!
\hbox{\mathbf c}_\alpha \hbox{\mathbf c}_\beta
\!-\!
{\displaystyle \frac{1}{32 \mbox{$\slashed{z}$}^4}} \!
\left( \!
\mbox{$\slashed{z}$}
\!-\!
{\displaystyle \frac{3}{\mbox{$\slashed{z}$}}} \!
\right) \!
x_\alpha x^{\mbox{\scriptsize T}}_\beta
\right. \\
\\[-10pt]
&\!\!&
\left.
~~~~~~~~~~~~~~~~~~~~~
-
{\displaystyle \frac{1}{8 \mbox{$\slashed{z}$}^2}} \!
\left( \!
\mbox{$\slashed{z}$}
\!-\!
{\displaystyle \frac{1}{\mbox{$\slashed{z}$}}} \!
\right) \!
\left(
x_\alpha
\hbox{\mathbf c}_\beta
\!+\!
x^{\mbox{\scriptsize T}}_\beta
\hbox{\mathbf c}_\alpha
\right)
\right] \!
U(G) \ket0 .
\EA
\right\}
\label{approxidentities3}
\ea\\[-12pt]
Using the set of
(\ref{approxidentities2}) and (\ref{approxidentities3}),
operations of the anticommutators of the fermion operators
$\hbox{\mathbf c}_\alpha$ and
$\hbox{\mathbf c}_\alpha^\dagger$ on the approximate
$\widetilde U(G) \ket0$ of the $U({\cal G}) \ket0$ are made as follows:\\[-16pt]
\ba
\!\!\!\!\!\!\!\!\!\!
\left.
\BA{ll}
&
\left(
\hbox{\mathbf c}_\alpha^\dagger \hbox{\mathbf c}_\beta
\!+\!
\hbox{\mathbf c}_\beta \hbox{\mathbf c}_\alpha^\dagger
\right) \!
\widetilde{U}(G) \ket0 
\!\cong\!
\left[
{\displaystyle \frac{1}{16 \mbox{$\slashed{z}$}^4}} \!\!
\left( \!
\mbox{$\slashed{z}$}
\!-\!
{\displaystyle \frac{3}{\mbox{$\slashed{z}$}}} \!
\right) \!
x^\star_\alpha x^{\mbox{\scriptsize T}}_\beta
\right.
\\
\\[-10pt]
&
\left.
+
{\displaystyle \frac{1}{2}} \!\!
\left( \!
\mbox{$\slashed{z}$}
\!+\!
{\displaystyle \frac{1}{\mbox{$\slashed{z}$}}} \!
\right) \!
\delta_{\alpha \beta }
\!+\!
\left( \!
\mbox{$\slashed{z}$}
\!-\!
{\displaystyle \frac{1}{\mbox{$\slashed{z}$}}} \!
\right) \!
\left\{ \!
\left( \!
1 \!-\! {\displaystyle \frac{1}{2 \mbox{$\slashed{z}$}^2}} \!
\right) \!
\delta_{\alpha \beta }
\!-\!\!
{\displaystyle \frac{1}{4 \mbox{$\slashed{z}$}^2}} \!
\left(
x^{\mbox{\scriptsize T}}_\beta \hbox{\mathbf c}^\dagger_\alpha
\!-\!
x^\star_\alpha \hbox{\mathbf c}_\beta
\right) \!
\right\}
\right] \!
U(G) \ket0 , \\
\\[-10pt]
&
\left(
\hbox{\mathbf c}_\alpha^\dagger \hbox{\mathbf c}_\beta^\dagger
\!+\!
\hbox{\mathbf c}_\beta^\dagger \hbox{\mathbf c}_\alpha^\dagger
\right) \!
\widetilde{U}(G) \ket0 \\
\\[-12pt]
&
\!\cong\!
\left[
-{\displaystyle \frac{1}{32 \mbox{$\slashed{z}$}^4}} \!
\left( \!
\mbox{$\slashed{z}$}
\!-\!
{\displaystyle \frac{3}{\mbox{$\slashed{z}$}}} \!
\right) \!
\left(
x^\star_\alpha x^\dagger_\beta
\!+\!
x^\star_\beta x^\dagger_\alpha
\right)
\!+\!
{\displaystyle \frac{1}{8 \mbox{$\slashed{z}$}^2}} \!
\left( \!
\mbox{$\slashed{z}$}
\!-\!
{\displaystyle \frac{1}{\mbox{$\slashed{z}$}}} \!
\right) \!
\left(
x^\star_\alpha \hbox{\mathbf c}^\dagger_\beta
\!+\!
x^\dagger_\beta
\hbox{\mathbf c}^\dagger_\alpha
\right)
\right] \!
U(G) \ket0 , \\
\\[-10pt]
&
\left(
\hbox{\mathbf c}_\alpha \hbox{\mathbf c}_\beta
\!+\!
\hbox{\mathbf c}_\beta \hbox{\mathbf c}_\alpha
\right) \!
\widetilde{U}(G) \ket0 \\
\\[-10pt]
&
\!\cong\!
\left[
-{\displaystyle \frac{1}{32 \mbox{$\slashed{z}$}^4}} \!
\left( \!
\mbox{$\slashed{z}$}
\!-\!
{\displaystyle \frac{3}{\mbox{$\slashed{z}$}}} \!
\right) \!
\left(
x_\alpha x^{\mbox{\scriptsize T}}_\beta
\!+\!
x_\beta x^{\mbox{\scriptsize T}}_\alpha
\right)
\!-\!
{\displaystyle \frac{1}{8 \mbox{$\slashed{z}$}^2}} \!
\left( \!
\mbox{$\slashed{z}$}
\!-\!
{\displaystyle \frac{1}{\mbox{$\slashed{z}$}}} \!
\right) \!
\left(
x^{\mbox{\scriptsize T}}_\alpha \hbox{\mathbf c}_\beta
\!+\!
x^{\mbox{\scriptsize T}}_\beta
\hbox{\mathbf c}_\alpha
\right)
\right] \!
U(G) \ket0 .
\EA \!\!
\right\}
\label{Opanticommu}
\ea\\[-10pt]
We should demand that the
expectation values of the anticommutators between
$\hbox{\mathbf c}_\alpha$
and
$\hbox{\mathbf c}^\dagger_\alpha$
by the approximate
$\widetilde{U}(G) \ket0$ of
$U({\cal G}) \ket0$ satifsy the following relations:\\[-18pt]
\ba
\left.
\BA{cc}
&\bra0
\widetilde{U}^\dagger(G) \!
\left(
\hbox{\mathbf c}_\alpha^\dagger \hbox{\mathbf c}_\beta
\!+\!
\hbox{\mathbf c}_\beta \hbox{\mathbf c}^\dagger_\alpha
\right) \!
\widetilde{U}(G)
\ket0
\!=\!
\delta_{\alpha \beta } , \\
\\[-10pt]
&\bra0
\widetilde{U}^\dagger(G) \!
\left(
\hbox{\mathbf c}_\alpha^\dagger \hbox{\mathbf c}_\beta^\dagger
\!+\!
\hbox{\mathbf c}_\beta^\dagger \hbox{\mathbf c}_\alpha^\dagger
\right) \!
\widetilde{U}(G)
\ket0
\!=\!
\bra0
\widetilde{U}^\dagger(G) \!
\left(
\hbox{\mathbf c}_\alpha \hbox{\mathbf c}_\beta
\!+\!
\hbox{\mathbf c}_\beta\hbox{\mathbf c}_\alpha
\right) \!
\widetilde{U}(G)
\ket0
\!=\!
0 ,
\EA
\right\}
\label{quasianticommurels}
\ea\\[-12pt]
from which and
(\ref{modifiedidentities}),
we can obtain the following approximate relations:\\[-16pt]
\ba
\!\!\!\!\!\!\!\!\!\!\!\!
\BA{ll}
&\bra0
\widetilde{U}^\dagger (G) \!
\left(
\hbox{\mathbf c}_\alpha^\dagger \hbox{\mathbf c}_\beta
\!+\!
\hbox{\mathbf c}_\beta \hbox{\mathbf c}_\alpha^\dagger
\right) \!
\widetilde{U}(G) \ket0 
\!=\!
{\displaystyle \frac{1}{2}} \!
\left( \!
\mbox{$\slashed{z}$}
\!+\!
{\displaystyle \frac{1}{\mbox{$\slashed{z}$}}} \!
\right) \!
\left[ \!
{\displaystyle \frac{1}{2}} \!
\left( \!
\mbox{$\slashed{z}$}
\!+\!
{\displaystyle \frac{1}{\mbox{$\slashed{z}$}}} \!
\right) \!
\delta_{\alpha \beta }
\!+\!
{\displaystyle \frac{1}{16 \mbox{$\slashed{z}$}^4}}
\left( \!
\mbox{$\slashed{z}$}
\!-\!
{\displaystyle \frac{3}{\mbox{$\slashed{z}$}}} \!
\right) \!
x^\star_\alpha
x^{\mbox{\scriptsize T}}_\beta
\right. \\
\\[-14pt]
&
\left.
\!+\!
\left( \!
\mbox{$\slashed{z}$}
\!-\!
{\displaystyle \frac{1}{\mbox{$\slashed{z}$}}} \!
\right) \!
\left\{ \!
\left( \!
1 \!\!-\! {\displaystyle \frac{1}{2 \mbox{$\slashed{z}$}^2}} \!
\right) \!
\delta_{\alpha \beta }
\!-\!
{\displaystyle \frac{1}{4 \mbox{$\slashed{z}$}^2}}
\left(
x^{\mbox{\scriptsize T}}_\beta \erw{c_\alpha }^{\!\!\!\!\!\!\star } ~
\!+\!
x^\star_\alpha \erw{c_\beta }
\right)
\right\}
\right]
\!\approx\!
\delta_{\alpha \beta } ,
\EA \!\!\!\!
\label{Opanticommu2}
\ea
\ba
\!\!\!\!\!\!\!\!\!\!\!\!
\left.
\BA{ll}
&\bra0
\widetilde{U}^\dagger (G) \!
\left(
\hbox{\mathbf c}_\alpha^\dagger \hbox{\mathbf c}_\beta^\dagger
\!+\!
\hbox{\mathbf c}_\beta^\dagger \hbox{\mathbf c}_\alpha^\dagger
\right)
\widetilde{U}(G) \ket0
\!=\!
{\displaystyle \frac{1}{2}} \!
\left( \!
\mbox{$\slashed{z}$}
\!+\!
{\displaystyle \frac{1}{\mbox{$\slashed{z}$}}} \!
\right) \!
\left[
-{\displaystyle \frac{1}{32 \mbox{$\slashed{z}$}^4}} \!
\left( \!
\mbox{$\slashed{z}$}
\!-\!
{\displaystyle \frac{3}{\mbox{$\slashed{z}$}}} \!
\right) \!
\left(
x^\star_\alpha x^\dagger_\beta
\!+\!
x^\star_\beta x^\dagger_\alpha
\right) \!
\right. \\
\\[-14pt]
&
\left.
-
{\displaystyle \frac{1}{8 \mbox{$\slashed{z}$}^2}} \!
\left( \!
\mbox{$\slashed{z}$}
\!-\!
{\displaystyle \frac{1}{\mbox{$\slashed{z}$}}} \!
\right) \!
\left(
x^\star_\alpha \erw{c_\beta }^{\!\!\!\!\!\!\star } ~~\!
\!+\!
x^\dagger_\beta
\erw{c_\alpha }^{\!\!\!\!\!\!\star } ~
\right)
\right] 
\!\approx\!
0 , \\
\\[-12pt]
&\bra0
\widetilde{U}^\dagger (G) \!
\left(
\hbox{\mathbf c}_\alpha \hbox{\mathbf c}_\beta
\!+\!
\hbox{\mathbf c}_\beta \hbox{\mathbf c}_\alpha
\right)
\widetilde{U}(G) \ket0
\!=\!
{\displaystyle \frac{1}{2}} \!
\left( \!
\mbox{$\slashed{z}$}
\!+\!
{\displaystyle \frac{1}{\mbox{$\slashed{z}$}}} \!
\right) \!
\left[
-{\displaystyle \frac{1}{32 \mbox{$\slashed{z}$}^4}} \!
\left( \!
\mbox{$\slashed{z}$}
\!-\!
{\displaystyle \frac{3}{\mbox{$\slashed{z}$}}} \!
\right) \!
\left(
x_\alpha x^{\mbox{\scriptsize T}}_\beta
\!+\!
x_\beta x^{\mbox{\scriptsize T}}_\alpha
\right)
\right. \\
\\[-14pt]
&
\left.
-
{\displaystyle \frac{1}{8 \mbox{$\slashed{z}$}^2}} \!
\left( \!
\mbox{$\slashed{z}$}
\!-\!
{\displaystyle \frac{1}{\mbox{$\slashed{z}$}}} \!
\right) \!
\left(
x^{\mbox{\scriptsize T}}_\alpha \erw{c_\beta }
\!+\!
x^{\mbox{\scriptsize T}}_\beta
\erw{c_\alpha }
\right)
\right]
\!\approx\!
0 ,
\EA
\label{Opanticommu2p}
\right\}
\ea\\[-12pt]
which are rewritten in compact forms as\\[-20pt]
\ba
\!\!\!\!
\left.
\BA{ll}
&
\left( \! 3 \mbox{$\slashed{z}$}^4
\!-\!
2 \mbox{$\slashed{z}$}^2
\!+\!
1 \! \right) \!
\delta_{\alpha \beta }
+\!
{\displaystyle \frac{1}{8 \mbox{$\slashed{z}$}^2}} \!
\left( \! \mbox{$\slashed{z}$}^2 \!-\! 3 \! \right) \!
x^\star_\alpha
x_\beta
\!=\!
{\displaystyle \frac{1}{2}} \!
\left( \! \mbox{$\slashed{z}$}^2 \!-\! 1 \! \right) \!
\left(
x_\beta \erw{c_\alpha }^{\!\!\!\!\!\!\star } ~
\!+\!
x^\star_\alpha \erw{c_\beta }
\right) , \\
\\[-12pt]
&
-{\displaystyle \frac{1}{2 \mbox{$\slashed{z}$}^2}} \!\!
\left( \!
\mbox{$\slashed{z}$}
\!-\!
{\displaystyle \frac{3}{\mbox{$\slashed{z}$}}} \!
\right) \!
x^\star_\alpha x^\star
_\beta
\!=\!
\left( \!
\mbox{$\slashed{z}$}
\!-\!
{\displaystyle \frac{1}{\mbox{$\slashed{z}$}}} \!
\right) \!\!
\left(
x^\star_\alpha \erw{c_\beta }^{\!\!\!\!\!\!\star } ~~\!
\!\!+\!\!
x^\star
_\beta
\erw{c_\alpha }^{\!\!\!\!\!\!\star } ~
\right) ~
\mbox{and the complex conjugate} .
\EA \!\!
\label{Opanticommu3}
\right\}
\ea
We call the demand mentioned above a
{\em quasi-anticommutation-relation approximation}
for
$\hbox{\mathbf c}_\alpha$ and $\hbox{\mathbf c}_\alpha^\dagger$,
proposed first in Ref.
\cite{Nishi.98}.
The quasi-anticommutation relation-approximation may work well
at the region very near $z \!=\! 1$, i.e.,
$\mbox{$\slashed{z}$}^2 \!=\! 1$.
To see more clearly the relation between SCF parameter $M$ and
Lagrange parameters $k$ and $l$,
we give another form of SCF condition for $M$
related directly to $k$ and $l$.
For instance,
multiplying by $k_{\beta \gamma }$ and $l_{\beta \gamma }$
the first equation of (\ref{Opanticommu3}) and its complex conjugation,
respectively and summing up with respect to $\gamma$ and
using the definition of $M_\alpha$ in
(\ref{SCFFandDandM}),
the conditions to determine the unknown parameters in the Lagrange multipliers for
$\mbox{$\slashed{z}$}^2 \!\approx\! 1$
are given as follows:\\[-20pt]
\ba
\!\!\!\!\!\!\!\!
\left.
\BA{ll}
&
2
{\displaystyle 
\frac{3 \mbox{$\slashed{z}$}^4 \!-\! 2 \mbox{$\slashed{z}$}^2 \!-\! 1}
{\mbox{$\slashed{z}$}^2 \!-\! 1}
}
\!\cdot\!
\left(
k_{\beta \alpha } x_\alpha
\!+\!
l_{\beta \alpha } x^\star_\alpha
\right)
\!+\!
{\displaystyle \frac{1}{4 \mbox{$\slashed{z}$}^2}}
{\displaystyle
\frac{\mbox{$\slashed{z}$}^2 \!-\! 3}
{\mbox{$\slashed{z}$}^2 \!-\! 1}
}
|x_\alpha|^2
\sum_\gamma
\left(
k_{\beta \gamma } x_\gamma
\!+\!
l_{\beta \gamma } x^\star_\gamma
\right) \\
\\[-12pt]
&-
\left(
x^\star_\alpha \erw{c_\alpha }
\right)
\sum_\gamma \!
l_{\beta \gamma } x_\gamma^\star
\!-\!
\left(
x_\alpha \erw{c_\alpha }^{\!\!\!\!\!\!\star } ~
\right)
\sum_\gamma
k_{\beta \gamma } x_\gamma 
\!=\!
|x_\alpha|^2 M_\beta
~~\hbox{(not summed for $\alpha$)} , \\
\\[-6pt]
&
-
\left\{
2 \mbox{$\slashed{z}$}^2 \!
\left(
\mbox{$\slashed{z}$}
\!-\!
{\displaystyle \frac{1}{\mbox{$\slashed{z}$}}}
\right)
\right\}^{-1} \!\!
\left(
\mbox{$\slashed{z}$}
\!-\!
{\displaystyle \frac{3}{\mbox{$\slashed{z}$}}}
\right)
\!\cdot\!
|x_\alpha|^2
\sum_\gamma
\left(
k_{\beta \gamma } x_\gamma
\!+\!
l_{\beta \gamma } x^\star_\gamma
\right) \\
\\[-12pt]
&
-
\left(
x^\star_\alpha \erw{c_\alpha }
\right)
\sum_\gamma k_{\beta \gamma } x_\gamma
\!-\!
\left(
x_\alpha \erw{c_\alpha }^{\!\!\!\!\!\!\star } ~
\right)
\sum_\gamma l_{\beta \gamma } x^\star_\gamma
\!=\!
|x_\alpha|^2 M_\beta
~~\hbox{(not summed for $\alpha$)} ,
\EA
\right\}
\label{Lagmultiplierkandl1}
\ea
\vspace{-0.4cm}
\ba
\!\!\!\!\!\!\!\!
\left.
\BA{ll}
&
2
{\displaystyle 
\frac{3 \mbox{$\slashed{z}$}^4 \!-\! 2 \mbox{$\slashed{z}$}^2 \!-\! 1}
{\mbox{$\slashed{z}$}^2 \!-\! 1}
}
\!\cdot\!
\left(
k^\star_{\alpha \beta } x^\star_\beta
\!+\!
l^\star_{\alpha \beta } x_\beta
\right)
\!+\!
{\displaystyle \frac{1}{4 \mbox{$\slashed{z}$}^2}}
{\displaystyle
\frac{\mbox{$\slashed{z}$}^2 \!-\! 3}
{\mbox{$\slashed{z}$}^2 \!-\! 1}
}
|x_\beta|^2
\sum_\gamma
\left(
k^\star_{\alpha \gamma } x^\star_\gamma
\!+\!
l^\star_{\alpha \gamma } x_\gamma
\right) \\
\\[-10pt]
&-
\left(
x^\star_\beta \erw{c_\beta }
\right)
\sum_\gamma \!
k^\star_{\alpha \gamma } x_\gamma^\star
\!-\!
\left(
x_\beta \erw{c_\beta }^{\!\!\!\!\!\!\star } ~
\right)
\sum_\gamma
l^\star_{\alpha \gamma } x_\gamma 
\!=\!
|x_\beta|^2 M^\star_\alpha
~~\hbox{(not summed for $\beta$)} , \\
\\[-6pt]
&
-
\left\{
2 \mbox{$\slashed{z}$}^2 \!
\left(
\mbox{$\slashed{z}$}
\!-\!
{\displaystyle \frac{1}{\mbox{$\slashed{z}$}}}
\right)
\right\}^{-1} \!\!
\left(
\mbox{$\slashed{z}$}
\!-\!
{\displaystyle \frac{3}{\mbox{$\slashed{z}$}}}
\right)
\!\cdot\!
|x_\beta|^2
\sum_\gamma
\left(
k_{\alpha \gamma } x_\gamma
\!+\!
l_{\alpha \gamma } x^\star_\gamma
\right) \\
\\[-10pt]
&
-
\left(
x^\star_\beta \erw{c_\beta }
\right)
\sum_\gamma k_{\alpha \gamma } x_\gamma
\!-\!
\left(
x_\beta \erw{c_\beta }^{\!\!\!\!\!\!\star } ~
\right)
\sum_\gamma l_{\alpha \gamma } x^\star_\gamma
\!=\!
|x_\beta|^2 M_\alpha
~~\hbox{(not summed for $\beta$)} ,
\EA
\right\}
\label{Lagmultiplierkandl2}
\ea
and their complex conjugations.
To write the conditions in symmetric forms with respect to
$\alpha$ and $\beta$,
we add
(\ref{Lagmultiplierkandl1}) and (\ref{Lagmultiplierkandl2}).
Then for
$\mbox{$\slashed{z}$}^2 \!\approx\! 1$
we have\\[-16pt]
\ba
\!\!\!\!\!\!\!\!
\BA{ll}
&
2
{\displaystyle 
\frac{3 \mbox{$\slashed{z}$}^4 \!-\! 2 \mbox{$\slashed{z}$}^2 \!-\! 1}
{\mbox{$\slashed{z}$}^2 \!-\! 1}
}
\!\cdot\!
\left\{
\left(
k_{\beta \alpha } x_\alpha
\!+\!
l_{\beta \alpha } x^\star_\alpha
\right)
\!+\!
\left(
k^\star_{\alpha \beta } x^\star_\beta
\!+\!
l^\star_{\alpha \beta } x_\beta
\right)
\right\} \\
\\[-8pt]
&
\!+\!
{\displaystyle \frac{1}{4 \mbox{$\slashed{z}$}^2}}
{\displaystyle
\frac{\mbox{$\slashed{z}$}^2 \!-\! 3}
{\mbox{$\slashed{z}$}^2 \!-\! 1}
}
\left\{
|x_\alpha|^2
\sum_\gamma \!
\left(
k_{\beta \gamma } x_\gamma
\!+\!
l_{\beta \gamma } x^\star_\gamma
\right)
\!+\!
|x_\beta|^2
\sum_\gamma \!
\left(
k^\star_{\alpha \gamma } x^\star_\gamma
\!+\!
l^\star_{\alpha \gamma } x_\gamma
\right) \!
\right\} \\
\\[-8pt]
&
\!-
x_\alpha \! \erw{c_\alpha }^{\!\!\!\!\!\!\star } ~
\sum_\gamma \!\!
k_{\beta \gamma } x_\gamma
\!-\!
x^\star_\alpha \! \erw{c_\alpha }
\sum_\gamma \!
l_{\beta \gamma } x_\gamma^\star
\!-\!
x_\beta \erw{c_\beta }^{\!\!\!\!\!\!\star } ~
\sum_\gamma \!\!
l^\star_{\alpha \gamma } x_\gamma
\!-\!
x^\star_\beta \erw{c_\beta }
\sum_\gamma \!\!
k^\star_{\alpha \gamma } x_\gamma^\star \\
\\[-4pt]
&
\!=\!
|x_\alpha|^2 M_\beta
\!+\!
|x_\beta|^2 M^\star_\alpha ,
~~~~~~~~~~~~~~~~~~~~~~~~~~~~~~~~~~~~~~~~~~~~~~~~~
\hbox{(not summed for $\alpha$ and $\beta$)} ,
\EA
\label{Lagmultiplierkandl3}
\ea
\ba
\!\!\!\!\!\!\!\!
\BA{ll}
&
-
{\displaystyle \frac{1}{2 \mbox{$\slashed{z}$}^2}}
{\displaystyle
\frac{\mbox{$\slashed{z}$}^2 \!-\! 3}
{\mbox{$\slashed{z}$}^2 \!-\! 1}
} \!
\left\{
|x_\alpha|^2
\sum_\gamma \!
\left(
k_{\beta \gamma } x_\gamma
\!+\!
l_{\beta \gamma } x^\star_\gamma
\right)
\!+\!
|x_\beta|^2
\sum_\gamma \!
\left(
k_{\alpha \gamma } x_\gamma
\!+\!
l_{\alpha \gamma } x^\star_\gamma
\right) \!
\right\} \\
\\[-10pt]
&
\!-
x^\star_\alpha \! \erw{c_\alpha }
\sum_\gamma \!\! k_{\beta \gamma } x_\gamma
\!-\!
x_\alpha \! \erw{c_\alpha }^{\!\!\!\!\!\!\star } ~
\sum_\gamma \! l_{\beta \gamma } x^\star_\gamma
\!-\!
x^\star_\beta \erw{c_\beta }
\sum_\gamma \!\! k_{\alpha \gamma } x_\gamma
\!-\!
x_\beta \erw{c_\beta }^{\!\!\!\!\!\!\star } ~
\sum_\gamma \! l_{\alpha \gamma } x^\star_\gamma \\
\\[-6pt]
&
\!=\!
|x_\alpha|^2 M_\beta
\!+\!
|x_\beta|^2 M_\alpha,
~~~~~~~~~~~~~~~~~~~~~~~~~~~~~~~~~~~~~~~~~~~~~~~~~
\hbox{(not summed for $\alpha$ and $\beta$)} .
\EA
\label{Lagmultiplierkandl4}
\ea\\[-12pt]
These conditions were first obtained in Ref.
\cite{Nishi.98}
but their expressions have been found to be wrong.
Now we have their correct forms.
Thus we could reach an ultimate goal to determine self-consistently
the Lagrange parameters $k_{\alpha \beta }$ and $l_{\alpha \beta }$.
This may play crucial roles for a unified description
of Bose-Fermi type collective excitations at a region very near
$\mbox{$\slashed{z}$}^2 \!\!\!=\!\!\! 1$,
since they are the necessary and sufficient conditions to determine
the unknown parameters in the Langrange multipliers.
To gain trust in this approximation, we should compare it with a
result of other Bose-Fermi theory in a concrete model.
Toward a possible determination of them for
$\mbox{$\slashed{z}$}^2 \!\approx\! 1$,
it is better to make rearrangements of
(\ref{Lagmultiplierkandl3})
and
(\ref{Lagmultiplierkandl4})
with respect to $k$ and $l$
in the following forms:\\[-22pt]
\ba
\!\!\!\!\!\!\!\!
\BA{ll}
&
2
{\displaystyle 
\frac{3 \mbox{$\slashed{z}$}^4 \!-\! 2 \mbox{$\slashed{z}$}^2 \!-\! 1}
{\mbox{$\slashed{z}$}^2 \!-\! 1}
}
\!\cdot\!
\left(
k_{\beta \alpha } x_\alpha
\!+\!
k^\star_{\alpha \beta } x^\star_\beta
\right) \\
\\[-12pt]
&
\!+\!
\left( \!
{\displaystyle \frac{1}{4 \mbox{$\slashed{z}$}^2}}
{\displaystyle
\frac{\mbox{$\slashed{z}$}^2 \!-\! 3}
{\mbox{$\slashed{z}$}^2 \!-\! 1}
}
|x_\alpha|^2
-
x_\alpha \erw{c_\alpha }^{\!\!\!\!\!\!\star }
\right) \!
\sum_\gamma
k_{\beta \gamma } x_\gamma
\!+\!
\left( \!
{\displaystyle \frac{1}{4 \mbox{$\slashed{z}$}^2}}
{\displaystyle
\frac{\mbox{$\slashed{z}$}^2 \!-\! 3}
{\mbox{$\slashed{z}$}^2 \!-\! 1}
}
|x_\beta|^2
\!-\!
x^\star_\beta \erw{c_\beta } \!
\right) \!
\sum_\gamma \!
k^\star_{\alpha \gamma } x_\gamma^\star \\
\\[-12pt]
&
+
2
{\displaystyle 
\frac{3 \mbox{$\slashed{z}$}^4 \!-\! 2 \mbox{$\slashed{z}$}^2 \!-\! 1}
{\mbox{$\slashed{z}$}^2 \!-\! 1}
}
\!\cdot\!
\left(
l_{\beta \alpha } x^\star_\alpha
\!+\!
l^\star_{\alpha \beta } x_\beta
\right) \\
\\[-10pt]
&
\!+\!
\left( \!
{\displaystyle \frac{1}{4 \mbox{$\slashed{z}$}^2}}
{\displaystyle
\frac{\mbox{$\slashed{z}$}^2 \!-\! 3}
{\mbox{$\slashed{z}$}^2 \!-\! 1}
}
|x_\alpha|^2
\!-\!
x^\star_\alpha \erw{c_\alpha } \!
\right) \!
\sum_\gamma
l_{\beta \gamma } x^\star_\gamma
\!+\!
\left( \!
{\displaystyle \frac{1}{4 \mbox{$\slashed{z}$}^2}}
{\displaystyle
\frac{\mbox{$\slashed{z}$}^2 \!-\! 3}
{\mbox{$\slashed{z}$}^2 \!-\! 1}
}
|x_\beta|^2
\!-\!
x_\beta \erw{c_\beta }^{\!\!\!\!\!\!\star }
\right) \!
\sum_\gamma
l^\star_{\alpha \gamma } x_\gamma \\
\\[-6pt]
&
\!=
|x_\alpha|^2 M_\beta
\!+\!
|x_\beta|^2 M^\star_\alpha ,
~~~~~~~~~~~~~~~~~~~~~~~~~~~~~~~~~~~~~~~~~~~~~~~~
\hbox{(not summed for $\alpha$ and $\beta$)} ,
\EA
\label{Lagmultiplierkandl5}
\ea
\vspace{-0.6cm}
\ba
\!\!\!\!\!\!\!\!
\BA{ll}
&
-
\left( \!
{\displaystyle \frac{1}{2 \mbox{$\slashed{z}$}^2}}
{\displaystyle
\frac{\mbox{$\slashed{z}$}^2 \!\!-\!\! 3}
{\mbox{$\slashed{z}$}^2 \!\!-\!\! 1}
}
|x_\alpha|^2 \!
\!-\!
x^\star_\alpha \erw{c_\alpha } \!
\right) \!
\sum_\gamma \!\! k_{\beta \gamma } x_\gamma
-
\left( \!
{\displaystyle \frac{1}{2 \mbox{$\slashed{z}$}^2}}
{\displaystyle
\frac{\mbox{$\slashed{z}$}^2 \!\!-\!\! 3}
{\mbox{$\slashed{z}$}^2 \!\!-\!\! 1}
}
|x_\beta|^2 \!
\!-\!
x^\star_\beta \erw{c_\beta } \!
\right) \!
\sum_\gamma \!\! k_{\alpha \gamma } x_\gamma \\
\\[-12pt]
&
-
\left( \!
{\displaystyle \frac{1}{2 \mbox{$\slashed{z}$}^2}}
{\displaystyle
\frac{\mbox{$\slashed{z}$}^2 \!\!-\!\! 3}
{\mbox{$\slashed{z}$}^2 \!\!-\!\! 1}
}
|x_\alpha|^2
\!-\!
x_\alpha \erw{c_\alpha }^{\!\!\!\!\!\!\star }
\right) \!
\sum_\gamma \! l_{\beta \gamma } x^\star_\gamma
\!-\!
\left( \!
{\displaystyle \frac{1}{2 \mbox{$\slashed{z}$}^2}}
{\displaystyle
\frac{\mbox{$\slashed{z}$}^2 \!\!-\!\! 3}
{\mbox{$\slashed{z}$}^2 \!\!-\!\! 1}
}
|x_\beta|^2
\!-\!
x_\beta \erw{c_\beta }^{\!\!\!\!\!\!\star }
\right) \!
\sum_\gamma \! l_{\alpha \gamma } x^\star_\gamma \\
\\[-6pt]
&
\!=
|x_\alpha|^2 M_\beta
\!+\!
|x_\beta|^2 M_\alpha
~~~~~~~~~~~~~~~~~~~~~~~~~~~~~~~~~~~~~~~~~~~~~~~~
\hbox{(not summed for $\alpha$ and $\beta$)} ,
\EA
\label{Lagmultiplierkandl6}
\ea\\[-12pt]
and their complex conjugations.
Owing to the symmetric forms with respect to $\alpha$ and $\beta$,
(\ref{Lagmultiplierkandl5}) and (\ref{Lagmultiplierkandl6})
can be separated into two parts. One is given as follows:\\[-22pt]
\ba
\!\!\!\!\!\!\!\!
\BA{ll}
&
2
{\displaystyle 
\frac{3 \mbox{$\slashed{z}$}^4 \!-\! 2 \mbox{$\slashed{z}$}^2 \!-\! 1}
{\mbox{$\slashed{z}$}^2 \!-\! 1}
}
\!\cdot\!
k_{\beta \alpha } x_\alpha
\!+\!
\left( \!
{\displaystyle \frac{1}{4 \mbox{$\slashed{z}$}^2}}
{\displaystyle
\frac{\mbox{$\slashed{z}$}^2 \!-\! 3}
{\mbox{$\slashed{z}$}^2 \!-\! 1}
}
|x_\alpha|^2
-
x_\alpha \erw{c_\alpha }^{\!\!\!\!\!\!\star }
\right) \!
\sum_\gamma
k_{\beta \gamma } x_\gamma \\
\\[-10pt]
&
+
2
{\displaystyle 
\frac{3 \mbox{$\slashed{z}$}^4 \!-\! 2 \mbox{$\slashed{z}$}^2 \!-\! 1}
{\mbox{$\slashed{z}$}^2 \!-\! 1}
}
\!\cdot\!
l_{\beta \alpha } x^\star_\alpha
\!+\!
\left( \!
{\displaystyle \frac{1}{4 \mbox{$\slashed{z}$}^2}}
{\displaystyle
\frac{\mbox{$\slashed{z}$}^2 \!-\! 3}
{\mbox{$\slashed{z}$}^2 \!-\! 1}
}
|x_\alpha|^2
\!-\!
x^\star_\alpha \erw{c_\alpha } \!
\right) \!
\sum_\gamma
l_{\beta \gamma } x^\star_\gamma \\
\\[-6pt]
&
\!=
|x_\alpha|^2 M_\beta ,
~~~~~~~~~~~~~~~~~~~~~~~~~~~~~~~~~~~~~~~~~~~~~~~~~~~~~~~~~~~
\hbox{(not summed for $\alpha$ and $\beta$)} ,
\EA
\label{Lagmultiplierkandl5r}
\ea
\vspace{-0.6cm}
\ba
\!\!\!\!\!\!\!\!
\BA{ll}
&
-
\left( \!
{\displaystyle \frac{1}{2 \mbox{$\slashed{z}$}^2}}
{\displaystyle
\frac{\mbox{$\slashed{z}$}^2 \!\!-\!\! 3}
{\mbox{$\slashed{z}$}^2 \!\!-\!\! 1}
}
|x_\alpha|^2 \!
\!-\!
x^\star_\alpha \erw{c_\alpha } \!
\right) \!
\sum_\gamma \!\! k_{\beta \gamma } x_\gamma
-
\left( \!
{\displaystyle \frac{1}{2 \mbox{$\slashed{z}$}^2}}
{\displaystyle
\frac{\mbox{$\slashed{z}$}^2 \!\!-\!\! 3}
{\mbox{$\slashed{z}$}^2 \!\!-\!\! 1}
}
|x_\alpha|^2
\!-\!
x_\alpha \erw{c_\alpha }^{\!\!\!\!\!\!\star }
\right) \!
\sum_\gamma \! l_{\beta \gamma } x^\star_\gamma \\
\\[-6pt]
&
\!=
|x_\alpha|^2 M_\beta
~~~~~~~~~~~~~~~~~~~~~~~~~~~~~~~~~~~~~~~~~~~~~~~~~~~~~~~~~~~~
\hbox{(not summed for $\alpha$ and $\beta$)} .
\EA
\label{Lagmultiplierkandl6r}
\ea\\[-12pt]
The other is obtained by taking the complex conjugation and
making the exchange between the indices $\alpha$ and
$\beta$ in
(\ref{Lagmultiplierkandl5r})
and
only by making the exchange betwenn the indices $\alpha$ and $\beta$ in
(\ref{Lagmultiplierkandl6r}),
respectively.

We are also able to consider another approximate
$SO(2N \!+\! 1)$ WF
$\widetilde{U}(G) \ket0$
at a region very near $z \!=\! 0$.
At that region the $SO(2N \!+\! 1)$ WF
has a peculiar feature that brings out the largest contributions
from the unpaired modes owing to the relation
$x_\alpha^\star x_\alpha \!+\! z^2 \!=\! 1$.
We have never experienced such a physical situation.
Then, an investigation of such a interesting problem must be meaninigful.
At the region very near $z \!\!=\!\! 0$,
the
$
\mbox{$\slashed{z}$}
$
is approximated to
$
{\displaystyle \frac{1}{\sqrt{2}}} \!
\left( \!
{\displaystyle \frac{1}{2}}
\!\!+\!\!
{\displaystyle \frac{1\!\!+\!\!z}{2}} \!
\right)
(z \!\!\approx\!\! 0)
$.
Introducing an approximate vacuum function
$
\widetilde{\Phi }_{00}^\star(G)
\!\!=\!\!
{\displaystyle \frac{1}{\sqrt{2}}} \!
\left\{ \!
{\displaystyle \frac{1}{2}}
\Phi^\star_{00}(g)
\!\!+\!\!
\mbox{$\slashed{z}$}
\Phi^\star_{00}(G) \!
\right\}
\!\!=\!\!
{\displaystyle \frac{1}{\sqrt{2}}} \!
\left( \!
\mbox{$\slashed{z}$}
\!\!+\!\!
{\displaystyle \frac{1}{2\mbox{$\slashed{z}$}}} \!
\right) \!
\Phi^\star_{00}(G) 
$,
we adopt an approximate
$SO(2N\!+\!1)$ WF expressed as\\[-20pt]
\ba
\widetilde{U}(G) \ket0
\!=\!
{\displaystyle
\frac{\widetilde{\Phi }^\star_{00}(G)}{\Phi^\star_{00}(G)}
}
U(G) \ket0
\!=\!
{\displaystyle \frac{1}{\sqrt{2}}} \!
\left( \!
\mbox{$\slashed{z}$}
\!+\!
{\displaystyle \frac{1}{2\mbox{$\slashed{z}$}}} \!
\right) \!
U(G) \ket0 .
\label{0approxwf}
\ea\\[-14pt]
Using
(\ref{0approxwf}) and (\ref{calphadiffrential}),
we get approximate identities\\[-18pt]
\ba
\!\!\!\!\!\!\!\!\!\!\!\!\!
\left.
\BA{ll}
&\hbox{\mathbf c}_\alpha \widetilde{U}(G) \! \ket0
\!\!\approx\!\!
{\displaystyle \frac{1}{\sqrt{2}}} \!\!
\left( \!\!
\mbox{$\slashed{z}$}
\!\!+\!\!
{\displaystyle \frac{1}{2\mbox{$\slashed{z}$}}} \!\!
\right) \!\!
\hbox{\mathbf c}_\alpha
U(G) \! \ket0
\!\!-\!\!
{\displaystyle \frac{1}{2\sqrt{2}}} \!
\left( \!\!
\mbox{$\slashed{z}$}
\!\!-\!\!
{\displaystyle \frac{1}{2\mbox{$\slashed{z}$}}} \!\!
\right) \!\!
\left\{ \!
r^{\!\mbox{\scriptsize T}} \! (1\!\!+\!\!q^\dagger q)^{\!-1}
\!\!+\!\!
r^\dagger \! q(1\!\!+\!\!q^\dagger q)^{\!-1} \!
\right\}_{\!\alpha } \!
U(G) \! \ket0 , \\
\\[-12pt]
&\hbox{\mathbf c}^\dagger_\alpha \widetilde{U}(G) \! \ket0
\!\!\approx\!\!
{\displaystyle \frac{1}{\sqrt{2}}} \!\!
\left( \!\!
\mbox{$\slashed{z}$}
\!\!+\!\!
{\displaystyle \frac{1}{2\mbox{$\slashed{z}$}}} \!\!
\right) \!\!
\hbox{\mathbf c}^\dagger_\alpha
U(G) \! \ket0
\!\!+\!\!
{\displaystyle \frac{1}{2\sqrt{2}}} \!\!
\left( \!\!
\mbox{$\slashed{z}$}
\!\!-\!\!
{\displaystyle \frac{1}{\mbox{2$\slashed{z}$}}} \!\!
\right) \!\!
\left\{ \!
r^\dagger \! (1\!\!+\!\!qq^\dagger )^{\!-1}
\!\!-\!\!
r^{\!\mbox{\scriptsize T}} \! q^\dagger (1\!\!+\!\!qq^\dagger )^{\!-1} \!
\right\}_{\!\alpha } \!\!
U(G) \! \ket0 .
\EA \!\!\!\!
\right\}
\label{0approxidentities}
\ea\\[-12pt]
With the aid of
(\ref{calphadiffrentialproof2}),
successive operations of the differential form
(\ref{Lieopratorc2})
for the fermion operators
$\hbox{\mathbf c}_\alpha$ and
$\hbox{\mathbf c}_\alpha^\dagger$ on the approximate identities
lead to the formulas
\\[-18pt]
\ba
\!\!\!\!\!\!\!\!\!\!
\left.
\BA{lll}
&\hbox{\mathbf c}^\dagger_\alpha
\hbox{\mathbf c}_\beta \widetilde{U}(G) \ket0
\!\!
&\!\!=\!
\left[
{\displaystyle \frac{1}{\sqrt{2}}} \!
\left( \!
\mbox{$\slashed{z}$}
\!+\!
{\displaystyle \frac{1}{2\mbox{$\slashed{z}$}}} \!
\right) \!
\hbox{\mathbf c}^\dagger_\alpha \hbox{\mathbf c}_\beta
\!\!+\!\!
{\displaystyle \frac{1}{\sqrt{2}}} \!
\left( \!
\mbox{$\slashed{z}$}
\!-\!
{\displaystyle \frac{1}{2\mbox{$\slashed{z}$}}} \!
\right) \!\!
\left( \!\!
1 \!\!-\!\! {\displaystyle \frac{1}{2 \mbox{$\slashed{z}$}^2}} \!\!
\right) \!
\delta_{\alpha \beta }
\!\!+\!\!
{\displaystyle \frac{1}
{16\sqrt{2} \mbox{$\slashed{z}$}^4}} \!\!
\left( \!
\mbox{$\slashed{z}$}
\!-\!
{\displaystyle \frac{3}{2\mbox{$\slashed{z}$}}} \!
\right) \!
x^\star_\alpha x^{\mbox{\scriptsize T}}_\beta
\right. \\
\\[-14pt]
&\!\!&
\left.
~~~~~~~~~~~~~~~~~~~~~
-
{\displaystyle \frac{1}{4\sqrt{2} \mbox{$\slashed{z}$}^2}} \!
\left( \!
\mbox{$\slashed{z}$}
\!-\!
{\displaystyle \frac{1}{2\mbox{$\slashed{z}$}}} \!
\right) \!
\left(
x^{\mbox{\scriptsize T}}_\beta
\hbox{\mathbf c}^\dagger_\alpha
\!-\!
x^\star_\alpha
\hbox{\mathbf c}_\beta
\right)
\right] \!
U(G) \ket0 , \\
\\[-12pt]
&\hbox{\mathbf c}_\beta
\hbox{\mathbf c}^\dagger_\alpha \widetilde{U}(G) \ket0
\!\!
&\!\!=\!
\left[
{\displaystyle \frac{1}{\sqrt{2}}} \!
\left( \!
\mbox{$\slashed{z}$}
\!+\!
{\displaystyle \frac{1}{2\mbox{$\slashed{z}$}}} \!
\right) \!
\hbox{\mathbf c}_\beta \hbox{\mathbf c}^\dagger_\alpha
\!\!+\!\!
{\displaystyle \frac{1}{\sqrt{2}}} \!
\left( \!
\mbox{$\slashed{z}$}
\!-\!
{\displaystyle \frac{1}{2\mbox{$\slashed{z}$}}} \!
\right) \!\!
\left( \!\!
1 \!\!-\!\! {\displaystyle \frac{1}{2 \mbox{$\slashed{z}$}^2}} \!\!
\right) \!
\delta_{\alpha \beta }
\!\!+\!\!
{\displaystyle \frac{1}
{16\sqrt{2} \mbox{$\slashed{z}$}^4}} \!\!
\left( \!
\mbox{$\slashed{z}$}
\!-\!
{\displaystyle \frac{3}{2\mbox{$\slashed{z}$}}} \!
\right) \!
x^\star_\alpha x^{\mbox{\scriptsize T}}_\beta
\right. \\
\\[-14pt]
&\!\!&
\left.
~~~~~~~~~~~~~~~~~~~~~
-
{\displaystyle \frac{1}{4\sqrt{2} \mbox{$\slashed{z}$}^2}} \!
\left( \!
\mbox{$\slashed{z}$}
\!-\!
{\displaystyle \frac{1}{2\mbox{$\slashed{z}$}}} \!
\right) \!
\left(
x^{\mbox{\scriptsize T}}_\beta
\hbox{\mathbf c}^\dagger_\alpha
\!-\!
x^\star_\alpha
\hbox{\mathbf c}_\beta
\right)
\right] \!
U(G) \ket0 ,
\EA \!\!
\right\}
\label{0approxidentities2}
\ea
\vspace{-0.5cm}
\ba
\!\!\!\!\!\!\!\!\!\!\!\!\!\!\!\!\!\!\!\!\!\!\!\!
\left.
\BA{lll}
&\hbox{\mathbf c}^\dagger_\alpha
\hbox{\mathbf c}^\dagger_\beta \widetilde{U}(G) \ket0
\!\!
&\!\!=\!
\left[
{\displaystyle \frac{1}{\sqrt{2}}} \!
\left( \!
\mbox{$\slashed{z}$}
\!+\!
{\displaystyle \frac{1}{2\mbox{$\slashed{z}$}}} \!
\right) \!
\hbox{\mathbf c}^\dagger_\alpha \hbox{\mathbf c}^\dagger_\beta
\!-\!
{\displaystyle \frac{1}{16\sqrt{2} \mbox{$\slashed{z}$}^4}} \!
\left( \!
\mbox{$\slashed{z}$}
\!-\!
{\displaystyle \frac{3}{2\mbox{$\slashed{z}$}}} \!
\right) \!
x^\star_\alpha x^\dagger_\beta
\right. \\
\\[-14pt]
&\!\!&
\left.
~~~~~~~~~~~~~~~~~~~~~
+
{\displaystyle \frac{1}{4\sqrt{2} \mbox{$\slashed{z}$}^2}} \!
\left( \!
\mbox{$\slashed{z}$}
\!-\!
{\displaystyle \frac{1}{2\mbox{$\slashed{z}$}}} \!
\right) \!
\left(
x^\star_\alpha
\hbox{\mathbf c}^\dagger_\beta
\!+\!
x^\dagger_\beta
\hbox{\mathbf c}^\dagger_\alpha
\right)
\right] \!
U(G) \ket0 , \\
\\[-12pt]
&\hbox{\mathbf c}_\alpha
\hbox{\mathbf c}_\beta \widetilde{U}(G) \ket0
\!\!
&\!\!=\!
\left[
{\displaystyle \frac{1}{\sqrt{2}}} \!
\left( \!
\mbox{$\slashed{z}$}
\!+\!
{\displaystyle \frac{1}{2\mbox{$\slashed{z}$}}} \!
\right) \!
\hbox{\mathbf c}_\alpha \hbox{\mathbf c}_\beta
\!-\!
{\displaystyle \frac{1}{16\sqrt{2} \mbox{$\slashed{z}$}^4}} \!
\left( \!
\mbox{$\slashed{z}$}
\!-\!
{\displaystyle \frac{3}{2\mbox{$\slashed{z}$}}} \!
\right) \!
x_\alpha x^{\mbox{\scriptsize T}}_\beta
\right. \\
\\[-14pt]
&\!\!&
\left.
~~~~~~~~~~~~~~~~~~~~~
-
{\displaystyle \frac{1}{4\sqrt{2} \mbox{$\slashed{z}$}^2}} \!
\left( \!
\mbox{$\slashed{z}$}
\!-\!
{\displaystyle \frac{1}{2\mbox{$\slashed{z}$}}} \!
\right) \!
\left(
x_\alpha
\hbox{\mathbf c}_\beta
\!+\!
x^{\mbox{\scriptsize T}}_\beta
\hbox{\mathbf c}_\alpha
\right)
\right] \!
U(G) \ket0 .
\EA
\right\}
\label{0approxidentities3}
\ea\\[-12pt]
Using the set of
(\ref{0approxidentities2})
and
(\ref{0approxidentities3}),
operations of the anticommutators of the fermion operators
$\hbox{\mathbf c}_\alpha$ and
$\hbox{\mathbf c}_\alpha^\dagger$ on the approximated
$\widetilde U(G) \ket0$
are made as follows:\\[-18pt]
\ba
\!\!\!\!\!\!\!\!\!\!
\left.
\BA{ll}
&
\left(
\hbox{\mathbf c}_\alpha^\dagger \hbox{\mathbf c}_\beta
\!+\!
\hbox{\mathbf c}_\beta \hbox{\mathbf c}_\alpha^\dagger
\right) \!
\widetilde{U}(G) \ket0 
\!\cong\!
\left[
{\displaystyle \frac{1}{8\sqrt{2} \mbox{$\slashed{z}$}^4}} \!\!
\left( \!
\mbox{$\slashed{z}$}
\!-\!
{\displaystyle \frac{3}{2\mbox{$\slashed{z}$}}} \!
\right) \!
x^\star_\alpha x^{\mbox{\scriptsize T}}_\beta
\right. \\
\\[-14pt]
&
\left.
+
{\displaystyle \frac{1}{\sqrt{2}}} \!\!
\left( \!
\mbox{$\slashed{z}$}
\!+\!
{\displaystyle \frac{1}{2\mbox{$\slashed{z}$}}} \!
\right) \!
\delta_{\alpha \beta }
\!+\!
\sqrt{2} \!
\left( \!
\mbox{$\slashed{z}$}
\!-\!
{\displaystyle \frac{1}{2\mbox{$\slashed{z}$}}} \!
\right) \!\!
\left\{ \!\!
\left( \!
1 \!-\! {\displaystyle \frac{1}{2 \mbox{$\slashed{z}$}^2}} \!
\right) \!
\delta_{\alpha \beta }
\!-\!
{\displaystyle \frac{1}{4 \mbox{$\slashed{z}$}^2}} \!
\left(
x^{\mbox{\scriptsize T}}_\beta \hbox{\mathbf c}^\dagger_\alpha
\!-\!
x^\star_\alpha \hbox{\mathbf c}_\beta
\right) \!
\right\} \!
\right] \!
U(G) \ket0 , \\
\\[-12pt]
&
\left( \!
\hbox{\mathbf c}_\alpha^\dagger \hbox{\mathbf c}_\beta^\dagger
\!+\!
\hbox{\mathbf c}_\beta^\dagger \hbox{\mathbf c}_\alpha^\dagger \!
\right) \!
\widetilde{U}(G) \ket0 \\
\\[-14pt]
&
\!\cong\!
\left[
-{\displaystyle \frac{1}{16\sqrt{2} \mbox{$\slashed{z}$}^4}} \!
\left( \!
\mbox{$\slashed{z}$}
\!-\!
{\displaystyle \frac{3}{2\mbox{$\slashed{z}$}}} \!
\right) \!\!
\left( \!
x^\star_\alpha x^\dagger_\beta
\!+\!
x^\star_\beta x^\dagger_\alpha \!
\right)
\!\!+\!\!
{\displaystyle \frac{1}{2\sqrt{2} \mbox{$\slashed{z}$}^2}} \!
\left( \!
\mbox{$\slashed{z}$}
\!-\!
{\displaystyle \frac{1}{2\mbox{$\slashed{z}$}}} \!
\right) \!\!
\left( \!
x^\star_\alpha \hbox{\mathbf c}^\dagger_\beta
\!+\!
x^\dagger_\beta
\hbox{\mathbf c}^\dagger_\alpha \!
\right)
\right] \!
U(G) \ket0  \\
\\[-10pt]
&\mbox{and the complex conjugate} .
\EA \!\!
\right\}
\label{0Opanticommu}
\ea\\[-12pt]
The expectation values of the anticommutators between
$\hbox{\mathbf c}_\alpha$
and
$\hbox{\mathbf c}^\dagger_\alpha$
by
$\widetilde{U}(G) \ket0$
should satifsy
(\ref{quasianticommurels}),
from which and
(\ref{modifiedidentities}),
we can obtain the following approximate relations:\\[-18pt]
\ba
\!\!\!\!\!\!\!\!\!\!\!\!
\left.
\BA{ll}
&\bra0
\widetilde{U}^\dagger (G) \!
\left(
\hbox{\mathbf c}_\alpha^\dagger \hbox{\mathbf c}_\beta
\!+\!
\hbox{\mathbf c}_\beta \hbox{\mathbf c}_\alpha^\dagger
\right) \!
\widetilde{U}(G) \ket0 
\!=\!
{\displaystyle \frac{1}{\sqrt{2}}} \!
\left( \!\!
\mbox{$\slashed{z}$}
\!+\!
{\displaystyle \frac{1}{2\mbox{$\slashed{z}$}}} \!
\right) \!\!
\left[ \!
{\displaystyle \frac{1}{\sqrt{2}}} \!
\left( \!\!
\mbox{$\slashed{z}$}
\!\!+\!\!
{\displaystyle \frac{1}{2\mbox{$\slashed{z}$}}} \!
\right) \!
\delta_{\alpha \beta }
\!\!+\!\!
{\displaystyle \frac{1}{8\sqrt{2} \mbox{$\slashed{z}$}^4}} \!
\left( \!\!
\mbox{$\slashed{z}$}
\!-\!
{\displaystyle \frac{3}{2\mbox{$\slashed{z}$}}} \!
\right) \!\!
x^\star_\alpha
x^{\mbox{\scriptsize T}}_\beta
\right. \\
\\[-14pt]
&
\left.
\!+\!
\sqrt{2} \!
\left( \!\!
\mbox{$\slashed{z}$}
\!-\!
{\displaystyle \frac{1}{2\mbox{$\slashed{z}$}}} \!
\right) \!\!
\left\{ \!\!
\left( \!
1 \!\!-\! {\displaystyle \frac{1}{2 \mbox{$\slashed{z}$}^2}} \!
\right) \!
\delta_{\alpha \beta }
\!-\!
{\displaystyle \frac{1}{4 \mbox{$\slashed{z}$}^2}}
\left(
x^{\mbox{\scriptsize T}}_\beta \erw{c_\alpha }^{\!\!\!\!\!\!\star } ~
\!+\!
x^\star_\alpha \erw{c_\beta }
\right) \!
\right\} \!
\right]
\!\approx\!
\delta_{\alpha \beta } , \\
\\[-12pt]
&\bra0
\widetilde{U}^\dagger (G) \!
\left(
\hbox{\mathbf c}_\alpha^\dagger \hbox{\mathbf c}_\beta^\dagger
\!+\!
\hbox{\mathbf c}_\beta^\dagger \hbox{\mathbf c}_\alpha^\dagger
\right)
\widetilde{U}(G) \ket0
\!=\!
{\displaystyle \frac{1}{\sqrt{2}}} \!
\left( \!
\mbox{$\slashed{z}$}
\!+\!
{\displaystyle \frac{1}{2\mbox{$\slashed{z}$}}} \!
\right) \!\!
\left[
-{\displaystyle \frac{1}{16\sqrt{2} \mbox{$\slashed{z}$}^4}} \!
\left( \!
\mbox{$\slashed{z}$}
\!-\!
{\displaystyle \frac{3}{\mbox{2$\slashed{z}$}}} \!
\right) \!
\left(
x^\star_\alpha x^\dagger_\beta
\!+\!
x^\star_\beta x^\dagger_\alpha
\right) \!
\right. \\
\\[-14pt]
&
\left.
-
{\displaystyle \frac{1}{2\sqrt{2} \mbox{$\slashed{z}$}^2}} \!
\left( \!
\mbox{$\slashed{z}$}
\!-\!
{\displaystyle \frac{1}{2\mbox{$\slashed{z}$}}} \!
\right) \!
\left(
x^\star_\alpha \erw{c_\beta }^{\!\!\!\!\!\!\star } ~~\!
\!+\!
x^\dagger_\beta
\erw{c_\alpha }^{\!\!\!\!\!\!\star } ~
\right)
\right] 
\!\approx\!
0~
\mbox{and the complex conjugate} ,
\EA \!\!\!\!
\label{0Opanticommu2}
\right\}
\ea\\[-10pt]
which are rewritten in compact forms as\\[-22pt]
\ba
\!\!\!\!\!\!\!\!
\left.
\BA{ll}
&
2 \!
\left( \!
6 \mbox{$\slashed{z}$}^4
\!-\!
3 \mbox{$\slashed{z}$}^2
\!+\!
1 \!
\right) \!
\delta_{\alpha \beta }
+\!
{\displaystyle \frac{1}{4 \mbox{$\slashed{z}$}^2}} \!
\left( \!
2\mbox{$\slashed{z}$}^2
\!-\!
3
\! \right) \!
x^\star_\alpha
x_\beta
\!=\!
\left( \!
2\mbox{$\slashed{z}$}^2
\!-\!
1
\! \right) \!
\left(
x_\beta \erw{c_\alpha }^{\!\!\!\!\!\!\star } ~
\!+\!
x^\star_\alpha \erw{c_\beta }
\right) , \\
\\[-14pt]
&
-{\displaystyle \frac{1}{8 \mbox{$\slashed{z}$}^2}} \!
\left( \!
2\mbox{$\slashed{z}$}^2
\!-\!
3 \!
\right) \!
x^\star_\alpha x^\star
_\beta
\!=\!
\left( \!
2\mbox{$\slashed{z}$}^2
\!-\!
1 \!
\right) \!
\left(
x^\star_\alpha \erw{c_\beta }^{\!\!\!\!\!\!\star } ~~\!
\!+\!
x^\star
_\beta
\erw{c_\alpha }^{\!\!\!\!\!\!\star } ~
\right) ~\!
\mbox{and the complex conjugate} ,
\EA \!\!\!
\label{0Opanticommu3}
\right\}
\ea\\[-14pt]
which is converted into the following conditions to determine the parameters
$k$ and $l$:\\[-14pt]
\ba
\!\!\!\!\!\!\!\!
\left.
\BA{ll}
&
2
{\displaystyle
\frac{
6 \mbox{$\slashed{z}$}^4
\!-\!
3 \mbox{$\slashed{z}$}^2
\!+\!
1 \!
}
{2\mbox{$\slashed{z}$}^2
\!-\!
1
}
}
\!\cdot\!
\left(
k_{\beta \alpha } x_\alpha
\!+\!
l_{\beta \alpha } x^\star_\alpha
\right)
\!+\!
{\displaystyle \frac{1}{4 \mbox{$\slashed{z}$}^2}}
{\displaystyle
\frac{2\mbox{$\slashed{z}$}^2
\!-\!
3}
{2\mbox{$\slashed{z}$}^2
\!-\!
1}
}
|x_\alpha|^2
\sum_\gamma
\left(
k_{\beta \gamma } x_\gamma
\!+\!
l_{\beta \gamma } x^\star_\gamma
\right) \\
\\[-10pt]
&-
\left(
x^\star_\alpha \erw{c_\alpha }
\right)
\sum_\gamma \!
l_{\beta \gamma } x_\gamma^\star
\!-\!
\left(
x_\alpha \erw{c_\alpha }^{\!\!\!\!\!\!\star } ~
\right)
\sum_\gamma
k_{\beta \gamma } x_\gamma 
\!=\!
|x_\alpha|^2 M_\beta
~~\hbox{(not summed for $\alpha$)} , \\
\\[-10pt]
&
-
{\displaystyle \frac{1}{8 \mbox{$\slashed{z}$}^2}}
{\displaystyle
\frac{2\mbox{$\slashed{z}$}^2
\!-\!
3}
{2\mbox{$\slashed{z}$}^2
\!-\!
1}
}
\!\cdot\!
|x_\alpha|^2
\sum_\gamma
\left(
k_{\beta \gamma } x_\gamma
\!+\!
l_{\beta \gamma } x^\star_\gamma
\right) \\
\\[-10pt]
&
-
\left(
x^\star_\alpha \erw{c_\alpha }
\right)
\sum_\gamma k_{\beta \gamma } x_\gamma
\!-\!
\left(
x_\alpha \erw{c_\alpha }^{\!\!\!\!\!\!\star } ~
\right)
\sum_\gamma l_{\beta \gamma } x^\star_\gamma
\!=\!
|x_\alpha|^2 M_\beta
~~\hbox{(not summed for $\alpha$)} ,
\EA
\right\}
\label{0Lagmultiplierkandl1}
\ea
\vspace{-0.4cm}
\ba
\!\!\!\!\!\!\!\!
\left.
\BA{ll}
&
2
{\displaystyle
\frac{
6 \mbox{$\slashed{z}$}^4
\!-\!
3 \mbox{$\slashed{z}$}^2
\!+\!
1 \!
}
{2\mbox{$\slashed{z}$}^2
\!-\!
1
}
}
\!\cdot\!
\left(
k^\star_{\alpha \beta } x^\star_\beta
\!+\!
l^\star_{\alpha \beta } x_\beta
\right)
\!+\!
{\displaystyle \frac{1}{4 \mbox{$\slashed{z}$}^2}}
{\displaystyle
\frac{2\mbox{$\slashed{z}$}^2
\!-\!
3}
{2\mbox{$\slashed{z}$}^2
\!-\!
1}
}
|x_\beta|^2
\sum_\gamma
\left(
k^\star_{\alpha \gamma } x^\star_\gamma
\!+\!
l^\star_{\alpha \gamma } x_\gamma
\right) \\
\\[-10pt]
&-
\left(
x^\star_\beta \erw{c_\beta }
\right)
\sum_\gamma \!
k^\star_{\alpha \gamma } x_\gamma^\star
\!-\!
\left(
x_\beta \erw{c_\beta }^{\!\!\!\!\!\!\star } ~
\right)
\sum_\gamma
l^\star_{\alpha \gamma } x_\gamma 
\!=\!
|x_\beta|^2 M^\star_\alpha
~~\hbox{(not summed for $\beta$)} , \\
\\[-10pt]
&
-
{\displaystyle \frac{1}{8 \mbox{$\slashed{z}$}^2}}
{\displaystyle
\frac{2\mbox{$\slashed{z}$}^2
\!-\!
3}
{2\mbox{$\slashed{z}$}^2
\!-\!
1}
}\!\cdot\!
|x_\beta|^2
\sum_\gamma
\left(
k_{\alpha \gamma } x_\gamma
\!+\!
l_{\alpha \gamma } x^\star_\gamma
\right) \\
\\[-10pt]
&
-
\left(
x^\star_\beta \erw{c_\beta }
\right)
\sum_\gamma k_{\alpha \gamma } x_\gamma
\!-\!
\left(
x_\beta \erw{c_\beta }^{\!\!\!\!\!\!\star } ~
\right)
\sum_\gamma l_{\alpha \gamma } x^\star_\gamma
\!=\!
|x_\beta|^2 M_\alpha
~~\hbox{(not summed for $\beta$)} ,
\EA
\right\}
\label{0Lagmultiplierkandl2}
\ea\\[-6pt]
and their complex conjugations.
Parallel to the way to get the conditions
(\ref{Lagmultiplierkandl3})
and
(\ref{Lagmultiplierkandl4}),
from
(\ref{0Lagmultiplierkandl1}) and
(\ref{0Lagmultiplierkandl2}),
we also have the symmetric conditions
with respect to
$\alpha$ and $\beta$.
Toward a possible determination of the unknown parameters for
$\mbox{$\slashed{z}$}^2 \!\approx\! \frac{1}{2}$,
we make again rearrangements of the conditions
with respect to $k$ and $l$
in the following forms:\\[-16pt]
\ba
\!\!\!\!\!\!\!\!
\BA{ll}
&
2 \!
{\displaystyle
\frac{
6 \mbox{$\slashed{z}$}^4
\!-\!
3 \mbox{$\slashed{z}$}^2
\!+\!
1 \!
}
{2\mbox{$\slashed{z}$}^2
\!-\!
1
}
}
\!\cdot\!
\left(
k_{\beta \alpha } x_\alpha
\!+\!
k^\star_{\alpha \beta } x^\star_\beta
\right) \\
\\[-12pt]
&
\!+\!
\left( \!
{\displaystyle \frac{1}{4 \mbox{$\slashed{z}$}^2}}
{\displaystyle
\frac{2\mbox{$\slashed{z}$}^2 \!-\! 3}
{\mbox{2$\slashed{z}$}^2 \!-\! 1}
}
|x_\alpha|^2
-
x_\alpha \erw{c_\alpha }^{\!\!\!\!\!\!\star }
\right) \!\!
\sum_\gamma
k_{\beta \gamma } x_\gamma
\!+\!
\left( \!
{\displaystyle \frac{1}{4 \mbox{$\slashed{z}$}^2}}
{\displaystyle
\frac{2\mbox{$\slashed{z}$}^2 \!-\! 3}
{2\mbox{$\slashed{z}$}^2 \!-\! 1}
}
|x_\beta|^2
\!-\!
x^\star_\beta \erw{c_\beta } \!
\right) \!\!
\sum_\gamma \!
k^\star_{\alpha \gamma } x_\gamma^\star \\
\\[-10pt]
&
+
2 \!
{\displaystyle
\frac{
6 \mbox{$\slashed{z}$}^4
\!-\!
3 \mbox{$\slashed{z}$}^2
\!+\!
1 \!
}
{2\mbox{$\slashed{z}$}^2
\!-\!
1
}
}
\!\cdot\!
\left(
l_{\beta \alpha } x^\star_\alpha
\!+\!
l^\star_{\alpha \beta } x_\beta
\right) \\
\\[-12pt]
&
\!+\!
\left( \!
{\displaystyle \frac{1}{4 \mbox{$\slashed{z}$}^2}}
{\displaystyle
\frac{2\mbox{$\slashed{z}$}^2 \!-\! 3}
{2\mbox{$\slashed{z}$}^2 \!-\! 1}
}
|x_\alpha|^2
\!-\!
x^\star_\alpha \erw{c_\alpha } \!
\right) \!\!
\sum_\gamma
l_{\beta \gamma } x^\star_\gamma
\!+\!
\left( \!
{\displaystyle \frac{1}{4 \mbox{$\slashed{z}$}^2}}
{\displaystyle
\frac{2\mbox{$\slashed{z}$}^2 \!-\! 3}
{\mbox{2$\slashed{z}$}^2 \!-\! 1}
}
|x_\beta|^2
\!-\!
x_\beta \erw{c_\beta }^{\!\!\!\!\!\!\star }
\right) \!\!
\sum_\gamma
l^\star_{\alpha \gamma } x_\gamma \\
\\[-8pt]
&
\!=
|x_\alpha|^2 M_\beta
\!+\!
|x_\beta|^2 M^\star_\alpha ,
~~~~~~~~~~~~~~~~~~~~~~~~~~~~~~~~~~~~~~~~~~~~~~~~
\hbox{(not summed for $\alpha$ and $\beta$)} ,
\EA
\label{0Lagmultiplierkandl5}
\ea
\vspace{-0.2cm}
\ba
\!\!\!\!\!\!\!\!
\BA{ll}
&
-
\left( \!
{\displaystyle \frac{1}{8 \mbox{$\slashed{z}$}^2}}
{\displaystyle
\frac{2\mbox{$\slashed{z}$}^2 \!\!-\!\! 3}
{2\mbox{$\slashed{z}$}^2 \!\!-\!\! 1}
}
|x_\alpha|^2 \!
\!-\!
x^\star_\alpha \erw{c_\alpha } \!
\right) \!\!
\sum_\gamma \!\! k_{\beta \gamma } x_\gamma
-
\left( \!
{\displaystyle \frac{1}{8 \mbox{$\slashed{z}$}^2}}
{\displaystyle
\frac{2\mbox{$\slashed{z}$}^2 \!\!-\!\! 3}
{2\mbox{$\slashed{z}$}^2 \!\!-\!\! 1}
}
|x_\beta|^2 \!
\!-\!
x^\star_\beta \erw{c_\beta } \!
\right) \!\!
\sum_\gamma \! k_{\alpha \gamma } x_\gamma \\
\\[-10pt]
&
-
\left( \!
{\displaystyle \frac{1}{8 \mbox{$\slashed{z}$}^2}}
{\displaystyle
\frac{2\mbox{$\slashed{z}$}^2 \!\!-\!\! 3}
{2\mbox{$\slashed{z}$}^2 \!\!-\!\! 1}
}
|x_\alpha|^2
\!-\!
x_\alpha \erw{c_\alpha }^{\!\!\!\!\!\!\star }
\right) \!\!
\sum_\gamma \! l_{\beta \gamma } x^\star_\gamma
\!-\!
\left( \!
{\displaystyle \frac{1}{8 \mbox{$\slashed{z}$}^2}}
{\displaystyle
\frac{2\mbox{$\slashed{z}$}^2 \!\!-\!\! 3}
{2\mbox{$\slashed{z}$}^2 \!\!-\!\! 1}
}
|x_\beta|^2
\!-\!
x_\beta \erw{c_\beta }^{\!\!\!\!\!\!\star }
\right) \!\!
\sum_\gamma \! l_{\alpha \gamma } x^\star_\gamma \\
\\[-8pt]
&
\!=
|x_\alpha|^2 M_\beta
\!+\!
|x_\beta|^2 M_\alpha
~~~~~~~~~~~~~~~~~~~~~~~~~~~~~~~~~~~~~~~~~~~~~~~~
\hbox{(not summed for $\alpha$ and $\beta$)} ,
\EA
\label{0Lagmultiplierkandl6}
\ea\\[-8pt]
which
can also be separated into two parts. One is given as follows:\\[-14pt]
\ba
\!\!\!\!\!\!\!\!
\BA{ll}
&
2 \!
{\displaystyle
\frac{
6 \mbox{$\slashed{z}$}^4
\!-\!
3 \mbox{$\slashed{z}$}^2
\!+\!
1 \!
}
{2\mbox{$\slashed{z}$}^2
\!-\!
1
}
}
k_{\beta \alpha } x_\alpha
\!+\!
\left( \!
{\displaystyle \frac{1}{4 \mbox{$\slashed{z}$}^2}}
{\displaystyle
\frac{2\mbox{$\slashed{z}$}^2 \!-\! 3}
{\mbox{2$\slashed{z}$}^2 \!-\! 1}
}
|x_\alpha|^2
-
x_\alpha \erw{c_\alpha }^{\!\!\!\!\!\!\star }
\right) \!\!
\sum_\gamma
k_{\beta \gamma } x_\gamma \\
\\[-12pt]
&
+
2 \!
{\displaystyle
\frac{
6 \mbox{$\slashed{z}$}^4
\!-\!
3 \mbox{$\slashed{z}$}^2
\!+\!
1 \!
}
{2\mbox{$\slashed{z}$}^2
\!-\!
1
}
}
l_{\beta \alpha } x^\star_\alpha
\!+\!
\left( \!
{\displaystyle \frac{1}{4 \mbox{$\slashed{z}$}^2}}
{\displaystyle
\frac{2\mbox{$\slashed{z}$}^2 \!-\! 3}
{2\mbox{$\slashed{z}$}^2 \!-\! 1}
}
|x_\alpha|^2
\!-\!
x^\star_\alpha \erw{c_\alpha } \!
\right) \!\!
\sum_\gamma
l_{\beta \gamma } x^\star_\gamma \\
\\[-10pt]
&
\!=
|x_\alpha|^2 M_\beta ,
~~~~~~~~~~~~~~~~~~~~~~~~~~~~~~~~~~~~~~~~~~~~~~~~~~~~~~~~~~~
\hbox{(not summed for $\alpha$ and $\beta$)} ,
\EA
\label{0Lagmultiplierkandl5r}
\ea
\vspace{-0.2cm}
\ba
\!\!\!\!\!\!\!\!
\BA{ll}
&
-
\left( \!
{\displaystyle \frac{1}{8 \mbox{$\slashed{z}$}^2}}
{\displaystyle
\frac{2\mbox{$\slashed{z}$}^2 \!\!-\!\! 3}
{2\mbox{$\slashed{z}$}^2 \!\!-\!\! 1}
}
|x_\alpha|^2 \!
\!-\!
x^\star_\alpha \erw{c_\alpha } \!
\right) \!\!
\sum_\gamma \!\! k_{\beta \gamma } x_\gamma
-
\left( \!
{\displaystyle \frac{1}{8 \mbox{$\slashed{z}$}^2}}
{\displaystyle
\frac{2\mbox{$\slashed{z}$}^2 \!\!-\!\! 3}
{2\mbox{$\slashed{z}$}^2 \!\!-\!\! 1}
}
|x_\alpha|^2
\!-\!
x_\alpha \erw{c_\alpha }^{\!\!\!\!\!\!\star }
\right) \!\!
\sum_\gamma \! l_{\beta \gamma } x^\star_\gamma \\
\\[-10pt]
&
\!=
|x_\alpha|^2 M_\beta
~~~~~~~~~~~~~~~~~~~~~~~~~~~~~~~~~~~~~~~~~~~~~~~~~~~~~~~~~~~~~
\hbox{(not summed for $\alpha$ and $\beta$)} .
\EA
\label{0Lagmultiplierkandl6r}
\ea\\[-8pt]
The other is also obtained by the same way as the previous one.
We are now in a stage to determine the unknown parameters
$k_{\alpha \beta }$ and $l_{\alpha \beta }$ in the Lagrange multipliers
under the quasi anti-commutation relation approximation for the fermions.
In the following section,
as an illustrative example, we will treat the superconducting toy-model 
used in the previous Sec. 4. 
The determination will be a great step to the completion of the present method.

\newpage

%%%%%%%%%%%%%%%%%%%%%%%%%%%%%%%%%%%%%%%%%%%%%%%%%%%%%%
%                                                    %
%  7  Solutions for unknown parameters $k$ and $l$   %
%                                                    %
%%%%%%%%%%%%%%%%%%%%%%%%%%%%%%%%%%%%%%%%%%%%%%%%%%%%%%

%%%%%%%%%%%%%%%%%%%%%%%%%%%%%%%%%%%%%%%%%%%%%%%%%%%%

\def\thesection{\arabic{section}}
\setcounter{equation}{0}
\renewcommand{\theequation}{\arabic{section}.\arabic{equation}}
\section{Solutions for unknown parameters $k$ and $l$}

%%%%%%%%%%%%%%%%%%%%%%%%%%%%%%%%%%%%%%%%%%%%%%%%%%%%
~~~First we give solutions at the region very near
$\mbox{$\slashed{z}$}^2 \!\!=\!\! 1$.
Using the Hartree eigenstate $w_{\alpha i}$, EHB
eigenvectors $u_i$ and $v_i$ and the solutions for the unpaired-mode amplitudes $x_i$ and $\bar x_i$,
which are already obtained in the previous Sec. 4,
(\ref{Lagmultiplierkandl5r}) and (\ref{Lagmultiplierkandl6r})
are rewritten as\\[-16pt]
\ba
\!\!\!\!\!\!\!\!\!\!\!\!
\BA{ll}
&
2
{\displaystyle 
\frac{3 \mbox{$\slashed{z}$}^4 \!-\! 2 \mbox{$\slashed{z}$}^2 \!-\! 1}
{\mbox{$\slashed{z}$}^2 \!-\! 1}
}
\!\cdot\!
k_{\beta \alpha } \sum_i x_i w^\star_{\alpha i}
\!+\!
2
{\displaystyle 
\frac{3 \mbox{$\slashed{z}$}^4 \!-\! 2 \mbox{$\slashed{z}$}^2 \!-\! 1}
{\mbox{$\slashed{z}$}^2 \!-\! 1}
}
\!\cdot\!
l_{\beta \alpha } \sum_i x^\star_i w_{\alpha i} \\
\\[-14pt]
&
\!+\!
\sum_{i^\prime j^\prime } \!\!
\left\{ \!\!
\left( \!
{\displaystyle \frac{1}{4 \mbox{$\slashed{z}$}^2}}
{\displaystyle
\frac{\mbox{$\slashed{z}$}^2 \!-\! 3}
{\mbox{$\slashed{z}$}^2 \!-\! 1}
}
\!-\!
{\displaystyle \frac{1}{2}}
\!+\!
|v_{i^\prime }|^2 \!
\right) \!
x^\star_{i^\prime } x_{j^\prime }
\!-\!
{\displaystyle \frac{1}{2}} \!
\left(
u^\star_{i^\prime }v_{i^\prime } \!+\! u_{i^\prime }v^\star_{i^\prime }\right) \!
\bar x^\star_{i^\prime } x_{j^\prime } \!\!
\right\} \!
w_{\alpha i^\prime }
w^\star_{\alpha j^\prime } \!
\sum_\gamma
k_{\beta \gamma } \sum_i x_i w^\star_{\gamma i} \\
\\[-14pt]
&
\!+\!
\sum_{i^\prime j^\prime } \!\!
\left\{ \!\!
\left( \!
{\displaystyle \frac{1}{4 \mbox{$\slashed{z}$}^2}}
{\displaystyle
\frac{\mbox{$\slashed{z}$}^2 \!-\! 3}
{\mbox{$\slashed{z}$}^2 \!-\! 1}
}
\!-\!
{\displaystyle \frac{1}{2}}
\!+\!
|v_{j^\prime }|^2 \!
\right) \!
x^\star_{i^\prime } x_{j^\prime }
-
{\displaystyle \frac{1}{2}} \!
\left(
u^\star_{j^\prime } v_{j^\prime } \!+\! u_{j^\prime }v^\star_{j^\prime }
\right) \!
x^\star_{i^\prime } \bar x_{j^\prime } \!\!
\right\} \!
w_{\alpha i^\prime }
w^\star_{\alpha j^\prime } \!
\sum_\gamma
l_{\beta \gamma } \sum_i x^\star_i w_{\gamma i} \\
\\[-12pt]
&
\!=
\sum_{i j} \!
w_{\alpha i}
w^\star_{\alpha j}
\sum_i M_i w^\star_{\beta i} ,
~~~~~~~~~~~~~~~~~~~~~~~~~~~~~~~~~~~~~~~~~~~~
\hbox{(not summed for $\alpha$ and $\beta$)} ,
\EA
\label{Lagmultiplierkandl5s}
\ea
\vspace{-0.3cm}
\ba
\!\!\!\!\!\!\!\!\!\!\!\!
\BA{ll}
&
-
\sum_{i^\prime j^\prime } \!\!
\left\{ \!
\left(
{\displaystyle \frac{1}{2 \mbox{$\slashed{z}$}^2}}
{\displaystyle
\frac{\mbox{$\slashed{z}$}^2 \!-\! 3}
{\mbox{$\slashed{z}$}^2 \!-\! 1}
}
\!-\!
{\displaystyle \frac{1}{2}}
\!+\!
|v_{j^\prime }|^2
\right) \!
x^\star_{i^\prime } x_{j^\prime }
-
{\displaystyle \frac{1}{2}} \!
\left(
u^\star_{j^\prime } v_{j^\prime } \!+\! u_{j^\prime }v^\star_{j^\prime }
\right) \!
x^\star_{i^\prime } \bar x_{j^\prime }
\right\} \!
w_{\alpha i^\prime }
w^\star_{\alpha j^\prime } \!
\sum_\gamma \! k_{\beta \gamma }  \sum_i x_i w^\star_{\gamma i} \\
\\[-14pt]
&
-
\sum_{i^\prime j^\prime } \!\!
\left\{ \!\!
\left( \!
{\displaystyle \frac{1}{2 \mbox{$\slashed{z}$}^2}}
{\displaystyle
\frac{\mbox{$\slashed{z}$}^2 \!-\! 3}
{\mbox{$\slashed{z}$}^2 \!-\! 1}
}
\!-\!
{\displaystyle \frac{1}{2}}
\!+\!
|v_{i^\prime }|^2 \!
\right) \!
x^\star_{i^\prime } x_{j^\prime }
\!-\!
{\displaystyle \frac{1}{2}} \!
\left(
u^\star_{i^\prime }v_{i^\prime } \!+\! u_{i^\prime }v^\star_{i^\prime }\right) \!
\bar x^\star_{i^\prime } x_{j^\prime } \!\!
\right\} \!
w_{\alpha i^\prime }
w^\star_{\alpha j^\prime } \!
\sum_\gamma \! l_{\beta \gamma } \sum_i x^\star_i w_{\gamma i} \\
\\[-12pt]
&
\!=
\sum_{i j} \!
w_{\alpha i}
w^\star_{\alpha j}
\sum_i M_i w^\star_{\beta i} .
~~~~~~~~~~~~~~~~~~~~~~~~~~~~~~~~~~~~~~~~~~~~~~~
\hbox{(not summed for $\alpha$ and $\beta$)} .
\EA
\label{Lagmultiplierkandl6s}
\ea\\[-8pt]
Subtracting
(\ref{Lagmultiplierkandl6s})
from (\ref{Lagmultiplierkandl5s}),
we can get a simple relation\\[-8pt]
\ba
\!\!\!\!\!\!\!\!
\BA{ll}
&
2
{\displaystyle 
\frac{3 \mbox{$\slashed{z}$}^4 \!-\! 2 \mbox{$\slashed{z}$}^2 \!-\! 1}
{\mbox{$\slashed{z}$}^2 \!-\! 1}
}
\!\cdot\!
\left(
k_{\beta \alpha } \! \sum_i x_i w^\star_{\alpha i} 
\!+\!
l_{\beta \alpha } \! \sum_i x^\star_i w_{\alpha i}
\right) \\
\\[-12pt]
&
\!+\!
\sum_{i^\prime \! j^\prime } \!\!
\left\{ \!\!
\left( \!\!
{\displaystyle \frac{3}{4 \mbox{$\slashed{z}$}^2}}
{\displaystyle
\frac{\mbox{$\slashed{z}$}^2 \!-\! 3}
{\mbox{$\slashed{z}$}^2 \!-\! 1}
}
\!\!-\!\!
1
\!\!+\!\!
|v_{i^\prime }|^2
\!\!+\!\!
|v_{\!j^\prime }|^2 \!\!
\right) \!\!
x^\star_{i^\prime } x_{\!j^\prime }
\!\!-\!\!
{\displaystyle \frac{1}{2}} \!
\left( \!
u^\star_{i^\prime }v_{i^\prime }
\!\!+\!\!
u_{i^\prime }v^\star_{i^\prime } \!
\right) \!
\bar x^\star_{i^\prime }x_{\!j^\prime }
\!\!-\!\!
{\displaystyle \frac{1}{2}} \!
\left( \!
u^\star_{\!j^\prime }v_{\!j^\prime }
\!\!+\!\!
u_{\!j^\prime }v^\star_{\!j^\prime } \!
\right) \!
x^\star_{i^\prime }\bar x_{\!j^\prime } \!\!
\right\} \!
w_{\alpha i^\prime }
w^\star_{\alpha \! j^\prime } \\
\\[-12pt]
&
\!\times\!
\left(
\sum_\gamma \!
k_{\beta \gamma } \! \sum_i x_i w^\star_{\gamma i} 
\!+\!
\sum_\gamma \!
l_{\beta \gamma } \! \sum_i x^\star_i w_{\gamma i} \!
\right)
\!=
0 ,
~~~~~~~~~~~~~~~~~~~~%~~~~~~~~~~~~~~~~~~
\hbox{(not summed for $\alpha$ and $\beta$)} ,
\EA
\label{Lagmultiplierkandl5c}
\ea
or adding (\ref{Lagmultiplierkandl5s})
to (\ref{Lagmultiplierkandl6s}),
we can reach another result\\[-16pt]
\ba
\!\!\!\!\!\!\!\!\!
\BA{ll}
&
2
{\displaystyle 
\frac{3 \mbox{$\slashed{z}$}^4 \!-\! 2 \mbox{$\slashed{z}$}^2 \!-\! 1}
{\mbox{$\slashed{z}$}^2 \!-\! 1}
}
\!\cdot\!
\left(
k_{\beta \alpha } \!\! \sum_i x_i w^\star_{\alpha i} 
\!+\!
l_{\beta \alpha } \!\! \sum_i x^\star_i w_{\alpha i}
\right) \\
\\[-14pt]
&
\!-
{\displaystyle \frac{1}{4 \mbox{$\slashed{z}$}^2}}
{\displaystyle
\frac{\mbox{$\slashed{z}$}^2 \!-\! 3}
{\mbox{$\slashed{z}$}^2 \!-\! 1}
}
\sum_{i^\prime \! j^\prime } \!
x^\star_{i^\prime } x_{\!j^\prime }
w_{\alpha \! i^\prime }
w^\star_{\alpha \! j^\prime } \!
\left( \!
\sum_\gamma \!
k_{\beta \gamma } \!\! \sum_i \! x_i w^\star_{\gamma i} 
\!+\!
\sum_\gamma \!
l_{\beta \gamma } \!\! \sum_i \! x^\star_i w_{\gamma i} \!
\right) \\
\\[-14pt]
&
\!+
\sum_{i^\prime \! j^\prime } \!
\left\{ \!
\left(
|v_{i^\prime }|^2
\!-\!
|v_{\!j^\prime }|^2
\right)
x^\star_{i^\prime } x_{\!j^\prime }
\!-\!
{\displaystyle \frac{1}{2}} \!
\left(
u^\star_{i^\prime }v_{i^\prime }
\!+\!
u_{i^\prime }v^\star_{i^\prime }
\right)
\bar x^\star_{i^\prime }x_{\!j^\prime }
\!+\!
{\displaystyle \frac{1}{2}} \!
\left(
u^\star_{\!j^\prime }v_{\!j^\prime }
\!+\!
u_{\!j^\prime }v^\star_{\!j^\prime }
\right)
x^\star_{i^\prime }\bar x_{\!j^\prime } \!
\right\} \!
w_{\alpha i^\prime }
w^\star_{\alpha \! j^\prime } \\
\\[-14pt]
&
\!\times\!\!
\left( \!
\sum_\gamma \!
k_{\beta \gamma } \!\! \sum_i \! x_i w^\star_{\gamma i} 
\!\!-\!\!
\sum_\gamma \!
l_{\beta \gamma } \!\! \sum_i \! x^\star_i w_{\gamma i} \!
\right)
\!\!=\!\!
2 \!
\sum_{i j} \!\!
w_{\alpha i}
w^\star_{\alpha j} \!
\sum_i \! M_i w^\star_{\beta i} ,~
\hbox{(not summed for $\alpha$ and $\beta$)} .
\EA
\label{Lagmultiplierkandl6c}
\ea
The other result is obtained by taking the complex conjugation and making the exchange between the indices $\alpha$ and $\beta$.

Next we give solutions at the region very near
$\mbox{$\slashed{z}$}^2 \!\!=\!\! \frac{1}{2}$.
(\ref{0Lagmultiplierkandl5r}) and (\ref{0Lagmultiplierkandl6r})
are rewritten as\\[-16pt]
\ba
\!\!\!\!\!\!\!\!
\BA{ll}
&
2 \!
{\displaystyle
\frac{
6 \mbox{$\slashed{z}$}^4
\!-\!
3 \mbox{$\slashed{z}$}^2
\!+\!
1 \!
}
{2\mbox{$\slashed{z}$}^2
\!-\!
1
}
}
\!\cdot\!
k_{\beta \alpha } \sum_i x_i w^\star_{\alpha i}
\!+\!
2 \!
{\displaystyle
\frac{
6 \mbox{$\slashed{z}$}^4
\!-\!
3 \mbox{$\slashed{z}$}^2
\!+\!
1 \!
}
{2\mbox{$\slashed{z}$}^2
\!-\!
1
}
}
l_{\beta \alpha } \sum_i x^\star_i w_{\alpha i}  \\
\\[-14pt]
&
\!+\!
\sum_{i^\prime j^\prime } \!\!
\left\{ \!
\left(
{\displaystyle \frac{1}{4 \mbox{$\slashed{z}$}^2}}
{\displaystyle
\frac{2 \mbox{$\slashed{z}$}^2 \!-\! 3}
{2 \mbox{$\slashed{z}$}^2 \!-\! 1}
}
\!-\!
{\displaystyle \frac{1}{2}}
\!+\!
|v_{i^\prime }|^2
\right) \!
x^\star_{i^\prime } x_{j^\prime }
\!-\!
{\displaystyle \frac{1}{2}} \!
\left(
u^\star_{i^\prime }v_{i^\prime } \!+\! u_{i^\prime }v^\star_{i^\prime }\right) \!
\bar x^\star_{i^\prime } x_{j^\prime } \!
\right\} \!
w_{\alpha i^\prime }
w^\star_{\alpha j^\prime } \!\!
\sum_\gamma \!
k_{\beta \gamma } \!
\sum_i x_i w^\star_{\gamma i} \\
\\[-14pt]
&
\!+\!
\sum_{i^\prime j^\prime } \!\!
\left\{ \!
\left(
{\displaystyle \frac{1}{4 \mbox{$\slashed{z}$}^2}}
{\displaystyle
\frac{2 \mbox{$\slashed{z}$}^2 \!-\! 3}
{2 \mbox{$\slashed{z}$}^2 \!-\! 1}
}
\!-\!
{\displaystyle \frac{1}{2}}
\!+\!
|v_{j^\prime }|^2
\right) \!
x^\star_{i^\prime } x_{j^\prime }
\!-\!
{\displaystyle \frac{1}{2}} \!
\left(
u^\star_{j^\prime } v_{j^\prime }
\!\!+\!\!
u_{j^\prime }v^\star_{j^\prime }
\right) \!
x^\star_{i^\prime } \bar x_{j^\prime } \!
\right\} \!
w_{\alpha i^\prime }
w^\star_{\alpha j^\prime } \!\!
\sum_\gamma \!
l_{\beta \gamma } \!
\sum_i x^\star_i w_{\gamma i} \\
\\[-8pt]
&
\!=
\sum_{i j} \!
w_{\alpha i}
w^\star_{\alpha j}
\sum_i M_i w^\star_{\beta i} ,
~~~~~~~~~~~~~~~~~~~~~~~~~~~~~~~~~~~~~~~~~~
\hbox{(not summed for $\alpha$ and $\beta$)} ,
\EA
\label{Lagmultiplierkandl7s}
\ea
\vspace{-0.3cm}
\ba
\!\!\!\!\!\!\!\!
\BA{ll}
&
-
\sum_{i^\prime j^\prime } \!\!
\left\{ \!
\left(
{\displaystyle \frac{1}{8 \mbox{$\slashed{z}$}^2}}
{\displaystyle
\frac{2 \mbox{$\slashed{z}$}^2 \!-\! 3}
{2 \mbox{$\slashed{z}$}^2 \!-\! 1}
}
\!-\!
{\displaystyle \frac{1}{2}}
\!+\!
|v_{j^\prime }|^2
\right) \!
x^\star_{i^\prime } x_{j^\prime }
\!-\!
{\displaystyle \frac{1}{2}} \!
\left(
u^\star_{j^\prime } v_{j^\prime }
\!\!+\!\!
u_{j^\prime }v^\star_{j^\prime }
\right) \!
x^\star_{i^\prime } \bar x_{j^\prime } \!
\right\} \!
w_{\alpha i^\prime }
w^\star_{\alpha j^\prime } \!\!
\sum_\gamma \!
k_{\beta \gamma } \!\!
\sum_i x_i w^\star_{\gamma i} \\
\\[-14pt]
&
-
\sum_{i^\prime j^\prime } \!\!
\left\{ \!
\left(
{\displaystyle \frac{1}{8 \mbox{$\slashed{z}$}^2}}
{\displaystyle
\frac{2 \mbox{$\slashed{z}$}^2 \!-\! 3}
{2 \mbox{$\slashed{z}$}^2 \!-\! 1}
}
\!-\!
{\displaystyle \frac{1}{2}}
\!+\!
|v_{i^\prime }|^2
\right) \!
x^\star_{i^\prime } x_{j^\prime }
\!-\!
{\displaystyle \frac{1}{2}} \!
\left(
u^\star_{i^\prime }v_{i^\prime } \!+\! u_{i^\prime }v^\star_{i^\prime }\right) \!
\bar x^\star_{i^\prime } x_{j^\prime } \!
\right\} \!
w_{\alpha i^\prime }
w^\star_{\alpha j^\prime } \!\!
\sum_\gamma \! 
l_{\beta \gamma } \!\!
\sum_i x^\star_i w_{\gamma i} \\
\\[-8pt]
&
\!=
\sum_{i j} \!
w_{\alpha i}
w^\star_{\alpha j}
\sum_i M_i w^\star_{\beta i} ,
~~~~~~~~~~~~~~~~~~~~~~~~~~~~~~~~~~~~~~~~~~
\hbox{(not summed for $\alpha$ and $\beta$)} .
\EA
\label{Lagmultiplierkandl8s}
\ea\\[-6pt]
Subtracting
(\ref{Lagmultiplierkandl8s})
from
(\ref{Lagmultiplierkandl7s}),
we can also get a simple relation\\[-18pt]
\ba
\!\!\!\!\!\!\!\!\!\!
\BA{ll}
&
2 \!
{\displaystyle
\frac{
6 \mbox{$\slashed{z}$}^4
\!-\!
3 \mbox{$\slashed{z}$}^2
\!+\!
1 \!
}
{2\mbox{$\slashed{z}$}^2
\!-\!
1
}
}
\!\cdot\!
\left(
k_{\beta \alpha } \! \sum_i x_i w^\star_{\alpha i} 
\!+\!
l_{\beta \alpha } \! \sum_i x^\star_i w_{\alpha i}
\right) \\
\\[-14pt]
&
\!+\!
\sum_{i^\prime \! j^\prime } \!\!
\left\{ \!\!
\left( \!\!
{\displaystyle \frac{3}{8 \mbox{$\slashed{z}$}^2}}
{\displaystyle
\frac{2 \mbox{$\slashed{z}$}^2 \!\!-\!\! 3}
{2 \mbox{$\slashed{z}$}^2 \!\!-\!\! 1}
}
\!\!-\!\!
1
\!\!+\!\!
|v_{i^\prime }|^2
\!\!+\!\!
|v_{\!j^\prime }|^2 \!\!
\right) \!\!
x^\star_{i^\prime } x_{\!j^\prime }
\!\!-\!\!
{\displaystyle \frac{1}{2}} \!
\left( \!
u^\star_{i^\prime }v_{i^\prime }
\!\!+\!\!
u_{i^\prime }v^\star_{i^\prime } \!
\right) \!
\bar x^\star_{i^\prime } x_{j^\prime }
\!\!-\!\!
{\displaystyle \frac{1}{2}} \!
\left( \!
u^\star_{\!j^\prime }v_{\!j^\prime }
\!\!+\!\!
u_{\!j^\prime }v^\star_{\!j^\prime } \!
\right) \!
x^\star_{i^\prime } \bar x_{\!j^\prime } \!\!
\right\} \!\!
w_{\alpha \! i^\prime }
w^\star_{\alpha \! j^\prime } \\
\\[-14pt]
&
\!\times\!
\left(
\sum_\gamma
k_{\beta \gamma } \sum_i x_i w^\star_{\gamma i}
\!+\!
\sum_\gamma
l_{\beta \gamma } \sum_i x^\star_i w_{\gamma i}
\right)
\!=
0 ,
~~~~~~~~~~~~~~~~~~%~~~~~~~~~~~~~~~~~~~~~~~~
\hbox{(not summed for $\alpha$ and $\beta$)} ,
\EA
\label{Lagmultiplierkandl7c}
\ea
or addting
(\ref{Lagmultiplierkandl7s})
to
(\ref{Lagmultiplierkandl8s}),
we can also reach another result\\[-18pt]
\ba
\!\!\!\!\!\!\!\!\!\!
\BA{ll}
&
2 \!
{\displaystyle
\frac{
6 \mbox{$\slashed{z}$}^4
\!-\!
3 \mbox{$\slashed{z}$}^2
\!+\!
1 \!
}
{2\mbox{$\slashed{z}$}^2
\!-\!
1
}
}
\!\cdot\!
\left(
k_{\beta \alpha } \! \sum_i x_i w^\star_{\alpha i} 
\!+\!
l_{\beta \alpha } \! \sum_i x^\star_i w_{\alpha i}
\right) \\
\\[-14pt]
&
\!+
{\displaystyle \frac{1}{8 \mbox{$\slashed{z}$}^2}}
{\displaystyle
\frac{2 \mbox{$\slashed{z}$}^2 \!-\! 3}
{2 \mbox{$\slashed{z}$}^2 \!-\! 1}
}
\sum_{i^\prime \! j^\prime } \!
x^\star_{i^\prime } x_{\!j^\prime }
w_{\alpha \! i^\prime }
w^\star_{\alpha \! j^\prime } \!
\left( \!
\sum_\gamma \!
k_{\beta \gamma } \!\! \sum_i \! x_i w^\star_{\gamma i} 
\!+\!
\sum_\gamma \!
l_{\beta \gamma } \!\! \sum_i \! x^\star_i w_{\gamma i} \!
\right) \\
\\[-14pt]
&
\!+
\sum_{i^\prime \! j^\prime } \!
\left\{ \!
\left(
|v_{i^\prime }|^2
\!-\!
|v_{\!j^\prime }|^2
\right)
x^\star_{i^\prime } x_{\!j^\prime }
\!-\!
{\displaystyle \frac{1}{2}} \!
\left(
u^\star_{i^\prime }v_{i^\prime }
\!+\!
u_{i^\prime }v^\star_{i^\prime }
\right)
\bar x^\star_{i^\prime } x_{j^\prime }
\!+\!
{\displaystyle \frac{1}{2}} \!
\left(
u^\star_{\!j^\prime }v_{\!j^\prime }
\!+\!
u_{\!j^\prime }v^\star_{\!j^\prime }
\right)
x^\star_{i^\prime } \bar x_{\!j^\prime } \!
\right\} \!
w_{\alpha \! i^\prime }
w^\star_{\alpha \! j^\prime } \\
\\[-14pt]
&
\!\times\!\!
\left( \!
\sum_\gamma \!
k_{\beta \gamma } \!\! \sum_i \! x_i w^\star_{\gamma i} 
\!\!-\!\!
\sum_\gamma \!
l_{\beta \gamma } \!\! \sum_i \! x^\star_i w_{\gamma i} \!
\right)
\!\!=\!\!
2 \!
\sum_{i j} \!\!
w_{\alpha i}
w^\star_{\alpha j} \!
\sum_i \! M_i w^\star_{\beta i} ,~
\hbox{(not summed for $\alpha$ and $\beta$)} .
\EA
\label{Lagmultiplierkandl8c}
\ea
The other is also obtained by the same way as the above one.

%%%%%%%%%%%%%%%%%%%%%%%%%%%%%%%%%%%%%%%%%%%%%%%

First we treat the region very near
$\mbox{$\slashed{z}$}^2 \!=\! 1$.
Subtracting and adding complex conjugation of
(\ref{Lagmultiplierkandl5c}) and (\ref{Lagmultiplierkandl6s})
from
(\ref{Lagmultiplierkandl5c}) and (\ref{Lagmultiplierkandl6s})
and to
(\ref{Lagmultiplierkandl5c}) and (\ref{Lagmultiplierkandl6s}),
respectively, we have\\[-18pt]
\ba
\!\!\!\!\!\!\!\!
\BA{ll}
&
2
{\displaystyle 
\frac{3 \mbox{$\slashed{z}$}^4 \!-\! 2 \mbox{$\slashed{z}$}^2 \!-\! 1}
{\mbox{$\slashed{z}$}^2 \!-\! 1}
}
\!\cdot\!
\left\{
\left( k_{\beta \alpha } \!\mp\! l^\star_{\beta \alpha } \right) \!
\sum_i x_i w^\star_{\alpha i} 
\!+\!
\left( l_{\beta \alpha } \!\mp\! k^\star_{\beta \alpha } \right) \!
\sum_i x^\star_i w_{\alpha i}
\right\} \\
\\[-12pt]
&
\!+\!
\sum_{i^\prime \! j^\prime } \!\!
\left\{ \!\!
\left( \!\!
{\displaystyle \frac{3}{4 \mbox{$\slashed{z}$}^2}}
{\displaystyle
\frac{\mbox{$\slashed{z}$}^2 \!\!-\!\! 3}
{\mbox{$\slashed{z}$}^2 \!\!-\!\! 1}
}
\!-\!
1
\!+\!
|v_{i^\prime }\!|^2
\!+\!
|v_{\!j^\prime }\!|^2 \!\!
\right) \!\!
x^\star_{i^\prime } x_{\!j^\prime }
\!\!-\!\!
{\displaystyle \frac{1}{2}} \!
\left( \!
u^\star_{\!j^\prime }v_{\!j^\prime }
\!\!+\!\!
u_{\!j^\prime }v^\star_{\!j^\prime } \!
\right) \!
x^\star_{i^\prime } \bar x_{\!j^\prime }
\!\!-\!\!
{\displaystyle \frac{1}{2}} \!
\left( \!
u^\star_{i^\prime }v_{i^\prime }
\!\!+\!\!
u_{i^\prime }v^\star_{i^\prime } \!
\right) \!
\bar x^\star_{i^\prime } x_{\!j^\prime } \!\!
\right\} \!\!
w_{\alpha i^\prime }
w^\star_{\alpha \! j^\prime } \\
\\[-12pt]
&
\!\times\!
\left\{
\sum_\gamma \!
\left( k_{\beta \gamma } \!\mp\! l^\star_{\beta \gamma } \right) \!
\sum_i x_i w^\star_{\gamma i}
\!+\!
\sum_\gamma \!
\left( l_{\beta \gamma } \!\mp\! k^\star_{\beta \gamma } \right) \!
\sum_i x^\star_i w_{\gamma i}
\right\}
\!=
0 , ~
\hbox{(not summed for $\alpha$ and $\beta$)} ,
\EA
\label{Lagmultiplierkandlminusplus1}
\ea
\vspace{-0.6cm}
\ba
\!\!\!\!\!\!\!\!
\BA{ll}
&
\sum_{i^\prime \! j^\prime } \!\!
\left\{ \!\!
\left( \!\!
-{\displaystyle \frac{1}{2 \mbox{$\slashed{z}$}^2}}
{\displaystyle
\frac{\mbox{$\slashed{z}$}^2 \!-\! 3}
{\mbox{$\slashed{z}$}^2 \!-\! 1}
}
\!+\!
{\displaystyle \frac{1}{2}}
\!-\!
|v_{\!j^\prime }|^2 \!\!
\right) \!
x^\star_{i^\prime } x_{\!j^\prime }
\!\!+\!\!
{\displaystyle \frac{1}{2}} \!
\left( \!
u^\star_{\!j^\prime }v_{\!j^\prime }
\!\!+\!\!
u_{j^\prime }v^\star_{\!j^\prime }\right) \!
x^\star_{i^\prime } \bar x_{j^\prime } \!\!
\right\} \!
w_{\alpha i^\prime }
w^\star_{\alpha \!j^\prime } \!
\sum_\gamma \!\!
\left( k_{\beta \gamma } \!\!\mp\!\! l^\star_{\beta \gamma } \right) \!\!
\sum_i x_i w^\star_{\gamma i} \\
\\[-12pt]
&
\!+\!
\sum_{i^\prime \! j^\prime } \!\!
\left\{ \!\!
\left( \!\!
-{\displaystyle \frac{1}{2 \mbox{$\slashed{z}$}^2}}
{\displaystyle
\frac{\mbox{$\slashed{z}$}^2 \!\!-\!\! 3}
{\mbox{$\slashed{z}$}^2 \!\!-\!\! 1}
}
\!+\!
{\displaystyle \frac{1}{2}}
\!-\!
|v_{i^\prime }|^2 \!\!
\right) \!
x^\star_{i^\prime } x_{j^\prime }
\!\!+\!\!
{\displaystyle \frac{1}{2}} \!
\left( \!
u^\star_{i^\prime } v_{i^\prime }
\!+\!
u_{i^\prime }v^\star_{i^\prime } \!
\right) \!
\bar x^\star_{i^\prime } x_{\!j^\prime } \!\!
\right\} \!
w_{\alpha i^\prime }
w^\star_{\alpha \! j^\prime } \!\!
\sum_\gamma \!\!
\left( l_{\beta \gamma } \!\!\mp\!\! k^\star_{\beta \gamma } \right) \!\!
\sum_i \! x^\star_i w_{\gamma i} \\
\\[-10pt]
&
\!=
\sum_{i j} \!
w_{\alpha i}
w^\star_{\alpha j} \!
\sum_i \!
\left( M_i w^\star_{\beta i} \!\mp\! M^\star_i w_{\beta i} \right) ,
~~~~~~~~~~~~~~~~~~~~~~~~~~~~~
\hbox{(not summed for $\alpha$ and $\beta$)} .
\EA
\label{Lagmultiplierkandlminusplus2}
\ea
To solve
(\ref{Lagmultiplierkandlminusplus1})
and
(\ref{Lagmultiplierkandlminusplus2}),
we make the following bold assumptions:\\[-14pt]
\ba
\!\!\!\!
\BA{l}
\sum_\gamma \!
\left( k_{\beta \gamma } \!\mp\! l^\star_{\beta \gamma } \right) \!
\sum_i x_i w^\star_{\gamma i}
\!=\!
\sum_i
M_i w^\star_{\beta i}, ~
\sum_\gamma \!
\left( l_{\beta \gamma } \!\mp\! k^\star_{\beta \gamma } \right) \!
\sum_i x_i w^\star_{\gamma i}
\!=\!
\mp
\sum_i
M^\star_i w_{\beta i} , \\
\\
~~~~~~~~~~~~~~~~~~~~~~~~~~~~~~~~~~~~~~~~~~~~~~~~~~~~~~~~~~~~~~~~~~~~~~~~~~~~~~~~~~
\hbox{(not summed for $\beta$)} , 
\EA
\label{boldassumption1}
\ea
\vspace{-0.8cm}
\ba
\!\!\!\!
\BA{ll}
&
\sum_{i^\prime \! j^\prime } \!\!
\left\{ \!\!
\left( \!\!
{\displaystyle \frac{1}{2 \mbox{$\slashed{z}$}^2}}
{\displaystyle
\frac{\mbox{$\slashed{z}$}^2 \!-\! 3}
{\mbox{$\slashed{z}$}^2 \!-\! 1}
}
\!-\!
{\displaystyle \frac{1}{2}}
\!+\!
|v_{j^\prime }|^2 \!\!
\right) \!
x^\star_{i^\prime } x_{j^\prime }
\!\!-\!\!
{\displaystyle \frac{1}{2}} \!
\left(
u^\star_{j^\prime }v_{j^\prime } \!+\! u_{j^\prime }v^\star_{j^\prime }\right) \!
x^\star_{i^\prime } \bar x_{j^\prime } \!\!
\right\} \!
w_{\alpha i^\prime }
w^\star_{\alpha j^\prime } \\
\\[-10pt]
&\!\!\!\!
\!=\!
\sum_{i^\prime \! j^\prime } \!\!
\left\{ \!\!
\left( \!\!
{\displaystyle \frac{1}{2 \mbox{$\slashed{z}$}^2}}
{\displaystyle
\frac{\mbox{$\slashed{z}$}^2 \!-\! 3}
{\mbox{$\slashed{z}$}^2 \!-\! 1}
}
\!-
{\displaystyle \frac{1}{2}}
\!+\!
|v_{i^\prime }|^2 \!\!
\right) \!
x^\star_{i^\prime } x_{j^\prime }
\!\!-\!\!
{\displaystyle \frac{1}{2}}
\left(
u^\star_{i^\prime } v_{i^\prime } \!+\! u_{i^\prime }v^\star_{i^\prime }
\right)
\bar x^\star_{i^\prime } x_{j^\prime } \!\!
\right\} \!
w_{\alpha i^\prime }
w^\star_{\alpha j^\prime } \\
\\[-10pt]
&\!\!\!\!
\!=\!
- \sum_{i j} \!
w_{\alpha i}
w^\star_{\alpha j} ,
~~~~~~~~~~~~~~~~~~~~~~~~~~~~~~~~~~~~~~~~~~~~~~~~~~~~~~~~~~
\hbox{(not summed for $\alpha$)} ,
\EA
\label{boldassumption2}
\ea\\[-10pt]
in which the second assumption
(\ref{boldassumption2})
is proved to hold
if the Hartree state is a real eigen state.
We, however, use this assumption even if the Hartree state is
a complex eigen state.
Substituting
(\ref{boldassumption1}) and (\ref{boldassumption2})
into
(\ref{Lagmultiplierkandlminusplus1}),
we reach the following equation:\\[-16pt]
\ba
\!\!\!\!\!\!\!\!\!\!
\BA{ll}
&
2
{\displaystyle 
\frac{3 \mbox{$\slashed{z}$}^4 \!-\! 2 \mbox{$\slashed{z}$}^2 \!-\! 1}
{\mbox{$\slashed{z}$}^2 \!-\! 1}
}
\!\cdot\!
\left\{
\left( k_{\beta \alpha } \!\mp\! l^\star_{\beta \alpha } \right) \!
\sum_i x_i w^\star_{\alpha i} 
\!+\!
\left( l_{\beta \alpha } \!\mp\! k^\star_{\beta \alpha } \right) \!
\sum_i x^\star_i w_{\alpha i}
\right\} \\
\\[-16pt]
&
\!=\!
\sum_{i j} \!
\left\{
2
\!+\!
{\displaystyle \frac{1}{4 \mbox{$\slashed{z}$}^2}}
{\displaystyle
\frac{\mbox{$\slashed{z}$}^2 \!\!-\!\! 3}
{\mbox{$\slashed{z}$}^2 \!\!-\!\! 1}
}
x^\star_{i } x_{j }
\right\}
w_{\alpha i}
w^\star_{\alpha j} \!
\sum_i \!
\left( M_i w^\star_{\beta i} \!\mp\! M^\star_i w_{\beta i} \right), ~~~
\hbox{(not summed for $\alpha$ and $\beta$)} .
\EA
\label{Lagmultiplierkandlminus3}
\ea\\[-12pt]
Equating each term in the l.h.s. and the r.h.s. of
(\ref{Lagmultiplierkandlminus3}),
respectively,
then we have\\[-18pt]
\ba
\!\!\!\!\!\!\!\!\!\!\!\!
\BA{ll}
&
2
{\displaystyle 
\frac{3 \mbox{$\slashed{z}$}^4 \!-\! 2 \mbox{$\slashed{z}$}^2 \!-\! 1}
{\mbox{$\slashed{z}$}^2 \!-\! 1}
}
\!\cdot\!
\left( k_{\beta \alpha } \!\mp\! l^\star_{\beta \alpha } \right) \!
\sum_i x_i w^\star_{\alpha i}
\!=\!
\sum_{i j} \!
\left\{
2
\!+\!
{\displaystyle \frac{1}{4 \mbox{$\slashed{z}$}^2}}
{\displaystyle
\frac{\mbox{$\slashed{z}$}^2 \!\!-\!\! 3}
{\mbox{$\slashed{z}$}^2 \!\!-\!\! 1}
}
x^\star_{i } x_{j }
\right\}
w_{\alpha i}
w^\star_{\alpha j} \!
\sum_i \!
M_i w^\star_{\beta i} , \\
\\[-14pt]
&~~~~~~~~~~~~~~~~~~~~~~~~~~~~~~~~~~~~~~~~~~~~~~~~~~~~~~~~~~~~~~~~~~~~~~~~~~~~
\hbox{(not summed for $\alpha$ and $\beta$)} ,
\EA
\label{Lagmultiplierkandlminus4}
\ea\\[-14pt]
in which making summation with respect to $\alpha$ and using
(\ref{boldassumption1}),
(\ref{Lagmultiplierkandlminus4})
can be converted to\\[-18pt]
\ba
\BA{l}
2
{\displaystyle 
\frac{3 \mbox{$\slashed{z}$}^4 \!-\! 2 \mbox{$\slashed{z}$}^2 \!-\! 1}
{\mbox{$\slashed{z}$}^2 \!-\! 1}
}
\!=\!
\sum_{i j} \!
\left\{
2
\!+\!
{\displaystyle \frac{1}{4 \mbox{$\slashed{z}$}^2}}
{\displaystyle
\frac{\mbox{$\slashed{z}$}^2 \!\!-\!\! 3}
{\mbox{$\slashed{z}$}^2 \!\!-\!\! 1}
}
x^\star_{i } x_{j }
\right\}
\delta_{ij} ,~
\left( \!
\mbox{$\slashed{z}$}
\!\equiv\! 
{\displaystyle \sqrt{\frac{1\!\!+\!\!z}{2}}} \!
\right) ,
\EA
\label{Lagmultiplierkandlminus5}
\ea\\[-16pt]
from which we can obtain the relation between the original variable $z$ and the number of single-particle state $N$ as\\[-22pt]
\ba
\BA{l}
z
\!=\!
{\displaystyle \frac{1}{7}}
\left\{
2N \!+\! 1
\!-\!
\left (2N \!+\! 1 \right) 
\sqrt{
1
\!-\!
{\displaystyle \frac{28N \!-\! 35}{\left (2N \!+\! 1 \right)^2}}
}
\right\}
\approx
1
\!-\!
{\displaystyle \frac{7}{2}\frac{1}{2N \!+\! 1}} ,
\EA
\label{Lagmultiplierkandlminus6}
\ea\\[-14pt]
fortunately, 
which is a little bit smaller than 1 for large $N$.
Further we study another case.

Contrasting to the above case,
subtraction and addition of complex conjugation of
(\ref{Lagmultiplierkandl8c})
from
(\ref{Lagmultiplierkandl8c})
and to
(\ref{Lagmultiplierkandl8c}),
lead to a very simple equation\\[-14pt]
\ba
\!\!\!\!\!\!\!\!
\BA{ll}
&
2
{\displaystyle 
\frac{3 \mbox{$\slashed{z}$}^4 \!-\! 2 \mbox{$\slashed{z}$}^2 \!-\! 1}
{\mbox{$\slashed{z}$}^2 \!-\! 1}
}
\!\cdot\!
\left\{
\left( k_{\beta \alpha } \!\mp\! l^\star_{\beta \alpha } \right) \!
\sum_i x_i w^\star_{\alpha i} 
\!+\!
\left( l_{\beta \alpha } \!\mp\! k^\star_{\beta \alpha } \right) \!
\sum_i x^\star_i w_{\alpha i}
\right\} \\
\\[-12pt]
&
-
{\displaystyle \frac{1}{2 \mbox{$\slashed{z}$}^2}}
{\displaystyle
\frac{\mbox{$\slashed{z}$}^2 \!-\! 3}
{\mbox{$\slashed{z}$}^2 \!-\! 1}
}
\sum_{i^\prime \! j^\prime } \!
x^\star_{i^\prime } x_{\!j^\prime }
w_{\alpha \! i^\prime }
w^\star_{\alpha \! j^\prime } \!
\left\{ \!
\sum_\gamma \!
\left( \!
k_{\beta \gamma } \!\mp\! l^\star_{\beta \gamma } \!
\right)
\sum_i \! x_i w^\star_{\gamma i} 
\!+\!
\sum_\gamma \!
\left( \!
l_{\beta \gamma } \!\mp\! k^\star_{\beta \gamma } \!
\right)
\sum_i \! x^\star_i w_{\gamma i} \!
\right\} \\
\\[-12pt]
&
\!=\!
2
\sum_{i j} \!
w_{\alpha i}
w^\star_{\alpha j} \!
\sum_i \!
\left( M_i w^\star_{\beta i} \!\mp\! M^\star_i w_{\beta i} \right) , 
~~~~~~~~~~~~~~~~~~~~~~~~~~~~
\hbox{(not summed for $\alpha$ and $\beta$)} .
\EA
\label{Lagmultiplierkandlminusplup0p}
\ea
In which we have used the trivial relation derived from
(\ref{boldassumption2}).
Substituting
(\ref{boldassumption1}) and (\ref{boldassumption2})
into
(\ref{Lagmultiplierkandlminusplup0p}),
we obtain\\[-16pt]
\ba
\!\!\!\!\!\!\!\!
\BA{ll}
&
2
{\displaystyle 
\frac{3 \mbox{$\slashed{z}$}^4 \!-\! 2 \mbox{$\slashed{z}$}^2 \!-\! 1}
{\mbox{$\slashed{z}$}^2 \!-\! 1}
}
\!\cdot\!
\left\{
\left( k_{\beta \alpha } \!\mp\! l^\star_{\beta \alpha } \right) \!
\sum_i x_i w^\star_{\alpha i} 
\!+\!
\left( l_{\beta \alpha } \!\mp\! k^\star_{\beta \alpha } \right) \!
\sum_i x^\star_i w_{\alpha i}
\right\} \\
\\[-12pt]
&
-
{\displaystyle \frac{1}{2 \mbox{$\slashed{z}$}^2}}
{\displaystyle
\frac{\mbox{$\slashed{z}$}^2 \!-\! 3}
{\mbox{$\slashed{z}$}^2 \!-\! 1}
}
\sum_{i^\prime \! j^\prime } \!
x^\star_{i^\prime } x_{\!j^\prime }
w_{\alpha \! i^\prime }
w^\star_{\alpha \! j^\prime } \!
\sum_i \!
\left( M_i w^\star_{\beta i} \!\mp\! M^\star_i w_{\beta i} \right) \\
\\[-8pt]
&
\!=\!
2
\sum_{i j} \!
w_{\alpha i}
w^\star_{\alpha j} \!
\sum_i \!
\left( M_i w^\star_{\beta i} \!\mp\! M^\star_i w_{\beta i} \right) , 
~~~~~~~~~~~~~~~~~~~~~~~~~~~~
\hbox{(not summed for $\alpha$ and $\beta$)} ,
\EA
\label{Lagmultiplierkandlminusplup0p1}
\ea
which is just identical with
(\ref{Lagmultiplierkandlminus3}).

Next we treat the region very near
$\mbox{$\slashed{z}$}^2 \!=\! \frac{1}{2}$.
Subtracting and adding complex conjugation of
(\ref{Lagmultiplierkandl7c}) and (\ref{Lagmultiplierkandl8s})
from
(\ref{Lagmultiplierkandl7c}) and (\ref{Lagmultiplierkandl8s})
and to
(\ref{Lagmultiplierkandl7c}) and (\ref{Lagmultiplierkandl8s}),
respectively, we have\\[-18pt]
\ba
\!\!\!\!\!\!\!\!
\BA{ll}
&
2
{\displaystyle 
\frac{6\mbox{$\slashed{z}$}^4 \!-\! 3 \mbox{$\slashed{z}$}^2 \!+\! 1}
{2 \mbox{$\slashed{z}$}^2 \!-\! 1}
}
\!\cdot\!
\left\{
\left( k_{\beta \alpha } \!\mp\! l^\star_{\beta \alpha } \right) \!
\sum_i x_i w^\star_{\alpha i} 
\!+\!
\left( l_{\beta \alpha } \!\mp\! k^\star_{\beta \alpha } \right) \!
\sum_i x^\star_i w_{\alpha i}
\right\} \\
\\[-12pt]
&
\!+\!
\sum_{i^\prime \! j^\prime } \!\!
\left\{ \!\!
\left( \!\!
{\displaystyle \frac{3}{8 \mbox{$\slashed{z}$}^2}}
{\displaystyle
\frac{2 \mbox{$\slashed{z}$}^2 \!\!-\!\! 3}
{2 \mbox{$\slashed{z}$}^2 \!\!-\!\! 1}
}
\!\!-\!\!
1
\!\!+\!\!
|v_{i^\prime }\!|^2
\!\!+\!\!
|v_{j^\prime }\!|^2 \!\!
\right) \!\!
x^\star_{i^\prime } x_{j^\prime }
\!\!-\!\!
{\displaystyle \frac{1}{2}} \!
\left( \!
u^\star_{j^\prime }v_{j^\prime }
\!\!+\!\!
u_{j^\prime }v^\star_{j^\prime } \!
\right) \!
x^\star_{i^\prime } \bar x_{j^\prime }
\!\!-\!\!
{\displaystyle \frac{1}{2}} \!
\left( \!
u^\star_{i^\prime }v_{i^\prime }
\!\!+\!\!
u_{i^\prime }v^\star_{i^\prime } \!
\right) \!
\bar x^\star_{i^\prime } x_{j^\prime } \!\!
\right\} \!\!
w_{\alpha i^\prime }
w^\star_{\alpha j^\prime } \\
\\[-12pt]
&
\!\times\!
\left\{
\sum_\gamma \!
\left( k_{\beta \gamma } \!\mp\! l^\star_{\beta \gamma } \right) \!
\sum_i x_i w^\star_{\gamma i}
\!+\!
\sum_\gamma \!
\left( l_{\beta \gamma } \!\mp\! k^\star_{\beta \gamma } \right) \!
\sum_i x^\star_i w_{\gamma i}
\right\}
\!=
0 , ~
\hbox{(not summed for $\alpha$ and $\beta$)} ,
\EA
\label{Lagmultiplierkandlminusplus1p}
\ea
\vspace{-0.6cm}
\ba
\!\!\!\!\!\!\!\!
\BA{ll}
&
\sum_{i^\prime \! j^\prime } \!\!
\left\{ \!\!
\left( \!\!
{\displaystyle \frac{1}{8 \mbox{$\slashed{z}$}^2}}
{\displaystyle
\frac{2 \mbox{$\slashed{z}$}^2 \!\!-\!\! 3}
{2 \mbox{$\slashed{z}$}^2 \!\!-\!\! 1}
}
\!+\!
{\displaystyle \frac{1}{2}}
\!-\!
|v_{j^\prime }|^2 \!\!
\right) \!
x^\star_{i^\prime } x_{j^\prime }
\!+\!
{\displaystyle \frac{1}{2}} \!
\left( \!
u^\star_{j^\prime }v_{j^\prime }
\!+\!
u_{j^\prime }v^\star_{j^\prime }\right) \!
x^\star_{i^\prime } \bar x_{j^\prime } \!\!
\right\} \!
w_{\alpha i^\prime }
w^\star_{\alpha j^\prime } \!
\sum_\gamma \!\!
\left( k_{\beta \gamma } \!\!\mp\!\! l^\star_{\beta \gamma } \right) \!\!
\sum_i x_i w^\star_{\gamma i} \\
\\[-12pt]
&
\!+\!
\sum_{i^\prime \! j^\prime } \!\!
\left\{ \!\!
\left( \!\!
{\displaystyle \frac{1}{8 \mbox{$\slashed{z}$}^2}}
{\displaystyle
\frac{2 \mbox{$\slashed{z}$}^2 \!\!-\!\! 3}
{2 \mbox{$\slashed{z}$}^2 \!\!-\!\! 1}
}
\!+\!
{\displaystyle \frac{1}{2}}
\!-\!
|v_{i^\prime }|^2 \!\!
\right) \!
x^\star_{i^\prime } x_{j^\prime }
\!+\!
{\displaystyle \frac{1}{2}} \!
\left( \!
u^\star_{i^\prime } v_{i^\prime }
\!+\!
u_{i^\prime }v^\star_{i^\prime } \!
\right) \!
\bar x^\star_{i^\prime } x_{j^\prime } \!\!
\right\} \!
w_{\alpha i^\prime }
w^\star_{\alpha j^\prime } \!
\sum_\gamma \!\!
\left( l_{\beta \gamma } \!\!\mp\!\! k^\star_{\beta \gamma } \right) \!\!
\sum_i x^\star_i w_{\gamma i} \\
\\[-10pt]
&
\!=
\sum_{i j} \!
w_{\alpha i}
w^\star_{\alpha j} \!
\sum_i \!
\left( M_i w^\star_{\beta i} \!\mp\! M^\star_i w_{\beta i} \right) ,
~~~~~~~~~~~~~~~~~~~~~~~~~~~~~
\hbox{(not summed for $\alpha$ and $\beta$)} .
\EA
\label{Lagmultiplierkandlminusplus2p}
\ea\\[-10pt]
Instead of the second assumption
(\ref{boldassumption2}),
we here make the third assumption\\[-18pt]
\ba
\!\!\!\!
\BA{ll}
&
\sum_{i^\prime \! j^\prime } \!\!
\left\{ \!\!
\left( \!\!
{\displaystyle \frac{1}{8 \mbox{$\slashed{z}$}^2}}
{\displaystyle
\frac{2 \mbox{$\slashed{z}$}^2 \!\!-\!\! 3}
{2 \mbox{$\slashed{z}$}^2 \!\!-\!\! 1}}
\!-\!
{\displaystyle \frac{1}{2}}
\!+\!
|v_{j^\prime }|^2 \!\!
\right) \!
x^\star_{i^\prime } x_{j^\prime }
\!\!-\!\!
{\displaystyle \frac{1}{2}} \!
\left(
u^\star_{j^\prime }v_{j^\prime } \!+\! u_{j^\prime }v^\star_{j^\prime }\right) \!
x^\star_{i^\prime } \bar x_{j^\prime } \!\!
\right\} \!
w_{\alpha i^\prime }
w^\star_{\alpha j^\prime } \\
\\[-10pt]
&\!\!\!\!
\!=\!
\sum_{i^\prime \! j^\prime } \!\!
\left\{ \!\!
\left( \!\!
{\displaystyle \frac{1}{8 \mbox{$\slashed{z}$}^2}}
{\displaystyle
\frac{2 \mbox{$\slashed{z}$}^2 \!\!-\!\! 3}
{2 \mbox{$\slashed{z}$}^2 \!\!-\!\! 1}}
\!-
{\displaystyle \frac{1}{2}}
\!+\!
|v_{i^\prime }|^2 \!\!
\right) \!
x^\star_{i^\prime } x_{j^\prime }
\!\!-\!\!
{\displaystyle \frac{1}{2}}
\left(
u^\star_{i^\prime } v_{i^\prime } \!+\! u_{i^\prime }v^\star_{i^\prime }
\right)
\bar x^\star_{i^\prime } x_{j^\prime } \!\!
\right\} \!
w_{\alpha i^\prime }
w^\star_{\alpha j^\prime } \\
\\[-10pt]
&\!\!\!\!
\!=\!
- \sum_{i j} \!
w_{\alpha i}
w^\star_{\alpha j} ,
~~~~~~~~~~~~~~~~~~~~~~~~~~~~~~~~~~~~~~~~~~~~~~~~~~~~~~~~~~
\hbox{(not summed for $\alpha$)} .
\EA
\label{boldassumption4}
\ea\\[-12pt]
Substituting
(\ref{boldassumption1})
into
(\ref{Lagmultiplierkandlminusplus1p}),
we obtain\\[-16pt]
\ba
\!\!\!\!\!\!\!\!\!\!
\BA{ll}
&
2
{\displaystyle 
\frac{6 \mbox{$\slashed{z}$}^4 \!-\! 3 \mbox{$\slashed{z}$}^2 \!+\! 1}
{2 \mbox{$\slashed{z}$}^2 \!-\! 1}
}
\!\cdot\!
\left\{
\left( k_{\beta \alpha } \!\mp\! l^\star_{\beta \alpha } \right) \!
\sum_i x_i w^\star_{\alpha i} 
\!+\!
\left( l_{\beta \alpha } \!\mp\! k^\star_{\beta \alpha } \right) \!
\sum_i x^\star_i w_{\alpha i}
\right\} \\
\\[-16pt]
&
\!=\!
\sum_{i j} \!
\left\{ \!
2
\!-\!
{\displaystyle \frac{1}{8 \mbox{$\slashed{z}$}^2}}
{\displaystyle
\frac{2 \mbox{$\slashed{z}$}^2 \!\!-\!\! 3}
{2 \mbox{$\slashed{z}$}^2 \!\!-\!\! 1}
}
x^\star_{i } x_{j } \!
\right\} \!
w_{\alpha i}
w^\star_{\alpha j} \!
\sum_i \!
\left( M_i w^\star_{\beta i} \!\mp\! M^\star_i w_{\beta i} \right), ~~~
\hbox{(not summed for $\alpha$ and $\beta$)} .
\EA
\label{Lagmultiplierkandlminus3p}
\ea\\[-12pt]
Equating each term in the l.h.s. and the r.h.s. of
(\ref{Lagmultiplierkandlminus3p}),
respectively,
then we have\\[-16pt]
\ba
\!\!\!\!\!\!\!\!\!\!
\BA{ll}
&
2
{\displaystyle 
\frac{6 \mbox{$\slashed{z}$}^4 \!-\! 3 \mbox{$\slashed{z}$}^2 \!+\! 1}
{2 \mbox{$\slashed{z}$}^2 \!-\! 1}
}
\!\cdot\!
\left( k_{\beta \alpha } \!\mp\! l^\star_{\beta \alpha } \right) \!
\sum_i x_i w^\star_{\alpha i}
\!=\!
\sum_{i j} \!
\left\{ \!
2
\!-\!
{\displaystyle \frac{1}{8 \mbox{$\slashed{z}$}^2}}
{\displaystyle
\frac{2 \mbox{$\slashed{z}$}^2 \!\!-\!\! 3}
{2 \mbox{$\slashed{z}$}^2 \!\!-\!\! 1}
}
x^\star_{i } x_{j } \!
\right\} \!
w_{\alpha i}
w^\star_{\alpha j} \!
\sum_i \!
M_i w^\star_{\beta i} , \\
\\[-14pt]
&~~~~~~~~~~~~~~~~~~~~~~~~~~~~~~~~~~~~~~~~~~~~~~~~~~~~~~~~~~~~~~~~~~~~~~~~~~
\hbox{(not summed for $\alpha$ and $\beta$)} ,
\EA
\label{Lagmultiplierkandlminus4p}
\ea\\[-12pt]
in which making summation with respect to $\alpha$ and using
(\ref{boldassumption1}),
(\ref{Lagmultiplierkandlminus4p})
can be converted to\\[-16pt]
\ba
\BA{l}
2
{\displaystyle 
\frac{6 \mbox{$\slashed{z}$}^4 \!-\! 3 \mbox{$\slashed{z}$}^2 \!+\! 1}
{2 \mbox{$\slashed{z}$}^2 \!-\! 1}
}
\!=\!
\sum_{i j} \!
\left\{ \!
2
\!-\!
{\displaystyle \frac{1}{8 \mbox{$\slashed{z}$}^2}}
{\displaystyle
\frac{2 \mbox{$\slashed{z}$}^2 \!\!-\!\! 3}
{2 \mbox{$\slashed{z}$}^2 \!\!-\!\! 1}
}
x^\star_{i } x_{j } \!
\right\} \!
\delta_{ij} ,
\EA
\label{Lagmultiplierkandlminus5p}
\ea\\[-14pt]
from which we acquire the relation similar to
(\ref{Lagmultiplierkandlminus6})
as\\[-16pt]
\ba
\BA{l}
z
\!=\!
{\displaystyle \frac{1}{22}}
\left\{ \!
8N \!-\! 15
\!-\!
\left (8N \!-\! 15 \right) 
\sqrt{
1
\!-\!
{\displaystyle \frac{264}{\left (8N \!-\! 15 \right)^2}}
}
\right\}
\approx
{\displaystyle \frac{6}{8N \!-\! 15}} ,
\EA
\label{Lagmultiplierkandlminus6p}
\ea\\[-12pt]
whose value is very near zero for large $N$.
We study another case.
Subtraction and addition of complex conjugation of
(\ref{Lagmultiplierkandl8c})
from
(\ref{Lagmultiplierkandl8c})
and to
(\ref{Lagmultiplierkandl8c}),
lead to a very simple equation\\[-18pt]
\ba
\!\!\!\!\!\!\!\!
\BA{ll}
&
2
{\displaystyle 
\frac{6 \mbox{$\slashed{z}$}^4 \!-\! 3 \mbox{$\slashed{z}$}^2 \!+\! 1}
{2 \mbox{$\slashed{z}$}^2 \!-\! 1}
}
\!\cdot\!
\left\{
\left( k_{\beta \alpha } \!\mp\! l^\star_{\beta \alpha } \right) \!
\sum_i x_i w^\star_{\alpha i} 
\!+\!
\left( l_{\beta \alpha } \!\mp\! k^\star_{\beta \alpha } \right) \!
\sum_i x^\star_i w_{\alpha i}
\right\} \\
\\[-12pt]
&
+
{\displaystyle \frac{1}{8 \mbox{$\slashed{z}$}^2}}
{\displaystyle
\frac{2 \mbox{$\slashed{z}$}^2 \!-\! 3}
{2 \mbox{$\slashed{z}$}^2 \!-\! 1}
}
\sum_{i^\prime \! j^\prime } \!
x^\star_{i^\prime } x_{\!j^\prime }
w_{\alpha \! i^\prime }
w^\star_{\alpha \! j^\prime } \!
\left\{ \!
\sum_\gamma \!
\left( \!
k_{\beta \gamma } \!\mp\! l^\star_{\beta \gamma } \!
\right)
\sum_i \! x_i w^\star_{\gamma i} 
\!+\!
\sum_\gamma \!
\left( \!
l_{\beta \gamma } \!\mp\! k^\star_{\beta \gamma } \!
\right)
\sum_i \! x^\star_i w_{\gamma i} \!
\right\} \\
\\[-8pt]
&
\!=\!
2
\sum_{i j} \!
w_{\alpha i}
w^\star_{\alpha j} \!
\sum_i \!
\left( M_i w^\star_{\beta i} \!\mp\! M^\star_i w_{\beta i} \right) , 
~~~~~~~~~~~~~~~~~~~~~~~~~~~~
\hbox{(not summed for $\alpha$ and $\beta$)} .
\EA
\label{Lagmultiplierkandlminusplup0pp}
\ea
In which we have used the trivial relation
which is also derived from
(\ref{boldassumption4}).
Substituting
(\ref{boldassumption1})
into
(\ref{Lagmultiplierkandlminusplup0p}),
we obtain the equation equivalent to
(\ref{Lagmultiplierkandlminus3p}).
This situation is quitely the same as the one of the previous case.
Using
(\ref{Lagmultiplierkandlminus3}) and (\ref{Lagmultiplierkandlminus3p}),
the Lagrange multipliers
$k_{\alpha \beta }$ and $l_{\alpha \beta }$
are determined in the following forms in which
the indices $\alpha$ and $\beta$ are not summed:\\[-18pt]
\ba
\!\!\!\!
\BA{l}
k_{\alpha \beta } \!\!\mp\!\! l^\star_{\beta \alpha }
\!\!=\!\!
c_1 \!\!
\sum_{i j} \!\!
\left(
2
\!\!+\!\!
c_2
x^\star_{i } x_{j }
\right) \!
w_{\beta i}
w^\star_{\beta j} \!
{\displaystyle 
\frac{\sum_i M_i w^\star_{\alpha i}}
{\sum_i x_i w^\star_{\beta i}}
} , 
k_{\beta \alpha } \!\!\mp\!\! l^\star_{\beta \alpha }
\!\!=\!\!
c_1 \!\!
\sum_{i j} \!\!
\left(
2
\!\!+\!\!
c_2
x^\star_{i } x_{j }
\right) \!
w_{\alpha i}
w^\star_{\alpha j} \!
{\displaystyle 
\frac{\sum_i M_i w^\star_{\beta i}}
{\sum_i x_i w^\star_{\alpha i}}
} , 
\EA
\label{Lagmultiplierkandldet1}
\ea\\[-14pt]
first of which is brought by exchange of the indices $\alpha$ and $\beta$
of the second which corresponds to
(\ref{Lagmultiplierkandlminus3}) and (\ref{Lagmultiplierkandlminus3p}).
The complex conjugation and exchange of the indices of
(\ref{Lagmultiplierkandldet1})
bring\\[-18pt]  
\ba
\!\!\!\!
\BA{l}
k_{\alpha \beta } \!\!\mp\!\! l_{\beta \alpha }
\!\!=\!\!
c_1 \!\!
\sum_{i j} \!\!
\left(
2
\!\!+\!\!
c_2
x^\star_{i } x_{j }
\right) \!
w_{\alpha i}
w^\star_{\alpha j} \!
{\displaystyle 
\frac{\sum_i M^\star_i w_{\beta i}}
{\sum_i x^\star_i w_{\alpha i}}
} , 
k_{ \beta \alpha } \!\!\mp\!\! l_{\beta \alpha }
\!\!=\!\!
c_1 \!\!
\sum_{i j} \!\!
\left(
2
\!\!+\!\!
c_2
x^\star_{i } x_{j }
\right) \!
w_{\beta i}
w^\star_{\beta j} \!
{\displaystyle 
\frac{\sum_i M^\star_i w_{\alpha i}}
{\sum_i x^\star_i w_{\beta i}}
} . 
\EA
\label{Lagmultiplierkandldet2}
\ea\\[-14pt]
We have used the properties
$k_{ \beta \alpha } \!\!=\!\! k^\star_{\alpha \beta }$
and
$l_{ \beta \alpha } \!\!=\!\! l_{\alpha \beta }$.
The coefficients $c_1$ and $c_2$ are given as\\[-18pt]
\ba
\left.
\BA{l}
c_1
\!=\!
8
\!-\!
{\displaystyle \frac{21}{2} \frac{1}{2N \!+\! 1}},~~
c_2
\!=\!
{\displaystyle \frac{2}{7}} (2N \!+\! 1) 
\left( \!
1
\!-\!
{\displaystyle \frac{7}{8}
\frac{1}{2N \!+\! 1 \!-\! {\displaystyle \frac{7}{4}}}
} \!
\right) , ~~
(\mbox{$\slashed{z}$}^2  \!\!\approx\!\! 1) , \\
\\[-4pt]
c_1
\!=\!
{\displaystyle \frac{8}{3} \frac{8N^2 \!-\! 21N \!+\! 18}{8N \!-\! 15}},~~
c_2
\!=\!
-
{\displaystyle \frac{1}{6} \frac{(8N \!-\! 15)(4N \!-\! 9)}{8N \!-\! 9}},~~
(\mbox{$\slashed{z}$}^2  \!\!\approx\!\! \frac{1}{2}) .
\EA
\right\}
\label{coefficients}
\ea\\[-12pt]
We here should notice that by only the equations
(\ref{Lagmultiplierkandldet1}) and (\ref{Lagmultiplierkandldet2})
the unknown parameters $k_{\alpha \beta }$ and $l_{\alpha \beta }$
cannot been determined.
Such a determination is possible if the equations
(\ref{Lagmultiplierkandldet1}) and (\ref{Lagmultiplierkandldet2})
are subtracted or added with each other.
Really, subtraction the second from the first in
(\ref{Lagmultiplierkandldet1})
and
(\ref{Lagmultiplierkandldet2})
leads to\\[-18pt]
\ba
\BA{l}
k_{\alpha \beta } \!-\! k_{\beta \alpha }
\!=\!
c_1 \!\!
\sum_{i j} \!\!
\left(
2
\!\!+\!\!
c_2
x^\star_{i } x_{j }
\right) \!
w_{\beta i}
w^\star_{\beta j} \!
{\displaystyle 
\frac{\sum_i M_i w^\star_{\alpha i}}
{\sum_i x_i w^\star_{\beta i}}
} 
\!-\!
c_1 \!\!
\sum_{i j} \!\!
\left(
2
\!\!+\!\!
c_2
x^\star_{i } x_{j }
\right) \!
w_{\alpha i}
w^\star_{\alpha j} \!
{\displaystyle 
\frac{\sum_i M_i w^\star_{\beta i}}
{\sum_i x_i w^\star_{\alpha i}}
} , \EA
\label{Lagmultiplierkandldet3}
\ea
\vspace{-0.7cm}
\ba
\BA{l}
k_{\alpha \beta } \!-\! k_{\beta \alpha }
\!=\!
c_1 \!\!
\sum_{i j} \!\!
\left(
2
\!\!+\!\!
c_2
x^\star_{i } x_{j }
\right) \!
w_{\alpha i}
w^\star_{\alpha j} \!
{\displaystyle 
\frac{\sum_i M^\star_i w_{\beta i}}
{\sum_i x^\star_i w_{\alpha i}}
} 
\!-\!
c_1 \!\!
\sum_{i j} \!\!
\left(
2
\!\!+\!\!
c_2
x^\star_{i } x_{j }
\right) \!
w_{\beta i}
w^\star_{\beta j} \!
{\displaystyle 
\frac{\sum_i M^\star_i w_{\alpha i}}
{\sum_i x^\star_i w_{\beta i}}
} ,
\EA
\label{Lagmultiplierkandldet4}
\ea\\[-14pt]
respectively.
From these relations,
we can determine the unknown parameter $k_{\alpha \beta }$ as\\[-18pt]
\ba
\!\!\!\!
\BA{l}
k_{\alpha \beta }
\!=\!
- {\displaystyle \frac{1}{2}}
c_1 \!\!
\sum_{i j} \!\!
\left(
2
\!\!+\!\!
c_2
x^\star_{i } x_{j }
\right) \!
w_{\alpha i}
w^\star_{\alpha j} \!
{\displaystyle 
\frac{\sum_i M_i w^\star_{\beta i}}
{\sum_i x_i w^\star_{\alpha i}}
} 
\!-\! {\displaystyle \frac{1}{2}}
c_1 \!\!
\sum_{i j} \!\!
\left(
2
\!\!+\!\!
c_2
x^\star_{i } x_{j }
\right) \!
w_{\beta i}
w^\star_{\beta j} \!
{\displaystyle 
\frac{\sum_i M^\star_i w_{\alpha i}}
{\sum_i x^\star_i w_{\beta i}}
}
\!=\!
k^\star_{\beta \alpha } .
\EA
\label{Lagmultiplierkalplhabeta}
\ea\\[-14pt]
On the other hand,
adding the second to the first in
(\ref{Lagmultiplierkandldet2})
yields\\[-18pt]
\ba
\!\!\!\!
\BA{l}
k_{\alpha \beta } \!\!+\!\! k_{\beta \alpha } \!\!\mp\!\! 2 l_{\beta \alpha }
\!\!=\!\!
c_1 \!\!
\sum_{i j} \!\!
\left(
2
\!\!+\!\!
c_2
x^\star_{i } x_{j }
\right) \!
w_{\alpha i}
w^\star_{\alpha j} \!
{\displaystyle 
\frac{\sum_i M^\star_i w_{\beta i}}
{\sum_i x^\star_i w_{\alpha i}}
} 
\!\!+\!\!
c_1 \!\!
\sum_{i j} \!\!
\left(
2
\!\!+\!\!
c_2
x^\star_{i } x_{j }
\right) \!
w_{\beta i}
w^\star_{\beta j} \!
{\displaystyle 
\frac{\sum_i M^\star_i w_{\alpha i}}
{\sum_i x^\star_i w_{\beta i}}
} , 
\EA
\label{Lagmultiplierkandldet5}
\ea\\[-14pt]
from which we are able to determine  the unknown parameter $l_{\alpha \beta }$ as\\[-18pt]
\ba
\!\!\!\!\!\!\!\!
\BA{l}
\mp l_{\beta \alpha }
\!\!=\!\!
{\displaystyle \frac{3}{4}} \!
c_1 \!\!
\sum_{i j} \!\!
\left(
2
\!\!+\!\!
c_2
x^\star_{i } x_{j }
\right) \!
w_{\alpha i}
w^\star_{\alpha j} \!
{\displaystyle 
\frac{\sum_i M^\star_i w_{\beta i}}
{\sum_i x^\star_i w_{\alpha i}}
} 
\!\!+\!\!
{\displaystyle \frac{3}{4}} \!
c_1 \!\!
\sum_{i j} \!\!
\left(
2
\!\!+\!\!
c_2
x^\star_{i } x_{j }
\right) \!
w_{\beta i}
w^\star_{\beta j} \!
{\displaystyle 
\frac{\sum_i M^\star_i w_{\alpha i}}
{\sum_i x^\star_i w_{\beta i}}
} \\
\\[-8pt]
~~~~~
+ {\displaystyle \frac{1}{4}}
c_1 \!\!
\sum_{i j} \!\!
\left(
2
\!\!+\!\!
c_2
x^\star_{i } x_{j }
\right) \!
w_{\alpha i}
w^\star_{\alpha j} \!
{\displaystyle 
\frac{\sum_i M_i w^\star_{\beta i}}
{\sum_i x_i w^\star_{\alpha i}}
} 
\!+\! {\displaystyle \frac{1}{4}}
c_1 \!\!
\sum_{i j} \!\!
\left(
2
\!\!+\!\!
c_2
x^\star_{i } x_{j }
\right) \!
w_{\beta i}
w^\star_{\beta j} \!
{\displaystyle 
\frac{\sum_i M_i w^\star_{\alpha i}}
{\sum_i x_i w^\star_{\beta i}}
}
\!=\!
\mp l_{\alpha \beta } .
\EA
\label{Lagmultiplierl}
\ea\\[-12pt]
To verify the validity of $k_{\alpha \beta }$
(\ref{Lagmultiplierkalplhabeta}),
we make the explicit form of $k_{\beta \alpha }$
from $k_{\alpha \beta }$
by exchanging the indices.
We calculate
$k_{\alpha \beta } \!-\! k_{\beta \alpha }$
and to the result of which further substitute
(\ref{Lagmultiplierkandldet1}) and (\ref{Lagmultiplierkandldet2}).
Then we can get
$k_{\alpha \beta } \!-\! k_{\beta \alpha }$
again.
This means that the expression for
$k_{\alpha \beta }$
is exactly valid and does not cause any inconsistency.
By the direct input of the explict expressions
(\ref{Lagmultiplierkalplhabeta}) and (\ref{Lagmultiplierl})
into the SCF parameter $M_\alpha$ difined by
(\ref{bosonizedSCFFandD}),
we could reach our ultimate goal to determine self-cosistently
the unknown Lagrange parameters
$k_{\alpha \beta }$ and $l_{\alpha \beta }$.

Finally summarizing shortly,
in the present manner
we could reach considerably good solutions for all unknown parameters
$k$ and $l$
at both the regions very near
$\mbox{$\slashed{z}$}^2 \!\!=\!\! 1$
and
$\mbox{$\slashed{z}$}^2 \!\!=\!\! \frac{1}{2}$.
We will apply the present approximation method
to concrte problems which should be solved.

\newpage

%%%%%%%%%%%%%%%%%%%%%%%%%%%%%%%%%%%%%%%%%%%%%%%%%%%%%%
%                                                    %
%  7  Concluding remarks and further perspectives    %
%                                                    %
%%%%%%%%%%%%%%%%%%%%%%%%%%%%%%%%%%%%%%%%%%%%%%%%%%%%%%

\def\thesection{\arabic{section}}
\setcounter{equation}{0}
\renewcommand{\theequation}{\arabic{section}.\arabic{equation}}
\section{Concluding remarks and further perspectives}

~~~In the present new description,
the extended HB (EHB) theory for a fermion system with $N$ single-particle states
has been derived from the extended TDHB (ETDHB) theory.
The EHB and ETDHB theories have been constructed
from a group theoretical deduction
starting with the fact that the fermion annihilation-creation and pair
operators form a Lie algebra of the $SO(2N\!\!+\!\!1)$ group.
The induced representations of the $SO(2N\!\!+\!\!1)$ group was found by means of a group extension of the $SO(2N)$ Bogoliubov transformation for the fermions
to the $SO(2N\!\!+\!\!1)$ tranformation group.
Embedding the $SO(2N\!\!+\!\!1)$ group into the $SO(2N\!+\!2)$ group and
using the boson images of the $SO(2N\!\!+\!\!2)$ Lie operators,
we have obtained the ETDHB equation from the Heisenberg
equation of motion for the boson operators.
We have expressed its final form through the variables
as the representatives of the paired modes and the unpaired modes.
From the ETDHB theory we have derived the static EHB theory
in which the paired modes and the unpaired modes are treated
in an equal manner.
The EHB theory applicable to both even and odd fermion systems is also the SCF
theory with the same capacity to provide
a mean field approximation as the usual HB theory for the even fermion systems.
We have obtained a new solution with unpaired-mode effects.
However, it includes the unknown parameters which originate from the Lagrange multipliers
involved in the image of the Hamiltonian in order to
select out the physical spinor subspace.
They cannot be determined in the classical limit only,
and a complete determination of them requires that the quantum mechanical fluctuations
are taken into account.
We have, instead, determined the 
parameters with the aid of the {\em quasi-anticommutation relation approximation}
for the differential form of the fermion operators.
Both the EHB eigenvalue equation and {\em quasi-anticommutation relation approximation} provide a group theoretically transparent kinematical frame,
which is able to 
describe both the paired and unpaired modes in many fermion dynamics.
They work well in both even and odd fermion systems with strong collective correlations
in which the effect of the unpaired modes is remarkable.
They play important and crucial roles for a unified SCF description of
the Bose-Fermi type collective excitations
in several fermion systems, e.g., quantum dots
\cite{YanLan.03}.
We stress again that
it is a very exciting problem to consider an approximate
$SO(2N\!\!+\!\!1)$ wave function very near $z \!\!=\!\! 0$.
We have never experienced such a physical situation
in which the largest contributions
from the unpaired-mode amplitudes occur
owing to the constraint
$x_\alpha^\star x_\alpha \!\!+\!\! z^2 \!\!=\!\! 1$
and consequently 
most physical contributions arise from the unpaired-mode effects.
This situation is in striking contrast to the situation in which we
have no effects due to the unpaired modes very near
$z \!\!=\!\! 1$.
A study of such a problem can be expected to open quite a new field for exploration of an exotic fermion dynamics
accompanying the stronger unpaired-mode effects.
Particularly, in a forthcoming paper,
we will investigate a concrete problem for such an object
using the present superconducting toy-model.
Then we will give detailed processes
to the EHB eigenvalue equation under
the {\em quas-anticommutation relation approximation} very near 
$z \!\!=\!\! 0$
and make analysess of solutions obtained by calculations.

Hitherto we have no effective method to
describe such Bose-Fermi type collective motions
except the previous theory proposed by Fukutome and
one of the present authors (S.N.) Ref.
\cite{FukuNishi.84}
in which the $SO(2N\!\!+\!\!1)$ density matrix plays
a crucial role.
To study the interrelation between the previous theory
and the present theory
will become a very interesting project.

Finally we will develop a group theoretical approach to
the formation of the Lax pair of the $SO(2N \!+\! 2)$ top
according to the ideas of Olshanetsky and Perelomov and
Reyman and Shansky
\cite{OlshanetskyPerelomov.81}-\cite{ReymanShansky.94}
in the near future.

\newpage

%%%%%%%%%%%%%%%%%
%               %
%    Appendix   %
%               %
%%%%%%%%%%%%%%%%%

\leftline{\large{\bf Appendix}}
\appendix

\vspace{-0.4cm}

%%%%%%%%%%%%%%%%%%%%%%%%%%%%%%%%%%%%%%%%%%%%%%%%%%%%%%%%%%%%%

\def\thesection{\Alph{section}}
\setcounter{equation}{0}
\renewcommand{\theequation}{\Alph{section}.\arabic{equation}}

\section{Differential SO(2N+2) Lie operators on coset space and operation of differential annihilation-creation operators on SO(2N+1) wave function $U({\cal G}) \ket0$}

%%%%%%%%%%%%%%%%%%%%%%%%%%%%%%%%%%%%%%%%%%%%%%%%%%%%%%%%%%%%%

\vspace{-0.1cm}

~~~Owing to
$
\bra 0 U(G) \ket0
\!=\!
\bra 0 U({\cal G}) \ket0
$,
we have
$\ket G \!=\! U({\cal G}) \ket 0 \!=\! \ket {\cal G}
({\cal G} \!\in\! SO(2N \!+\! 2))$.
Following
Refs.
\cite{Fu.77} and \cite{Fu.81},
expressions for
$\hbox{\mathbf c}_{\alpha }$ and $\hbox{\mathbf c}^\dagger_\alpha$
are given
in terms of the variables $q_{\alpha \beta }$ and $r_\alpha$:\\[-26pt]
\ba
\BA{ll}
&\hbox{\mathbf c}_{\alpha }
\!=\!
{\displaystyle \frac{\partial }{\partial r^\star_\alpha }}
\!+\!
r^\star_\xi
{\displaystyle \frac{\partial }{\partial q^\star_{\alpha \xi }}}
\!+\!
(r_{\alpha } r_\xi \!-\! q_{\alpha \xi })
{\displaystyle \frac{\partial }{\partial r_\xi }}
\!-\!
q_{\alpha \xi } r_\eta
{\displaystyle \frac{\partial }{\partial q_{\xi \eta }}}
\!+\!
ir_{\alpha }
{\displaystyle \frac{\partial }{\partial \tau }} ,~~
\hbox{\mathbf c}^\dagger_\alpha
\!=\!
-\hbox{\mathbf c}^\star_{\alpha } .
\label{Lieopratorc2}
\EA
\ea\\[-28pt]

Using
(\ref{Lieopratorc2})
we prove the identities
(\ref{identities})
as follows:\\[-16pt]
\ba
\!\!\!\!\!\!\!\!\!\!
\left.
\BA{ll}
&\hbox{\mathbf c}_{\alpha } U({\cal G}) \ket0
\!=\!
\hbox{\mathbf c}_{\alpha } U(G) \ket0
\!=\!
\hbox{\mathbf c}_{\alpha } \!
\left\{ \!
\Phi^\star_{00}(G)
( 1 \!+\! r_\beta c_\beta^\dagger)
e^{\frac{1}{2}q_{\gamma \delta }c^\dagger_\gamma c^\dagger_\delta }
\ket0 \!
\right\} \\
\\[-14pt]
&
\!=\!
-r_\alpha U(G) \ket0
\!+\!
\Phi^\star_{00}(G) \!
\left\{ \!\!
(r_{\alpha } r_\xi \!\!-\!\! q_{\alpha \xi })
c_\xi ^\dagger
e^{\frac{1}{2}q_{\gamma \delta }c^\dagger_\gamma c^\dagger_\delta }
\ket0
\!\!-\!
( 1 \!\!+\!\! r_\beta c_\beta^\dagger )
q_{\alpha \xi } r_\eta \!
{\displaystyle \frac{\partial }{\partial q_{\xi \eta }}}
e^{\frac{1}{2}q_{\gamma \delta }c^\dagger_\gamma c^\dagger_\delta }
\ket0 \!\!
\right\} \\
\\[-14pt]
&
\!=\!
-
r_\alpha U(G) \ket0
\!+\!
\left\{
r_{\alpha } r_{\xi } c^\dagger_{\xi }
(1 \!+\! r_\eta c^\dagger_\eta
\!-\! r_\eta c^\dagger_\eta)
\!-\!
q_{\alpha \xi } c^\dagger_{\xi } \!
(1 \!+\! r_\eta c^\dagger_\eta
\!-\! r_\eta c^\dagger_\eta)
\right\} \!
\Phi^\star_{00}(G)
e^{\frac{1}{2}q_{\gamma \delta }c^\dagger_\gamma c^\dagger_\delta }
\ket0 \! \\
\\[-10pt]
&
~~~~~~~~~~~~~~~~~~~~~~~~~~~~~~~~~~~~~~~~~~~~~~~~~~~~\!
-
q_{\alpha \xi } r_\eta c^\dagger_{\xi } c^\dagger_\eta\!
\Phi^\star_{00}(G)
e^{\frac{1}{2}q_{\gamma \delta }c^\dagger_\gamma c^\dagger_\delta }
\ket0 \! \\
\\[-8pt]
&
\!=\!
-
r_{\alpha } U(G) \ket0
\!+\!
r_{\alpha } r_\xi c^\dagger_\xi U(G) \ket0
\!-\!
q_{\alpha \xi } c^\dagger_\xi U(G) \ket0  \\
\\[-10pt]
&
\!=\!
\left( \!
-
r_{\alpha }
\!+\!
r_{\alpha } r_\xi c^\dagger_\xi
\!-\!
q_{\alpha \xi } c^\dagger_\xi \!
\right)
\!\cdot\!
U(G) \ket0 ,
\EA \!\!\!
\right\}
\label{calphaoperation}
\ea\\[-12pt]
where we have used
$r_\xi c^\dagger_\xi r_\eta c^\dagger_\eta = 0$.
Thus we have given the proof of the first identity of
(\ref{identities}).
Further using
$r_\beta c^\dagger_\beta r_\xi c^\dagger_\xi \!=\! 0$,
the second identity is also derived as follows:\\[-18pt]
\ba
\!\!\!\!\!\!
\left.
\BA{ll}
&\hbox{\mathbf c}^\dagger_{\alpha } U\!({\cal G}) \! \ket0
\!=\!
-\hbox{\mathbf c}^\star_{\alpha } \!
\left\{ \!
\Phi^\star_{00}(G)
( 1 \!+\! r_\beta c_\beta^\dagger)
e^{\frac{1}{2}q_{\gamma \delta }c^\dagger_\gamma c^\dagger_\delta }
\ket0 \!
\right\} \\
\\[-14pt]
&
\!=\!
-\Phi^\star_{00}(G) \!
\left\{ \!
{\displaystyle \frac{\partial }{\partial r_\alpha }}
(1 \!+\! r_\beta c_\beta^\dagger)
e^{\frac{1}{2}q_{\gamma \delta }c^\dagger_\gamma c^\dagger_\delta }
\ket0
\!+\!
(1 \!+\! r_\beta c_\beta^\dagger)
r_{\xi } \!
{\displaystyle \frac{\partial }{\partial q_{\alpha \xi }}}
e^{\frac{1}{2}q_{\gamma \delta }c^\dagger_\gamma c^\dagger_\delta }
\ket0 \!
\right\} \\
\\[-12pt]
&
\!=\!
-c^\dagger_{\alpha } \Phi^\star_{00}(G) \!
\left( \!
1
\!+\!
r_{\xi } c^\dagger_{\xi }
\!-\!
r_\beta r_{\xi } c_\beta^\dagger c^\dagger_{\xi } \!
\right)
e^{\frac{1}{2}q_{\gamma \delta }c^\dagger_\gamma c^\dagger_\delta }
\ket0
\!=\!
-c^\dagger_{\alpha }
\!\cdot
U\!(G) \! \ket0 .
\EA \!\!
\right\}
\label{calphaoperation2}
\ea\\[-10pt]
By successive use of
(\ref{identities}) and (\ref{vacfunc}),
we get the formulas below\\[-14pt]
\ba
\!\!\!\!\!\!\!\!\!\!
\left.
\BA{ll}
&
\hbox{\mathbf c}^\dagger_\alpha \hbox{\mathbf c}_\beta
U\!({\cal G}) \! \ket0
\!\!=\!\!
\left( \!
\delta_{\alpha \beta }
\!-\!
r_\beta c^\dagger_\alpha
\!+\!
r_\beta r_\xi c^\dagger_\alpha c_\xi^\dagger
\!-\!
q_{\beta \xi } c^\dagger_\alpha c_\xi^\dagger \!
\right)
\!\cdot\! U\!(G) \! \ket0 , \\
\\[-12pt]
&
\hbox{\mathbf c}_\beta \hbox{\mathbf c}^\dagger_\alpha
U\!({\cal G}) \! \ket0
\!\!=\!\!
\left( \!
r_\beta c^\dagger_\alpha
\!-\!
r_\beta r_\xi c^\dagger_\alpha c_\xi^\dagger
\!+\!
q_{\beta \xi } c^\dagger_\alpha c_\xi^\dagger \!
\right)
\!\cdot\! U\!(G) \! \ket0 ,
\hbox{\mathbf c}_\alpha^\dagger \hbox{\mathbf c}_\beta^\dagger
U\!({\cal G}) \! \ket0
\!\!=\!\!
-c_\alpha^\dagger c_\beta^\dagger \!\cdot\! U\!(G) \! \ket0 ,\\
\\[-12pt]
&
\hbox{\mathbf c}_\alpha \hbox{\mathbf c}_\beta
U\!({\cal G}) \! \ket0
\!\!=\!\!
\left\{ \!
q_{\alpha \beta }
\!-\!
q_{\alpha \xi } q_{\beta \eta } c_\xi^\dagger c^\dagger_\eta
\!\!-\!\!
\left( q_{\alpha \xi } r_\beta
\!\!-\!\!
r_\alpha q_{\beta \xi }
\right) \!
c^\dagger_\xi
\!\!+\!\!
\left( q_{\alpha \xi } r_\beta
\!\!-\!\!
r_\alpha q_{\beta \xi }
\right) \!
r_\eta 
c_\xi^\dagger c^\dagger_\eta
\right\}
\!\cdot\! U\!(G) \! \ket0 .
\EA \!\!\!
\right\}
\label{calphaoperation3}
\ea\\[-10pt]
The first and last equations of
(\ref{calphaoperation3})
are proved as follows:\\[-14pt]
\ba
\!\!\!\!\!\!
\left.
\BA{ll}
&
\hbox{\mathbf c}^\dagger_{\alpha }
\hbox{\mathbf c}_{\beta }
U({\cal G}) \ket0
\!=\!
\left\{ \!
\hbox{\mathbf c}^\dagger_{\alpha } \!
\left( \!
-
r_{\beta }
\!+\!
r_{\beta } r_\xi c^\dagger_\xi
\!-\!
q_{\beta \xi } c^\dagger_\xi
\right)
\!\!+\!\!
\left( \!
-
r_{\beta }
\!+\!
r_{\beta } r_\xi c^\dagger_\xi
\!-\!
q_{\beta \xi } c^\dagger_\xi
\right) \!\!
\left( - c^\dagger_\alpha \right) \!
\right\}
\!\cdot\!
U(G) \ket0 \\
\\[-14pt]
&~~~~~
\!=\!
\left\{ \!
{\displaystyle \frac{\partial r_{\beta }}{\partial r_\alpha }}
\!-\!
{\displaystyle \frac{\partial r_{\beta }}{\partial r_\alpha }}
r_\xi c^\dagger_\xi
\!-\!
r_{\beta } 
{\displaystyle \frac{\partial r_\xi }{\partial r_\alpha }}
c^\dagger_\xi
\!+\!
r_\eta
{\displaystyle \frac{\partial q_{\beta \xi } }
{\partial q_{\alpha \eta }}}
c^\dagger_\xi
\!+\!
r_{\beta } c^\dagger_\alpha
\!+\!
r_{\beta } c^\dagger_\alpha r_\xi c^\dagger_\xi
\!-\!
q_{\beta \xi } c^\dagger_\xi c^\dagger_\alpha \!
\right\}
\!\cdot\!
U(G) \ket0 \\
\\[-12pt]
&~~~~~
\!=\!
\left(
\delta_{\alpha \beta }
\!-\!
r_\beta c^\dagger_\alpha
\!+\!
r_\beta r_\xi c^\dagger_\alpha c_\xi^\dagger
\!-\!
q_{\beta \xi } c^\dagger_\alpha c_\xi^\dagger
\right)
\!\cdot\! U(G) \ket0 , \\
\\[-12pt]
&
\hbox{\mathbf c}_{\alpha }
\hbox{\mathbf c}_{\beta }
U({\cal G}) \ket0
\!=\!
\left\{ \!
\left[
(r_{\alpha } r_\xi \!-\! q_{\alpha \xi })
{\displaystyle \frac{\partial }{\partial r_\xi }}
\!-\!
q_{\alpha \xi } r_\eta
{\displaystyle \frac{\partial }{\partial q_{\xi \eta }}} \!
\right] \!
\left(
-
r_{\beta }
\!+\!
r_{\beta } r_\xi c^\dagger_\xi
\!-\!
q_{\beta \xi } c^\dagger_\xi
\right)
\right. \\
\\[-16pt]
&
\left. 
~~~~~~~~~~~~~~~~~~~~~~
\!+\!
\left( \!
-r_{\beta }
\!+\!
r_{\beta } r_\xi c^\dagger_\xi
\!-\!
q_{\beta \xi } c^\dagger_\xi \!
\right) \!\!
\left( \!
-r_{\alpha }
\!+\!
r_{\alpha } r_\xi c^\dagger_\xi
\!-\!
q_{\alpha \xi } c^\dagger_\xi \!
\right) \!\!\!\!\!
{}^{^{^{^{^{^{.}}}}}}
\right\}
\!\cdot\!
U(G) \ket0 \\
\\[-14pt]
&
\!=\!
\left\{ \!
q_{\alpha \beta }
\!-\!
q_{\alpha \xi } q_{\beta \eta } c_\xi^\dagger c^\dagger_\eta
\!-\!
\left( q_{\alpha \xi } r_\beta
\!-\!
r_\alpha q_{\beta \xi }
\right)
c^\dagger_\xi
\!+\!
\left( q_{\alpha \xi } r_\beta
\!-\!
r_\alpha q_{\beta \xi }
\right)
r_\eta 
c_\xi^\dagger c^\dagger_\eta \!
\right\}
\!\cdot\!
U(G) \ket0 .
\EA \!\!
\right\}
\label{calphaoperation4}
\ea
\newpage
Using another form of
$\mbox{$\slashed{z}$}$, i.e.,
$
\mbox{$\slashed{z}$}
\!=\!
(1 \!+\! r^\dagger \chi r)^{-\frac{1}{2}}
(\chi
\!=\!
(1 \!+\! qq^\dagger)^{-1})
$,
we give useful formulas to evaluate
expectation values of the anticommutators of the operators
\hbox{\mathbf c}$_{\alpha }$ and
\hbox{\mathbf c}$_{\alpha }^\dagger$
in differential forms
by an approximate
$\widetilde{U}(G) \ket0$ of the $SO(2N \!\!+\!\! 1)$ WF
$U(G) \ket0$,
i.e.,
(\ref{approxwf}):\\[-16pt]
\ba
\left.
\BA{l}
\hbox{\mathbf c}_\alpha \mbox{$\slashed{z}$}
\!=\!
-{\displaystyle \frac{1}{2}}
\mbox{$\slashed{z}$}
\left\{
r^{\mbox{\scriptsize T}} (1\!+\!q^\dagger q)^{-1}
\!+\!
r^\dagger q(1\!+\!q^\dagger q)^{-1}
\right\}_\alpha
\!\equiv\!
\left\{
r^{\mbox{\scriptsize T}}\!\oplus\!r^\dagger
\right\}_\alpha
\!=\!
-{\displaystyle
\frac{x_\alpha^{\mbox{\scriptsize T}}}{4 \mbox{$\slashed{z}$}}
} , \\
\\[-16pt]
\hbox{\mathbf c}^\dagger_\alpha \mbox{$\slashed{z}$}
\!=\!
~~
{\displaystyle \frac{1}{2}}
\mbox{$\slashed{z}$}
\left\{
r^\dagger( 1 + qq^\dagger )^{-1}
\!-\!\!
r^{\mbox{\scriptsize T}} \! q^\dagger \! ( 1 \!+\! qq^\dagger )^{-1}
\right\}_\alpha
\!\equiv\!
\left\{
r^\dagger\!\ominus\!r^{\mbox{\scriptsize T}}
\right\}_\alpha
\!=\!~~
{\displaystyle 
\frac{x_\alpha^\dagger }{4 \mbox{$\slashed{z}$}}
} , 
\EA 
\mbox{$\slashed{z}$}
\!\equiv\!
\sqrt{{\displaystyle \frac{1 \!\!+\!\! z}{2}}} .
\right\}
\label{calphadiffrential}
\ea\\[-12pt]
The proof of the first equation in
(\ref{calphadiffrential})
is given as follows:\\[-18pt]
\ba
\left.
\BA{ll}
&
{\displaystyle \frac{\partial }{\partial r^\star_\alpha }}
\mbox{$\slashed{z}$}
\!=\!
-{\displaystyle \frac{1}{2}}
{\displaystyle \frac{1}{1\!+\!r^\dagger \chi r}}
\sqrt{{\displaystyle \frac{1}{1\!+\!r^\dagger \chi r}}}
{\displaystyle \frac{\partial }{\partial r^\star_\alpha }}
\left(
1 \!+\! r^\star _\beta \chi_{\beta \gamma } r_\gamma
\right)
\!=\!
-{\displaystyle \frac{1}{2}} \mbox{$\slashed{z}$}^3
\left( \chi r \right)_\alpha , \\
\\[-12pt]
&
r^\star_\xi
{\displaystyle \frac{\partial }{\partial q^\star_{\alpha \xi }}}
\mbox{$\slashed{z}$}
\!=\!
-{\displaystyle \frac{1}{2}} \mbox{$\slashed{z}$}^3
r^\star_\xi
r^\star _\beta
{\displaystyle \frac{\partial }{\partial q^\star_{\alpha \xi }}}
\left\{(1\!+\!qq^\dagger)^{-1}\right\}_{\beta \gamma } \! r_\gamma \\
\\[-12pt]
&~~~~~~~~~~~
\!=\!
{\displaystyle \frac{1}{2}} \mbox{$\slashed{z}$}^3
\left\{
r^\dagger \chi q r^\star \!\cdot\! (\chi r)
\!-\!
r^\dagger \chi r \!\cdot\! (r^\star \chi q)
\right\}_\alpha
\!=\!
-{\displaystyle \frac{1}{2}} \mbox{$\slashed{z}$}^3
r^\dagger \chi r \!\cdot\! (r^\dagger \chi q)_\alpha , \\
\\[-12pt]
&
(r_{\alpha } r_\xi \!\!-\!\! q_{\alpha \xi })
{\displaystyle \frac{\partial }{\partial r_\xi }}
\mbox{$\slashed{z}$}
\!=\!
-{\displaystyle \frac{1}{2}} \mbox{$\slashed{z}$}^3
(r_{\alpha } r_\xi \!\!-\!\! q_{\alpha \xi })
{\displaystyle \frac{\partial }{\partial r_\xi }} \!
\left( \!
1 \!+\! r^\star _\beta \chi_{\beta \gamma } r_\gamma \!
\right)
\!=\!
-{\displaystyle \frac{1}{2}} \mbox{$\slashed{z}$}^3 \!
\left\{ \!
r^\dagger \chi r \!\cdot\! r
\!-\!
\left( q \chi^\star r^\star \right) \!
\right\}_\alpha \! , \\
\\[-12pt]
&
\!-\!
q_{\alpha \xi } r_\eta
{\displaystyle \frac{\partial }{\partial q_{\xi \eta }}}
\mbox{$\slashed{z}$}
\!=\!
{\displaystyle \frac{1}{2}} \mbox{$\slashed{z}$}^3
q_{\alpha \xi } r_\eta
r^\star _\beta
{\displaystyle \frac{\partial }
{\partial (1\!+\!qq^\dagger)_{\delta \epsilon }}}
\left\{(1\!+\!qq^\dagger)^{-1}\right\}_{\beta \gamma }\!
{\displaystyle \frac{\partial (1\!+\!qq^\dagger)_{\delta \epsilon }}
{\partial q_{\xi \eta }}}
r_\gamma \\
\\[-12pt]
&~~~~~~~~~~~~~~~~\!
\!=\!
-{\displaystyle \frac{1}{2}} \mbox{$\slashed{z}$}^3
\left\{
r^{\mbox{\scriptsize T}} q^\dagger \chi q r \!\cdot\! ( q r^\star \chi)
\!-\!
r^\dagger \chi r \!\cdot\! ( q q^\dagger \chi r )
\right\}_\alpha
\!=\!
{\displaystyle \frac{1}{2}} \mbox{$\slashed{z}$}^3
r^\dagger \chi r \!\cdot\! (r \!-\! \chi r)_\alpha .
\EA
\right\}
\label{calphadiffrentialproof}
\ea\\[-10pt]
Gathering these results,
the first equation is obtained.
Using the relation
$
\hbox{\mathbf c}^\dagger_\alpha
\!=\!
-\hbox{\mathbf c}^\star_{\alpha }
$,
the last equation of
(\ref{Lieopratorc2}),
the second equation is also proved.
With the use of the same simbols as the ones introduced in
(\ref{calphadiffrential}),
we also give the following useful formulas first
presented in Ref.
\cite{Nishi.98}:\\[-22pt]
\ba
\left.
\BA{ll}
&\hbox{\mathbf c}_\alpha
\left\{
r^{\mbox{\scriptsize T}}\!\oplus\!r^\dagger
\right\}_\beta
\!=\!
{\displaystyle
\frac{x_\alpha x^{\mbox{\scriptsize T}}_\beta }{4 \mbox{$\slashed{z}$}^2}
} , ~
\hbox{\mathbf c}_\alpha^\dagger
\left\{
r^{\mbox{\scriptsize T}}\!\oplus\!r^\dagger
\right\}_\beta
\!=
-
2
\left( \!
1 \!-\! {\displaystyle \frac{1}{2 \mbox{$\slashed{z}$}^2}} \!
\right)
\!\cdot\!
\delta_{\alpha \beta }
-
{\displaystyle
\frac{x_\alpha^\star x^{\mbox{\scriptsize T}}_\beta }
{4 \mbox{$\slashed{z}$}^2}
} ,\\
\\[-14pt]
&\hbox{\mathbf c}_\beta
\left\{
r^\dagger\!\ominus\!r^{\mbox{\scriptsize T}}
\right\}_\alpha
\!=
2
\left( \!
1 \!-\! {\displaystyle \frac{1}{2 \mbox{$\slashed{z}$}^2}} \!
\right)
\!\cdot\!
\delta_{\alpha \beta }
+
{\displaystyle
\frac{x_\alpha^\star x^{\mbox{\scriptsize T}}_\beta }
{4 \mbox{$\slashed{z}$}^2}
} .
\EA
\right\}
\label{calphadiffrentialproof2}
\ea\\[-12pt]
The first and second equations of
(\ref{calphadiffrentialproof2})
are proved as follows:\\[-20pt]
\ba
\!\!\!\!\!\!\!\!\!\!\!\!
\left.
\BA{ll}
&
{\displaystyle \frac{\partial }{\partial r^\star_\alpha }}
\left\{
r^{\mbox{\scriptsize T}}\!\oplus\!r^\dagger
\right\}_\beta
\!=\!
\left\{
q (1 \!+\! q^\dagger q )^{-1}
\right\}_{\alpha \beta }, \\
\\[-14pt]
&
( r_{\alpha } r_\xi \!-\! q_{\alpha \xi } )
{\displaystyle \frac{\partial }{\partial r_\xi }}
\left\{
r^{\mbox{\scriptsize T}}\!\oplus\!r^\dagger
\right\}_\beta
\!=
r_\alpha
\left\{
r^{\mbox{\scriptsize T}} ( 1 \!+\! q^\dagger q )^{-1}
\right\}_\beta
\!-\!
\left\{
q(1\!+\!q^\dagger q)^{-1}
\right\}_{\alpha \beta }, \\
\\[-14pt]
&
r^\star_\xi
{\displaystyle \frac{\partial }{\partial q^\star_{\alpha \xi }}}
\left\{
r^{\mbox{\scriptsize T}}\!\oplus\!r^\dagger
\right\}_\beta
\!=
r^\star_\xi
r^{\mbox{\scriptsize T}}_\epsilon
{\displaystyle
\frac{\partial }
{\partial (1 \!+\! q^\dagger q )_{\gamma \gamma^\prime }}
} \!
\left\{( 1 \!+\! q^\dagger q )^{-1}\right\}_{\epsilon \beta } \!
{\displaystyle
\frac{\partial (1 \!+\! q^\dagger q )_{\gamma \gamma^\prime }}
{\partial q^\star_{\alpha \xi }}
} \\
\\[-14pt]
&
~~~~~~~~~~~~~~~~~~~~~~~
+\!
r^\star_\xi
r^\dagger_\epsilon
q_{\epsilon \delta }
{\displaystyle
\frac{\partial }
{\partial (1 \!+\! q^\dagger q )_{\gamma \gamma^\prime }}
} \!
\left\{(1 \!+\! q^\dagger q )^{-1}\right\}_{\delta \beta } \!
{\displaystyle
\frac{\partial (1 \!+\! q^\dagger q )_{\gamma \gamma^\prime }}
{\partial q^\star_{\alpha \xi }}
} \\
\\[-8pt]
&
=
-
r^\dagger ( 1 \!+\! qq^\dagger ) r \!
\left\{
q( 1 \!+\! q^\dagger q )^{-1} \!
\right\}_{\alpha \beta }
\!+\!\!
\left\{
r^{\mbox{\scriptsize T}}( 1 \!+\! q^\dagger q )^{-1}
\!+\!
r^\dagger q( 1 \!+\! q^\dagger q )^{-1} \!\!
\right\}_{\alpha } \!
\left\{
r^\dagger q( 1 \!+\! q^\dagger q )^{-1} \!
\right\}_{\beta } , \\
\\[-12pt]
&
-
q_{\alpha \xi } r_\eta
{\displaystyle \frac{\partial }{\partial q_{\xi \eta }}} \!
\left\{
r^{\mbox{\scriptsize T}}\!\oplus\!r^\dagger
\right\}_\beta
\!=\!
-
q_{\alpha \xi } r_\eta
r^{\mbox{\scriptsize T}}_\epsilon
{\displaystyle \frac{\partial }
{\partial (1 \!+\! q^\dagger q )_{\gamma \gamma^\prime }}} \!
\left\{( 1 \!+\! q^\dagger q )^{-1}\right\}_{\epsilon \beta } \!\!
{\displaystyle
\frac{\partial ( 1 \!+\! q^\dagger q )_{\gamma \gamma^\prime }}
{\partial q_{\xi \eta }}
} \\
\\[-14pt]
&
-
q_{\alpha \xi } r_\eta
r^{\mbox{\scriptsize T}}_\epsilon
{\displaystyle \frac{\partial q_{\epsilon \delta }}
{\partial q_{\xi \eta }}}
\left\{( 1 \!+\! q^\dagger q )^{-1}\right\}_{\delta \beta }
\!-\!
q_{\alpha \xi } r_\eta
r^\star_\xi
r^\dagger_\epsilon
q_{\epsilon \delta } \!
{\displaystyle
\frac{\partial }
{\partial ( 1 \!+\! q^\dagger q )_{\gamma \gamma^\prime }}
}
\left\{( 1 \!+\! q^\dagger q )^{-1}\right\}_{\delta \beta } \!\!
{\displaystyle
\frac{\partial ( 1\!+\! q^\dagger q )_{\gamma \gamma^\prime }}
{\partial q_{\xi \eta }}
} \\
\\[-6pt]
&
=
-
r_\alpha \!
\left\{
r^{\mbox{\scriptsize T}} ( 1 \!+\! q^\dagger q )^{-1}
\right\}_\beta
\!+\!
r^\dagger ( 1 \!+\! qq^\dagger ) r \!
\left\{
q(1\!+\!q^\dagger q)^{-1} \!
\right\}_{\alpha \beta } \\
\\[-6pt]
&
~~~~~~~~~~~~~~~~~~~~~~~~~~~~~~
\!+\!
\left\{
r^{\mbox{\scriptsize T}}( 1 \!+\! q^\dagger q )^{-1}
\!+\!
r^\dagger q( 1 \!+\! q^\dagger q )^{-1} \!
\right\}_{\alpha } \!
\left\{
r^{\mbox{\scriptsize T}} ( 1 \!+\! q^\dagger q )^{-1} \!
\right\}_{\beta } .
\EA \!\!
\right\}
\label{calphadiffrentialproof3}
\ea
On the other hand, we have\\[-16pt]
\ba
\!\!\!\!\!\!\!\!\!\!\!\!\!\!
\left.
\BA{ll}
&
-{\displaystyle \frac{\partial }{\partial r_\alpha }} \!\!
\left\{
r^{\mbox{\scriptsize T}}\!\oplus\!r^\dagger
\right\}_{\!\beta }
\!=\!
-\!
\left\{
( \! 1 \!+\! q^\dagger q \! )^{-1}
\right\}_{\!\alpha \beta } , \\
\\[-12pt]
&
-(r^\star_{\alpha } r^\star_\xi \!-\! q^\star_{\alpha \xi })
{\displaystyle \frac{\partial }{\partial r^\star_\xi }} \!
\left\{ 
r^{\mbox{\scriptsize T}}\!\oplus\!r^\dagger
\right\}_{\!\beta }
\!=\!
-
r^\star_\alpha
\left\{
r^\dagger \! ( \! 1\!+\!q^\dagger q \! )^{-1}
\right\}_{\!\beta }
\!-\!
\delta_{\alpha \beta }
\!+\!
\left\{
( \! 1 \!+\!q^\dagger q \! )^{-1}
\right\}_{\!\alpha \beta } , \\
\\[-12pt]
&
-
r_\xi
{\displaystyle \frac{\partial }{\partial q_{\alpha \xi }}}
\left\{
r^{\mbox{\scriptsize T}}\!\oplus\!r^\dagger
\right\}_\beta
\!=\!
-
r_\xi
r^{\mbox{\scriptsize T}}_\epsilon
{\displaystyle \frac{\partial }
{\partial ( 1 \!+\! q^\dagger q )_{\gamma \gamma^\prime }}}
\left\{( 1 \!+\! q^\dagger q )^{-1}\right\}_{\epsilon \beta }
{\displaystyle
\frac{\partial ( 1 \!+\! q^\dagger q )_{\gamma \gamma^\prime }}
{\partial q_{\alpha \xi }}
} \\
\\[-12pt]
&
~~~~~~~~~~~~~~-\!
r_\xi
r^\dagger_\epsilon
{\displaystyle \frac{\partial q_{\epsilon \delta }}
{\partial q_{\alpha \xi }}} \!\!
\left\{\!( 1 \!+\! q^\dagger q )^{-1}\!\right\}_{\delta \beta }
\!\!-\!
r_\xi
r^\dagger_\epsilon
q_{\epsilon \delta }
{\displaystyle \frac{\partial }
{\partial ( 1 \!+\! q^\dagger q )_{\gamma \gamma^\prime }}} \!
\left\{( \! 1 \!+\! q^\dagger q )^{-1}\!\right\}_{\delta \beta } \!
{\displaystyle
\frac{\partial ( 1 \!+\! q^\dagger q )_{\gamma \gamma^\prime }}
{\partial q_{\alpha \xi }}
} \\
\\[-8pt]
&
=
r^\dagger (1\!+\!qq^\dagger) r
\left\{
( 1 \!+\! q^\dagger q )^{-1} \!
\right\}_{\alpha \beta }
\!-\!
\left\{
r^\dagger ( 1 \!+\! qq^\dagger )^{-1}
\!-\!
r^{\mbox{\scriptsize T}} q^\dagger ( 1 \!+\! qq^\dagger )^{-1} \!
\right\}_{\alpha } \!
\left\{
r^{\mbox{\scriptsize T}} ( 1 \!+\! q^\dagger q )^{-1} \!
\right\}_{\beta } , \\
\\[-10pt]
&
q^\star_{\alpha \xi } r^\star_\eta
{\displaystyle \frac{\partial }{\partial q^\star_{\xi \eta }}} \!
\left\{
r^{\mbox{\scriptsize T}}\!\oplus\!r^\dagger
\right\}_\beta
\!=\!
q^\star_{\alpha \xi } r^\star_\eta
r^{\mbox{\scriptsize T}}_\epsilon
{\displaystyle \frac{\partial }
{\partial ( 1 \!+\! q^\dagger q )_{\gamma \gamma^\prime }}} \!
\left\{(1 \!+\! q^\dagger q )^{-1}\right\}_{\epsilon \beta } \!\!
{\displaystyle
\frac{\partial ( 1 \!+\! q^\dagger q )_{\gamma \gamma^\prime }}
{\partial q^\star_{\xi \eta }}
} \\
\\[-12pt]
&
~~~~~~~~~~~~~~~~~~~~~~~~~
+
q^\star_{\alpha \xi } r^\star_\eta
r^\star_\xi
r^\dagger_\epsilon
q_{\epsilon \delta }
{\displaystyle \frac{\partial }
{\partial ( 1 \!+\! q^\dagger q )_{\gamma \gamma^\prime }}}
\left\{( 1 \!+\! q^\dagger q )^{-1}\right\}_{\delta \beta } \!
{\displaystyle
\frac{\partial ( 1 \!+\! q^\dagger q )_{\gamma \gamma^\prime }}
{\partial q^\star_{\xi \eta }}
} \\
\\[-6pt]
&
=
r^\dagger ( 1 \!+\! qq^\dagger ) r
\!\cdot\!
\delta_{\alpha \beta }
\!-\!
r^\dagger ( 1 \!+\! qq^\dagger ) r \!
\left\{
q ( 1 \!+\! q^\dagger q )^{-1} \!
\right\}_{\alpha \beta }
\!+\!
r^\dagger_\alpha \!
\left\{
r^\dagger q ( 1 \!+\! q^\dagger q )^{-1}
\right\}_\beta \\
\\[-6pt]
&
~~~~~~~~~~~~~~~~~~~~~~~~~~~~~~
\!+\!
\left\{
r^{\mbox{\scriptsize T}}(1\!+\!q^\dagger q)^{-1} q^\dagger
\!-\!
r^\dagger ( 1 \!+\! q^\dagger q )^{-1} \!
\right\}_{\alpha } \!
\left\{
r^\dagger q (1 \!+\! q^\dagger q )^{-1} \!
\right\}_{\beta } .
\EA \!\!
\right\}
\label{calphadiffrentialproof4}
\ea\\[-16pt]
Gathering results
(\ref{calphadiffrentialproof3}) and 
(\ref{calphadiffrentialproof4}),
the first and second equations of
(\ref{calphadiffrentialproof2})
are obtained.
By using the relation
$
\hbox{\mathbf c}^\dagger_\alpha
\!=\!
-\hbox{\mathbf c}^\star_{\alpha }
$
and
the second equations of
(\ref{calphadiffrentialproof2}),
the last equation of
(\ref{calphadiffrentialproof2})
is also proved.
Using the identities
(\ref{identities}) and the two important relations
$\erw{c_\xi }^{\!\!\!\!\!\!\star } ~~\! r_\xi \!=\! 0$ and
$q_{\alpha \xi } \! \erw{c_\xi }^{\!\!\!\!\!\!\star } ~
=\!
-{\displaystyle \frac{1}{2}} \!
\left(
x
\!+\!
q x^\star \!
\right)_\alpha \!
\!+\!
\erw{c_\alpha }^{\!\!\!\!\!\!\star }
$~,
the following expectation values of the annihilation and creation operators,
$\hbox{\mathbf c}_\alpha$
and
$\hbox{\mathbf c}^\dagger_\alpha$,
are derived:\\[-16pt]
\ba
\!\!\!\!\!\!\!\!\!\!\!\!
\left.
\BA{ll}
&
\bra0
\widetilde{U}^\dagger({\cal G}) \!
\hbox{\mathbf c}_\alpha
U(G) \ket0
\!=\!
-{\displaystyle \frac{1}{2}} \!\!
\left( \!
\mbox{$\slashed{z}$}
\!+\!
{\displaystyle \frac{1}{\mbox{$\slashed{z}$}}} \!
\right) \!\!
\left\{ \!
\erw{c_\alpha }
\!+\!
{\displaystyle \frac{1 \!-\! \mbox{$\slashed{z}$}^2}
{2 \mbox{$\slashed{z}$}^2}} \!
\left( \!
x
\!+\!
q x^\star \!
\right)_\alpha \!
\right\}
\!\!\approx\!\!
-{\displaystyle \frac{1}{2}} \!\!
\left( \!
\mbox{$\slashed{z}$}
\!+\!
{\displaystyle \frac{1}{\mbox{$\slashed{z}$}}} \!
\right) \!
\erw{c_\alpha } , \\
\\[-10pt]
&
\bra0
\widetilde{U}^\dagger({\cal G}) \!
\hbox{\mathbf c}^\dagger_\alpha
U(G) \ket0
\!=\!
-{\displaystyle \frac{1}{2}} \!\!
\left( \!
\mbox{$\slashed{z}$}
\!+\!
{\displaystyle \frac{1}{\mbox{$\slashed{z}$}}} \!
\right) \!
\erw{c_\alpha }^{\!\!\!\!\!\!\star } ~, ~
(0 \!\ll\! \mbox{$\slashed{z}$}^2 \!\approx\! 1) ,
\EA \!\!
\right\}
\label{modifiedidentities}
\ea\\[-16pt]
which have been used to compute the expectation values of the anticommutators between
$\hbox{\mathbf c}_\alpha$
and
$\hbox{\mathbf c}^\dagger_\alpha$
by
$\widetilde{U}(G) \ket0$
(\ref{0Opanticommu2}).

\newpage
%\vspace{-0.5cm}

%%%%%%%%%%%%%%%%%%%%%%
%                    %
%  Acknowledgements  %
%                    %
%%%%%%%%%%%%%%%%%%%%%%

%\vskip1.5cm
\begin{center}
{\bf Acknowledgements}
\end{center}
~~~~S. N. would like to
express his sincere thanks to 
Professor Manuel Fiolhais for kind and
warm hospitality extended to
him at the Centro de F\'\i sica Computacional,
Universidade de Coimbra, Portugal.
This work was supported by FCT (Portugal) under the project
CERN/FP/83505/2008.
The authors thank the Yukawa Institute for Theoretical Physics
at Kyoto University. 
Discussions during the YITP workshop
YITP-W-10-02 on 
``Development of Quantum Fields and String Theory 2010''
were useful to complete this work.

\newpage
%\vspace{-0.4cm}

%%%%%%%%%%%%%%%%
%              %
%  References  %
%              %
%%%%%%%%%%%%%%%%


\begin{thebibliography}{999}
\bibitem{Bogo.58}
N.N. Bogoliubov,
{\it Soviet Phys. JETP} {\bf 7}, 41 (1958).
\bibitem{RS.80}
P. Ring and P. Schuck,
{\it The nuclear many-body problem},
Springer, Berlin, 1980.
\bibitem{BlaizotRipka.86}
J.P. Blaizot and G. Ripka,
{\it Quantum Theory of Finite Systems},
MIT Press, Cambridge, MA, 1986.
\bibitem{Bogo.59}
N.N. Bogoliubov,
{\it Usp. Fiz. Nauk} {\bf 67}, 549 (1959).
\bibitem{Nishi.81}
S. Nishiyama,
{\it Prog. Theor. Phys.} {\bf 66}, 348 (1981).
\bibitem{FYN.77}
H. Fukutome, M. Yamamura and S. Nishiyama,
{\it Prog. Theor. Phys.} {\bf 57}, 1554 (1977).
\bibitem{Sch.65}
J. Schwinger,
in {\it Quantum Theory of Angular Momentum},
L. Biedenharn and H. van Dam (eds.),
Academic Press, New York, 1965, 229.
\bibitem{YN.76}
M. Yamamura and S. Nishiyama,
{\it Prog. Theor. Phys.} {\bf 56}, 124 (1976).
\bibitem{Nishi.98}
S. Nishiyama,
{\it Int. J. Mod. Phys.} {\bf E7}, 677 (1998).
\bibitem{SJCF.08}
S. Nishiyama, J. da Provid\^{e}ncia,
C. Provid\^{e}ncia and F. Cordeiro,
{\it Nucl. Phys.} {\bf B802}, 121 (2008).
\bibitem{SJCF.11}
S. Nishiyama, J. da Provid\^{e}ncia,
C. Provid\^{e}ncia and F. Cordeiro,
{\it J. High Energy Phys.} {\bf02}, 093-1 (2011).
\bibitem{Ba.61}
M. Baranger,
{\it Phys. Rev.} {\bf122}, 992 (1961).
\bibitem{Sawada.57}
K. Sawada,
{\it Phys. Rev.} {\bf106}, 372 (1957).
\bibitem{Sawada.60} 
K. Sawada,
{\bf119}, 2090 (1960).
\bibitem{Nishi.96}
S. Nishiyama,
in the Proceedings of the XXI International Colloquium on Group
Theoretical Methods in Physics, 
edited by H.-D. Doebner, W. Scherer and C. Schulte,
World Scientific, Singapore, 1998, pp. 786-790. 
\bibitem{TripodiLima.97}
L. Tripodi and C.L. Lima,
{\it Phys. Lett.} {\bf B412}, 7 (1977).
\bibitem{AlginArik.01}
A. Algin and M. Arik,
{\it Eur. Phys. J.} {\bf C19, 583} (2001).
\bibitem{AlginArikAlikan.02}
A. Algin, M. Arik and A.S. Arikan,
{\it Phys. Rev.} {\bf E65}, 026140 (2002).
\bibitem{BonatsosDaskoloyannis.99}
D. Bonatsos and C. Daskoloyannis,
{\it Prog. Part. Nucl. Phys.} {\bf 43}, 537 (1999) 618.
\bibitem{OlshanetskyPerelomov.81}
M.A. Olshanetsky and A.M. Perelomov,
{\it Phys. Rep.} {\bf 71} 
(1981) 313-400.
\bibitem{OlshanetskyPerelomov.94}
M.A. Olshanetsky and A.M. Perelomov,
in {\em Dynamical Systems VII}, 
edited by V.I. Arnol'd and S.P. Novikov,
Springer, 
1994, pp. 87-116.
\bibitem{ReymanShansky.94}
A.G. Reyman and M.A. Semenov-Tian-Shansky, 
{\em Dynamical Systems VII}, 
edited by V.I. Arnol'd and S.P. Novikov,
Springer, 
1994, pp. 116-225.
\bibitem{Fu.77}
H. Fukutome,
{\it Prog. Theor. Phys.} {\bf 58}, 1692 (1977).
\bibitem{Fu.81}
H. Fukutome,
{\it Prog. Theor. Phys.} {\bf 65}, 809 (1981).
\bibitem{Doba.81}
J. Dobaczewski,
{\it Nucl. Phys.} {\bf A369}, 213 (1981).
\bibitem{Doba2.81}
J. Dobaczewski,
{\it Nucl. Phys.} {\bf A369}, 237  (1981).
\bibitem{Doba.82}
J. Dobaczewski,
{\it Nucl. Phys.} {\bf A380}, 1 (1982).
\bibitem{FukuNishiFuku.92}
S. Nishiyama and H. Fukutome,
{\it J. Phys. G: Nucl. Part. Phys.} {\bf 18}, 317 (1992).
\bibitem{BCS.80}
J. Bardeen, L.N. Cooper and J.R. Schrieffer,
{\it Phys. Rev.} {\bf 108}, 1175 (1957).
\bibitem{NIFY.80}
S. Nishiyama, M. Iwasaki, H. Fukutome and M. Yamamura,
{\it Prog. Theor. Phys.} {\bf 64}, 558 (1980) 558.
\bibitem{NIFY2.80}
S. Nishiyama, M. Iwasaki, H. Fukutome and M. Yamamura,
{\it Prog. Theor. Phys.} {\bf 64}, 911 (1980).
\bibitem{MNY.77}
Y. Mizobuchi, S. Nishiyama and M. Yamamura,
{\it Prog. Theor. Phys.} {\bf 57}, 96 (1977).
\bibitem{YanLan.03}
C. Yannouleas and U. Landman,
{\it Phys. Rev.} {\bf B68}, 035325 (2003).
\bibitem{FukuNishi.84}
H. Fukutome and S. Nishiyama,
{\it Prog. Theor. Phys.} {\bf 72}, 239 (1984).
\end{thebibliography}
\end{document}